\documentclass[12pt]{report}

\begin{document}

\setcounter{page}{0} \topmargin 0pt \oddsidemargin 5mm
\renewcommand{\thefootnote}{\arabic{footnote}}
\newpage
\setcounter{page}{0}

\begin{titlepage}

\vspace{1cm}

\begin{center}\huge
BOUNDARY EFFECTS IN TWO-DIMENSIONAL CRITICAL AND OFF-CRITICAL SYSTEMS
\end{center}

\vspace{1cm}

\begin{center}
{\Large Valentina Riva$^{1,2}$} \\
\vspace{0.5cm} {\em $^{1}$Department of chemical, physical and mathematical sciences,
University of Insubria, Como, Italy}\\
\vspace{0.3cm} {\em $^{2}$International School for Advanced Studies (SISSA), Trieste, Italy }
\end{center}

\vspace{2cm}

\begin{center}\Large
Degree Thesis
\end{center}

\vspace{1cm}

\begin{flushleft}\large
Supervisor:

Prof. Giuseppe Mussardo \quad (\textit{SISSA})

\vspace{0.5cm}

Internal supervisor:

Prof. Vittorio Gorini \quad (\textit{University of Insubria})
\end{flushleft}

\end{titlepage}

\tableofcontents

\newpage

\chapter{Introduction}

Statistical mechanical systems undergoing a continuous (or second order) phase transition are called critical, and
are characterized by the divergence of the correlation length, which leads to scaling invariance. Thanks to the
Polyakov theorem, which states that local field theories that are scaling invariant are also conformally
invariant, the behaviour of these systems can be described by conformal field theories (CFT), i.e. quantum field
theories that are invariant under conformal symmetry. The microscopic dynamics may vary considerably from one
statistical system to another, but the scaling behavior near a second order phase transition falls into
universality classes, each of which is described by a CFT.

In a generic space-time dimension $d$ the conformal group possesses $(d+1)(d+2)/2$ free parameters, while if the
space-time is two-dimensional this group enlarges to an infinite set of transformations, and this makes it
possible to exactly solve the dynamics of critical systems. All correlators, in fact, can be computed assuming
conformal invariance and the existence of an algebra of scaling fields defined by a short-distance operator
product expansion (OPE) for the fields.

In a critical system with boundary, conformal transformations must map the boundary onto itself and preserve the
boundary conditions (BC), with the consequence that only half of the conformal generators remain. In
two-dimensions, the prototype of a system with boundary is the upper half plane; in this case it is possible to
study exactly the surface critical behaviour, i.e. to determine the correlation functions near the boundary when
the bulk is critical, because the $n$-point correlators can be expressed in terms of the $2n$-point ones on the
infinite plane. The underlying technique, called method of images, entails in a natural way the existence of
scaling fields living on the boundary, which appear in the OPE between a bulk scaling field close to the boundary
and its mirror image. The insertion of these boundary fields at a certain point has the effect of changing the BC,
which are explicitly related the bulk operator content of the theory.

We have performed an explicit calculation of some two-point correlation functions in the presence of boundary in a
class of CFT called minimal models (MM), characterized by the possibility of organizing the operators in a finite
number of families. These correlators are expressed in terms of hypergeometric functions, and can be completely
fixed.

\vspace{0.5cm}

Dynamics in the vicinity of second order phase transitions can be described by CFT perturbed by the addition of
operators that break the conformal symmetry and introduce a mass scale in the system. For opportune choices of the
perturbing operator the off-critical massive field theory can be integrable, i.e. it can possess an infinite set
of commuting integrals of motion. In this case, the corresponding scattering theory is purely elastic and the
$n$-particle S-matrix is factorized into a product of $n(n-1)/2$ two-particles S-matrices. In two dimensions,
these two-particles scattering amplitudes satisfy a number of constraints which are restrictive enough to
calculate them exactly, up to the so-called CDD-ambiguity. In the case of distinguishable particles, these
constraints are expressed by unitarity, crossing and bootstrap equations. The bootstrap approach consists in
identifying the S-matrix bound states with some of the particles appearing as asymptotic states, so that the
spectrum of the theory is encoded in the analytic structure of the S-matrix. In two-dimensional elastic
scattering, since there's only one independent Mandelstam variable, it is convenient to introduce a
parametrization of the momenta in terms of the so-called rapidity variable $\theta$:
\begin{displaymath}
p^{0}=m\cosh\theta\,,\qquad\qquad p^{1}=m\sinh\theta \,.
\end{displaymath}
The S-matrix expressed in this variable has just pole singularities. Simple poles are naturally identified with
the bound states which appear in the intermediate channels, while higher order poles demand an explanation in
terms of elementary collision processes. It can be shown that odd-order poles provide a new mechanism to produce
bound states, while even-order poles correspond to multiple scattering processes without the creation of new
intermediate bound states (Coleman-Thun mechanism).

Another class of integrable quantum field theories, in addition to perturbed CFT, is given by the so-called affine
Toda field theories (ATFT), which are defined by a Lagrangian built up from some characteristic quantities of an
affine Lie algebra $\hat{g}$. The corresponding scattering theory contains a number of particles equal to the
rank of $\hat{g}$, and the S-matrix, depending on the choice of $\hat{g}$, coincides with the one of a specific
perturbed minimal model \lq\lq dressed\rq\rq \, with some CDD-factors (related to the ambiguity mentioned above),
which depend on the coupling constant present in the Lagrangian. Since the CDD-factors don't introduce new poles
with physical meaning, the ATFT and the corresponding perturbed MM share the same spectrum of bound states.

In the presence of a boundary, assuming that the BC are compatible with integrability, the scattering theory is
still elastic and factorized, meaning that every process can be expressed in terms of two-particles bulk
scattering amplitudes and single-particle boundary reflection factors. There is a set of equations, analogous to
the unitarity, crossing and bootstrap equations of the bulk, which relate the reflection matrix to the bulk
S-matrix. Hence, if the bulk scattering theory of a system is solved, it is possible to investigate which are the
compatible reflection amplitudes, expecting as many different solutions as the number of possible integrable
boundary conditions, related one to the other by CDD-factors due to an ambiguity of the same kind as the bulk
one. The analytic structure of the reflection matrix encodes the boundary spectrum of the theory, in the light of
a bootstrap approach analogous to the bulk one. In fact, the boundary can exist in several stable states, and the
presence of poles in the reflection amplitude of a particle indicates the possibility for this particle to excite
it.

A complete analysis of the boundary scattering for many systems as in the bulk case has not jet been performed,
and many aspects of the corresponding bound states structure are still obscure.

Starting from two kinds of known reflection amplitudes, we have performed a detailed study of the boundary bound
states structure for the three ATFT related to the exceptional affine Lie algebras $E_{6}$, $E_{7}$ and $E_{8}$
and for the corresponding perturbed MM, which describe the universality classes of particularly important
statistical systems, respectively the thermal tricritical 3-state Potts model, the thermal tricritical Ising
model and the Ising model in a magnetic field. A typical feature of these systems is a very rich analytical
structure of the S-matrix, and it is equally interesting the structure of the reflection amplitudes. In the bulk
case, in order to iterate consistently the bootstrap procedure, one considers all odd-order poles with positive
residue, which correspond to bound states in the direct intermediate channel. In the presence of a boundary,
however, it is sometimes possible to describe also odd-order poles with pure multiple scattering processes,
without involving any new bound state. The existence of these generalized Coleman-Thun mechanisms leads to the
possibility of excluding the creation of some excited boundary states, and in the case of perturbed MM this is
essential in order to avoid an infinite cascade of such states. In fact, the \lq\lq dressing\rq\rq \,
CDD-factors, relating the reflection matrix of ATFT to the one of perturbed MM, as in the bulk case don't
introduce physical poles, but now can change the residue's sign of the existing ones. This leads to the presence
of a number of odd-order poles with positive residue much smaller in the ATFT reflection matrices then in the
perturbed MM ones, with the consequence that the number of boundary states is finite just in the first case. The
condition of having a finite number of bound states, essential for the consistence of the bootstrap procedure,
can be recovered in the case of perturbed MM only admitting an intensive use of generalized Coleman-Thun
mechanisms, with the hypothesis that their existence excludes the boundary excitation process. In the case of the
Potts model, however, neither with this mechanisms was it possible to obtain a consistent bootstrap; this is an
indication that the two analyzed reflection amplitudes don't have a physical meaning, i.e. they don't correspond
to any integrable BC. Using the CDD-ambiguity mentioned, we have then constructed another reflection matrix which
gives rise to a finite number of boundary bound states.

The described on-shell techniques allow a detailed analysis of the boundary states structure of various systems,
even if they leave many problems unsolved, most of all the one of associating the reflection amplitudes to the
corresponding BC. In order to face these problems it will be necessary to adopt off-shell methods, such as the
thermodynamic Bethe ansatz and the form factors approach, which are already known and widely applied in the bulk
case but still require improvement in the presence of a boundary.

\chapter{Conformal invariance}

\section{Critical phenomena}

A statistical mechanical system is said to be critical when its correlation length $\xi$, defined as the typical
distance over which the order parameters are statistically correlated, increases infinitely
($\xi\rightarrow\infty$). Correspondingly, length scales lose their relevance, and scale invariance emerges. This
is peculiar of the continuous (or second order) phase transitions, which consist in a sudden change of the
macroscopic properties of the system as some parameters (e.g. the temperature) are varied, without finite jumps
in the energy (characteristic of first order transitions).

A remarkable property of these systems is that the fine details of their microscopic structure become
unimportant, and the various possible critical behaviours are organized in universality classes, which depend
only on the space dimensionality and on the underlying symmetry. This allows a description of the order parameter
fluctuations in the language of a field theory, which is invariant under the global scale transformations
\begin{equation}\label{scaletr}
x^{\mu}\rightarrow x'\,^{\mu}=\lambda x^{\mu} \,,
\end{equation}
provided that the fields transform as
\begin{equation}\label{scalingdim}
\Phi(x)\rightarrow \Phi'(x')=\lambda^{-\Delta}\Phi(x) \,,
\end{equation}
where $\Delta$ is called the scaling dimension of the field $\Phi$.

The use of conformal invariance to describe statistical mechanical systems at criticality is motivated by a
theorem, due to Polyakov, which states that local field theories which are scaling invariant are also conformally
invariant (\cite{pol}). Hence, every universality class of critical behaviour can be identified with a conformal
field theory (CFT), i.e. a quantum field theory that is invariant under conformal symmetry. This way of studying
critical systems started with a pioneering paper by Belavin, Polyakov and Zamolodchikov (\cite{bpz}), and is now
systematically presented in many review articles and text books (see for instance
\cite{ginsp},\cite{zamrev},\cite{dif}).

\vspace{1cm}

\section{The conformal group}

An infinitesimal coordinate transformation $x^{\mu}\rightarrow x^{\mu}+\xi^{\mu}(x)$ is called conformal if it
leaves the metric tensor $g_{\mu\nu}$ invariant up to a local scale factor, i.e.
\begin{equation}\label{metr}
g_{\mu\nu}(x)\rightarrow \varrho(x)g_{\mu\nu}(x).
\end{equation}
In particular, these transformations preserve the angle between two vectors. The definition implies the following
condition:
\begin{equation}\label{csi}
\partial_{\mu}\xi_{\nu}+\partial_{\nu}\xi_{\mu}=\frac{2}{d}\,\eta_{\mu\nu}(\partial\cdot\xi),
\end{equation}
where $d$ is the dimension of space-time.

The finite version of these transformations is given by:

\begin{eqnarray}
&& x'_{i}=\Lambda_{ik}x_{k}+b_{i} \qquad\quad\textrm{Poincar\'e transformations} \\
&& x'_{i}=\lambda x_{i}\qquad\qquad\quad\;\;\;\textrm{Dilatations} \\
&& \frac{x'_{i}}{(x')^{2}}=\frac{x_{i}}{(x)^{2}}+a_{i}\qquad\textrm{Special transformations}
\end{eqnarray}
and constitutes the $SO(d+1,1)$ group, with $(d+1)(d+2)/2$ free parameters.

\subsubsection{Conformal invariance in quantum field theory}

A spinless field $\phi(x)$ is called quasi-primary if, under a conformal transformation $x\rightarrow x'$,
transforms as
\begin{equation}
\phi(x)\rightarrow\phi'(x')=\left|\frac{\partial x'}{\partial x}\right|^{-\Delta/d}\phi(x),
\end{equation}
where $\Delta$ is the scaling dimension of the field, and the Jacobian of the conformal transformation is related
to the scale factor of eq. (\ref{metr}) by
\begin{equation}\label{jac}
\left|\frac{\partial x'}{\partial x}\right|=\varrho(x)^{-d/2}.
\end{equation}

Conformal invariance fixes the form of the correlators of two and three quasi-primary fields up to a
multiplicative constant:

\begin{equation}
\langle\phi_{1}(x_{1})\phi_{2}(x_{2})\rangle=\left\{
\begin{array}{ll}
\frac{C_{12}}{x_{12}^{2\Delta_{1}}} &
\quad  \textrm{if}\;\;\Delta_{1}=\Delta_{2} \medskip \\
0 & \quad \textrm{if}\;\;\Delta_{1}\neq\Delta_{2}\,\,\, \label{2ptd}
\end{array}
\right. \,\, ,
\end{equation}

\begin{equation}\label{3ptd}
\langle\phi_{1}(x_{1})\phi_{2}(x_{2})\phi_{3}(x_{3})\rangle=\frac{C_{123}}{x_{12}^{\Delta_{1}+\Delta_{2}-\Delta_{3}}x_{23}^{\Delta_{2}+\Delta_{3}-\Delta_{1}}x_{13}^{\Delta_{3}+\Delta_{1}-\Delta_{2}}},
\end{equation}
where $x_{ij}=|x_{i}-x_{j}|$.

Higher correlators cannot be fixed by analogous considerations, because with four points (or more) it is possible
to construct some conformal invariants, called anharmonic ratios, as for instance:
\begin{equation}\label{ratiosd}
\frac{x_{12}x_{34}}{x_{13}x_{24}}, \quad \frac{x_{12}x_{34}}{x_{23}x_{14}}.
\end{equation}
The $n$-point functions may have an arbitrary dependence on these ratios; in particular, the four-point function
may take the form
\begin{equation}\label{4ptd}
\langle\phi_{1}(x_{1})\phi_{2}(x_{2})\phi_{3}(x_{3})\phi_{4}(x_{4})\rangle=f\left(\frac{x_{12}x_{34}}{x_{13}x_{24}},
 \frac{x_{12}x_{34}}{x_{23}x_{14}}\right)\prod_{i<j}^{4}x_{ij}^{\Delta/3-\Delta_{i}-\Delta_{j}},
\end{equation}
where $\Delta=\sum_{i=1}^{4}\Delta_{i}$.

\vspace{1cm}

\section{Two-dimensional conformal field theory}

In two dimensions, since the conformal group enlarges to an infinite set of transformations, it is possible to
solve exactly the dynamics of a critical system, assuming conformal invariance and a short-distance operator
product expansion (OPE) for the fluctuating fields.

\subsection{Conformal coordinate transformations}

In euclidean two-dimensional space-time, eq.(\ref{csi}) specializes to the Cauchy-Riemann equations for
holomorphic functions. This motivates the use of complex coordinates $z$ and $\bar{z}$, with
\begin{eqnarray}\label{z}
z=x^{0}+ix^{1} && \quad \bar{z}=x^{0}-ix^{1}\\
\partial=\frac{1}{2}(\partial_{0}-i\partial_{1}) && \quad \bar{\partial}=\frac{1}{2}(\partial_{0}+i\partial_{1})
\end{eqnarray}

With this notation, solutions of eq.(\ref{csi}) are holomorphic or anti-holomorphic transformations,
$z\rightarrow f(z)$ and $\bar{z}\rightarrow \bar{f}(\bar{z})$, such that $\bar{\partial}f=\partial\bar{f}=0$.
These functions admit the Laurent expansion
\begin{equation}\label{lau}
f(z)=\sum_{n=-\infty}^{\infty}a_{n}z^{n+1} \qquad \qquad
\bar{f}(\bar{z})=\sum_{n=-\infty}^{\infty}a'_{n}\bar{z}^{n+1}
\end{equation}
which has an infinite number of parameters. In this way, the conformal group enlarges to an infinite set of
transformations, even if the only invertible mappings of the whole complex plane into itself are the so-called
Moebious transformations, defined as
\begin{equation}\label{moebious}
z\rightarrow w(z)=\frac{az+b}{cz+d}\,,\quad \textrm{with} \quad ad-bc=1 \,.
\end{equation}
These are the global conformal transformations under which the theory is invariant, while a generic holomorphic
or antiholomorphic transformation will have anomalies encoded in the so-called Ward identities.

\vspace{0.5cm}

Given an holomorphic infinitesimal transformation
\begin{equation}\label{lauinf}
z\rightarrow z+\epsilon(z) \qquad \qquad \epsilon(z)=\sum_{n=-\infty}^{\infty}c_{n}z^{n+1},
\end{equation}
and its antiholomorphic counterpart, the corresponding generators
\begin{equation}\label{gencl}
\ell_{n}=-z^{n+1}\partial_{z}\qquad\qquad \bar{\ell}_{n}=-\bar{z}^{n+1}\partial_{\bar{z}}
\end{equation}
obey the following commutation relations:
\begin{eqnarray}\label{vircl}
&& [\ell_{n},\ell_{m}]=(n-m)\ell_{n+m}\\
&& [\bar{\ell}_{n},\bar{\ell}_{m}]=(n-m)\bar{\ell}_{n+m}\\
&& [\ell_{n},\bar{\ell}_{m}]=0
\end{eqnarray}
Thus the conformal algebra is a direct sum of two infinite-dimensional algebras, one in the holomorphic and the
other in the antiholomorphic sector, and this makes it convenient to regard $z$ and $\bar{z}$ as independent
variables, remembering that the physical surface is given by the condition $\bar{z}=z^{*}$.

Each of these two algebras contains a finite subalgebra generated by $\ell_{-1}$, $\ell_{0}$ and $\ell_{1}$,
obtained demanding regularity of the transformations on the whole complex plane, and associated to the global
conformal group. In particular, $\ell_{0}+\bar{\ell}_{0}$ generates dilatations on the real surface and
$i(\ell_{0}-\bar{\ell}_{0})$ generates rotations.

\subsection{Primary fields and their correlators}

In two dimensions the concept of quasi-primary field can be extended to fields with spin $s$ generally non zero
(\cite{bpz}), defining the holomorphic and antiholomorphic comformal dimensions $h$ and $\bar{h}$ as
\begin{equation}\label{h}
h=\frac{1}{2}(\Delta+s)\qquad\qquad \bar{h}=\frac{1}{2}(\Delta-s),
\end{equation}
where $\Delta$ is the scaling dimension. Under a global conformal transformation (\ref{moebious}), a
quasi-primary field transforms as
\begin{equation}
\phi'(w,\bar{w})=\left(\frac{dw}{dz}\right)^{-h}\left(\frac{d\bar{w}}{d\bar{z}}\right)^{-\bar{h}}\phi(z,\bar{z}).
\end{equation}
If the same is true under any local conformal transformation $z\rightarrow f(z)$, the field is called primary.

Conformal invariance forces the two- and three-point functions of quasi primary fields to have the following form:
\begin{equation}
\langle\phi_{1}(z_{1},\bar{z}_{1})\phi_{2}(z_{2},\bar{z}_{2})\rangle=\left\{
\begin{array}{ll}
\frac{C_{12}}{z_{12}^{2h}\bar{z}_{12}^{2\bar{h}}} &
\quad  \textrm{if}\;\;h_{1}=h_{2}=h \;\;\textrm{and}\;\;\bar{h}_{1}=\bar{h}_{2}=\bar{h}\medskip \\
0 & \quad \textrm{otherwise}\,\,\, \label{2pt}
\end{array}
\right. \,\, ,
\end{equation}

\begin{equation}\label{3pt}
\langle\phi_{1}(z_{1},\bar{z}_{1})\phi_{2}(z_{2},\bar{z}_{2})\phi_{3}(z_{3},\bar{z}_{3})\rangle=C_{123}\frac{1}{z_{12}^{h_{1}+h_{2}-h_{3}}z_{23}^{h_{2}+h_{3}-h_{1}}z_{13}^{h_{3}+h_{1}-h_{2}}}
\end{equation}
\begin{displaymath}
\qquad\qquad\qquad\qquad\times\frac{1}{\bar{z}_{12}^{\bar{h}_{1}+\bar{h}_{2}-\bar{h}_{3}}\bar{z}_{23}^{\bar{h}_{2}+\bar{h}_{3}-\bar{h}_{1}}\bar{z}_{13}^{\bar{h}_{3}+\bar{h}_{1}-\bar{h}_{2}}}\,,
\end{displaymath}
where $z_{ij}=z_{i}-z_{j}$ and $\bar{z}_{ij}=\bar{z}_{i}-\bar{z}_{j}$.

As in generic dimension, higher correlators cannot be fixed because of the existence of anharmonic ratios, but
now the number of independent ratios is reduced, since the four points are forced to lie in the same plane, and
we have:
\begin{equation}\label{ratios}
\eta=\frac{z_{12}z_{34}}{z_{13}z_{24}} \qquad 1-\eta=\frac{z_{14}z_{23}}{z_{13}z_{24}} \qquad
\frac{\eta}{1-\eta}=\frac{z_{12}z_{34}}{z_{23}z_{14}}
\end{equation}
The four-point function may then have the form:
\begin{equation}\label{4pt}
\langle\phi_{1}(z_{1},\bar{z}_{1})\phi_{2}(z_{2},\bar{z}_{2})\phi_{3}(z_{3},\bar{z}_{3})\phi_{4}(z_{4},\bar{z}_{4})\rangle=G\left(\eta,\bar{\eta}\right)\prod_{i<j}^{4}z_{ij}^{h/3-h_{i}-h_{j}}\bar{z}_{ij}^{\bar{h}/3-\bar{h}_{i}-\bar{h}_{j}},
\end{equation}
where $h=\sum_{i=1}^{4}h_{i}$ and $\bar{h}=\sum_{i=1}^{4}\bar{h}_{i}$.

\subsection{Stress-energy tensor and conformal Ward identity}

In a two-dimensional field theory, the variation of the action under a transformation of coordinates
$x^{\mu}\rightarrow x^{\mu}+\epsilon^{\mu}(x)$ is given by
\begin{equation}\label{action}
\delta S=-\frac{1}{2\pi}\int d^{2}x \,T^{\mu\nu}(x)\partial_{\mu}\epsilon_{\nu},
\end{equation}
where $T^{\mu\nu}$ is the stress-energy tensor. Conformal invariance is equivalent to the vanishing of this
quantity under the condition (\ref{csi}), and it is guaranteed by the tracelessness of the stress-energy tensor.
Together with translation and rotation invariance ($\partial_{\mu}T^{\mu\nu}=0$), the condition $T^{\mu}_{\mu}=0$
is expressed in complex coordinates as
\begin{equation}\label{T}
\bar{\partial} T=0 \quad \textrm{and}\quad \partial\bar{T}=0,
\end{equation}
where $T(z)=T_{11}-T_{22}+2iT_{12}$ and $\bar{T}(\bar{z})=T_{11}-T_{22}-2iT_{12}$. Thus the stress-energy tensor
splits into a holomorphic and an antiholomorphic part (\cite{bpz}).

It is possible to deduce the following Ward identity for the variation of a correlator of $n$ primary fields
$\langle X\rangle=\langle\phi_{1}(z_{1},\bar{z}_{1})...\phi_{n}(z_{n},\bar{z}_{n})\rangle$ under a local conformal
transformation $z\rightarrow z+\epsilon(z)$, $\bar{z}\rightarrow \bar{z}+\bar{\epsilon}(\bar{z})$:

\begin{center}
\begin{tabular}{l}
$\frac{1}{2\pi i}\oint_{C} d z
\,\epsilon(z)\sum_{i=1}^{n}\left[\frac{h_{i}}{(z-z_{i})^{2}}+\frac{1}{z-z_{i}}\partial_{i}\right]\langle
X\rangle-\frac{1}{2\pi i}\oint_{C} d\bar{z}
\,\bar{\epsilon}(\bar{z})\sum_{i=1}^{n}\left[\frac{\bar{h}_{i}}{(\bar{z}-\bar{z}_{i})^{2}}+\frac{1}{\bar{z}-\bar{z}_{i}}\bar{\partial}_{i}\right]\langle
X\rangle=$
\end{tabular}
\end{center}

\begin{equation}\label{ward}
=-\,\delta_{\epsilon,\bar{\epsilon}}\langle X\rangle=\frac{1}{2\pi i}\oint_{C} d z \,\epsilon(z)\langle T(z)
X\rangle-\frac{1}{2\pi i}\oint_{C} d \bar{z} \,\bar{\epsilon}(\bar{z})\langle \bar{T}(\bar{z}) X\rangle
\end{equation}
$C$ is a counterclockwise contour that includes all the positions $(z_{i},\bar{z}_{i})$ of the fields contained
in $X$.

\subsection{Operator product expansion}

It is typical of correlation functions to have singularities when the positions of two or more fields coincide.
The operator product expansion (OPE) is the representation of a product of operators (at positions $x$ and $y$ in
a $d$-dimensional space-time) by a sum of terms involving single operators multiplied by functions of $x$ and $y$,
possibly diverging as $x\rightarrow y$. This expansion has a weak sense, being valid within correlation
functions, and leads to the construction of an algebra of scaling fields defined by
\begin{equation}\label{alg}
A_{i}(x)A_{j}(y)=\sum_{k}\hat{C}_{ij}^{k}(x,y)A_{k}(y),
\end{equation}
where $\hat{C}_{ij}^{k}(x,y)$ are the structure constants. Translation and scaling invariance forces these
functions to have the following form:
\begin{equation}
\hat{C}_{ij}^{k}(x,y)=\frac{C_{ij}^{j}}{|x-y|^{\Delta_{i}+\Delta_{j}-\Delta_{k}}},
\end{equation}
where $C_{ij}^{j}$ are exactly the undetermined multiplicative constants of the tree-point correlators (see
eq.(\ref{3ptd})).

\vspace{0.5cm}

In two dimensions, from the Ward identity (\ref{ward}), since $\epsilon$ and $\bar{\epsilon}$ are arbitrary it is
possible to deduce the following OPE for the stress-energy tensor and a primary field of dimension $(h,\bar{h})$:
\begin{equation}\label{Tprim}
T(z)\phi(w,\bar{w})=\frac{h}{(z-w)^{2}}\phi(w,\bar{w})+\frac{1}{z-w}\partial_{w}\phi(w,\bar{w})\:\textrm{+ regular
terms },
\end{equation}
\begin{equation}\label{Tbprim}
\bar{T}(\bar{z})\phi(w,\bar{w})=
\frac{\bar{h}}{(\bar{z}-\bar{w})^{2}}\phi(w,\bar{w})+\frac{1}{\bar{z}-\bar{w}}\partial_{\bar{w}}\phi(w,\bar{w})\:\textrm{+
regular terms }.
\end{equation}

\subsection{Central charge and Virasoro algebra}

It is possible to show that the OPE of the stress-energy tensor with itself has the form:
\begin{equation}\label{TT}
T(z)T(w)=\frac{c/2}{(z-w)^{4}}+\frac{2}{(z-w)^{2}}T(w)+\frac{1}{z-w}\partial T(w)\:\textrm{+ regular terms },
\end{equation}
where the constant $c$, called central charge, depends on the specific model. A similar expression holds for the
antiholomorphic component, and from now on, when its form is obvious, we will write explicitly just the
holomorphic part.

The conformal Ward identity (\ref{ward}) with $X=T$ implies the following expression for the infinitesimal
variation of the stress-energy tensor (\cite{bpz}):
\begin{equation}\label{dT}
\delta_{\epsilon}
T(w)=-\left(2\partial\epsilon(w)+\epsilon(w)\partial\right)T(w)-\frac{c}{12}\partial^{3}\epsilon(w).
\end{equation}
In the case of a finite transformation of the form (\ref{moebious}), this corresponds to
\begin{equation}\label{T'}
T'(w)=\left(\frac{dw}{dz}\right)^{-2}\left[T(z)-\frac{c}{12}\{w;z\}\right],
\end{equation}
where $\{w;z\}$, called Schwarzian derivative, is defined as
\begin{equation}\label{schw}
\{w;z\}=\frac{w'''}{w'}-\frac{3}{2}\left(\frac{w''}{w'}\right)^{2}
\end{equation}
(with the \lq prime\rq \, symbol we mean $d/dz$).

It is straightforward to verify that the Schwarzian derivative of a global conformal map (\ref{moebious})
vanishes, hence the stress-energy tensor is a quasi-primary field, even if it is not primary.

An immediate application of the transformation property (\ref{T'}) shows the physical meaning of $c$, as a
measure of the response of the system to the introduction of a macroscopic scale
(\cite{casimir1},\cite{casimir2}). In fact, if we map the complex plane to an infinite cylinder of circumference
$L$ by
\begin{equation}\label{cyl}
z\rightarrow w(z)=\frac{L}{2\pi}\ln z\,,
\end{equation}
we get
\begin{equation}
T_{cyl.}(w)=\left(\frac{2\pi}{L}\right)^{2}\left[T_{pl.}(z)z^{2}-\frac{c}{24}\right].
\end{equation}
If we assume that the vacuum energy density $\langle T_{pl.}\rangle$ vanishes on the plane, we see that it is non
zero on the cylinder:
\begin{equation}\label{cas}
\langle T_{cyl.}\rangle=-\frac{c\pi^{2}}{6L^{2}}.
\end{equation}
The central charge is then proportional to the Casimir energy, which naturally goes to zero as the macroscopic
scale $L$ goes to infinity.

\vspace{0.5cm}

The holomorphic and antiholomorphic components of the stress-energy tensor can be expanded in Laurent series
respectively on modes $L_{n}$ and $\bar{L}_{n}$, which are the quantum generators of the local conformal
transformations:
\begin{equation}\label{Ln}
T(z)=\sum_{n=-\infty}^{\infty}\frac{L_{n}}{z^{n+2}}\qquad\qquad
\bar{T}(\bar{z})=\sum_{n=-\infty}^{\infty}\frac{\bar{L}_{n}}{\bar{z}^{n+2}}
\end{equation}
These generators obey the Virasoro algebra
\begin{eqnarray}\label{vir}
&& [L_{n},L_{m}]=(n-m)L_{n+m}+\frac{c}{12}n(n^{2}-1)\delta_{n+m,0}\\
&& [\bar{L}_{n},\bar{L}_{m}]=(n-m)\bar{L}_{n+m}+\frac{c}{12}n(n^{2}-1)\delta_{n+m,0}\\
&& [L_{n},\bar{L}_{m}]=0
\end{eqnarray}
Comparing definition (\ref{Ln}) with the OPE (\ref{Tprim}), we can deduce the action of some generators on a
primary field:
\begin{eqnarray}\label{Lprim}
\left(L_{0}\phi\right)(z)&=&h\,\phi(z)\\
\left(L_{-1}\phi\right)(z)&=&\partial\phi(z)\\
\left(L_{n}\phi\right)(z)&=&0\quad\textrm{if}\quad n\geq 1
\end{eqnarray}
The relation $[L_{0},L_{n}]=-nL_{n}$ leads to the interpretation of generators $L_{n}$ with $n>0$ as destruction
operators and with $n<0$ as creation operators. Hence primary fields define highest weight representations of the
Virasoro algebra, being annihilated by all destruction operators. The action of creation operators on these
fields is encoded in the regular part of the OPE (\ref{Tprim}), and defines the so-called descendant fields
\begin{equation}\label{desc}
\phi^{(n_{1},n_{2},...,n_{k})}=\left(L_{-n_{1}}L_{-n_{2}}...L_{-n_{k}}\right)\phi \,,
\end{equation}
which are again eigenvectors of $L_{0}$:
\begin{equation}\label{hdesc}
L_{0}\left[\phi^{(n_{1},n_{2},...,n_{k})}\right]=\left(h+\sum_{i=1}^{k}n_{i}\right)\phi^{(n_{1},n_{2},...,n_{k})}\,.
\end{equation}
The number $N=\sum_{i=1}^{k}n_{i}$ is called level of the descendant. As an example, the stress-energy tensor is
a level two descendant of the identity $(T=L_{-2}1)$. The set $[\phi]$ constituted by all the descendant fields
of a primary operator $\phi$ is called conformal family (\cite{bpz}).

It is possible to show that every correlation function involving descendant fields can be computed acting with a
linear differential operator on the correlation function of the corresponding primary fields. In fact, given $n$
primary fields $\phi_{i}$ of dimensions $h_{i}$, the following relation holds for $k\geq 2$:
\begin{equation}\label{corrdesc}
\langle\phi_{1}(w_{1})...\phi_{n-1}(w_{n-1})\left(L_{-k}\phi_{n}\right)(z)\rangle={\cal
L}_{-k}\langle\phi_{1}(w_{1})...\phi_{n-1}(w_{n-1})\phi_{n}(z)\rangle,
\end{equation}
where
\begin{equation}\label{diffop}
{\cal
L}_{-k}=-\sum_{i=1}^{n-1}\left[\frac{(1-k)h_{i}}{(w_{i}-z)^{k}}+\frac{1}{(w_{i}-z)^{k-1}}\partial_{w_{i}}\right].
\end{equation}
This means that, if we indicate by $\phi_{p}^{\{k\}}$ a level $k$ descendant of a primary field $\phi_{p}$, we can
write the OPE of two primary fields in the following way
\begin{equation}
\phi_{i}(z_{1})\phi_{j}(z_{2})=\sum_{p,k}C_{ijp}^{\{k\}}(z_{1}-z_{2})^{h_{p}-h_{i}-h_{j}+k}\phi_{p}^{\{k\}} \,,
\end{equation}
and the structure constants $C_{ijp}^{\{k\}}$ can be determined algebraically from the structure constant
$C_{ijp}$ of the primary field $\phi_{p}$.

\vspace{1cm}

\section{Minimal models}

\subsection{Verma modules}\label{Verma modules}

In order to construct the Hilbert space of a conformal field theory, we define the vacuum state $|0\rangle$ by
the requirement:
\begin{equation}\label{vacuum}
L_{n}|0\rangle=\bar{L}_{n}|0\rangle=0\qquad\qquad\textrm{for}\;n\geq -1
\end{equation}
In particular, this condition implies invariance of the vacuum under global conformal transformations.

On the infinite plane, it is convenient to adopt the so-called radial quantization, which consists in choosing
the space direction along concentric cycles centered at the origin, and the time direction orthogonal to space.
This choice looks natural if we initially consider our theory defined on an infinite cylinder with canonical
quantization, i.e. with time $t$ going to $-\infty$ to $+\infty$ along the flat direction of the cylinder, and
space being compactified with a coordinate $x$ going from $0$ to $L$. In the Euclidean space, the cylinder is
described by a single complex coordinate $w=t+ix$, and if we map it onto the infinite plane by the inverse of
transformation (\ref{cyl}), the remote past ($t\rightarrow -\infty$) is situated at the origin $z=0$, the remote
future ($t\rightarrow \infty$) lies on the point at infinity on the Riemann sphere, and equal time surfaces
($t$=const) are mapped onto concentric circumferences constituted by complex numbers of equal modulus.

The asymptotic \lq\lq in\rq\rq\, states are then obtained acting on the vacuum by operators situated at $z=0$.
Primary fields create the so-called highest weight states
\begin{equation}\label{hwst}
|h\rangle\equiv \phi(0)|0\rangle,
\end{equation}
which satisfy
\begin{equation}
L_{0}|h\rangle=h|h\rangle\qquad\textrm{and}\qquad L_{n}|h\rangle=0\quad\textrm{if}\quad n>0 \,.
\end{equation}
Acting on $|h\rangle$ with the creation operators, we obtain the descendant states, whose set
\begin{equation}\label{verma}
V(c,h)=\left\{L_{-n_{1}}...L_{-n_{k}}|h\rangle:\;n_{i}\geq 0\right\}
\end{equation}
is called Verma module and is a subset of the Hilbert space invariant under the Virasoro algebra.

The same construction holds for the antiholomorphic states $|\bar{h}\rangle\equiv \phi(\bar{0})|0\rangle$ and
their descendants, obtained acting on them with the generators $\bar{L}_{-n}$ and constituting the Verma module
$\bar{V}(c,\bar{h})$. The total Hilbert space is then the direct sum
\begin{equation}\label{hilbert}
{\cal H}=\sum_{h,\bar{h}}V(c,h)\otimes \bar{V}(c,\bar{h}),
\end{equation}
where $h,\bar{h}$ run over all conformal dimensions occurring in the theory.

The structure of a Verma module is encoded in the associated Virasoro character, defined as
\begin{equation}\label{char}
\chi _{c,h }(q)=\textrm{Tr}\:q^{L_{0}-c/24}=\sum_{N =0}^{\infty }\dim (N )q^{h +N -c/24}\;,
\end{equation}
where $\dim (N)$ is the number of linearly independent descendant states at level $N$ and $q$ is a complex
variable. The characters are generating functions for the level degeneracy $\dim (N)$.

\subsection{The Kac determinant}

To every Verma module it is possible to associate the Gram matrix $M(c,h)$ of inner products between all basis
states. This matrix is block diagonal, and a generic element of the block $M^{(N)}(c,h)$ (corresponding to states
of level $N$) is
\begin{equation}
\left\langle h |L_{m_{1}}\cdots L_{m_{l}}L_{-n_{1}}\cdots L_{-n_{k}}|h \right\rangle \;,\quad n_{i},m_{i}\geq
0,\;\sum_{i=1}^{k}n_{i}=\sum_{i=1}^{l}m_{i}=N
\end{equation}
(we have defined the Hermitian conjugate $L_{n}^{\dagger}=L_{-n}$).

States with negative norm are present in a Verma module if and only if $M(c,h)$ has one or more negative
eigenvalues. The requirement for a representation of the Virasoro algebra to be unitary, equivalent to the
absence of negative norm states, imposes then some constraints on the parameters $c$ and $h$. It is easy to
verify that all representations with $c<0$ or $h<0$ are non unitary.

Presence of null vectors with zero norm (equivalent to the existence of zero eigenvalues) indicates instead that
the representation is reducible, and irreducible representations can be constructed quotienting out of $V(c,h)$
the submodules generated by these vectors.

There is a general formula for the determinant of the Gram matrix, called Kac determinant:
\begin{equation}\label{Kdet}
\det M^{(N )}(c,h)=\alpha_{N }\prod_{rs\leq N }\left( h -h _{r,s}(c)\right) ^{p(N -rs)}
\end{equation}
where $r,s\geq 1$, $p(N -rs)$ is the number of partitions of the integer $N -rs$, and $\alpha_{N }$ is a positive
constant independent of $h$ or $c$. The functions $h _{r,s}$ can be parametrized as
\begin{equation}\label{hrs}
h_{r,s}(m)=\frac{\left[ (m+1)r-ms\right] ^{2}-1}{4m(m+1)}\qquad \textrm{with}\qquad m=-\frac{1}{2}\pm
\frac{1}{2}\sqrt{\frac{25-c}{1-c}} \,.
\end{equation}
Using this explicit expression, it is possible to show that all representations with $c\geq 1$ and $h\geq 0$ are
unitary. For $0<c<1$, instead, only a discrete set of possible unitary theories exists, characterized by a
central charge of the form
\begin{equation}\label{min}
c=1-\frac{6}{m(m+1)}\qquad \textrm{with}\qquad m=3,4,...
\end{equation}
and by conformal dimensions parametrized by (\ref{hrs}) with $1\leq r<m,\,1\leq s\leq r$. These theories, called
unitary minimal models, possess only a finite number of primary fields.

\subsection{Minimal models}

In general, the presence of null vectors in a Verma module imposes constraints on the three point functions and
then on the operator algebra. An example of null vector at level 2 is
\begin{equation}\label{nullvect}
|\chi\rangle=\left[L_{-2}-\frac{3}{2(2h+1)}L_{-1}^{2}\right]|h\rangle \qquad\textrm{with}\qquad
h=\frac{1}{16}\left\{5-c\pm\sqrt{(c-1)(c-25)}\right\} \,.
\end{equation}
Another possible parametrization of the functions $h_{r,s}$ in (\ref{Kdet}) is
\begin{eqnarray}\label{hrs2}
&& h_{r,s}(c)=h_{0}+\frac{1}{4}(r\alpha_{+}+s\alpha_{-})^{2}\\
&& h_{0}=\frac{1}{24}(c-1)\\
&& \alpha_{\pm}=\frac{\sqrt{1-c}\pm\sqrt{25-c}}{\sqrt{24}}
\end{eqnarray}
In this notation the above null vector corresponds to $h=h_{1,2}$ or $h=h_{2,1}$. It follows from
(\ref{corrdesc}) that every correlator involving the primary field $\phi$ associated to $|h\rangle$ satisfies the
following differential equation:
\begin{equation}
\left\{{\cal L}_{-2}-\frac{3}{2(2h+1)}{\cal L}_{-1}^{2}\right\}\langle\phi(z)X\rangle=0\,.
\end{equation}
Parametrizing the conformal dimensions as $h(\alpha)\equiv h_{0}+\frac{1}{4}\alpha^{2}$, it is easy to see that
the three point correlators $\langle\phi(z)\phi_{1}(z_{1})\phi_{2}(z_{2})\rangle$ vanish unless
\begin{eqnarray}
\alpha_{2}=\alpha_{1}\pm \alpha_{+} && (h=h_{2,1}) \\
\alpha_{2}=\alpha_{1}\pm \alpha_{-} && (h=h_{1,2})
\end{eqnarray}
It then follows that the corresponding operator algebra will assume the symbolic form
\begin{eqnarray}
\label{21OPE} && \phi_{(2,1)}\times\phi_{(\alpha)}=\phi_{(\alpha-\alpha_{+})}+\phi_{(\alpha+\alpha_{+})} \\
\label{12OPE} && \phi_{(1,2)}\times\phi_{(\alpha)}=\phi_{(\alpha-\alpha_{-})}+\phi_{(\alpha+\alpha_{-})}
\end{eqnarray}
which means that the OPE of fields belonging to the families $\left[\phi_{(2,1)}\right]$ and
$\left[\phi_{(\alpha)}\right]$ may contain terms belonging only to the conformal families
$\left[\phi_{(\alpha-\alpha_{+})}\right]$ and $\left[\phi_{(\alpha+\alpha_{+})}\right]$. The implicit coefficient
multiplying the families are the structure constants of the algebra, and they may vanish. The conditions under
which a conformal family occurs in the OPE of two conformal fields are called the fusion rules of the theory, and
are represented as
\begin{equation}\label{fusion}
\phi_{i}\times\phi_{j}=\sum_{k}{\cal N}_{ij}^{k}\,\phi_{k} \qquad\textrm{with}\qquad {\cal N}_{ij}^{k}\in\{0,1\}.
\end{equation}

\vspace{0.5cm}

A minimal model ${\cal M}(p,p')$ is a conformal theory defined by the existence of two coprime integers $p$ and
$p'$ such that
\begin{equation}
p\,\alpha_{-}+p'\,\alpha_{+}=0 \; ,
\end{equation}
implying the periodicity property
\begin{equation}
h_{r,s}=h_{r+p',s+p}\,.
\end{equation}
In terms of these two integers, the central charge and the Kac formula become
\begin{eqnarray}
&& c=1-6\frac{(p-p')^{2}}{p\, p'}\\
&& h_{r,s}=\frac{(p\, r-p'\,s)^{2}-(p-p')^{2}}{4 p \,p'}
\end{eqnarray}
In this case, in every Verma module $V_{r,s}$ there is an infinite number of null vectors, whose effect is a
truncation of the operator algebra, yielding a finite set of conformal families with $h_{r,s}$ delimited by
\begin{equation}
1\leq r< p'\qquad\textrm{and}\qquad 1\leq s< p \; .
\end{equation}
These conformal dimensions are organized in a rectangle in the $(r,s)$ plane, called Kac table. The symmetry
$h_{r,s}=h_{p'-r,p-s}$ makes half of this rectangle redundant, and the number of distinct fields is
$(p-1)(p'-1)/2$.

It can be shown that minimal models are unitary only if $|p-p'|=1$, and their list coincides with the one of
unitary representations given in (\ref{min}), fixing $p'=m$ and $p=m+1$.

\newpage

\subsection{Examples}

If not differently specified, we will always assume that the considered fields have no spin, hence their
holomorphic and antiholomorphic dimensions are equal $(h=\bar{h})$.

\subsubsection{${\cal M}(5,2)$: the Yang-Lee model}

This minimal model, studied by Cardy in \cite{M52}, is non unitary, has central charge $c=-\frac{22}{5}$ and
contains only two primary fields: $\phi_{(1,1)}$ of dimension $h_{1,1}=0$ (the identity operator, present in all
models) and $\phi_{(1,2)}$ of dimension $h_{1,2}=-\frac{1}{5}$.

\vspace{0.5cm}

\subsubsection{${\cal M}(4,3)$: the Ising model}

This model, which has central charge $c=\frac{1}{2}$, has been analyzed in \cite{bpz}. The Kac table is

\vspace{2cm}

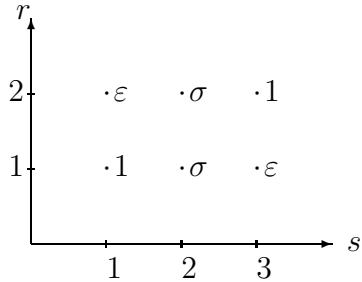
\setlength{\unitlength}{0.01cm}
\begin{picture}( 1000,300)(0,100)

\put(500,200){\vector( 1, 0){400}} \put(500,200){\vector(0,1){300}}
\put(600,195){\line(0,1){10}}\put(700,195){\line(0,1){10}}\put(800,195){\line(0,1){10}}
\put(495,300){\line(1,0){10}}\put(495,400){\line(1,0){10}}

\put(595,290){$\cdot$}\put(695,290){$ \cdot$}\put(795,290){$ \cdot$} \put(595,390){$ \cdot$}\put(695,390){$
\cdot$}\put(795,390){$ \cdot$}

\put(920,190){$s$}\put(480,500){$r$}\put(600,155){$1$}\put(700,155){$2$}\put(800,155){$3$}
\put(470,290){$1$}\put(470,390){$2$}

\put(610,290){$1$}\put(710,290){$\sigma$}\put(810,290){$\varepsilon$}
\put(810,390){$1$}\put(710,390){$\sigma$}\put(610,390){$\varepsilon$}

\end{picture}

This field theory is in the same universality class as the lattice Ising model, defined by the usual
configuration energy
\begin{equation}\label{ising}
E[\sigma]=-J\sum_{\langle i,j\rangle}\sigma_{i}\sigma_{j}-h\sum_{i}\sigma_{i} \, \qquad \sigma_{i}\in\{-1,1\} \,.
\end{equation}
The operator $\phi_{1,2}=\sigma$, with conformal dimension $h_{\sigma}=\frac{1}{16}$, is the continuum version of
the lattice spin $\sigma_{i}$, while $\phi_{1,3}=\varepsilon$, with $h_{\varepsilon}=\frac{1}{2}$, corresponds to
the interaction energy $\sigma_{i}\sigma_{i+1}$. The fusion rules are
\begin{eqnarray}\label{fusis}
&&\sigma\times\sigma=1+\varepsilon\\
&&\sigma\times\varepsilon=\sigma\\
&&\varepsilon\times\varepsilon=1
\end{eqnarray}
and are compatible with the $Z_{2}$ symmetry $\sigma_{i}\rightarrow -\sigma_{i}$ of the Ising model.

\vspace{0.5cm}

\subsubsection{${\cal M}(5,4)$: the tricritical Ising model (TIM)}

The Kac table of this model, which has central charge $c=\frac{7}{10}$, is

\vspace{2cm}

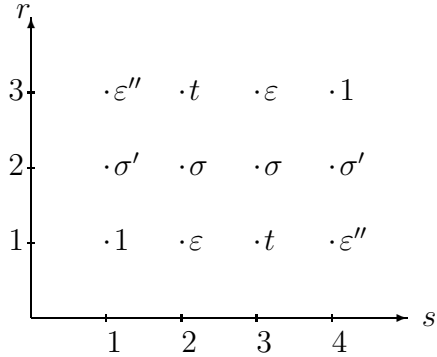
\setlength{\unitlength}{0.01cm}
\begin{picture}( 1000,300)(0,100)

\put(500,100){\vector( 1, 0){500}} \put(500,100){\vector(0,1){400}}
\put(600,95){\line(0,1){10}}\put(700,95){\line(0,1){10}}\put(800,95){\line(0,1){10}}\put(900,95){\line(0,1){10}}
\put(495,200){\line(1,0){10}}\put(495,300){\line(1,0){10}}\put(495,400){\line(1,0){10}}

\put(595,190){$\cdot$}\put(695,190){$ \cdot$}\put(795,190){$ \cdot$}\put(895,190){$ \cdot$} \put(595,290){$
\cdot$}\put(695,290){$ \cdot$}\put(795,290){$ \cdot$}\put(895,290){$ \cdot$}\put(595,390){$ \cdot$}\put(695,390){$
\cdot$}\put(795,390){$ \cdot$}\put(895,390){$ \cdot$}

\put(1020,90){$s$}\put(480,500){$r$}\put(600,55){$1$}\put(700,55){$2$}\put(800,55){$3$}\put(900,55){$4$}
\put(470,190){$1$}\put(470,290){$2$}\put(470,390){$3$}

\put(610,190){$1$}\put(710,190){$\varepsilon$}\put(810,190){$t$}\put(910,190){$\varepsilon''$}
\put(610,290){$\sigma'$}\put(710,290){$\sigma$}\put(810,290){$\sigma$}\put(910,290){$\sigma'$}
\put(910,390){$1$}\put(810,390){$\varepsilon$}\put(710,390){$t$}\put(610,390){$\varepsilon''$}

\end{picture}

\vspace{1cm}

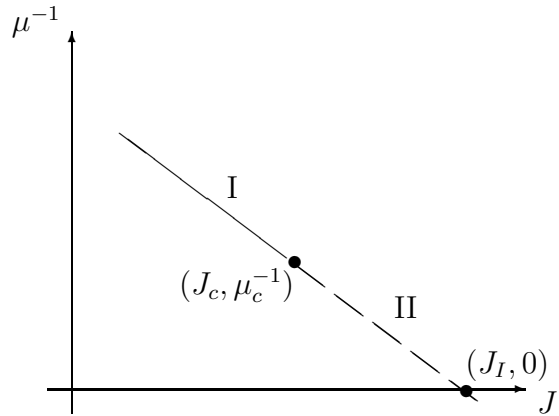
\begin{figure}[h]
\setlength{\unitlength}{0.0125in}
\begin{picture}(40,90)(60,420)
\put(210,420){\vector(1,0){200}} \put(220,410){\vector(0,1){160}} \put(415,410){$J$} \put(195,570){$\mu^{-1}$}

\put(390,415){\line(-4,3){12}} \put(374,427){\line(-4,3){12}} \put(358,439){\line(-4,3){12}}
\put(342,451){\line(-4,3){12}} \put(326,463){\line(-4,3){12}}

\put(310,475){\line(-4,3){70}}

\put(285,500){I}\put(355,450){II}

\put(310,470){$ \bullet$}  \put(265,460){$(J_{c},\mu_{c}^{-1})$}

\put(382,416){$ \bullet$} \put(385,427){$(J_{I},0)$}
\end{picture}
\caption{Phase diagram of the TIM} \label{TIMphasediagr}
 \end{figure}

It was recognized in \cite{M54} that the lattice model associated with this conformal field theory is the dilute
Ising model at its tricritical fixed point, defined by
\begin{equation}\label{TIM}
E[\sigma,t]=-J\sum_{\langle i,j\rangle}\sigma_{i}\sigma_{j}t_{i}t_{j}-\mu\sum_{i}(t_{i}-1) \, \qquad
\sigma_{i}\in\{-1,1\},\,t_{i}\in\{0,1\} \,,
\end{equation}
where $\mu$ is the chemical potential and $t_{i}$ is the vacancy variable. The corresponding phase diagrams is
drawn in Figure \ref{TIMphasediagr}, where I and II denote respectively a first and second order phase transition,
and the point $(J_{I},0)$ represents the Ising model, with all lattice's site occupied.

\vspace{2.5cm}

\vspace{0.5cm}

The field $\phi_{1,2}=\varepsilon$, with $h_{\varepsilon}=\frac{1}{10}$, corresponds to the energy density, while
$\phi_{1,3}=t$, with $h_{t}=\frac{6}{10}$, is the vacancy (or subleading energy) operator. The leading and
subleading magnetization fields are respectively $\phi_{2,2}=\sigma$ and $\phi_{2,1}=\sigma'$, with
$h_{\sigma}=\frac{3}{80}$ and $h_{\sigma'}=\frac{7}{16}$. The remaining field $\phi_{1,4}=\varepsilon''$ has
conformal dimension $h_{\varepsilon''}=\frac{3}{2}$. Dividing the operators in even and odd with respect to the
$Z_{2}$ symmetry of the model, we can list the fusion rules in the following way (\cite{TIMscaling}):

\begin{center}
\begin{tabular}{|l l l l l|} \hline
$\textrm{even}\times \textrm{even}$ & \hspace{0.5cm} & $\textrm{even} \times \textrm{odd}$ & \hspace{0.5cm} &  $\textrm{odd} \times \textrm{odd}$ \\
\hline
$\varepsilon \times \varepsilon=1+t$ &                  & $\varepsilon \times \sigma=\sigma+\sigma'$ &    & $\sigma\times \sigma= 1+\varepsilon+t+\varepsilon''$\\
$\varepsilon \times t=\varepsilon+\varepsilon''$  &     & $\varepsilon \times \sigma'=\sigma$ &           & $\sigma\times \sigma'= \varepsilon+t$ \\
$\varepsilon \times \varepsilon''=t$ &                  & $t \times \sigma=\sigma+\sigma'$ &              & $\sigma' \times \sigma'= 1+\varepsilon''$\\
$t \times t=1+t$ &                                      & $t \times \sigma'=\sigma$ &                     &\\
$t \times \varepsilon''=\varepsilon$  &                 & $\varepsilon'' \times \sigma=\sigma$ &          &\\
$\varepsilon'' \times \varepsilon''=1$  &               & $\varepsilon'' \times \sigma'=\sigma'$ &        &\\
\hline
\end{tabular}
\end{center}

\vspace{0.5cm}

\subsubsection{${\cal M}(6,5)$: the three-state Potts model}

The $Q$-state Potts model is defined in terms of a spin variable $\sigma_{i}$ taking $Q$ different values:
\begin{equation}\label{potts}
E[\sigma]=-\sum_{\langle i,j\rangle}\delta_{\sigma_{i}\sigma_{j}} \,.
\end{equation}
The case $Q=2$ is equivalent to the Ising model, while for $Q=3$ Dotsenko has shown (\cite{M65}) that the critical
point is described by a subset of the ten scaling fields contained in the ${\cal M}(6,5)$ minimal model, with
central charge $c=\frac{4}{5}$. The combination of the holomorphic and antiholomorphic parts can now give some
fields with non zero spin, i.e. with $h\neq\bar{h}$.

The tricritical version of this model, defined by the possibility of having empty sites in the lattice
realization, is described by a subset of the scaling fields contained in the ${\cal M}(7,6)$ minimal model.

\vspace{0.5cm}

\subsection{Four-point correlation functions}

In a minimal model, the presence of null vectors implies certain differential equations for the four point
functions (\ref{4pt}). We will focus our attention on the case where one of the fields is $\phi_{1,2}$, referring
to the parametrization  (\ref{hrs2}). Global conformal invariance makes it possible to fix the positions of three
of the four points to be $0,1,\infty$; we will call $z$ the remaining free coordinate.

The function $G(z)$ in the correlator
\begin{equation}\label{G}
\langle\phi_{r_{1},s_{1}}(0)\phi_{1,2}(z)\phi_{r_{3},s_{3}}(1)\phi_{r_{4},s_{4}}(\infty)\rangle=z^{h/3-h_{r_{1},s_{1}}-h_{1,2}}(1-z)^{h/3-h_{1,2}-h_{r_{3},s_{3}}}G\left(z\right)
\end{equation}
satisfies an ordinary hypergeometric differential equation, and the same holds for the antiholomorphic part.
Employing the so-called Coulomb-gas formalism, developed by Dotsenko and Fateev (\cite{dosfat1},\cite{dosfat2}),
it is possible to show that the general solution is
\begin{equation}
G(z,\bar{z})=\sum_{i,j=1,2}X_{ij}I_{i}(z)\overline{I_{j}(z)}\, ,
\end{equation}
where $I_{1,2}$ are expressed in terms of the hypergeometric function $F(\lambda,\mu,\nu;z)$ as
\begin{displaymath}
I_{1}(a,b,c;z)=\frac{\Gamma(-a-b-c-1)\Gamma(b+1)}{\Gamma(-a-c)}F(-c,-a-b-c-1,-a-c;z)\\
\end{displaymath}
\begin{equation}
I_{2}(a,b,c;z)=z^{1+a+c}\,\frac{\Gamma(a+1)\Gamma(c+1)}{\Gamma(a+c+2)}F(-b,a+1,a+c+2;z) \label{I12}
\end{equation}
and the parameters are defined by
\begin{equation}
a=2\alpha_{-}\alpha_{r_{1},s_{1}}\qquad b=2\alpha_{-}\alpha_{r_{3},s_{3}}\qquad c=2\alpha_{-}\alpha_{1,2}
\end{equation}
with
\begin{equation}\label{alphars}
\alpha_{r,s}=\frac{1}{2}(1-r)\alpha_{+}+\frac{1}{2}(1-s)\alpha_{-}\,.
\end{equation}

The coefficients $X_{ij}$ are determined by enforcing the monodromy invariance of the function $G$, due to the
fact that a physical correlator must not be affected by analytical continuation along contours surrounding
singular points. Up to an overall normalization the examined correlator is then
\begin{equation}
\langle\phi_{r_{1},s_{1}}(0)\phi_{1,2}(z)\phi_{r_{3},s_{3}}(1)\phi_{r_{4},s_{4}}(\infty)\rangle\sim
\end{equation}
\begin{displaymath}
|z|^{4\alpha_{1,2}\alpha_{r_{1},s_{1}}}|1-z|^{4\alpha_{1,2}\alpha_{r_{3},s_{3}}}\left[\frac{s(b)s(a+b+c)}{s(a+c)}|I_{1}(z)|^{2}+\frac{s(a)s(c)}{s(a+c)}|I_{2}(z)|^{2}\right]
\end{displaymath}
where $s(x)=\sin(\pi x)$.

\vspace{0.5cm}

From the explicit expressions of four-point correlators it is possible to recover the structure constants of the
theory, inserting the operator product expansions of two couples of fields in the opportune limits, and comparing
the coefficients of the singular terms. The standard normalization is $C_{r,s;r,s}^{1,1}=1$, which corresponds to
normalizing to one the two-point functions.

\vspace{1cm}

\section{Modular invariance}

The requirement of modular invariance in conformal field theory was first analyzed by Cardy (\cite{cardymod}), who
studied the minimal models in a finite geometry and derived constraints on their possible operator content. The
decoupling of the holomorphic and antiholomorphic sectors exists only in the infinite plane, while physical
constraints on the left-right content of the theory are imposed by the geometry of the space in which it is
defined.

Let us consider a torus with complex periods $\omega_{1}$ and $\omega_{2}$, calling modular parameter their ratio
$\tau=\frac{\omega_{2}}{\omega_{1}}$. The partition function of the system is
\begin{equation}
Z(\omega_{1},\omega_{2})=\textrm{Tr}\,e^{-\{H\,\footnotesize\textrm{Im}\normalsize\,\omega_{2}-iP\,\footnotesize\textrm{Re}\normalsize\,\omega_{2}\}}\,
,
\end{equation}
where $H$ and $P$ are respectively the hamiltonian and the total momentum of the theory. It can be shown that the
canonical quantization procedure on a cylinder of circumference $L$ (discussed in section \ref{Verma modules})
leads to the following expression for the hamiltonian and the momentum in terms of the Virasoro generators:
\begin{equation}\label{hamcyl}
H=\frac{2\pi}{L}\left(L_{0}+\bar{L}_{0}-\frac{c}{12}\right) \qquad \qquad P=\frac{2\pi i
}{L}\left(L_{0}-\bar{L}_{0}\right)
\end{equation}
The torus can be regarded as a cylinder with periodic conditions at its ends, and this corresponds to the choice
$\omega_{1}=L$. In this way, defining the parameters $q=e^{2\pi i \tau}$ and $\bar{q}=e^{-2\pi i \bar{\tau}}$,
the partition function takes the form
\begin{equation}\label{partfct}
Z(\tau)=\textrm{Tr}\left(q^{L_{0}-\frac{c}{24}}\bar{q}^{\bar{L}_{0}-\frac{c}{24}}\right)\,,
\end{equation}
which depends on the periods $\omega_{1,2}$ only through their ratio $\tau$. From the definition (\ref{char}) it
follows that
\begin{equation}
Z(\tau)=\sum_{h,\bar{h}}{\cal N}_{h,\bar{h}}\,\chi _{c,h }(\tau)\bar{\chi} _{c,\bar{h} }(\bar{\tau})\,,
\end{equation}
where ${\cal N}_{h,\bar{h}}$ is the multiplicity of occurrence of $V(c,h)\otimes \bar{V}(c,\bar{h})$ in ${\cal
H}$ (see eq.(\ref{hilbert})).

The requirement of modular invariance consists in imposing that the partition function is independent of the
choice of periods $\omega_{1,2}$ for a given torus, i.e. it doesn't vary defining new periods as integer
combinations of $\omega_{1}$ and $\omega_{2}$. This translates in transformations of the modular parameter of the
form
\begin{equation}
\tau\rightarrow\frac{a\tau+b}{c\tau+d}\qquad\textrm{with}\qquad a,b,c,d\in Z \,,\quad ad-bc=1
\end{equation}
which constitute the modular group $SL(2,Z)/Z_{2}$ (the signs of all parameters may be simultaneously changed
without affecting the transformation).

It can be proved that the two transformations
\begin{equation}\label{T,S}
{\cal T}:\;\tau\rightarrow\tau+1\qquad\qquad\textrm{and}\qquad\qquad {\cal S}:\;\tau\rightarrow -\frac{1}{\tau}
\end{equation}
generate the whole modular group.

In a minimal model, it is possible to show that the matrix elements of these transformations on the basis of
minimal characters, defined as
\begin{equation}
\chi_{r,s}(\tau+1)=\sum_{(\varrho,\sigma)}{\cal
T}_{rs;\varrho\sigma}\chi_{\varrho,\sigma}(\tau)\qquad\textrm{and}\qquad
\chi_{r,s}(-\frac{1}{\tau})=\sum_{(\varrho,\sigma)}{\cal S}_{rs;\varrho\sigma}\chi_{\varrho,\sigma}(\tau)
\end{equation}
have the form:
\begin{eqnarray}
{\cal T}_{rs;\varrho\sigma}&=& \delta_{r\varrho}\delta_{s\sigma}e^{2\pi i\left(h_{r,s}-\frac{c}{24}\right)}\\
{\cal S}_{rs;\varrho\sigma}&=&
2\sqrt{\frac{2}{p\,p'}}(-1)^{1+s\varrho+r\sigma}\sin\left(\pi\frac{p}{p'}r\varrho\right)
\sin\left(\pi\frac{p'}{p}s\sigma\right) \label{modmatr}
\end{eqnarray}

The classification of modular invariant partition functions for minimal models was performed by Cappelli,
Itzykson and Zuber in \cite{ciz}. One of the main features is that, except for $p$ or $p'=2,4$, there is always
more than one modular-invariant theory at a given value of the central charge $c=1-6(p-p')^{2}/p\,p'$, i.e. one
can find different operator algebras, closed under OPE, built out of the same set of primary fields.

The simplest possibility is a diagonal theory, where each field of the Kac table appears in the partition
function exactly once and in a spinless left-right combination:
\begin{equation}
Z_{diag}=\sum_{(r,s)}|\chi _{r,s }|^{2}\,.
\end{equation}

An example of non diagonal theory is the already mentioned three-state Potts model, whose (left or right) field
content is a subset of the ${\cal M}(6,5)$ minimal model. The partition function is
\begin{equation}
Z_{Potts\;3}=\sum_{r=1,2}\left\{|\chi _{r,1}+\chi _{r,5}|^{2}+2|\chi _{r,3}|^{2}\right\}\,.
\end{equation}
The multiplicity $2$ of the operators $\phi_{r,3}$ is reflected in a non trivial structure of the fusion rules,
and the asymmetric left-right combinations $\phi_{r,1}\otimes\bar{\phi}_{r,5}$ (with their complex conjugates)
have a non vanishing spin $\pm (2r-5)$.

Finally, we present a remarkable result, called Verlinde formula (\cite{verl}), which relates the fusion number
${\cal N}_{rs,mn}^{kl}$ of the minimal theories (see eq.(\ref{fusion})) to the ${\cal S}$ matrix elements
(\ref{modmatr}):
\begin{equation}\label{verlform}
{\cal N}_{rs,mn}^{kl}=\sum_{(i,j)}\frac{{\cal S}_{rs,ij}{\cal S}_{mn,ij}{\cal S}_{ij,kl}}{{\cal S}_{11,ij}}\;.
\end{equation}

\chapter{Critical systems with boundary}

In a two-dimensional system with a boundary, the study of the surface critical behaviour is the determination of
correlation functions near the boundary when the bulk is critical. If the model has to preserve some form of
conformal symmetry at criticality, conformal transformations must map the boundary onto itself and preserve the
boundary conditions. This has the consequence that holomorphic and antiholomorphic fields no longer decouple, and
only half of the conformal generators remain.

The main applications of conformal invariance to systems with boundary are due to Cardy, who discussed in
\cite{cardy0},\cite{cardy1} surface critical behaviour and the relation between boundary conditions and the
operator content of a theory.

\vspace{1cm}

\section{Method of images}

The prototype of a two-dimensional system with boundary is the upper half plane. Infinitesimal local conformal
transformations of the form $z\rightarrow z+\epsilon (z)$ map the real axis onto itself if and only if $\epsilon
(\bar{z})=\bar{\epsilon} (z)$, i.e. $\epsilon$ is real on the real axis. This constraint eliminates half of the
conformal generators.

Conformal invariance implies that the boundary conditions must be homogeneous, as for instance:
\begin{equation}
\phi|_{B}=0 \; , \;\phi|_{B}=\infty \; , \;\frac{\partial\phi}{\partial n}|_{B}=0
\end{equation}
due to the fact that the transformations laws of the primary fields are multiplicative.

The condition $\phi|_{B}=0$ is called \lq\lq free\rq\rq\, boundary condition, and a critical system obeying such a
boundary condition is said to undergo an ordinary transition. On the other hand the condition $\phi|_{B}=\infty$
refers to the extraordinary transition, when the boundary orders before the bulk.

\vspace{0.5cm}

The bulk Ward identity (\ref{ward}) is in fact a pair of identities giving the independent variations
$\delta_{\epsilon}\langle X\rangle$ and $\delta_{\bar{\epsilon}}\langle X\rangle$, because the infinitesimal
variations $\epsilon(z)$ and $\bar{\epsilon}(\bar{z})$ are independent.

On the upper half-plane this identity is still applicable, except that the integration contour $C$ must lie
entirely in the upper half plane, and the coordinate variation $\bar{\epsilon}$ is the complex conjugate of
$\epsilon$: we no longer have a decoupling into holomorphic and antiholomorphic identities.

To proceed further, we can regard the dependence of the correlators on antiholomorphic coordinates $\bar{z}_{i}$
on the upper half-plane as a dependence on holomorphic coordinates $z_{i}^{*}=\bar{z}_{i}$ on the lower
half-plane. We thus introduce a mirror image of the system, and we extend the definition of $T(z)$ into the lower
half-plane according to the following relation:
\begin{equation}
T(z^{*})=\overline{T}(z).
\end{equation}
Such an extension is compatible with the boundary conditions, because
\begin{equation}
T|_{B}=\overline{T}|_{B},
\end{equation}
which in cartesian coordinates means
\begin{equation}
T_{xy}|_{B}=0
\end{equation}
i.e. there is no energy or momentum flux across the surface.

We now have
\begin{equation}
\delta_{\epsilon,\epsilon^{*}}\langle X\rangle=-\frac{1}{2\pi i}\oint_{C}dz\,\epsilon(z)\langle T(z)X'\rangle
-\frac{1}{2\pi i}\oint_{\bar{C}}dz\,\epsilon(z)\langle T(z)X'\rangle,
\end{equation}
where $\bar{C}$ is the mirror image of the contour $C$ in the lower half-plane, and
\begin{equation}
X'=\phi_{h_{1}}(z_{1})\bar{\phi}_{\bar{h}_{1}}(z_{1}^{*})\ldots
\phi_{h_{n}}(z_{n})\bar{\phi}_{\bar{h}_{n}}(z_{n}^{*})\,.
\end{equation}

\vspace{2cm}

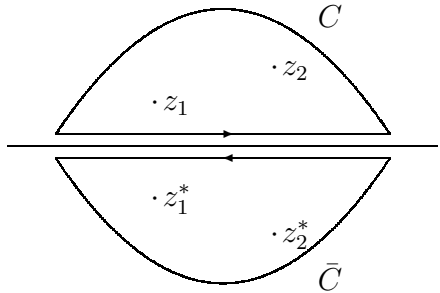
\setlength{\unitlength}{0.0125in}
\begin{picture}(40,90)(60,420)
\put(210,490){\line(1,0){180}}

\qbezier(230,495)(300,600)(370,495) \put(230,495){\vector(1,0){75}} \put(300,495){\line(1,0){70}}

\qbezier(230,485)(300,380)(370,485) \put(230,485){\line(1,0){70}}\put(370,485){\vector(-1,0){70}}

\put(270,505){$\cdot$} \put(275,505){$z_{1}$}\put(320,520){$\cdot$} \put(325,520){$z_{2}$}

\put(270,465){$\cdot$} \put(275,465){$z_{1}^{*}$}\put(320,450){$\cdot$} \put(325,450){$z_{2}^{*}$}

\put(340,540){$C$} \put(340,430){$\bar{C}$}

\end{picture}

\vspace{0.5cm}

Since $\overline{T}=T$ on the real axis, the two disjoint contours may be fused into one, their horizontal parts
canceling each other, and we end up with a single contour circling around twice the number of points:
\begin{equation}
\delta_{\epsilon}\langle X\rangle=-\frac{1}{2\pi i}\oint_{C}dz\epsilon(z)\langle T(z)X'\rangle .
\end{equation}

We thus conclude that the correlator $\langle X\rangle$ on the upper half-plane, as a function of the $2n$
variables $z_{1},\bar{z}_{1},...,z_{n},\bar{z}_{n}$, satisfies the same differential equation as the correlator
$\langle X'\rangle$ on the entire plane, regarded as a function of the $2n$ holomorphic variables
$z_{1},...,z_{2n}$ where $z_{n+i}=z_{i}^{*}$. We have effectively converted the antiholomorphic degrees of freedom
on the upper half-plane into holomorphic ones on the lower half-plane. A $n$-point function on the upper
half-plane is replaced here by a $2n$-point function on the infinite plane. The role of the boundary is simulated
by the interaction between mirror images of the same holomorphic field.

The simplest application of the method of images consists in the determination of the order parameter profile
near the boundary. Assume that in the bulk $\langle\phi(z)\rangle=0$. The one point function
$\langle\phi(z,\bar{z})\rangle$ in the upper half-plane is given by the two point function on the infinite plane
$\langle\phi(z)\phi(z^{*})\rangle=(z-z^{*})^{-2h}$. Thus, if $y$ is the distance from the real axis, the order
parameter profile is
\begin{equation}\label{bvev}
\langle\phi(y)\rangle_{\alpha}=A_{\phi}^{\alpha} \frac{1}{2y^{2h}},
\end{equation}
where $\alpha$ labels the boundary condition and $A_{\phi}^{\alpha}$ is a universal amplitude.

\vspace{1cm}

\section{Boundary operators}

The existence of scaling fields living on the boundary appears naturally within the method of images. If we bring
a bulk scaling field $\phi(z)$ on the upper half-plane closer and closer to the boundary (the real axis), it
interacts with its mirror image $\phi(z^{*})$, and can be replaced by the OPE with its image:
\begin{equation}\label{bOPE}
\phi(x,y)\sim\phi(z)\phi(z^{*})\sim\sum_{i}(z-z^{*})^{h_{i}-2h}C_{\phi\psi_{i}}^{\alpha}\psi_{i}(x)\,,
\end{equation}
where $x=(z+z^{*})/2$. The $\psi_{i}(x)$ are boundary operators with scaling dimension $h_{i}$, normalized so
that $\langle\psi_{i}(x)\psi_{i}(0)\rangle_{\alpha}=x^{-2h_{i}}$ (note that this boundary scaling dimension is
half of the corresponding one for a bulk spinless field).

The bulk-boundary structure constants $C_{\phi\psi_{i}}^{\alpha}$ depend on the type of boundary condition and on
the bulk and boundary operators. Taking the expectation values of both sides of eq.(\ref{bOPE}) we can see that
the coupling of any field to the boundary identity operator is exactly the amplitude of eq.(\ref{bvev})\,
($C_{\phi 1}^{\alpha}=A_{\phi}^{\alpha}$), and that $C_{\phi\psi _{i}}^{\alpha}\langle\psi _{i} \rangle
_{\alpha}=0$ for $\psi _{i}\neq 1$.

As we shall see, the boundary fields, when inserted at a point on the boundary, have the effect of changing the
boundary conditions.

\vspace{2cm}

\begin{figure}[h]
\setlength{\unitlength}{0.0125in}
\begin{picture}(40,90)(60,420)

\put(200,440){\line(0,1){100}} \put(250,440){\line(0,1){100}}
\qbezier(200,440)(225,425)(250,440)\qbezier(200,440)(225,455)(250,440)
\qbezier(200,540)(225,525)(250,540)\qbezier(200,540)(225,555)(250,540)

\put(225,420){$\alpha$} \put(225,555){$\beta$} \put(248,425){$T$} \put(258,480){$L$}
\put(177,490){$H_{\alpha\beta}$} \qbezier(210,490)(225,480)(240,490) \put(210,486){\line(0,1){8}}
\put(238,488){\vector(3,2){6}} \put(225,475){$t$}

\put(380,440){\line(0,1){100}} \put(430,440){\line(0,1){100}}
\qbezier(380,440)(405,425)(430,440)\qbezier(380,440)(405,455)(430,440)
\qbezier(380,540)(405,525)(430,540)\qbezier(380,540)(405,555)(430,540)

\put(405,420){$|\,\alpha\rangle$} \put(405,555){$|\,\beta\rangle$} \put(438,467){$H$}
\put(440,480){\line(1,0){8}} \put(444,480){\vector(0,1){30}} \put(450,490){$t$}

\end{picture} \caption{Two different quantization schemes}
 \end{figure}
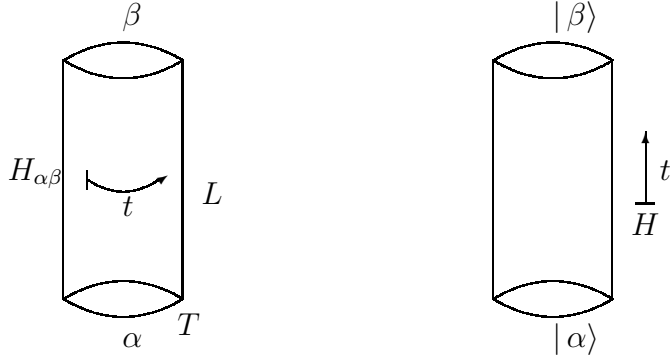

Let us consider a conformal field theory defined on a finite cylinder of circumference $T$ and length $L$, with
boundary conditions $\alpha$ and $\beta$ imposed on the two edges. There are two equivalent quantization schemes,
one in which time flows around the cylinder, another one in which it flows along the cylinder. In the first
scheme, the hamiltonian $H_{\alpha\beta}$ depends on the boundary conditions on the edges, while in the second
one the boundary conditions are embodied in initial and final states $|\alpha\rangle$ and $|\beta\rangle$ and the
hamiltonian $H$ is obtained directly from the whole complex plane. In this second scheme, Cardy and Lewellen have
shown (\cite{cardylew}) that the amplitude in eq.(\ref{bvev}) is given by
\begin{equation}\label{Aphi}
A_{\phi}^{\alpha} =\frac{\langle\phi|\alpha\rangle}{\langle 0|\alpha\rangle}\,,
\end{equation}
where $|0\rangle$ is the ground state of $H$.

In the first scheme the partition function is
\begin{equation}
Z_{\alpha\beta}(q)=Tr\, q^{H_{\alpha\beta}}\qquad\qquad \textrm{with}\qquad q= e^{2\pi
i\tau},\;\tau=\frac{iT}{2L}\,.
\end{equation}
Local conformal invariance implies that the spectrum of $H_{\alpha\beta}$ falls into irreducible representations
of the Virasoro algebra:
\begin{equation}
Z_{\alpha\beta}(q)=\sum_{i}n_{\alpha\beta}^{i}\chi_{i}(q)\,,
\end{equation}
where $n_{\alpha\beta}^{i}$ is the number of copies of the representation labeled $i$ occurring in the spectrum.
This is a linear, and not bilinear, combination of characters because the full theory resides on the holomorphic
sector only.

In a minimal model, under a modular transformation $\tau \rightarrow -1/\tau$ the holomorphic characters transform
as follows (see (\ref{modmatr})):
\begin{equation}
\chi_{i}(q)=\sum_{i}S_{ij}\chi_{j}(\tilde{q})\qquad \tilde{q}\equiv e^{-2\pi i/\tau}\,,
\end{equation}
hence
\begin{equation}\label{schemes}
Z_{\alpha\beta}(q)=\sum_{i}n_{\alpha\beta}^{i}S_{ij}\chi_{j}(\tilde{q})\,.
\end{equation}
Such a modular transformation interchanges the roles of $L$ and $T$, i.e. it interchanges the two quantization
schemes.

In the second scheme the partition function is expressed as
\begin{equation}
Z_{\alpha\beta}(q)=\langle\alpha|e^{LH}|\beta\rangle=\langle\alpha|\left(\tilde{q}^{1/2}\right)^{L_{0}+\bar{L}_{0}-\frac{c}{12}}|\beta\rangle
\,,
\end{equation}
where the last equality is due to the bulk hamiltonian expression (\ref{hamcyl}).

For all boundary conditions we must have
\begin{equation}
T_{cyl.}(0,t)=\overline{T}_{cyl.}(0,t)\qquad \textrm{and}\qquad T_{cyl.}(L,t)=\overline{T}_{cyl.}(L,t)
\end{equation}
In terms of the Virasoro generators acting on the boundary state $|\alpha\rangle$, this constraint becomes
\begin{equation}
\left(L_{n}-\bar{L}_{-n}\right)|\alpha\rangle=0\,,
\end{equation}
whose solutions are the so-called Ishibashi states
\begin{equation}
|j\rangle\rangle\equiv\sum_{N}|j;N\rangle\otimes U |\overline{j;N}\rangle\,,
\end{equation}
where $|j;N\rangle$ is a holomorphic state belonging to the irreducible quotient of the Verma module $j$,
$|\overline{j;N}\rangle$ is the corresponding antiholomorphic state, and $U$ is an antiunitary operator such that
\begin{equation}
U|\overline{j;0}\rangle=|\overline{j;0}\rangle^{*}\qquad U\bar{L}_{n}=\bar{L}_{n}U\,.
\end{equation}
The boundary states $|\alpha\rangle$ and $|\beta\rangle$ will then be linear combinations of Ishibashi states
associated with different Verma modules. Assuming that the states $|j\rangle\rangle$ have been normalized, we may
write
\begin{equation}
Z_{\alpha\beta}(q)=\sum_{i,j}\langle\alpha|i\rangle\rangle\langle\langle i|
\left(\tilde{q}^{1/2}\right)^{L_{0}+\bar{L}_{0}-\frac{c}{12}}|j\rangle\rangle\langle\langle j|\beta\rangle\,.
\end{equation}
If the theory is diagonal, this can be rewritten as
\begin{equation}
Z_{\alpha\beta}(q)=\sum_{j}\langle\alpha|j\rangle\rangle\langle\langle j|\beta\rangle\chi_{j}(\tilde{q})\,,
\end{equation}
and comparing the above result with eq.(\ref{schemes}) leads to the relation
\begin{equation}
\sum_{i}S_{ij}n_{\alpha\beta}^{i}=\langle\alpha|j\rangle\rangle\langle\langle j|\beta\rangle\,.
\end{equation}

It follows from this equation that a boundary state $|\tilde{0}\rangle$ such that the only representation
occurring in the hamiltonian $H_{\tilde{0}\tilde{0}}$ is the identity
$\left(n_{\tilde{0}\tilde{0}}^{i}=\delta_{0}^{i}\right)$ has to satisfy $|\langle\tilde{0}|j\rangle|^{2}=S_{0j}$.
In a unitary model $S_{0j}$ can be shown to be positive, and therefore this state exists and can be taken as
\begin{equation}
|\tilde{0}\rangle=\sum_{j}\sqrt{S_{0j}}|j\rangle\rangle\,.
\end{equation}
Likewise, we define a state
\begin{equation}\label{bstates}
|\tilde{l}\rangle=\sum_{j}\frac{S_{lj}}{\sqrt{S_{0j}}}|j\rangle\rangle\,,
\end{equation}
such that $n_{\tilde{0}\tilde{l}}^{i}=\delta_{l}^{i}$: only the representation $l$ propagates in
$H_{\tilde{0}\tilde{l}}$. With this identification we have a 1:1 correspondence between physical boundary states
(i.e. conformal boundary conditions) and the primary operators of the bulk theory. Furthermore, we have
\begin{equation}
\sum_{i}S_{ij}n_{\tilde{k}\tilde{l}}^{i}=\langle\tilde{k}|j\rangle\rangle\langle\langle
j|\tilde{l}\rangle=\frac{S_{ki}S_{lj}}{S_{0j}}\,,
\end{equation}
and from the Verlinde formula (\ref{verlform}) we conclude that
\begin{equation}
n_{\tilde{k}\tilde{l}}^{i}=N_{kl}^{i}\,,
\end{equation}
i.e. the number of times representation $i$ occurs in the Hamiltonian $H_{\tilde{k}\tilde{l}}$ is precisely the
fusion coefficient $N_{kl}^{i}$. This result warrants the interpretation that boundary conditions may be changed
by inserting a local operator on the boundary.

Using the explicit form (\ref{bstates}) for the boundary states, eq.(\ref{Aphi}) gives the remarkable result
\begin{equation}\label{Aexpl}
A_{\phi}^{\tilde{k}}=\frac{S_{k\phi}}{S_{k0}}\sqrt{\frac{S_{00}}{S_{0\phi}}}\,.
\end{equation}

\newpage

\section{An explicit calculation}

We will now explicitly calculate two-point functions involving the field $\phi_{1,2}$ in minimal models with
boundary, which can be expressed in terms of the bulk holomorphic four-point functions as
\begin{equation}
\langle \phi _{n,m}\left(z_{1},\overline{z}_{1} \right)\phi _{1,2}\left(z_{2},\overline{z}_{2} \right)\rangle
_{\alpha}= \langle \phi _{n,m}\left(z_{1} \right)\phi _{1,2}\left(z_{2} \right)\phi _{1,2}\left(z^{*}_{2}
\right)\phi _{n,m}\left(z^{*}_{1} \right)\rangle
\end{equation}
Defining
\begin{center}
\begin{tabular}{c|c}
$z_{1}=x_{1}+iy_{1}$ & $z_{2}=x_{2}+iy_{2}$ \\
$z_{4}=x_{1}-iy_{1}$ & $z_{3}=x_{2}-iy_{2}$ \\
\end{tabular}
\end{center}
and following the parametrization (\ref{hrs2}),(\ref{alphars}), we have
\begin{equation}\label{explcalc}
\langle \phi _{n,m}\left(z_{1} \right)\phi _{1,2}\left(z_{2} \right)\phi _{1,2}\left(z_{3} \right)\phi
_{n,m}\left(z_{4} \right)\rangle=\left(\prod_{i<j}z_{ij}^{\gamma _{ij}} \right)\eta^{2\alpha
_{n,m}\alpha_{1,2}-\gamma _{12}} \left(1-\eta  \right)^{2\alpha _{1,2}^{2}-\gamma _{23}} Y\left(\eta  \right)
\end{equation}
where $\eta =\frac{z_{12}z_{34}}{z_{13}z_{24}}$, $\gamma _{ij}=\frac{1}{3}\left(\sum _{k}h_{k}
\right)-h_{i}-h_{j}$, and $Y\left(\eta  \right)$ is a linear combination of the functions $I_{1}\left(a,b,c,\eta
\right)$ and $I_{2}\left(a,b,c,\eta  \right)$ defined in (\ref{I12}):
\begin{equation}
Y\left(\eta  \right)=A I_{1}\left(a,b,c,\eta  \right)+B I_{2}\left(a,b,c,\eta  \right) \,.
\end{equation}
This combination in linear (and not bilinear) because we are left with just the holomorphic sector of the theory,
and the coefficients $A$ and $B$ are no more determined enforcing monodromy invariance, but imposing the correct
behaviour of the correlator in certain limits.

\subsubsection{$x_{12}\rightarrow \infty$ limit}

\vspace{2cm}

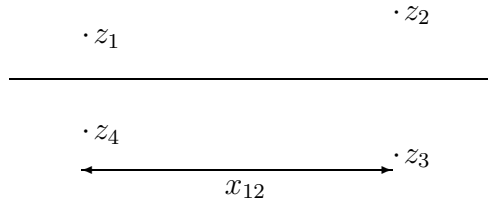
\setlength{\unitlength}{0.0125in}
\begin{picture}(40,90)(60,420)
\put(200,490){\line(1,0){200}}

\put(230,505){$\cdot$} \put(235,505){$z_{1}$}\put(360,515){$\cdot$} \put(365,515){$z_{2}$}

\put(230,465){$\cdot$} \put(235,465){$z_{4}$}\put(360,455){$\cdot$} \put(365,455){$z_{3}$}

\put(295,452){\vector(1,0){65}} \put(295,452){\vector(-1,0){65}} \put(290,442){$x_{12}$}
\end{picture}

\vspace{0.5cm}

We start considering the limit in which the two fields are very far from each other, in comparison to their
distance from the boundary. The expected behaviour of the correlator is
\begin{eqnarray}
\langle \phi _{n,m}\left(z_{1},\overline{z}_{1} \right)\phi _{1,2}\left(z_{2},\overline{z}_{2} \right)\rangle
_{\alpha} &\rightarrow&  \langle \phi _{n,m}\left(z_{1},\overline{z}_{1} \right)\rangle _{\alpha}\langle \phi
_{1,2}\left(z_{2},\overline{z}_{2} \right)\rangle _{\alpha}\\
&& =A_{nm}^{\alpha}A_{12}^{\alpha}\left(2y_{1} \right)^{-2h_{nm}}\left(2y_{2} \right)^{-2h_{12}}
\end{eqnarray}
where the constants $A_{nm}^{\alpha}$ have the form (\ref{Aexpl}).

Being
\begin{equation}
\eta =\frac{x_{12}^{2}+\left(y_{1}-y_{2} \right)^{2}}{x_{12}^{2}+\left(y_{1}+y_{2} \right)^{2}} \qquad 1-\eta
=\frac{4y_{1}y_{2}}{x_{12}^{2}+\left(y_{1}+y_{2} \right)^{2}} \,,
\end{equation}
this limit corresponds to $ \eta \rightarrow 1$. In order to extract the corresponding behaviour of $I_{1,2}$
using the property $F(\lambda,\mu,\nu;0)=1$ of the hypergeometric function, we write these functions in the
$1-\eta$ basis:
\begin{equation}
I_{1}\left(a,b,c,\eta  \right)=\frac{s(a)}{s(b+c)}\tilde{I_{1}}\left(b,a,c,1-\eta
\right)-\frac{s(c)}{s(b+c)}\tilde{I_{2}}\left(b,a,c,1-\eta  \right)\\
\end{equation}
\begin{displaymath}
I_{2}\left(a,b,c,\eta  \right)=-\frac{s(a+b+c)}{s(b+c)}\tilde{I_{1}}\left(b,a,c,1-\eta
\right)-\frac{s(b)}{s(b+c)}\tilde{I_{2}}\left(b,a,c,1-\eta  \right)
\end{displaymath}
where $s\left(x \right)=sin\left(\pi x \right)$. In this way we obtain

\begin{tabular}{c}
$Y\left(\eta  \right)\rightarrow \frac{\left[As\left(2\alpha _{-}\alpha _{nm} \right)-Bs\left(2\alpha _{-}\alpha
_{nm} -2\alpha _{-}^{2}\right)\right]\Gamma\left(2\alpha _{-}^{2}-2\alpha _{-}\alpha _{nm}-1\right) \Gamma
\left(2\alpha _{-}\alpha _{nm}+1\right)}{s\left(-2\alpha _{-}^{2}\right)\Gamma \left(2\alpha _{-}^{2}\right)}$
\end{tabular}
\begin{equation}
 -(A+B)\frac{s\left(\alpha _{-}^{2}\right)}{s\left(2\alpha _{-}^{2}\right)}\frac{\Gamma^{2} \left(1-\alpha
_{-}^{2}\right)}{\Gamma\left(2-2\alpha _{-}^{2}\right) }\left[\left(4y_{1}y_{2} \right)x_{12}^{-2}
\right]^{1-2\alpha _{-}^{2}}
\end{equation}
The prefactor which multiplies $Y(\eta)$ in eq.(\ref{explcalc}) goes like
\begin{equation}\label{prefactor}
i^{-\frac{1}{6}\left[\left(n^{2}-1 \right)\alpha _{+}^{2}+\left(m^{2}+2 \right)\alpha _{-}^{2}-2nm
\right]}x_{12}^{2-4\alpha _{-}^{2}}\left(2y_{1} \right)^{\frac{1}{2}\alpha_{-}^{2}+2h_{12}+2h_{nm} }\left(2y_{2}
\right)^{\frac{1}{2}\alpha_{-}^{2} }
\end{equation}
The corresponding phase $ i^{-\frac{1}{6}\left[\left(n^{2}-1 \right)\alpha _{+}^{2}+\left(m^{2}+2 \right)\alpha
_{-}^{2}-2nm \right]}$ can be fixed equal to zero by multiplying the whole correlator by an opportune constant $
\lambda $. However, it is possible to see that it simplifies in the calculation of the bulk-boundary structure
constants.

The exponent $ 2-4\alpha _{-}^{2}$ is negative for $ -2< c< 7$, hence we have the condition
\begin{equation}\label{A+B}
A+B=-\frac{s\left(2\alpha _{-}^{2} \right)}{s\left(\alpha _{-}^{2} \right)}\frac{\Gamma \left(2-2\alpha _{-}^{2}
\right)}{\Gamma ^{2}\left(1-\alpha _{-}^{2} \right)}A_{nm}^{\alpha}A_{12}^{\alpha}
\end{equation}

\subsubsection{$ \mid z_{1}-z_{2}\mid \ll y_{1},y_{2}$  limit}

\vspace{3cm}

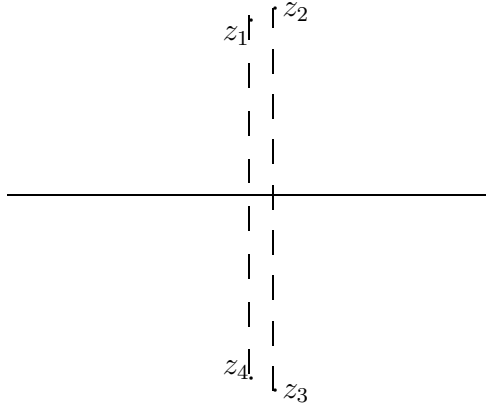
\setlength{\unitlength}{0.0125in}
\begin{picture}(40,90)(60,420)
\put(200,510){\line(1,0){200}}

\put(301,435){\line(0,1){10}} \put(301,455){\line(0,1){10}} \put(301,475){\line(0,1){10}}
\put(301,495){\line(0,1){10}} \put(301,515){\line(0,1){10}} \put(301,535){\line(0,1){10}}
\put(301,555){\line(0,1){10}} \put(301,575){\line(0,1){10}}

\put(311,428){\line(0,1){10}} \put(311,447){\line(0,1){10}} \put(311,466){\line(0,1){10}}
\put(311,485){\line(0,1){10}} \put(311,504){\line(0,1){10}} \put(311,523){\line(0,1){10}}
\put(311,542){\line(0,1){10}} \put(311,561){\line(0,1){10}} \put(311,580){\line(0,1){8}}

 \put(300,580){$\cdot$} \put(290,575){$z_{1}$}\put(310,585){$\cdot$}
\put(315,585){$z_{2}$}

\put(300,430){$\cdot$} \put(290,435){$z_{4}$}\put(310,425){$\cdot$} \put(315,425){$z_{3}$}
\end{picture}

We consider the limit in which $ \mid z_{1}-z_{2}\mid\rightarrow 0$ and $ y_{1}\rightarrow y_{2}\equiv y$, so
that the operators are placed deep in the bulk, far from the boundary, and we can use their bulk OPE's.

The bulk expansion
\begin{equation}
\phi _{n,m}\left(z_{1},\overline{z}_{1} \right)\phi _{1,2}\left(z_{2},\overline{z}_{2} \right)\sim
\frac{C_{nm,12}^{n,m-1}\phi _{n,m-1}\left(z_{2},\overline{z}_{2} \right)}{\mid z_{1}-z_{2}\mid
^{2\left(h_{nm}+h_{12}-h_{n,m-1} \right)}}+\frac{C_{nm,12}^{n,m+1}\phi _{n,m+1}\left(z_{2},\overline{z}_{2}
\right)}{\mid z_{1}-z_{2}\mid ^{2\left(h_{nm}+h_{12}-h_{n,m+1} \right)}}
\end{equation}
forces the correlator to have the behaviour:
\begin{displaymath}
\langle \phi _{n,m}\left(z_{1},\overline{z}_{1} \right)\phi _{1,2}\left(z_{2},\overline{z}_{2} \right)\rangle
_{\alpha}\rightarrow A_{n,m-1}^{\alpha}C_{nm,12}^{n,m-1}\mid z_{1}-z_{2}\mid ^{-2\left(h_{nm}+h_{12}-h_{n,m-1}
\right)} \left(2y\right)^{-2h_{n,m-1}}+
\end{displaymath}
\begin{equation}
+A_{n,m+1}^{\alpha}C_{nm,12}^{n,m+1}\mid z_{1}-z_{2}\mid ^{-2\left(h_{nm}+h_{12}-h_{n,m+1} \right)}
\left(2y\right)^{-2h_{n,m+1}}
\end{equation}
Since $\eta =\frac{\mid z_{1}-z_{2}\mid ^{2}}{4y^{2}}$, this limit correspond to $\eta\rightarrow 0$, and we have
\begin{equation}
Y\left(\eta  \right)\rightarrow A\frac{\Gamma\left(2\alpha _{-}^{2}-2\alpha _{-}\alpha _{nm}-1\right) \Gamma
\left(1-\alpha _{-}^{2}\right)}{\Gamma \left(\alpha _{-}^{2}-2\alpha _{-}\alpha _{nm}\right)}+
\end{equation}
\begin{displaymath}
+B\frac{\Gamma \left(2\alpha _{-}\alpha _{nm}+1 \right)\Gamma \left(1-\alpha
_{-}^{2}\right)}{\Gamma\left(2+2\alpha _{-}\alpha _{nm}-\alpha _{-}^{2}\right) }\left[\mid z_{1}-z_{2}\mid
\left(2y \right)^{-1} \right]^{2+4\alpha _{-}\alpha _{nm}-2\alpha _{-}^{2}}
\end{displaymath}
The prefactor goes like
\begin{equation}
i^{-\frac{1}{6}\left[\left(n^{2}-1 \right)\alpha _{+}^{2}+\left(m^{2}+2 \right)\alpha _{-}^{2}-2nm \right]}\mid
z_{1}-z_{2}\mid ^{-2\alpha _{-}\alpha _{nm}}\left(2y \right)^{2\alpha_{-}\alpha _{nm}-2h_{12}-2h_{nm} }
\end{equation}
and comparing the two channels, we get the conditions
\begin{equation}\label{A}
A=\frac{\Gamma \left(1-n+m\alpha _{-}^{2} \right)}{\Gamma \left(-n+(1+m)\alpha _{-}^{2} \right)\Gamma
\left(1-\alpha _{-}^{2} \right)}\,C_{nm,12}^{n,m+1}A_{n,m+1}^{\alpha}
\end{equation}
\begin{equation}\label{B}
B=\frac{\Gamma \left(1+n-m\alpha _{-}^{2} \right)}{\Gamma \left(n+(1-m)\alpha _{-}^{2} \right)\Gamma
\left(1-\alpha _{-}^{2} \right)}\,C_{nm,12}^{n,m-1}A_{n,m-1}^{\alpha}
\end{equation}

As a check, we have to verify if this values are compatible with the one previously found for $A+B$.

\subsubsection{Back to the $ x_{12}\rightarrow \infty $  limit}

In this limit $\left(\mid z_{1}-z_{2}\mid \gg y_{1},y_{2} \right)$ we can use the expansion (\ref{bOPE}), which
implies the behaviour
\begin{equation}
\langle \phi _{n,m}\left(z_{1},\overline{z}_{1} \right)\phi _{1,2}\left(z_{2},\overline{z}_{2} \right)\rangle
_{\alpha}\rightarrow
\end{equation}
\begin{displaymath}
\sum _{i}\left(2y_{1} \right)^{h_{i}-2h_{nm}}\left(2y_{2}
\right)^{h_{i}-2h_{12}}\left[C_{(n,m)\psi _{i}}^{\alpha} \right]\left[C_{(1,2)\psi _{i}}^{\alpha}
\right]\left(x_{1}-x_{2} \right)^{-2h_{i}}
\end{displaymath}
where the sum runs over the operators common to the expansions $ \phi _{1,2}\times\phi _{1,2}$ and  $ \phi
_{n,m}\times\phi _{n,m}$, i.e. it cannot involve other boundary operators in addition to $\psi _{1,1}$ and $\psi
_{1,3}$ (see (\ref{12OPE})). As in the other limit, comparing the two channels we have the relations
\begin{equation}\label{bstructconst11}
\left[C_{(n,m)\psi _{1,1}}^{\alpha} \right]\left[C_{(1,2)\psi _{1,1}}^{\alpha} \right] =-\frac{s\left(\alpha
_{-}^{2} \right)}{s\left(2\alpha _{-}^{2} \right)}\frac{\Gamma^{2} \left(1-\alpha _{-}^{2} \right)}{\Gamma
\left(2-2\alpha _{-}^{2} \right)}\left(A+B \right)= A_{n,m}^{\alpha}A_{1,2}^{\alpha}
\end{equation}
and
\begin{eqnarray*}\label{bstructconst13}
\left[C_{(n,m)\psi _{1,3}}^{\alpha} \right]\left[C_{(1,2)\psi _{1,3}}^{\alpha} \right] &=&\frac{\Gamma
\left(-n+(1+m)\alpha
_{-}^{2} \right)\Gamma \left(n+(1-m)\alpha _{-}^{2} \right)}{\Gamma \left(2\alpha _{-}^{2} \right)} \times \\
&& \times \frac{As\left(n-1+(1-m)\alpha _{-}^{2} \right)-Bs\left(n-1-(1+m)\alpha _{-}^{2} \right)}{s\left(-2\alpha
_{-}^{2} \right)}=
\end{eqnarray*}
\begin{eqnarray*}
&& =\frac{s\left(n-1+(1-m)\alpha _{-}^{2} \right)}{s\left(-2\alpha _{-}^{2} \right)}\frac{\Gamma
\left(n+(1-m)\alpha _{-}^{2} \right)\Gamma \left(1-n+m\alpha _{-}^{2} \right)}{\Gamma\left(2\alpha _{-}^{2}
\right) \Gamma \left(1-\alpha _{-}^{2} \right)}C_{nm,12}^{n,m+1}A_{n,m+1}^{\alpha}+ \\
&& -\frac{s\left(n-1-(1+m)\alpha _{-}^{2} \right)}{s\left(-2\alpha _{-}^{2} \right)}\frac{\Gamma
\left(-n+(1+m)\alpha _{-}^{2} \right)\Gamma \left(1+n-m\alpha _{-}^{2} \right)}{\Gamma\left(2\alpha _{-}^{2}
\right) \Gamma \left(1-\alpha _{-}^{2} \right)}C_{nm,12}^{n,m-1}A_{n,m-1}^{\alpha}
\end{eqnarray*}
Note that if we had kept the phase in (\ref{prefactor}), it would have simplified in these last identities. The
result obtained for $C_{(n,m)\psi _{1,1}}^{\alpha}$ is exactly what we expected from (\ref{bOPE}).

\vspace{0.5cm}

The described procedure can be repeated in the computation of the correlator
\begin{equation}
\langle \phi _{n,m}\left(z_{1},\overline{z}_{1} \right)\phi _{2,1}\left(z_{2},\overline{z}_{2} \right)\rangle
_{\alpha}= \langle \phi _{n,m}\left(z_{1} \right)\phi _{2,1}\left(z_{2} \right)\phi _{2,1}\left(z^{*}_{2}
\right)\phi _{n,m}\left(z^{*}_{1} \right)\rangle \,,
\end{equation}
with $\alpha _{-}$ replaced by $ \alpha _{+}$, and the roles of $n$ and $m$ interchanged. In this way it is
possible to determine also the bulk-boundary structure constants $C_{(n,m)\psi _{3,1}}^{\alpha}$.

\vspace{0.5cm}

The explicit values of the constants $A$, $B$ and the bulk-boundary structure constants $C_{(n,m)\psi
_{1,1}}^{\alpha}$, $C_{(n,m)\psi _{1,3}}^{\alpha}$ and $C_{(n,m)\psi _{3,1}}^{\alpha}$ are listed in appendix
\ref{confcalc} for the Ising model and the tricritical Ising model.

\newpage

\chapter{Off-critical systems}

Dynamics in the vicinity of second order phase transitions can be described by CFT perturbed by the addition of
operators that break the conformal symmetry and introduce a mass scale in the system. A complete review of this
subject can be found in (\cite{mussrev}).

\vspace{1cm}

\section{Scaling region near the critical points}

The specific values of the parameters for which a statistical system is critical are associated to fixed points of
the renormalization group (RG) flow. A RG trajectory flowing away from a fixed point is obtained by combinations
of the relevant scalar operators $\Phi _{i}$ present in the corresponding CFT, which have anomalous dimensions
$x_{i}=2 h _{i}< 2$. The corresponding off-critical action is given by
\begin{equation}\label{action}
A=A_{CFT}+\sum _{i}\lambda _{i}\int\Phi _{i}(x)d^{2}x.
\end{equation}
The coupling constants $ \lambda _{i}$, having mass dimension $ 2\left(1-h _{i} \right)$, introduce a
characteristic length in the system. It is then possible to have non vanishing vacuum expectation values of the
fields, proportional to $ \lambda ^{\frac{2h }{2\left(1-h _{i} \right)}}$. The relevant operators don't affect
the behavior of the system at short distances (they are of superrenormalizable type with respect to the
ultraviolet divergences), but they do change it at large distance scales. The RG trajectories can reach another
critical point (as an example, the minimal model ${\cal M}_{p,p-1}$ perturbed by $\phi _{(1,3)}$ flows to ${\cal
M}_{p-1,p-2}$) or end at a non-critical fixed point, corresponding to a massive quantum field theory. We will
restrict our attention on this second case, described by a relativistic scattering theory that is completely
defined specifying the $S$-matrix. The CFT data contain the information about the short distance (UV) properties
of the field theory, while the $S$-matrix data display the information about the long distance (IR) behaviour.

The main result of this approach is the Zamolodchikov c-theorem (\cite{zamcth}), which states that, for quantum
field theories which possess rotational invariance, reflection positivity and conservation of the stress-energy
tensor, there is a function $C\left(\lambda _{i} \right)$ of the coupling constants $\lambda _{i}$ which is
non-increasing along the RG trajectories and is stationary only at the fixed points, where it coincides with the
central charge $c$ of the corresponding CFT.

Thus, if two critical points are linked by a RG trajectory, then the values $c_{0}$ and $c_{1}$ of the central
charge in the conformal theories obey the inequality $c_{0}> c_{1}$, and this makes it possible to give a RG
meaning to the \lq\lq ordering\rq\rq\, of the CFT solutions by the magnitude of the central charge $c$.

In the simplest case of a deformation of CFT achieved by perturbing only with one relevant field $\Phi$, Cardy
(\cite{Cardysum}) established the following sum rule for the total change in $c$ from short to large distances:
\begin{equation}\label{Cardysum}
\Delta c=3\pi \lambda ^{2}\left(2-2 h \right)^{2}\int d^{2}x\,|x| ^{2}\langle\Phi (x)\Phi (0)\rangle \,.
\end{equation}

\vspace{1cm}

\section{Conservation laws and integrable models}

An integrable model is characterized by the presence of an infinite set of conserved currents, which in
two-dimensional systems satisfy the equations
\begin{equation}
\partial _{\overline{z}}J_{z,z,...}+\partial _{z}J_{\overline{z},\overline{z},...}=0 \,.
\end{equation}
In CFT, any operator in the conformal family of the stress-energy tensor, being an analytic function, trivially
satisfies these equations, while in the deformed theory this is generally destroyed. However, Zamolodchikov
(\cite{sbrind}) discovered the so-called integrable deformations of CFT, with a corresponding QFT that possesses
an infinite set of conserved charges $P_{s}$ in involution, which permit to solve the theory non-perturbatively.
This theories are called Minimal Integrable Models (MIM).

For integrable models originating from a perturbation of a CFT, the integrals of motion can be interpreted as
deformations of the conformal conservation laws. Let $T_{s+1}$ be the quasi-primary descendants of the
stress-energy tensor, with spin $s+1$. The conformal conservation laws are
\begin{equation}
\partial _{\overline{z}}T_{s+1}=0 \,,
\end{equation}
and the problem is to find the spins $s$ and the local fields $\Theta _{s-1}$ such that we have the off-critical
conservation laws
\begin{equation}\label{offcritcons}
\partial _{\overline{z}}T_{s+1}=\partial _{z}\Theta _{s-1}\,,
\end{equation}
which give rise to the conserved charges
\begin{equation}
P_{s}=\int\left(T_{s+1}\, dz -\Theta _{s-1} \, d\overline{z} \right)\,.
\end{equation}

\vspace{0.5cm}

We will now show which are the conditions under which it is possible to write eq.(\ref{offcritcons}). Let
$C_{s}(z)$ be a conserved current with spin $s$ in the conformal minimal model ${\cal M}_{p,p'}$, local with
respect to the perturbing field $\Phi _{lk}\left(z,\overline{z}\right)=\phi
_{lk}\left(z\right)\overline{\phi}_{lk}\left(\overline{z} \right)$:
\begin{equation}
C_{s}(z)\Phi _{lk}\left(w,\overline{w} \right)=\sum _{n=2}^{m}\frac{d_{lk}^{(n)}}{\left(z-w \right)^{n}}\Phi
_{lk}^{(n)}\left(w,\overline{w} \right)+\frac{1}{z-w}B_{lk}\left(w,\overline{w} \right)+...
\end{equation}
($n$ is an integer, $ \Phi _{lk}^{(n)}$ and $ B_{lk}$ are descendants of $ \Phi _{lk}$ and $ d_{lk}^{(n)}$ are
some constants). Referring to the action (\ref{action}), the deformed Ward identities for $C_{s}(z,\overline{z})$
can be written in terms of the conformal ones as
\begin{equation}
\langle C_{s}\left(z,\overline{z} \right)\cdots \rangle =\langle C_{s}\left(z \right)\cdots \rangle_{0}+\lambda
\int dw\, d\overline{w}\,\langle C_{s}\left(z\right)\Phi _{lk}\left(w,\overline{w} \right)\cdots
\rangle_{0}+O\left(\lambda ^{2} \right)\,.
\end{equation}
These relations, together with the identity
\begin{equation}
\partial _{\overline{z}}\,\frac{1}{z-w+i\varepsilon }=\delta \left(z-w \right)\delta \left(\overline{z}-\overline{w} \right),
\end{equation}
lead to the following equation, valid to the first order in $\lambda$:
\begin{equation}
\partial _{\overline{z}}C_{s}\left(z,\overline{z} \right)=\lambda \left[B_{lk}\left(z,\overline{z} \right)-d_{lk}^{(2)}\partial _{z}\Phi_{lk}^{(2)} \left(z,\overline{z} \right) \right].
\end{equation}
Hence, the existence of the off-critical conservation law depends on whether $B_{lk}$ is a total derivative with
respect to $z$.

\vspace{0.5cm}

The simplest example is the energy-momentum conservation: $C_{s}=T$. In this case we have
\begin{equation}
\partial _{\overline{z}}T\left(z,\overline{z} \right)=-\frac{1}{4}\partial _{z}\Theta ,
\end{equation}
where $\Theta=4T_{z\overline{z}}=-4\lambda\left(1-h\right)\Phi_{lk}\left(z,\overline{z}\right)$ is the trace
component of $T^{\mu\nu}$ in terms of the perturbation $\Phi_{lk}$. The corresponding conserved charge, present
for every choice of the perturbation, is the momentum:
\begin{equation}
C_{1}\equiv P=\int\left(T dz+\frac{1}{4}\Theta d \overline{z}\right)\,.
\end{equation}

\vspace{0.5cm}

An interesting example, involving higher integrals of motion, is given by the QFT defined by a $\Phi_{13}$
deformation of the minimal model ${\cal M}_{p,p+1}$. Let $C_{s}=T_{4}=:T^{2}:$ be the quasi-primary field of spin
4 in the conformal family of the identity operator. We have
\begin{equation}
B_{13}=\lambda(h-1)\left[2L_{-1}L_{-2}-2L_{-3}+\frac{1}{6}(h-3)L_{-1}^{3}\right]\Phi_{13}\left(z,\overline{z}\right)\,.
\end{equation}
The second term may spoil the existence of the conservation law, but the field $\Phi_{13}$ satisfies the
third-level null-vector equation
\begin{equation}\label{3levnull}
\left(L_{-3}-\frac{2}{h+2}L_{-1}L_{-2}+\frac{1}{(h+1)(h+2)}L_{-1}^{3}\right)\Phi_{1,3}=0 \,,
\end{equation}
and consequently there exist a conserved charge with spin 3:
\begin{equation}
P_{3}=\int\left(T_{4} dz+\Theta_{2} d \overline{z}\right) \,,
\end{equation}
where
\begin{equation}
\Theta_{2}=\lambda\frac{h-1}{h+2}\left(2h
L_{-2}+\frac{\left(h-2\right)\left(h-1\right)\left(h+3\right)}{6\left(h+1\right)}L_{-1}^{2}\right)\Phi_{1,3} \, .
\end{equation}

\vspace{1cm}

\section{Counting argument}

The counting argument, introduced by Zamolodchikov (\cite{sbrind}), is a sufficient criterion for the existence of
non-trivial conservation laws. We restrict to the case of conserved currents originated from the conformal family
of the identity operator.

Let us consider the space of linearly independent descendants of the identity operator at level $s+1$, i.e. the
factor space
\begin{equation}
\hat{T}_{s+1}=T_{s+1}/\partial_{z}T_{s}\,,
\end{equation}
and the analogous factor space at level $s$ for the perturbing field $\Phi$:
\begin{equation}
\hat{\Phi}_{s}=\Phi_{s}/\partial_{z}\Phi_{s-1}\,.
\end{equation}
If the mapping
\begin{equation}
\partial_{\overline{z}}:\hat{T}_{s+1}\rightarrow\lambda \hat{\Phi}_{s}
\end{equation}
has a non vanishing kernel, i.e. if
\begin{equation}\label{countdim}
\textrm{dim}\,\hat{T}_{s+1}>\textrm{dim}\, \hat{\Phi}_{s}\,,
\end{equation}
then there are some fields $T_{s+1}\left(z,\overline{z}\right)\in\hat{T}_{s+1}$ and
$\Phi_{s-1}\left(z,\overline{z}\right)\in\hat{\Phi}_{s-1}$ such that
\begin{equation}
\partial_{\overline{z}}T_{s+1}\left(z,\overline{z}\right)=\lambda
\partial_{z}\Phi_{s-1}\left(z,\overline{z}\right),
\end{equation}
i.e. there is a conserved charge with spin $s$.

The condition (\ref{countdim}) can be checked using the character formulae (see (\ref{char})):
\begin{equation}
\sum_{s=0}^{\infty}q^{s}\textrm{dim}\hat{T}_{s}=\left(1-q\right)\tilde{\chi}_{1,1}(q)+q
\end{equation}
and
\begin{equation}
\sum_{s=0}^{\infty}q^{s}\textrm{dim}\left(\hat{\Phi}_{k,l}\right)_{s}=\left(1-q\right)\tilde{\chi}_{k,l}(q)\,,
\end{equation}
where
\begin{equation}
\tilde{\chi}_{r,s}(q)=q^{c/24-h_{r,s}}\chi_{r,s}(q)\,.
\end{equation}
In the case of minimal models, these relations immediately give the values of the lowest conserved spins, because
the characters can be explicitly expanded in power series.

\vspace{0.5cm}

With this technique Zamolodchikov has proved that the operators $\Phi_{1,3}$, $\Phi_{1,2}$ and $\Phi_{2,1}$ always
define an integrable deformation of the conformal minimal models.

In the case of the $\Phi_{1,3}$ perturbation, for example, thanks to the third-level null-vector equation
(\ref{3levnull}) we have
\begin{equation}
\textrm{dim}\left(\hat{\Phi}_{1,3}\right)_{3}=0\,,
\end{equation} and the mapping
\begin{equation}
\partial_{\overline{z}}:\hat{T}_{4}\rightarrow\lambda \hat{\Phi}_{3}
\end{equation}
has a non vanishing kernel. In a similar way, it is possible to see that $\partial_{\overline{z}}$ has non
vanishing kernels also for the next few odd $s$, and it is natural to conjecture that the perturbed QFT possesses
conserved charges at each odd value of $s$.

\vspace{1cm}

\section{Scattering theory}

\subsection{Elasticity and factorization}

In the case of deformations which give rise to massive integrable models, the corresponding scattering theory is
purely elastic and factorized (\cite{Svecchio}). In fact, the presence of an infinite number of conserved charges
$P_{s}$ (with spin $s$) implies the conservation of all the $s$ powers of the momenta, and this is equivalent to
the absence of particle production and the equality of the sets of initial and final momenta. If the mass spectrum
is not degenerate, the $S$-matrix is completely diagonal, while, if the system presents multiplets of degenerate
particles, there can be a redistribution of the momenta among the particles with the same masses.

Furthermore, an arbitrary $n$-particle collision process becomes factorized into the product of $n(n-1)/2$
elastic pair collisions.

\vspace{2.5cm}

\begin{figure}[h]
\setlength{\unitlength}{0.0125in}
\begin{picture}(40,90)(60,420)
\put(260,420){\line(-3,4){100}} \put(160,420){\line(3,4){100}} \put(210,420){\line(0,1){130}}
\put(430,420){\line(-3,4){100}} \put(330,420){\line(3,4){100}} \put(400,420){\line(0,1){130}} \put(290,490){$=$}
\end{picture}
\caption{Factorization} \label{figfactor}
 \end{figure}
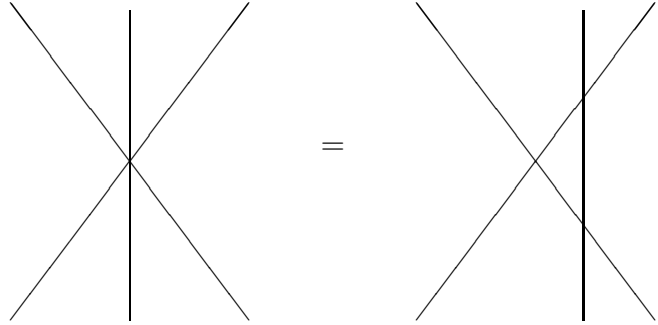

\vspace{0.5cm}

This corresponds to the commutativity of the processes shown in Figure \ref{figfactor}, where the lines drawn are
exactly the physical lines of the particles in the two-dimensional space-time, and can be explained by a
particle-displacement argument. In fact, using the conserved charges $P_{s}$, we can define the operators
$T_{s}(a)=\exp \left(iaP_{s}\right)$. $T_{1}$ is constructed with the momentum operator and, applied to any state
of the system, only produces a uniform translation in space-time. But applying any other operator $T_{s}$ to the
wave packets that describe the particles, we can move them by an amount which depends on their momentum. Hence,
with a fine-tuning combination of these operators, we can arbitrarily shift the points of interaction in any
scattering process, obtaining the same amplitude because the conserved charges commute with the Hamiltonian.

Hence the $n$-particle $S$-matrix factorizes into a product of $n(n-1)/2$ elastic two-particle $S$-matrices, which
satisfy, in addition to the usual requirements of unitarity and crossing symmetry, the star-triangle (or
Yang-Baxter) equations, and are linked among themselves by the bootstrap equations.

The two-particle $S$-matrix elements are defined by
\begin{equation}
|A_{i}\left(p_{1}\right)A_{j}\left(p_{2}\right)\rangle_{in}=S_{ij}^{kl}\:|A_{k}\left(p_{3}\right)A_{l}\left(p_{4}\right)\rangle_{out}\,,
\end{equation}
where $A_{i}\left(p_{1}\right)$ and $A_{j}\left(p_{2}\right)$ denote the incoming particles (with 2-momenta
$p_{1}^{\mu}$ and $p_{2}^{\mu}$), and $A_{k}\left(p_{3}\right)$ and $A_{l}\left(p_{4}\right)$ the outgoing
states.

\vspace{1cm}

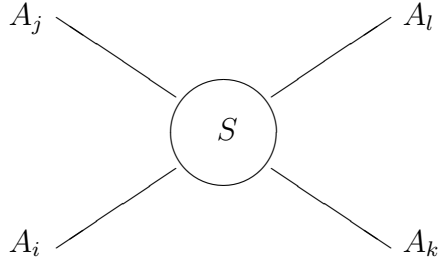
\begin{figure}[h]
\setlength{\unitlength}{0.0125in}
\begin{picture}(40,90)(60,420)
\put(292,467){$S$} \put(295,470){\circle{55}} \put(315,485){\line(3,2){50}} \put(315,455){\line(3,-2){50}}
\put(275,485){\line(-3,2){50}} \put(275,455){\line(-3,-2){50}}
\put(205,420){$A_{i}$}\put(205,515){$A_{j}$}\put(370,515){$A_{l}$}\put(370,420){$A_{k}$}
\end{picture}
\caption{Two-particle $S$-matrix}
 \end{figure}

\vspace{0.5cm}

Lorentz invariance fixes the two body $S$-matrix to be a function of the Mandelstam variables
$s=\left(p_{1}+p_{2}\right)^{2}$, $t=\left(p_{1}-p_{3}\right)^{2}$ and $u=\left(p_{1}-p_{4}\right)^{2}$, which
satisfy the relation $s+t+u=\sum_{i=1}^{4}m_{i}^{2}$. Since in (1+1) dimensions and for elastic scattering only
one of these variables is independent, it is convenient to introduce a parametrization of the momenta in terms of
the so-called rapidity variable $\theta$:
\begin{equation}
p_{i}^{0}=m_{i}\cosh\theta_{i}\,,\qquad\qquad p_{i}^{1}=m_{i}\sinh\theta_{i}\,,
\end{equation}
which corresponds to the following expression for the Mandelstam variable $s$:
\begin{equation}\label{rapidity}
s(\theta)=\left(p_{1}+p_{2}\right)^{2}=m_{i}^{2}+m_{j}^{2}+2m_{i}m_{j}\cosh\theta_{ij}\,,
\end{equation}
with $\theta_{ij}=\theta_{i}-\theta_{j}$. The functions $S_{ij}^{kl}$ will then depend only on the rapidity
difference of the involved particles:
\begin{equation}
|A_{i}\left(\theta_{1}\right)A_{j}\left(\theta_{2}\right)\rangle_{in}=S_{ij}^{kl}\left(\theta_{12}\right)|
A_{k}\left(\theta_{2}\right)A_{l}\left(\theta_{1}\right)\rangle_{out}\,.
\end{equation}

The elastic $S$-matrices are analytic functions in the complex plane of $s$, with square branch cut singularities
at $\left(m_{i}-m_{j}\right)^{2}$ and $\left(m_{i}+m_{j}\right)^{2}$. From (\ref{rapidity}) it follows that the
functions $S_{ij}^{kl}\left(\theta\right)$ are meromorphic in $\theta$, and real at $\textrm{Re}(\theta)=0$. The
physical sheet of the $s$ plane is mapped into the strip $0\leq \textrm{Im}(\theta)\leq\pi$, and the $S$-matrix
poles are mapped into the imaginary axis. The structure in the $\theta$ plane repeats with periodicity $2\pi i$.

The functions $S_{ij}^{kl}\left(\theta\right)$ satisfy the unitarity equations
\begin{equation}\label{unitbulk}
\sum_{n,m}S_{ij}^{nm}\left(\theta\right)S_{nm}^{kl}\left(-\theta\right)=\delta_{i}^{k}\delta_{j}^{l}\,,
\end{equation}
and the crossing symmetry is expressed by
\begin{equation}\label{crossbulk}
S_{ik}^{lj}\left(\theta\right)=S_{ij}^{kl}\left(i\pi-\theta\right)\,,
\end{equation}
because the analytic continuation from the $s$-channel to the $t$-channel corresponds to the substitution
$\theta\rightarrow i\pi-\theta$.

The two-particle amplitudes have also to satisfy the star-triangle or Yang-Baxter equations:
\begin{equation}\label{YBbulk}
S_{i_{1}i_{2}}^{k_{1}k_{2}}\left(\theta_{12}\right)S_{k_{1}k_{3}}^{j_{1}j_{3}}\left(\theta_{13}\right)S_{k_{2}i_{3}}^{j_{2}k_{3}}\left(\theta_{23}\right)=S_{i_{1}i_{3}}^{k_{1}k_{3}}\left(\theta_{13}\right)S_{k_{1}k_{2}}^{j_{1}j_{2}}\left(\theta_{12}\right)S_{i_{2}k_{3}}^{k_{2}j_{3}}\left(\theta_{23}\right)\,,
\end{equation}
with an implicit sum on the intermediate indices.

\vspace{2.5cm}

\begin{figure}[h]
\setlength{\unitlength}{0.0125in}
\begin{picture}(40,90)(60,420)
\put(260,420){\line(-3,4){100}} \put(160,420){\line(3,4){100}} \put(185,420){\line(0,1){130}}
\put(155,410){$i_{1}$}\put(185,410){$i_{2}$}\put(155,560){$i_{3}$}\put(265,560){$j_{1}$}\put(185,560){$j_{2}$}
\put(265,410){$j_{3}$}\put(195,455){$k_{1}$}\put(170,490){$k_{2}$}\put(195,515){$k_{3}$}
\put(290,490){$=$}

\put(430,420){\line(-3,4){100}} \put(330,420){\line(3,4){100}} \put(405,420){\line(0,1){130}}
\put(335,410){$i_{1}$}\put(405,410){$i_{2}$}\put(335,560){$i_{3}$}\put(435,560){$j_{1}$}\put(405,560){$j_{2}$}
\put(435,410){$j_{3}$}\put(385,455){$k_{3}$}\put(410,490){$k_{2}$}\put(385,515){$k_{1}$}
\end{picture}
\caption{Yang-Baxter equations} \label{figYBbulk}
 \end{figure}
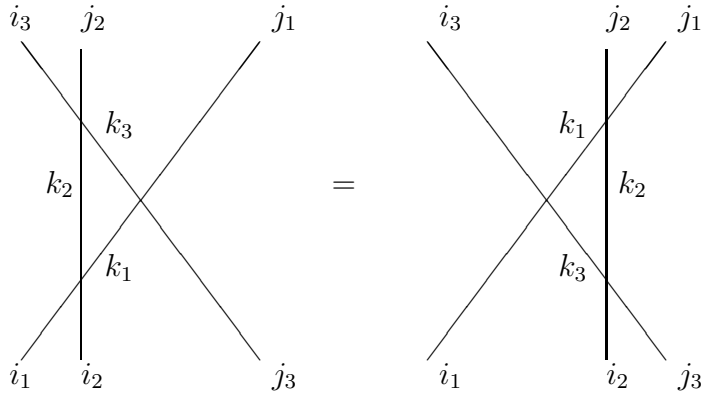

These equations correspond to the commutativity of the processes shown in figure \ref{figYBbulk}, and can be
justified by a particle-displacement argument analogous to the one described for the factorization property.

The system of equations (\ref{unitbulk}), (\ref{crossbulk}) and (\ref{YBbulk}) for the two-particle $S$-matrix is
in many cases sufficient to determine a consistent solution, up to a so-called CDD ambiguity, which consists in
multiplying a given solution by factors that alone satisfy the same equations (we will comment on the role of
these factors at the end of section \ref{sectCDD}).

\subsection{Analytic structure of the $S$-matrix}

The bound states of a theory correspond to singularities of the $S$-matrix. The bootstrap approach consists in
identifying the bound states with some of the particles appearing as asymptotic states, so that the spectrum of
the theory is encoded in the analytic structure of the S-matrix. Stable bound states are usually associated to
simple poles with positive residues which lie on the imaginary axis of the physical strip, but this assumption
may be generalized to the case of poles with negative residues (in massive theories obtained perturbing
non-unitary CFT) and to the case of odd higher order poles.

If an $S$-matrix with initial particle states $A_{i}$ and $A_{j}$ has a simple pole in the $s$-channel at
$\theta=iu_{ij}^{n}$ ($A_{n}$ is the associated intermediate bound state), in the vicinity of this singularity we
have
\begin{equation}
S_{ij}^{kl}(\theta)\sim\frac{iR_{ij}^{n}}{\left(\theta-iu_{ij}^{n}\right)},
\end{equation}
and the residue $R_{ij}^{n}$ is related to the on mass-shell coupling constants of the underlying quantum field
theory by $R_{ij}^{n}=f_{ij}^{n}f_{kl}^{n}$.

\vspace{4cm}

\begin{figure}[h]
\setlength{\unitlength}{0.0125in}
\begin{picture}(40,0)(60,470)

\put(250,510){\line(1,0){90}} \put(205,555){\line(2,-3){22}} \put(205,465){\line(2,3){22}}
\put(385,555){\line(-2,-3){22}} \put(385,465){\line(-2,3){22}}

\put(235,510){\circle{30}} \put(355,510){\circle{30}} \put(230,508){$f_{ij}^{n}$} \put(350,508){$f_{kl}^{n}$}

\put(290,515){$A_{n}$} \put(185,465){$A_{i}$} \put(185,550){$A_{j}$} \put(395,465){$A_{l}$} \put(395,550){$A_{k}$}

\end{picture}
 \caption{First-order pole}
 \end{figure}
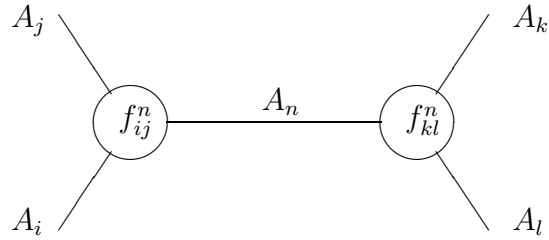

\vspace{0.5cm}

Remembering that the corresponding singularity in the $s$ variable is of the form $\left(s-m_{n}^{2}\right)^{-1}$
and using relation (\ref{rapidity}), we get the following expression for the mass of the bound state:
\begin{equation}
m_{n}^{2}=m_{i}^{2}+m_{j}^{2}+2m_{i}m_{j}\cos u_{ij}^{n}.
\end{equation}
This formula describes a triangle with sides of lengths $m_{i}$, $m_{j}$ and $m_{n}$. The existence of a non-zero
coupling constant $f_{ij}^{n}$ implies a pole singularity in the amplitudes $S_{in}$ and $S_{jn}$ as well, due to
the intermediate bound states $A_{j}$ and $A_{i}$, respectively.  Hence, the location of the three poles is
restricted by the identity
\begin{equation}
u_{ij}^{n}+u_{in}^{j}+u_{jn}^{i}=2\pi.
\end{equation}

\vspace{2.5cm}

\begin{figure}[h]
\setlength{\unitlength}{0.0125in}
\begin{picture}(40,0)(60,470)

\put(305,465){$m_{n}$} \put(275,510){$m_{i}$} \put(380,500){$m_{j}$}

\put(200,460){\line(1,0){200}} \put(200,460){\line(2,1){150}} \put(400,460){\line(-2,3){50}}

\qbezier(240,480)(250,475)(250,460) \put(228,465){$\bar{u}_{in}^{j}$}

\qbezier(365,460)(365,480)(383,485) \put(375,465){$\bar{u}_{jn}^{i}$}

\qbezier(320,520)(340,510)(360,518) \put(340,521){$\bar{u}_{ij}^{n}$}

\end{picture}
\caption{Mass triangle, with $\bar{u}_{ij}^{k}=\pi -u_{ij}^{k}.$}
 \end{figure}
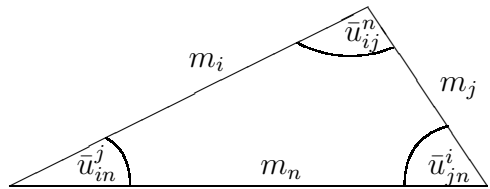

\subsection{Distinguishable particles case} \label{sectCDD}

\vspace{0.5cm}

In the case when the system has all distinguishable particles, the $S$-matrix is diagonal and the Yang-Baxter
equations are trivially satisfied. This can happen when the mass spectrum is non degenerate or when particles of
the same mass can be distinguished by their higher charge eigenvalues.

\vspace{0.5cm}

The unitarity and crossing equations become
\begin{equation}\label{disting}
S_{ab}\left(\theta\right)S_{ab}\left(-\theta\right)=1,\qquad
S_{ab}\left(i\pi-\theta\right)=S_{\bar{a}b}\left(\theta\right).
\end{equation}
These equations imply that the $S_{ab}(\theta)$ are $2\pi i$-periodic functions of $\theta$, and their most
general solution in the space of $2\pi i$-periodic meromorphic functions is of the form
$S_{ab}(\theta)=\prod_{\left\{x\right\}}s_{x}(\theta)$, with
\begin{equation}
s_{x}(\theta)=\frac{\sinh\left[\frac{1}{2}(\theta+i\pi x)\right]}{\sinh\left[\frac{1}{2}(\theta-i\pi x)\right]}\,.
\end{equation}
Due to the $2\pi i$-periodicity, it is possible to choose the parameters $x$ in the range $-1\leq x\leq 1$. These
functions have a simple pole at $\theta=i\pi x$ and a simple zero at $\theta=-i\pi x$ in the range
$-\pi\leq\textrm{Im}\,\theta\leq\pi$, and they satisfy the following properties:
\begin{eqnarray}
&& s_{x}(\theta)s_{x}(-\theta)=s_{x}(\theta)s_{-x}(\theta)=1\,,\\
&& s_{x}(\theta)=s_{x+2}(\theta)\,,\\
&& s_{0}(\theta)=-s_{1}(\theta)=1\,,\\
&& s_{x}(i\pi-\theta)=-s_{1-x}(\theta)\,.
\end{eqnarray}

\vspace{0.5cm}

If all the particles are self-conjugate, the general solution becomes
\begin{equation}
S_{ab}(\theta)=\prod_{\left\{x\right\}}f_{x}(\theta)=\prod_{\left\{x\right\}}s_{x}(\theta)s_{x}(i\pi-\theta)=\prod_{\left\{x\right\}}\frac{\tanh\left[\frac{1}{2}(\theta+i\pi
x)\right]}{\tanh\left[\frac{1}{2}(\theta-i\pi x)\right]}\,.
\end{equation}
The simple poles of these functions are located at the crossing symmetric points $\theta=i\pi x$ and $\theta=i\pi
(1-x)$, and the zeros at $\theta=-i\pi x$ and $\theta=-i\pi(1-x)$. Important properties of $f_{x}(\theta)$ are
\begin{equation}
f_{x}(\theta)=f_{x}(i\pi-\theta)=f_{1-x}(\theta),\qquad f_{x}(-\theta)=f_{-x}(\theta)=1/f_{x}(\theta).
\end{equation}

\vspace{0.5cm}

In this way the kinematic problem is solved. In order to determine the dynamics of the system, i.e. to find the
values of the parameters $\left\{x\right\}$, we have to implement the bootstrap principle, which leads to the
so-called bootstrap equations (\cite{sbrind})
\begin{equation}\label{bootbulk}
S_{i\bar{l}}(\theta)=S_{ij}(\theta+i\bar{u}_{jl}^{k})S_{ik}(\theta-i\bar{u}_{lk}^{j})\,.
\end{equation}

\vspace{2cm}

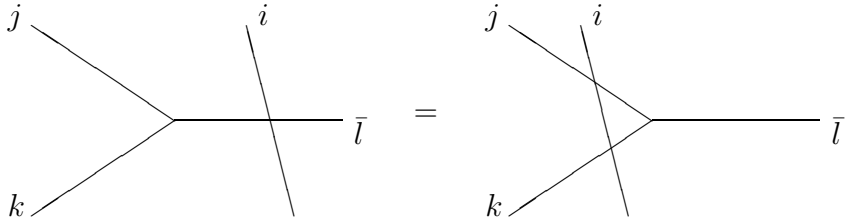
\begin{figure}[h]
\setlength{\unitlength}{0.0125in}
\begin{picture}(40,90)(60,420)
\put(210,490){\line(-3,2){60}} \put(210,490){\line(-3,-2){60}} \put(210,490){\line(1,0){70}}
\put(260,450){\line(-1,4){20}}

\put(245,530){$i$}\put(140,530){$j$}\put(140,450){$k$}\put(285,480){$\bar{l}$}

\put(310,490){$=$}

\put(410,490){\line(-3,2){60}} \put(410,490){\line(-3,-2){60}} \put(410,490){\line(1,0){70}}
\put(400,450){\line(-1,4){20}}

\put(385,530){$i$}\put(340,530){$j$}\put(340,450){$k$}\put(485,480){$\bar{l}$}
\end{picture}
\caption{Bootstrap equations}
 \end{figure}

\vspace{0.5cm}

These equations are justified by a particle displacement argument analogous to the one previously mentioned. To
solve them, it is often necessary to find an ansatz for the two-particles bootstrap \lq\lq fusions\rq\rq\, of some
minimal subset of particles (called \lq\lq fundamental\rq\rq ), which should be consistent with the conservation
laws and the symmetries of the model. Usually the \lq\lq fundamental\rq\rq \, particle is the one with the
lightest mass, and its $S$-matrix will have a minimal pole structure, whereas the $S$-matrices of the other
particles will present higher order poles, supporting the interpretation of the mass of a particle as a dynamical
parameter related to the complexity of its interactions.

The CDD ambiguity consists in multiplying a given solution by factors which satisfy equations (\ref{disting}) and
(\ref{bootbulk}), and don't have poles in the physical strip $0\leq \textrm{Im}\,\theta\leq\pi$, hence they don't
change the spectrum of the theory. These factors give the possibility to link $S$-matrices related to different
models which share the same symmetry and the same IR properties, but have different UV behaviours. However, some
aspects of the role of CDD factors have still to be clarified. For example, an important consequence of their
insertion can be a change from a fermionic to a bosonic-type $S$-matrix, where these properties refer to the sign
of the two-particle amplitudes $S_{aa}$ evaluated at zero rapidity difference (an $S$-matrix is said to be of
fermionic-type if $S_{aa}(0)=-1$ and of bosonic-type if $S_{aa}(0)=1$). At the moment, only fermionic-type
scattering theories are clearly understood. The bosonic situation was firstly faced in \cite{musssim}, where it
was shown that the corresponding UV behaviour is characterized by the divergence of the central charge, which can
be explained by the presence of a certain set of irrelevant perturbing operators.

\subsection{Bootstrap consistency equations}

From the bootstrap principle, knowing the $S$-matrix of a theory it is possible to derive a system of consistency
equations for the eigenvalues of the conserved charges $P_{s}$, and to find a restriction on the possible values
of their spin $s$ (\cite{sbrind}).

Choosing the asymptotic states $\mid A_{a}(\theta)\rangle$ as eigenstates of $P_{s}$,
\begin{equation}
P_{s}\mid A_{a}(\theta)\rangle =\omega_{s}^{a}(\theta) \mid A_{a}(\theta)\rangle\,,
\end{equation}
Lorentz invariance fixes the functional form of the corresponding eigenvalues to be
\begin{equation}
\omega_{s}^{a}(\theta)=\chi_{s}^{a}e^{s\theta}\,,
\end{equation}
where $\chi_{s}^{a}$ are constants, and $\chi_{1}^{a}=m_{a}$. Locality implies
\begin{equation}
P_{s}\mid A_{a_{1}}(\theta_{1})\cdots A_{a_{k}}(\theta_{k})\rangle
=\left(\omega_{s}^{a_{1}}(\theta_{1})+\ldots+\omega_{s}^{a_{k}}(\theta_{k})\right) \mid
A_{a_{1}}(\theta_{1})\cdots A_{a_{k}}(\theta_{k})\rangle.
\end{equation}

If $u_{ab}^{\bar{c}}$ is the imaginary value of the rapidity at which particles $a$ and $b$ \lq\lq fuse\rq\rq\,
together and create the bound state $\bar{c}$, we have:
\begin{equation}
\lim_{\varepsilon\rightarrow 0}\,\varepsilon\mid
A_{a}\left(\theta+i\bar{u}_{ac}^{b}-\frac{1}{2}\varepsilon\right)A_{b}\left(\theta-i\bar{u}_{bc}^{a}+\frac{1}{2}\varepsilon\right)\rangle
=\mid A_{\bar{c}}\left(\theta\right)\rangle .
\end{equation}
Applying $P_{s}$ to both sides of this equation we get the following infinite system of linear equations for the
$\chi_{s}^{a}$s:
\begin{equation}\label{consisteq}
\chi_{s}^{a}\exp\left(is\bar{u}_{ac}^{b}\right)+\chi_{s}^{b}\exp\left(-is\bar{u}_{bc}^{a}\right)=\chi_{s}^{\bar{c}}.
\end{equation}
Non-trivial solutions of these equations are obtained for special sets of resonance angles $u_{ab}^{c}$ and for
specific values of the spin $s$. For example, if $a=b$ and $\chi_{s}^{a}\neq 0$, they reduce to
\begin{equation}
2\cos\left(s\bar{u}_{ac}^{a}\right)=\chi_{s}^{\bar{c}}/\chi_{s}^{a}\,.
\end{equation}
In the case $a=b=c$, the above equation has an unique solution
\begin{equation} \label{phi3}
\bar{u}_{aa}^{a}=\frac{1}{3}\pi, \qquad s=1,5 \quad (\textrm{mod}\;6).
\end{equation}
The corresponding $S$-matrix presents the so-called $\Phi^{3}$ property, i.e. the particle $A_{a}$ appears as
bound state of itself.

We will now consider the following examples, involving two different particles:
\begin{eqnarray}
(i) && \qquad A_{a}\times A_{a}\rightarrow A_{b}\,,\qquad\qquad\:\: A_{b}\times A_{b}\rightarrow A_{a}\,,\\
(ii)&& \qquad A_{a}\times A_{a}\rightarrow A_{a}+A_{b}\,,\qquad A_{b}\times A_{b}\rightarrow A_{a}\,.
\end{eqnarray}
The consistency conditions for the process (i) are
\begin{equation}
2\chi_{s}^{a}\cos\left(s\bar{u}_{ab}^{a}\right)=\chi_{s}^{b}\,,\qquad
2\chi_{s}^{b}\cos\left(s\bar{u}_{ab}^{b}\right)=\chi_{s}^{a}\,,
\end{equation}
and for $\chi_{s}^{a,b}\neq 0$ they reduce to
\begin{equation}
\cos\left(s\bar{u}_{ab}^{a}\right)\cos\left(s\bar{u}_{ab}^{b}\right)=\frac{1}{4}\,.
\end{equation}
This equation admits two solutions:
\begin{eqnarray}
&& \bar{u}_{ab}^{a}=\frac{1}{12}\pi\,, \quad \bar{u}_{ab}^{b}=\frac{5}{12}\pi\,, \quad s=1,4,5,7,8,11 \:(\textrm{mod}\:12)\,,\\
\label{isingfusang}&& \bar{u}_{ab}^{a}=\frac{1}{5}\pi\,, \quad\;\: \bar{u}_{ab}^{b}=\frac{2}{5}\pi\,, \quad \;\:
s=1,3,7,9 \:(\textrm{mod}\:10)\,.
\end{eqnarray}
For self-conjugate particles the even spins do not exist, and in the first solution we have $s=1,5,7,11
\:(\textrm{mod}\:12)$.

For the process (ii), we have to take for the spin $s$ the common solution of the process (i) and $A_{a}\times
A_{a}\rightarrow A_{a}$. For instance, in the case of the second solution of (i) we have
\begin{equation}\label{isingspin}
s=1,7,11,13,17,19,23,29 \;(\textrm{mod}\,30)\,.
\end{equation}

\subsection{Higher-order poles}

The multiple pole structure of the $S$-matrices is an unavoidable consequence of the iterative application of the
bootstrap equations. The simple poles naturally correspond to the bound states which appear in the intermediate
channels, while higher-order poles are a peculiar feature of (1+1)-dimensional systems (in four dimensions, these
anomalous threshold singularities would be branch cuts). As we will see, odd-order poles provide a new mechanism
to produce bound states, while even-order poles describe purely multiple scattering processes without the
creation of intermediate bound states. Consequently, all odd-order poles must be taken into account to iterate
consistently the bootstrap procedure by means of eq.(\ref{bootbulk}).

A complete analysis of higher order pole singularities has been performed in \cite{chrmuss1}, \cite{dur1},
\cite{dur2}; here, we will just explain the basic ideas in the case of double and triple poles.

Let us start with second-order poles, firstly analyzed by Coleman and Thun in the sine-Gordon model
(\cite{colthun}). It can be shown that an order $p$ singularity of the $S$-matrix, which has the form
\begin{equation}
S_{AB}(\theta)\sim\frac{g^{2p}R_{p}}{(\theta-\theta_{0})^{p}}\,,
\end{equation}
originates from a Feynman diagram with $P$ propagators and $L$ loops, with the condition $p=P-2L$. Hence, a
second order pole can be described by the following on-shell diagram:

\vspace{4cm}

\begin{figure}[h]
\setlength{\unitlength}{0.01in}
\begin{picture}(40,0)(60,470)
\put(360,540){\line(-2,1){60}} \put(360,540){\line(-2,-1){60}} \put(300,510){\line(0,1){60}}
\put(300,570){\line(-2,3){30}} \put(300,510){\line(-2,-3){30}} \put(290,540){$c$} \put(255,465){$A$}
\put(255,610){$B$} \put(330,560){$b$} \put(330,515){$a$} \put(360,540){\line(2,1){60}}
\put(360,540){\line(2,-1){60}} \put(420,510){\line(0,1){60}} \put(420,570){\line(2,3){30}}
\put(420,510){\line(2,-3){30}} \put(430,540){$c$} \put(455,465){$B$} \put(455,610){$A$} \put(390,560){$a$}
\put(385,515){$b$}
\end{picture}
 \caption{Second-order pole}
 \end{figure}
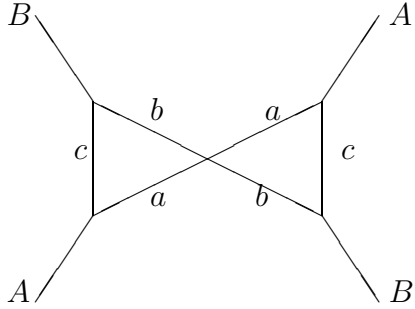

The diagram can be drawn if there are particles $a$, $b$ and $c$ with the opportune values of the resonance
angles $u_{ij}^{k}$, which are exactly the angles in the figure. The double pole of $S_{AB}$ is located at
\begin{equation}
\theta_{AB}=u_{Bc}^{b}+u_{Ac}^{a}-\pi \,,
\end{equation}
and $S_{ab}$ has to be regular at
\begin{equation}
\theta_{ab}=\pi-u_{ac}^{A}-u_{bc}^{B} \,.
\end{equation}
From this discussion it follows that the scattering amplitude $S_{11}$ of the lightest particle cannot have higher
order poles, because the resonance angles of two heavier particles with it are greater than $\pi/2$ and therefore
it is impossible to draw this diagram with $A=B=1$.

\vspace{0.5cm}

If we now assume that $\theta_{ab}$ is a simple pole for $S_{ab}$, the order of the diagram becomes three, and we
have an intermediate propagator:

\vspace{4cm}

\begin{figure}[h]
\setlength{\unitlength}{0.01in}
\begin{picture}(40,0)(60,470)
\put(330,540){\line(-2,1){60}} \put(330,540){\line(-2,-1){60}}

\put(270,510){\line(0,1){60}} \put(270,570){\line(-2,3){30}} \put(270,510){\line(-2,-3){30}}

\put(260,540){$c$} \put(225,465){$A$} \put(225,610){$B$} \put(300,560){$b$} \put(300,515){$a$}

\put(330,540){\line(1,0){60}} \put(360,545){$C$}

\put(390,540){\line(2,1){60}} \put(390,540){\line(2,-1){60}} \put(450,510){\line(0,1){60}}
\put(450,570){\line(2,3){30}} \put(450,510){\line(2,-3){30}}

\put(460,540){$c$} \put(485,465){$B$} \put(485,610){$A$} \put(420,560){$a$} \put(415,515){$b$}
\end{picture}
 \caption{Third-order pole}
 \end{figure}
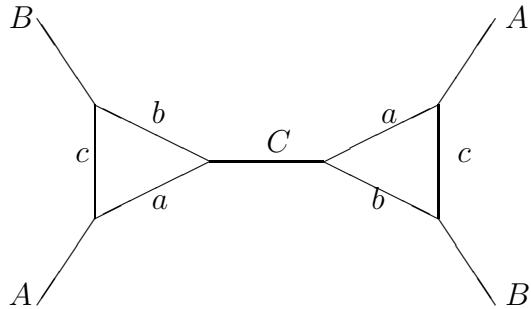

These are the basic mechanisms involved in the explanation of higher order singularities. In fact, fourth-order
poles arises when $S_{ab}$ has a double pole at $\theta_{ab}$, fifth-order ones when the pole is triple, and so
on.

\vspace{4cm}

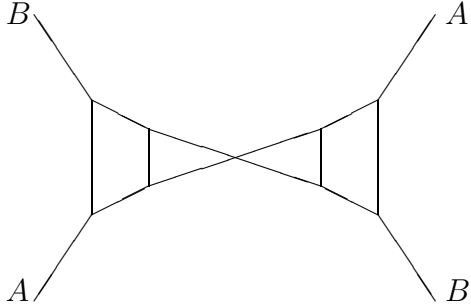
\begin{figure}[h]
\setlength{\unitlength}{0.01in}
\begin{picture}(40,0)(60,470)
\put(280,570){\line(2,-1){30}} \put(280,510){\line(2,1){30}} \put(310,525){\line(0,1){30}}
\put(280,510){\line(0,1){60}}

\put(280,570){\line(-2,3){30}} \put(280,510){\line(-2,-3){30}}

\put(235,465){$A$} \put(235,610){$B$}

\put(310,525){\line(3,1){90}} \put(310,555){\line(3,-1){90}}

\put(430,570){\line(-2,-1){30}} \put(430,510){\line(-2,1){30}} \put(430,510){\line(0,1){60}}
\put(400,525){\line(0,1){30}}

\put(430,570){\line(2,3){30}} \put(430,510){\line(2,-3){30}}

\put(465,465){$B$} \put(465,610){$A$}
\end{picture}
 \caption{Fourth-order pole}
 \end{figure}

In the light of these considerations, it is possible to classify all massive integrable systems taking the
bootstrap equations as basic entities, restricted only by the consistency equations of the conserved spins and by
a multi-scattering interpretation of higher order poles. Any odd-order pole is assumed to be in correspondence
with a bound state, and the whole set of bound states should form a complete set of states which describes the
asymptotic particles as well.

\vspace{1cm}

\section{Elastic $S$-matrices of perturbed minimal models}

Knowing the first spins of the conserved currents away from criticality, on the basis of symmetry and analyticity
arguments alone it is possible to fix the minimal form of the $S$-matrix of a system, i.e. the solution of
equations (\ref{disting}) and (\ref{bootbulk}) with the minimum number of poles and zeroes in the physical strip.

\subsection{The Yang-Lee model perturbed by $\phi_{1,2}$}

The $S$-matrix of this model has been found by Cardy and Mussardo in \cite{SYL}. Using the counting argument it is
possible to establish the existence of conserved currents with spins
\begin{equation}
s=1,5,7,11,13,17,19,23\,.
\end{equation}
It is then natural to conjecture property (\ref{phi3}), which allows the existence of a particle $A$ that appears
as a bound state of itself, with a corresponding $s$-channel pole at $\theta=i\frac{2}{3}\pi$ and a related
$t$-channel pole at $\theta=i\frac{1}{3}\pi$. The minimal solution of the bootstrap equation
\begin{equation}
S_{AA}(\theta)=S_{AA}\left(\theta-i\frac{1}{3}\pi\right)S_{AA}\left(\theta+i\frac{1}{3}\pi\right)
\end{equation}
is given by $S_{AA}(\theta)=f_{2/3}(\theta)$, hence $A$ is the only particle present in the theory. It is
interesting to note that the residue at the pole $\theta=i\frac{2}{3}\pi$ has negative sign, meaning that the
$S$-matrix is unitary in the sense of eq.(\ref{unitbulk}), i.e. it preserves the probability, but it is not
one-particle unitary.

\subsection{The magnetic perturbation of the Ising model}

The $S$-matrix of the Ising model perturbed by the magnetic operator $\phi_{1,2}=\sigma$ has been found by Fateev
and Zamolodchikov in \cite{fatzam}. Applying the counting argument, it is possible to see that the lowest
conserved spins are
\begin{equation}
s=1,7,11,13,17,19\,.
\end{equation}
The absence of multiples of 3 can be explained by the existence of a fundamental particle $A_{1}$ with the
$\Phi^{3}$ property (see (\ref{phi3})), while the absence of multiples of 5 by the existence of a second particle
$A_{2}$ that gives rise to the bootstrap fusions (see (\ref{isingspin})):
\begin{equation}
A_{1}\times A_{1}\rightarrow A_{1}+A_{2}\,,\qquad A_{2}\times A_{2}\rightarrow A_{1}\,.
\end{equation}
Correspondingly, the scattering amplitude $S_{11}$ has poles with positive residue at the resonance angles
$u_{11}^{1}=i\frac{2}{3}\pi$ and $u_{11}^{2}=i\frac{2}{5}\pi$ (see (\ref{isingfusang})). However, since it is
impossible to satisfy the bootstrap equation
\begin{equation}
S_{11}(\theta)=S_{11}\left(\theta-i\frac{1}{3}\pi\right)S_{11}\left(\theta+i\frac{1}{3}\pi\right)
\end{equation}
with only these poles, Zamolodchikov (\cite{sbrind}) conjectured the following minimal $S$-matrix for the
fundamental particle:
\begin{equation}
S_{11}(\theta)=f_{1/3}(\theta)f_{2/5}(\theta)f_{1/15}(\theta)\,.
\end{equation}
The bootstrap generated by this amplitude closes within eight particles, whose mass ratios can be determined by
equations (\ref{consisteq}) with $s=1$.

As we will see in section \ref{Toda}, the values of the conserved spins are related to an underlying Lie
algebraic symmetry of the theory; in fact, they are exactly the exponents of the Lie algebra $E_{8}$, and it is
natural to conjecture that their infinite set is given by
\begin{equation}
s=1,7,11,13,17,19,23,29\quad(\textrm{mod}\, 30)\,,
\end{equation}
where $30$ is the dual Coxeter number of $E_{8}$ (for the definitions of these quantities, see section
\ref{prelLiealg}). The $S$-matrix of this model coincides with the minimal part of the $S$-matrix of the $E_{8}$
affine Toda field theory, whose elements are listed in appendix \ref{Smatrlist}. Also the number of particles and
the mass ratios of the two theories are the same.

\subsection{The thermal perturbation of the tricritical Ising model}

The scaling region of the tricritical Ising model has been studied in \cite{TIMscaling}. The $S$-matrix of this
model perturbed by the energy operator $\phi_{1,2}=\varepsilon$ has been found in \cite{fatzam} and
\cite{chrmuss1}. In this case, applying the counting argument it is possible to see that the lowest conserved
spins are
\begin{equation}
s=1,5,7,9,11,13,17\,.
\end{equation}
The presence of the spin $s=9$ forbids the $\Phi^{3}$ property for the fundamental particle, and arguments
analogous to the ones described for the other systems lead to a bootstrap closing on seven particles.

The conserved spins coincide with the exponents of the Lie algebra $E_{7}$, whose dual Coxeter number is $18$. The
natural conjecture is then
\begin{equation}
s=1,5,7,9,11,13,17 \quad (\textrm{mod}\, 18)\,,
\end{equation}
and the $S$-matrix is equal to the minimal part of the one of the $E_{7}$ affine Toda field theory, presented in
appendix \ref{Smatrlist}.

\subsection{The thermal perturbation of the tricritical three-state Potts model}

The $S$-matrix of the tricritical three-state Potts model perturbed by the energy operator
$\phi_{1,2}=\varepsilon$, with conformal weight $(h,\bar{h})=\left(\frac{1}{7},\frac{1}{7}\right)$, has been
found by Sotkov and Zhu in \cite{E6}. The lowest conserved spins coincide with the exponents of the Lie algebra
$E_{6}$, so that it is natural to conjecture that their infinite set will be given by
\begin{equation}
s=1,4,5,7,8,11 \quad (\textrm{mod}\, 12)\,.
\end{equation}
The complete $S$-matrix is listed in appendix \ref{Smatrlist}, as the minimal part of the one of the $E_{6}$
affine Toda field theory. The bootstrap closes on six particles, two of which are neutral, while the other four
are organized in two couples $(a,\bar{a})$ of conjugate particles with degenerate mass.

\vspace{1cm}

\section{Affine Toda field theories (ATFT)} \label{Toda}

Affine Toda field theories (ATFT) are integrable massive field theories defined by a Lagrangian built up from some
characteristic quantities of an affine Lie algebra $\hat{g}$. The corresponding scattering theory contains a
number of particles equal to the rank of $\hat{g}$, and the S-matrix, depending on the choice of $\hat{g}$,
coincides with the one of a specific perturbed minimal model \lq\lq dressed\rq\rq \, with some CDD-factors which
depend on the coupling constant present in the Lagrangian. The main results on this subject have been found in
\cite{arfatzam}, \cite{chrmuss1}, \cite{chrmuss2}, \cite{dur1} and \cite{dur2}.

\subsection{Preliminaries on Lie algebras}\label{prelLiealg}

For a general reference on this subject, see \cite{wyb}.

\subsubsection{Simple Lie algebras}

A Lie algebra $g$ is a vector space equipped with an antisymmetric binary operation $[\;,\,]$, called a
commutator, mapping $g\times g$ into $g$, and constrained to satisfy the Jacobi identity
\begin{equation}
[X,[Y,Z]]+[Z,[X,Y]]+[Y,[Z,X]]=0 \qquad \textrm{for} \:X,Y,Z\in g\,.
\end{equation}
A Lie algebra can be specified by a set of generators $\left\{J^{a}\right\}$, whose number is the dimension of the
algebra, and their commutation relations
\begin{equation}
[J^{a},J^{b}]=\sum_{c}if^{ab}_{c}J^{c}\,,
\end{equation}
where $f^{ab}_{c}$ are the structure constants. A Lie algebra is simple if it contains no proper subset of
generators $\left\{L^{a}\right\}$ such that $[L^{a},J^{b}]\in\left\{L^{a}\right\}$ for any $J^{b}$.

A representation of an algebra is the association of every element of $g$ to a linear operator acting on some
vector space $V$, which respects the commutation relations of the algebra. The dimension of $V$ is the dimension
of the representation. Relative to a given basis, each element of $g$ can thus be represented by a square matrix,
and the commutator correspond to the usual matrix commutation. A representation is said to be irreducible if the
matrices representing the elements of $g$ cannot all be brought in a block-diagonal form by a change of basis.
The adjoint representation is the one in which the Lie algebra itself serves as the vector space on which the
generators act. A matrix realization of this representation in the basis $\left\{J^{a}\right\}$ is given by
$\left(J^{a}\right)_{bc}=-if_{abc}$.

In order to construct the so-called Cartan-Weyl basis, we first identify the maximal set of commuting Hermitian
generators $H^{i}$, $i=1,..,r$ ($r$ is the rank of the algebra):
\begin{equation}
[H^{i},H^{j}]=0\,.
\end{equation}
This set of generators forms the Cartan subalgebra. The remaining generators are chosen to be particular
combinations of the $J^{a}$'s that satisfy the eigenvalue equation:
\begin{equation}\label{root}
[H^{i},E^{\alpha}]=\alpha^{i}E^{\alpha}\,,
\end{equation}
where $\alpha=\left(\alpha_{1},...,\alpha_{r}\right)$ is called a root, and $\Delta$ denotes the set of all roots.

In order to define a scalar product for a Lie algebra, one defines the Killing form
\begin{equation}
K(X,Y)=\frac{1}{2h}\textrm{Tr}(\textrm{ad}X\,\textrm{ad}Y)\,,
\end{equation}
where $h$ is a constant called the dual Coxeter number of the algebra. The standard basis $\left\{J^{a}\right\}$
and the Cartan subalgebra are orthonormal with respect to $K$:
\begin{equation}
K\left(J^{a},J^{b}\right)=\delta^{ab}\,, \qquad\qquad K\left(H^{i},H^{j}\right)=\delta^{ij}\,.
\end{equation}

The Killing form defines a scalar product in the root space:
\begin{equation}
\left(\alpha,\beta\right)=K\left(H^{\beta},H^{\alpha}\right)\,,
\end{equation}
with $H^{\alpha}=\alpha\cdot H=\sum_{i}\alpha^{i}H^{i}$; we will use the notation
$\left(\alpha,\alpha\right)=|\alpha|^{2}$.

It can be shown that in the set of roots of a simple Lie algebra, at most two lengths (long and short) are
possible, and the standard normalization convention is to set the square length of the long roots equal to two.
When all the roots have the same length, the algebra is said to be simply laced.

In the adjoint representation, the eigenvalues of the Cartan generators are the roots. For an arbitrary
representation, a basis $\left\{|\lambda\rangle\right\}$ can always be found such that
\begin{equation}
H^{i}|\lambda\rangle=\lambda^{i}|\lambda\rangle,
\end{equation}
and the vector $\lambda=\left(\lambda^{1},...,\lambda^{r}\right)$ is called a weight. Scalar product between
weights is fixed by the Killing form. Eq.(\ref{root}) shows that $E^{\alpha}$ changes the eigenvalue of a state
by $\alpha$:
\begin{equation}
H^{i}E^{\alpha}|\lambda\rangle=\left(\lambda^{i}+\alpha^{i}\right)E^{\alpha}|\lambda\rangle
\end{equation}
so that $E^{\alpha}|\lambda\rangle$, if nonzero, must be proportional to a state $|\lambda+\alpha\rangle$, and
$E^{\alpha}$ is called ladder operator. The triplet of generators $E^{\alpha}$, $E^{-\alpha}$ and $\alpha\cdot
H/|\alpha|^{2}$ forms a $su(2)$ subalgebra.

Fixing a basis $\left\{\beta_{1},...,\beta_{r}\right\}$ in the $r$-dimensional space of roots, any root can be
expanded as
\begin{equation}
\alpha=\sum_{i=1}^{r}k_{i}\beta_{i}.
\end{equation}
A root $\alpha$ is said to be positive $\left(\alpha\in\Delta_{+}\right)$ if the first nonzero number in the
sequence $\left(k_{1},...,k_{r}\right)$ is positive. The set of negative roots $\Delta_{-}$ is defined in the
analogous way. A root $\alpha_{i}$ is called simple if it cannot be written as a sum of two positive roots. There
are necessarily $r$ simple roots, and their set $\left\{\alpha_{1},...,\alpha_{r}\right\}$ provides the most
convenient basis for the space of roots.

The scalar product of simple roots define the Cartan matrix:
\begin{equation}
A_{ij}=\frac{2\left(\alpha_{i},\alpha_{j}\right)}{|\alpha_{j}|^{2}}\,.
\end{equation}
The entries of this matrix, which is not in general symmetric, are necessarily integers. The diagonal elements are
all equal to 2, while the off-diagonal ones can only be $0,-1,-2$ or $-3$.

The Cartan matrix can be written in the compact form
\begin{equation}
A_{ij}=\left(\alpha_{i},\alpha_{j}^{\vee}\right),
\end{equation}
where $\alpha_{i}^{\vee}=2\alpha_{i}/|\alpha_{i}|^{2}$ is the coroot associated with the root $\alpha_{i}$.

The highest root $\theta$ is the unique root for which, in the expansion $\sum k_{i}\alpha_{i}$, the sum $\sum
k_{i}$ is maximized. The standard normalization is fixed by setting $|\theta|^{2}=2$. The coefficients of the
decomposition of $\theta$ in the bases $\left\{\alpha_{i}\right\}$ and $\left\{\alpha_{i}^{\vee}\right\}$ are
called marks $\left(m_{i}\in N\right)$ and comarks $\left(n_{i}\in N\right)$:
\begin{equation}
\theta=\sum_{i=1}^{r}m_{i}\alpha_{i}=\sum_{i=1}^{r}n_{i}\alpha_{i}^{\vee}\,,
\end{equation}
and are related by
\begin{equation}
m_{i}=n_{i}\frac{2}{|\alpha_{i}|^{2}}\,.
\end{equation}
The dual Coxeter number is defined as
\begin{equation}
h=\sum_{i=1}^{r}n_{i}+1\,.
\end{equation}

All the information on the structure of the algebra $g$ is contained in the Cartan matrix and can be encoded in a
simple planar diagram, the so-called Dynkin diagram, constructed associating a node to every simple root
$\alpha_{i}$ and joining the nodes $i$ and $j$ with $A_{ij}A_{ji}$ lines. The classification of simple Lie
algebras reduces then to a classification of Dynkin diagrams.

The convenient basis for weights is the one dual to the simple coroot basis, i.e. the set
$\left\{\omega_{i}\right\}$ such that
\begin{equation}
\left(\omega_{i},\alpha_{j}^{\vee}\right)=\delta_{ij}\,.
\end{equation}
The $\omega_{i}$ are called fundamental weights. The expansion coefficient $\lambda_{i}$ of a weight $\lambda$ in
the fundamental weight basis are called Dynkin labels:
\begin{equation}
\lambda=\sum_{i=1}^{r}\lambda_{i}\omega_{i}\qquad\Longleftrightarrow\qquad\lambda_{i}=\left(\lambda,\alpha_{j}^{\vee}\right)\,.
\end{equation}
The Dynkin labels of weights in finite-dimensional irreducible representations are always integers. The elements
of the Cartan matrix are the Dynkin labels of the simple roots:
\begin{equation}
\alpha_{i}=\sum_{j=1}^{r}A_{ij}\omega_{j}\,.
\end{equation}

The Weyl vector is defined as the weight for which all Dynkin labels are unity:
\begin{equation}
\rho=\sum_{i=1}^{r}\omega_{i}=(1,1,...,1)\,,
\end{equation}
and it can be shown that
\begin{equation}
\rho=\frac{1}{2}\sum_{\alpha\in\Delta_{+}}\alpha\,.
\end{equation}

Any finite-dimensional representation has a unique highest-weight state $|\lambda\rangle$, for which the sum of
the coefficient expansion in the basis of simple roots is maximal (the highest weight of the adjoint
representation is $\theta$). It follows that, for any $\alpha>0$, $\lambda+\alpha$ cannot be a weight in the
representation, hence
\begin{equation}
E^{\alpha}|\lambda\rangle=0\qquad\forall\alpha>0\,.
\end{equation}
Representations are often specified by their highest weights.

Starting from the highest weight $|\lambda\rangle$, all the states in the representation space (or irreducible
module) $L_{\lambda}$ can be obtained by the action of the lowering operators of $g$ as
\begin{equation}
E^{-\beta}E^{-\gamma}\cdots E^{-\eta}|\lambda\rangle\qquad \textrm{for} \; \beta,\gamma,...,\eta\in\Delta_{+}\,.
\end{equation}
The set of eigenvalues of all the states in $L_{\lambda}$ is the weight system $\Omega_{\lambda}$. Any weight
$\lambda'\in\Omega_{\lambda}$ is such that $\lambda-\lambda'\in\Delta_{+}$. Hence $\lambda'$ is necessarily of
the form $\lambda-\sum k_{i}\alpha_{i}$, with $k_{i}\in Z_{+}$, and  $\sum k_{i}$ is called the level of the
weight $\lambda'$ in the representation $\lambda$.

Casimir operators are defined by their commutativity with all the generators of the algebra $g$. If $\{J^{a}\}$
is an orthonormal basis, a quadratic Casimir operator is
\begin{equation}
{\cal Q}=\sum_{a}J^{a}J^{a}\,.
\end{equation}
In the case of $su(2)$ this is the only possible Casimir operator, but for higher-rank algebras there are Casimir
operators of higher degree. Their degrees minus one are called the exponents of the algebra.

For simply laced Lie algebras, the exponents can also be defined through the adjacency matrix
$G_{ij}=2\delta_{ij}-A_{ij}$, whose eigenvalues are of the form $2\cos(k_{i}\pi/h)$, where the $k_{i}$'s are the
exponents.

\vspace{0.5cm}

\subsubsection{Affine Lie algebras}

To every (finite) Lie algebra $g$, we associate an affine extension $\hat{g}$ by adding to the Dynkin diagram of
$g$ an extra node, related to the highest root $\theta$. As a result of this insertion, highest-weight
representations of the algebra become infinite dimensional, and are organized in terms of a new parameter, called
the level. The level of a weight is the sum of all its Dynkin labels, that are now $r+1$, each multiplied by the
corresponding comark.

We start considering the so-called loop algebra $\tilde{g}$,
\begin{equation}
\tilde{g}=g\otimes C\left[t,t^{-1}\right]\,,
\end{equation}
i.e. the generalization of $g$ in which the elements of the algebra are also Laurent polynomials in some variable
$t$, with generators $J_{n}^{a}\equiv J^{a}\otimes t^{n}$ such that
\begin{equation}\label{currentalg}
[J_{n}^{a},J_{m}^{b}]=\sum_{c}if^{ab}_{c}J_{n+m}^{c}+\hat{k}n\delta_{ab}\delta_{n+m,0}\,,
\end{equation}
where $\hat{k}$ is a central element which commutes with all $J^{a}$'s:
\begin{equation}
[J_{n}^{a},\hat{k}]=0\,.
\end{equation}

The set of generators $\left\{H_{0}^{1},...,H_{0}^{r},\hat{k}\right\}$ is abelian, but it is not a maximal
abelian subalgebra; hence, it must be augmented by the addition of a new grading operator $L_{0}$, defined as
\begin{equation}
L_{0}=-t\frac{d}{dt}\,,
\end{equation}
and such that
\begin{equation}
[L_{0},J_{n}^{a}]=-nJ_{n}^{a}\,.
\end{equation}
The maximal Cartan subalgebra is generated by $\left\{H_{0}^{1},...,H_{0}^{r},\hat{k},L_{0}\right\}$, and the
other generators $E_{n}^{\alpha}$ (for any $n$) and $H_{n}^{i}$ (for $n\neq 0$) play the role of ladder operators.
The resulting affine Lie algebra is
\begin{equation}
\hat{g}=\tilde{g}\oplus C\hat{k}\oplus C L_{0}\,,
\end{equation}
which is infinite dimensional ($g$ will be referred as the corresponding finite algebra).

Extending the definition of the Killing form from $g$ to $\hat{g}$ we get
\begin{equation}
K\left(J_{n}^{a},J_{m}^{b}\right)=\delta^{ab}\delta_{n+m,0}\,,
\end{equation}
\begin{eqnarray}
K\left(J_{n}^{a},\hat{k}\right)=0\qquad &\textrm{and}& \qquad K\left(\hat{k},\hat{k}\right)=0\,,\\
K\left(J_{n}^{a},L_{0}\right)=0\qquad &\textrm{and}& \qquad K\left(L_{0},\hat{k}\right)=-1\,,
\end{eqnarray}
and by convention we choose
\begin{equation}
K\left(L_{0},L_{0}\right)=0\,.
\end{equation}
Let the components of the vector $\hat{\lambda}$ be the eigenvalues of a state that is a simultaneous eigenvector
of all the generators of the Cartan subalgebra:
\begin{equation}
\hat{\lambda}=\left(\hat{\lambda}(H_{0}^{1}),\hat{\lambda}(H_{0}^{2}),...,\hat{\lambda}(H_{0}^{r});\hat{\lambda}(\hat{k});\hat{\lambda}(-L_{0})\right)=(\lambda;k_{\lambda};n_{\lambda})
\end{equation}
($\hat{\lambda}$ is called an affine weight). The scalar product induced by the extended Killing form is
\begin{equation}
(\hat{\lambda},\hat{\mu})=(\lambda,\mu)+k_{\lambda}n_{\mu}+k_{\mu}n_{\lambda}.
\end{equation}
Weights in the adjoint representation are called roots; since $\hat{k}$ commutes with all the generators of
$\hat{g}$, its eigenvalue on the states of the adjoint representation is equal to zero. Hence, affine roots are
of the form
\begin{equation}
\hat{\beta}=(\beta;0;n)\,,
\end{equation}
and their scalar product is exactly the same as in the finite case
\begin{equation}
(\hat{\beta},\hat{\alpha})=(\beta,\alpha)\,.
\end{equation}
If we define
\begin{equation}
\alpha\equiv(\alpha;0;0)\qquad \textrm{and} \qquad \delta=(0;0;1)\,,
\end{equation}
the roots can be reexpressed as $\hat{\alpha}=\alpha+n\delta$.

A basis of simple roots for the affine Lie algebra must contain $r+1$ elements, $r$ of which are necessarily the
finite simple roots $\alpha_{i}$. The extra simple root is
\begin{equation}
\alpha_{0}\equiv (-\theta;0;1)=-\theta+\delta\,.
\end{equation}

The extended Cartan matrix is defined as
\begin{equation}
\widehat{A}_{ij}=\left(\alpha_{i},\alpha_{j}^{\vee}\right)\qquad 0\leq i,j\leq r\,,
\end{equation}
where affine coroots are given by
\begin{equation}
\hat{\alpha}^{\vee}=\left(\alpha^{\vee};0;\frac{2}{|\alpha|^{2}}n\right)\,.
\end{equation}
Compared to the finite Cartan matrix, $\widehat{A}_{ij}$ contains an extra row and column, and these additional
entries are easily calculated as
\begin{equation}
\left(\alpha_{0},\alpha_{j}^{\vee}\right)=-\left(\theta,\alpha_{j}^{\vee}\right)=-\sum_{i=1}^{r}m_{i}\left(\alpha_{i},\alpha_{j}^{\vee}\right).
\end{equation}

The extended Dynkin diagram is obtained from that of $g$ by the addition of an extra node, representing
$\alpha_{0}$, linked to the $\alpha_{i}$-nodes by $\widehat{A}_{0i}\widehat{A}_{i0}$ lines. Extended Dynkin
diagrams obviously have more symmetry than their finite version.

The zeroth mark $m_{0}$ is defined to be 1, and since $|\alpha_{0}|^{2}=2$, the zeroth comark is also 1. The dual
Coxeter number then reads
\begin{equation}\label{dualCox}
h=\sum_{i=0}^{r}n_{i}\,.
\end{equation}

As in the finite case, the fundamental weights $\left\{\hat{\omega}_{i}\right\}$, $0\leq i\leq r$, are defined to
be the elements of the basis dual to the simple coroots, and now are assumed to be eigenstates of $L_{0}$ with
zero eigenvalue:
\begin{eqnarray}
\hat{\omega}_{i}=\left(\omega_{i};n_{i};0\right) &\qquad& (i\neq 0)\,,\\
\hat{\omega}_{0}=\left(0;1;0\right)&\qquad &(\textrm{basic fundamental weight})\,.
\end{eqnarray}
With $\omega_{i}\equiv(\omega_{i};0;0)$, we have
\begin{equation}
\hat{\omega}_{i}=n_{i}\hat{\omega}_{0}+\omega_{i}.
\end{equation}
Affine weights can thus be expanded in terms of the affine fundamental weights and $\delta$ as
\begin{equation}
\hat{\lambda}=\sum_{i=0}^{r}\lambda_{i}\hat{\omega}_{i}+\ell\delta\qquad\ell\in R.
\end{equation}
Since each fundamental weight contributes to the $\hat{k}$ eigenvalue by a factor $n_{i}$, we have
\begin{equation}
k\equiv\hat{\lambda}(\hat{k})=\sum_{i=0}^{r}n_{i}\lambda_{i},
\end{equation}
and $k$ is called the level. It follows from this relation that the zeroth Dynkin label $\lambda_{0}$ is related
to the finite Dynkin labels and the level by
\begin{equation}
\lambda_{0}=k-\sum_{i=1}^{r}n_{i}\lambda_{i}=k-(\lambda,\theta)\,.
\end{equation}

Affine weights are generally given in terms of Dynkin labels under the form
\begin{equation}
\hat{\lambda}=[\lambda_{0},\lambda_{1},...,\lambda_{r}],
\end{equation}
and for simple roots these are the rows of the affine Cartan matrix:
\begin{equation}
\alpha_{i}=[\widehat{A}_{io},\widehat{A}_{i1},...,\widehat{A}_{ir}].
\end{equation}

Highest-weight representations are characterized by a unique highest weight state $|\hat{\lambda}\rangle$
annihilated by the ladder operators:
\begin{equation}
E_{0}^{\alpha}|\hat{\lambda}\rangle=E_{n}^{\pm\alpha}|\hat{\lambda}\rangle=H_{n}^{i}|\hat{\lambda}\rangle=0\qquad\textrm{for}\quad
n>0,\alpha>0\,.
\end{equation}

A representation is called integrable if it decompose into finite irreducible representations of $su(2)$ and can
further be written as a direct sum of finite dimensional weight spaces. This requires in particular that
\begin{equation}
\lambda_{0}=k-(\lambda,\theta)\in Z_{+}\,.
\end{equation}
The adjoint representation, although not a highest-weight representation, is integrable.

\subsection{WZW models and coset construction}

Without entering the technical details, we will give the basic ideas of the general discussion presented in
\cite{dif}.

A conformal field theory with Lie-algebraic symmetry can be formulated in terms of the so-called
Wess-Zumino-Witten action
\begin{equation}\label{WZWaction}
{\cal A}^{WZW}=\frac{k}{16\pi}\int d^{2}x\,\textrm{Tr}\left(\partial^{\mu}g^{-1}\partial_{\mu}g\right)+k\Gamma\,,
\end{equation}
where $g(x)$ is a matrix bosonic field living on the group manifold $G$ associated to the Lie algebra $g$, $k$ is
a positive integer, and $\Gamma$ is a topological term necessary to preserve conformal invariance at quantum
level.

The study of conformal aspects of WZW models was initiated by Knizhnik and Zamolodchikov in \cite{knizam}.
Starting from the conserved currents
\begin{equation}
J=-k\partial_{z}g\,g^{-1}\qquad\qquad \bar{J}=k\,g^{-1}\partial_{\bar{z}}g \,,
\end{equation}
one defines the quantities $J^{a}$ and $\bar{J}^{a}$ as
\begin{equation}
J=\sum_{a}J^{a}t^{a}\qquad\qquad \bar{J}=\sum_{a}\bar{J}^{a}t^{a}\,,
\end{equation}
where $t^{a}$ are the Lie algebra generators; introducing the modes $J_{n}^{a}$ and $\bar{J}_{n}^{a}$ from the
Laurent expansions
\begin{equation}
J^{a}(z)=\sum_{n\in Z}z^{-n-1}J_{n}^{a}\qquad\qquad \bar{J}^{a}(z)=\sum_{n\in Z}\bar{z}^{-n-1}\bar{J}_{n}^{a}\,,
\end{equation}
it is possible to show that they generate two independent copies of the so-called current algebra, with the
commutation relations (\ref{currentalg}) characteristic of the affine Lie algebra $\hat{g}$ at level $k$.

In the so-called Sugawara construction, the stress-energy tensor is defined as
\begin{equation}
T(z)=\frac{1}{2(k+h)}\sum_{a}:J^{a}J^{a}:(z)\,,
\end{equation}
where $:\,:$ stands for the normal ordering, and $h$ is the dual Coxeter number (\ref{dualCox}). The
multiplicative constant is fixed requiring the correct form for the OPE of $T$ with itself, and the corresponding
central charge is
\begin{equation}\label{cWZW}
c\equiv c\left(\hat{g}_{k}\right)=\frac{k\,\textrm{dim}g}{k+h}\geq 1\,.
\end{equation}

The stress-energy tensor modes can be shown to be given by
\begin{equation}\label{WZWVirmodes}
L_{n}=\frac{1}{2(k+h)}\sum_{a}\sum_{m}:J_{m}^{a}J_{n-m}^{a}:\,,
\end{equation}
and the complete affine Lie and Virasoro algebra is
\begin{eqnarray}
&& [L_{n},L_{m}]= (n-m)L_{n+m}+\frac{c}{12}(n^{3}-n)\delta_{n+m,0} \\
&& [L_{n},J_{m}^{a}]= -m J_{n+m}^{a} \\
&& [J_{n}^{a},J_{m}^{b}]= \sum_{c}if^{ab}_{c}J_{n+m}^{c}+k n\delta_{ab}\delta_{n+m,0}
\end{eqnarray}

In the case of a direct sum of Lie algebras $g=\oplus_{i}g_{i}$, the stress-energy tensor is the sum of the
Sugawara stress-energy tensors associated to each $g_{i}$. Since all these tensors commute, the total central
charge is the sum of the central charges of the contributing pieces.

A field is said to be WZW primary if it transforms covariantly with respect to some representation specified by a
highest weight $\lambda$ in the holomorphic sector and by a highest weight $\mu$ in the antiholomorphic one; it
can be characterized by the OPE
\begin{eqnarray}
J^{a}(z)\phi_{\lambda,\mu}(w,\bar{w})&\sim& \frac{-t_{\lambda}^{a}\phi_{\lambda,\mu}(w,\bar{w})}{z-w}\\
\bar{J}^{a}(\bar{z})\phi_{\lambda,\mu}(w,\bar{w})&\sim&
\frac{\phi_{\lambda,\mu}(w,\bar{w})t_{\mu}^{a}}{\bar{z}-\bar{w}}
\end{eqnarray}
where $t_{\lambda}^{a}$ and $t_{\mu}^{a}$ are the matrix $t^{a}$ in the representations $\lambda$ and $\mu$,
respectively. In terms of the modes and concentrating on the holomorphic sector, this translates into
\begin{eqnarray}
J_{0}^{a}|\phi_{\lambda}\rangle &=& -t_{\lambda}^{a}|\phi_{\lambda}\rangle\\
J_{n}^{a}|\phi_{\lambda}\rangle &=& 0\qquad\qquad\textrm{for}\; n>0
\end{eqnarray}

It follows from (\ref{WZWVirmodes}) that WZW primary fields are also Virasoro primary fields with conformal weight
\begin{equation}\label{WZWVirh}
h_{\lambda}=\frac{(\lambda,\lambda+2\rho)}{2(k+h)}\,,
\end{equation}
but the inverse is not true.

\vspace{1cm}

A coset $\hat{g}/\hat{p}$ is a quotient of two WZW models, related to affine Lie algebras $\hat{g}$ and $\hat{p}$,
with $\hat{p}$ subalgebra of $\hat{g}$. If the level of $\hat{g}$ is $k$, that of $\hat{p}$ is given by $x_{e}k$,
with $x_{e}$ (the so-called embedding index) given by
\begin{equation}
x_{e}=\frac{|P\theta_{g}|^{2}}{|\theta_{p}|^{2}}\,,
\end{equation}
where $\theta_{g}$ and $\theta_{p}$ are respectively the highest roots of $g$ and $p$, and $P$ gives the
projection of every weight of $g$ onto a weight of $p$.

The stress-energy tensor modes
\begin{equation}
L_{m}^{(g/p)}\equiv L_{m}^{(g)}-L_{m}^{(p)}
\end{equation}
satisfy the Virasoro algebra with central charge given by the difference of the central charges of the WZW
components:
\begin{equation}
c\left(\hat{g}_{k}/\hat{p}_{x_{e}k}\right)=\frac{k\,\textrm{dim}g}{k+h_{q}}-\frac{x_{e}k\,\textrm{dim}p}{x_{e}k+h_{p}}\,.
\end{equation}
Hence, models with central charge lower than one may be represented by the coset construction.

In diagonal cosets, which are of the form
\begin{equation}
\frac{\hat{g}_{k_{1}}\oplus\hat{g}_{k_{2}}}{\hat{g}_{k_{1}+k_{2}}}\:,
\end{equation}
the embedding index is equal to 1, and the central charge is
\begin{equation}
c=\textrm{dim}g \left(\frac{k_{1}}{k_{1}+h}+\frac{k_{2}}{k_{2}+h}-\frac{k_{1}+k_{2}}{k_{1}+k_{2}+h}\right)\,.
\end{equation}
In this case, it can be shown that the field identifications take a simple form. In fact, primary fields are
specified by three weights $\{\hat{\lambda},\hat{\mu};\hat{\nu}\}$, which are integrable highest weights at
respective levels $k_{1}$, $k_{2}$ and $k_{1}+k_{2}$, selected by the condition
\begin{equation}\label{cosetfields}
\lambda+\mu-\nu\in Q \,,
\end{equation}
where $Q$ is the so-called root lattice of $g$, i.e. the set of all points whose expansion coefficients on the
basis of simple roots of $g$ are integers: $Q=Z\alpha_{1}+...+Z\alpha_{r}$. The fractional part of the
corresponding conformal weight $h_{\{\hat{\lambda},\hat{\mu};\hat{\nu}\}}$ is given by
$h_{\lambda}+h_{\mu}-h_{\nu}$ (see (\ref{WZWVirh})).

Furthermore, if we consider the group ${\cal O}(\hat{g})$ of outer automorphisms of $\hat{g}$, i.e. the set of
symmetry transformations of the Dynkin diagram of $\hat{g}$ which are not symmetry transformations of the Dynkin
diagram of $g$, the primary fields are identified according to
\begin{equation}
\{\hat{\lambda},\hat{\mu};\hat{\nu}\}\sim \{A\hat{\lambda},A\hat{\mu};A\hat{\nu}\}\qquad\forall A\in{\cal
O}(\hat{g})\,.
\end{equation}

\vspace{0.5cm}

It can be shown that the diagonal coset
\begin{equation}
\frac{\widehat{su}(2)_{k}\oplus\widehat{su}(2)_{1}}{\widehat{su}(2)_{k+1}}
\end{equation}
describes the series of unitary minimal models (\ref{min}) with $k+2=m\geq 3$. As an immediate check, remembering
that $\textrm{dim}(su(2))=3$ and $h=2$, we can easily compute the corresponding central charge:
\begin{equation}
c=\frac{3k}{k+2}+1-\frac{3(k+1)}{k+3}=1-\frac{6}{(k+2)(k+3)}\,.
\end{equation}

\vspace{0.5cm}

Other very interesting realizations are given by the cosets
\begin{equation}
\frac{(\widehat{E}_{n})_{1}\oplus(\widehat{E}_{n})_{1}}{(\widehat{E}_{n})_{2}}\qquad n=6,7,8\,.
\end{equation}
The properties of the corresponding Lie algebras are listed in appendix \ref{Liealg}. These three theories
describe respectively the tricritical three-state Potts model ($n=6$), the tricritical Ising model ($n=7$) and
the Ising model ($n=8$).

For example, in the case of $E_{7}$ we easily see that the central charge is $c=7/10$. At level 1, the only two
integrable representations are given by $\hat{\omega}_{0}$ and $\hat{\omega}_{6}$, and the corresponding primary
operators have conformal weights $h_{\hat{\omega}_{0}}=0$ and $h_{\hat{\omega}_{6}}=\frac{3}{4}$. At level 2 the
integrable representations are $2\hat{\omega}_{0}$, $\hat{\omega}_{0}+\hat{\omega}_{6}$, $\hat{\omega}_{1}$ (the
adjoint representation), $\hat{\omega}_{5}$ and $\hat{\omega}_{7}$, with weights $h_{2\hat{\omega}_{0}}=0$,
$h_{\hat{\omega}_{0}+\hat{\omega}_{6}}=\frac{57}{80}$, $h_{\hat{\omega}_{1}}=\frac{9}{10}$,
$h_{\hat{\omega}_{5}}=\frac{7}{5}$ and $h_{\hat{\omega}_{7}}=\frac{21}{16}$. The coset fields determined by the
rule (\ref{cosetfields}) are exactly the ones of the tricritical Ising model, and are associated to the
integrable representations in the following way:
\begin{eqnarray*}
&&(\hat{\omega}_{0})_{1}\times(\hat{\omega}_{0})_{1}=(1)_{\tiny\textrm{TIM}}\times(2\hat{\omega}_{0})_{2}+(\varepsilon)_{\tiny\textrm{TIM}}\times(\hat{\omega}_{1})_{2}+(t)_{\tiny\textrm{TIM}}\times(\hat{\omega}_{5})_{2}\\
&&(\hat{\omega}_{0})_{1}\times(\hat{\omega}_{6})_{1}=(\sigma)_{\tiny\textrm{TIM}}\times(\hat{\omega}_{0}+\hat{\omega}_{6})_{2}+(\sigma')_{\tiny\textrm{TIM}}\times(\hat{\omega}_{7})_{2}\\
&&(\hat{\omega}_{6})_{1}\times(\hat{\omega}_{6})_{1}=(\varepsilon'')_{\tiny\textrm{TIM}}\times(2\hat{\omega}_{0})_{2}+(t)_{\tiny\textrm{TIM}}\times(\hat{\omega}_{1})_{2}+(\varepsilon)_{\tiny\textrm{TIM}}\times(\hat{\omega}_{5})_{2}
\end{eqnarray*}
Note the identification of $t$ and $\varepsilon$ in the third row with the ones in the first, by means of the
$Z_{2}$ outer automorphism of the extended Dynkin diagram.

\subsection{Away from criticality}

The possibility of describing a perturbed CFT by a Lagrangian theory related to an affine Lie algebra $\hat{g}$
of rank $r$ was analyzed in \cite{egyang} and \cite{hollmansf}.

The idea is to start with a CFT described by a coset $\hat{g}_{k}\oplus\hat{g}_{1}/\hat{g}_{k+1}$ and identify as
the perturbing field $\Phi$ the one with conformal weight $h_{\Phi}=\frac{k+1}{k+1+h}$. It is possible to see
that the conserved currents of this theory have spins given by the exponents of $g$ modulo the dual Coxeter
number $h$ (the lowest values can be checked using the counting argument).

The affine Toda field theory related to the affine Lie algebra $\hat{g}$ is defined by the Lagrangian:
\begin{equation}
{\cal
L}=\frac{1}{2}\sum_{j=1}^{r}\left(\partial_{\mu}\phi^{j}\right)^{2}-\frac{A}{\beta^{2}}\left[\sum_{i=1}^{r}e^{\beta\alpha_{i}\cdot\phi}+e^{\beta\alpha_{0}\cdot\phi}\right]-B\,,
\end{equation}
where $\{\phi^{1},...,\phi^{r}\}$ are $r$ free bosonic fields, $\{\alpha_{0},...,\alpha_{r}\}$ are the simple
roots of $\hat{g}$, and $A,B,\beta\in R$. This Lagrangian presents a minimum at the point $\phi_{0}$, defined by
$\alpha_{i}\cdot \phi_{0}=\frac{1}{\beta}\ln (n_{i}N)$, with $N=\prod_{k=1}^{r}n_{k}^{-n_{k}/h}$ ($n_{i}$ are the
comarks of $\hat{g}$). If we redefine the set of bosonic fields as shifted from this minimum
($\phi\rightarrow\phi-\phi_{0}$), the Lagrangian becomes
\begin{equation}\label{todalagr}
{\cal
L}=\frac{1}{2}\sum_{j=1}^{r}\left(\partial_{\mu}\phi^{j}\right)^{2}-\frac{m^{2}}{\beta^{2}}\sum_{i=0}^{r}n_{i}\left[e^{\beta\alpha_{i}\cdot\phi}-1\right]\,,
\end{equation}
where $m$ is a real parameter which sets the mass scale of the system. The classically conserved currents of this
theory have spins given by the exponents of $g$ modulo the dual Coxeter number $h$, which are exactly the degrees
of the conserved currents in the $\hat{g}_{k}\oplus\hat{g}_{1}/\hat{g}_{k+1}$ coset perturbed by $\Phi$, and this
is at the basis of the correspondence between the two theories.

The formalism of this correspondence is given by the so-called Feigin-Fuchs construction, which consists in
representing the coset fields in terms of the $r$ free bosonic fields $\{\phi^{1},...,\phi^{r}\}$ by means of
vertex operators. The basic difference between affine Toda field theories and perturbed minimal models is that
the first ones are usual Lagrangian theories with real coupling constants, while the second ones have imaginary
coupling constants, which imply some difficulties in the Lagrangian approach. The rigorous way to treat this
problem is given by the so-called quantum group reduction, whose details are beyond our interests and have been
studied in \cite{smirnov}.

Expanding the potential in the Lagrangian (\ref{todalagr}) in a power series, we have
\begin{equation}
V(\phi)=m^{2}\frac{1}{2}\sum_{i=0}^{r}n_{i}\alpha_{i}^{a}\alpha_{i}^{b}\phi^{a}\phi^{b}+m^{2}\beta\frac{1}{6}\sum_{i=0}^{r}n_{i}\alpha_{i}^{a}\alpha_{i}^{b}\alpha_{i}^{c}\phi^{a}\phi^{b}\phi^{c}+...
\end{equation}
The quadratic term corresponds to the mass matrix
\begin{equation}
M_{ab}^{2}=m^{2}\sum_{i=0}^{r}n_{i}\alpha_{i}^{a}\alpha_{i}^{b}\,,
\end{equation}
whose eigenvalues $\{m_{1}^{2},...,m_{r}^{2}\}$ give the classical mass spectrum. In the case of simply laced
theories the masses can be organized in a vector $m=(m_{1},...,m_{r})$, which is the so-called Perron-Frobenius
eigenvector of $g$, i.e. the only eigenvector of the adjacency matrix of $g$ whose entries are all positive. In
this way we have a correspondence between the particle with mass $m_{i}$ and the relative representation of $g$.
In appendix \ref{Smatrlist} we list the mass spectra for the affine Toda field theories related to the $E_{n}$
affine Lie algebras, associating each mass to the corresponding node of the Dynkin diagram. The mass spectrum is
degenerate whether the group of automorphisms of the non-affine Dynkin diagram is non trivial. This is the case of
$E_{6}$, whose non-affine diagram possesses a $Z_{2}$ symmetry. The corresponding theory presents two couples of
degenerate particles, which however are distinguishable, because they have different eigenvalues with respect to
the $Z_{3}$ symmetry of the extended diagram.

The three particle couplings are given by
\begin{equation}
f^{abc}=m^{2}\beta\sum_{i=0}^{r}n_{i}\alpha_{i}^{a}\alpha_{i}^{b}\alpha_{i}^{c}\,,
\end{equation}
and, when different from zero, are proportional to the area ${\cal A}^{abc}$ of the triangle constructed with the
values of the masses $m_{a}$, $m_{b}$ and $m_{c}$. If $g$ is simply laced we have
\begin{equation}
|f^{abc}|=\frac{4\beta}{\sqrt{h}}{\cal A}^{abc}\,.
\end{equation}

\vspace{1cm}

The scattering theory of affine Toda field theories based on simply laced Lie algebras presents universal
properties which lead to a unified description. In this case, the known classical masses, three-particle
couplings and \lq\lq fusing angles\rq\rq\, provide a solution to the bootstrap equations for the $S$-matrix of
the quantized theory, which has the form
\begin{equation}\label{Stoda}
S_{ab}(\theta,\beta)=S_{ab}^{min}(\theta)Z_{ab}(\theta,b(\beta))\,.
\end{equation}
The minimal part $S_{ab}^{min}(\theta)$ only depends on the \lq\lq fusing angles\rq\rq\, $u_{ab}^{c}$ (it is
independent of the coupling constant $\beta$) and encodes the information on the mass spectrum, having all the
poles necessary to identify the physical bound states. The term $Z_{ab}$ is a CDD factor which depends on the
coupling constant $\beta$ through the function
\begin{equation}
b(\beta)=\left(\frac{1}{2\pi h}\right)\frac{\beta^{2}}{1+\beta^{2}/4\pi}\,.
\end{equation}
The full $S_{ab}$-matrix (\ref{Stoda}) must have the same pole structure of $S_{ab}^{min}$ in the physical strip,
hence $Z_{ab}$ cannot have poles with $0\leq\textrm{Im}(\theta)\leq\pi$. Furthermore, $Z_{ab}$ has to satisfy
unitarity, crossing and bootstrap equations, and in the limit $\beta\rightarrow 0$ it must reduce to the inverse
of the minimal part
\begin{equation}
Z_{ab}(\theta,b(0))=\left[S_{ab}^{min}(\theta)\right]^{-1}\,,
\end{equation}
because the full $S$-matrix reduces to the identity operator.

Defining $B=hb$, it has been found that the $S$-matrix elements of all simply laced ATFT have the form
$S_{ab}=\prod_{x}\{x\}_{\theta}$, with
\begin{equation}\label{todablocks}
\{x\}_{\theta}=\frac{s_{\frac{x+1}{h}}(\theta)s_{\frac{x-1}{h}}(\theta)}{s_{\frac{x+1-B}{h}}(\theta)s_{\frac{x-1+B}{h}}(\theta)}
, \qquad s_{\alpha}(\theta)=\frac{\sinh \left[\frac{1}{2}\left(\theta+i\pi \alpha\right)\right]}{\sinh
\left[\frac{1}{2}\left(\theta-i\pi\alpha\right)\right]}\,.
\end{equation}
The parameter $B$ varies in $[0,2]$ as $\beta$ varies in $[0,\infty]$, and these functions present the so-called
strong-weak duality
\begin{equation}
S_{ab}(B)=S_{ab}(2-B)\qquad\Longleftrightarrow\qquad  S_{ab}(\beta)=S_{ab}(4\pi/\beta)\,.
\end{equation}
Starting from a Toda $S$-matrix of the form (\ref{todablocks}), where $B$ is real and varies in $[0,2]$, we can
recover its minimal part performing the so-called \lq\lq roaming limit\rq\rq, which consists in taking $B=1+iC$
and sending the real quantity $C$ to infinity.

\vspace{0.5cm}

The specific values of the parameters $x$ for the ATFT related to the affine Lie algebras $E_{n}$ are listed in
appendix \ref{Smatrlist}. The minimal parts of these $S$-matrices exactly correspond to the $S$ matrices,
respectively, of the tricritical three-state Potts model perturbed by $\varepsilon$ ($n=6$), the tricritical
Ising model perturbed by $\varepsilon$ ($n=7$) and the Ising model perturbed by $\sigma$ ($n=8$).

\chapter{Scattering off the boundary}

As we have seen, the presence of an infinite set of conserved currents makes it possible to determine the exact
$S$-matrix of many two-dimensional quantum field theories. In the presence of a boundary, the knowledge of the
bulk scattering amplitudes allows us to find exact expressions also for the boundary reflection factors. However,
a complete analysis of the boundary scattering for many systems as in the bulk case has not jet been performed,
and many aspects of the corresponding bound states structure are still obscure.

In \cite{io}, starting from two kinds of known reflection amplitudes, we have performed a detailed study of the
boundary bound states structure for the three ATFT related to the affine Lie algebras $E_{6}$, $E_{7}$ and
$E_{8}$ and for the corresponding perturbed minimal models.

\vspace{1cm}

\section{General principles}

A two-dimensional quantum field theory with boundary can be basically defined in two ways (\cite{ghoszam}). Let us
consider the semi-infinite euclidean plane, $x\in (-\infty,0]$, $y\in (-\infty,\infty)$, where the $y$-axis
represents the boundary. In the Lagrangian approach one writes the action in the form
\begin{equation}\label{lagr}
{\cal A}=\int_{-\infty}^{\infty}dy\int_{-\infty}^{0}dx\:
a\left(\varphi,\partial_{\mu}\varphi\right)+\int_{-\infty}^{\infty}dy
\:b\left(\varphi_{B},\frac{d}{dy}\varphi_{B}\right),
\end{equation}
where $a$ and $b$ are local functions respectively of the bulk and boundary \lq\lq fundamental fields\rq\rq \,
$\varphi,\varphi_{B}$ ($\varphi_{B}(y)=\varphi(x,y)|_{x=0}$). In the \lq\lq perturbed conformal field theory\rq\rq
\, approach one writes the symbolic action
\begin{equation}\label{pert}
{\cal A}={\cal A}_{CFT+CBC}+\int_{-\infty}^{\infty}dy\int_{-\infty}^{0}dx\: \Phi(x,y)+\int_{-\infty}^{\infty}dy
\:\Phi_{B}(y),
\end{equation}
where ${\cal A}_{CFT+CBC}$ is the action of a conformal field theory (CFT) on the semi-infinite plane with certain
conformal boundary conditions (CBC), and $\Phi,\Phi_{B}$ are specific bulk and boundary relevant fields.

As we have seen, if a \lq\lq bulk\rq\rq \, theory is integrable the scattering processes are purely elastic, and
the $S$-matrix factorizes into two-particle amplitudes. This can be generalized to the case with boundary
defining, in addition to the bulk $S$-matrix, an opportune boundary reflection matrix. Assuming that the effect
of the boundary is to reverse the momentum and preserve the energy of a particle which scatters off it, we define
the reflection amplitude $K(\theta)$ as the proportionality factor which relates a single-particle \lq out\rq \,
state to the \lq in\rq \, state with reversed momentum:
\begin{equation}
|a,\theta\rangle_{out}=K_{a}\left(\theta\right)|a,-\theta\rangle_{in}.
\end{equation}

\vspace{5cm}

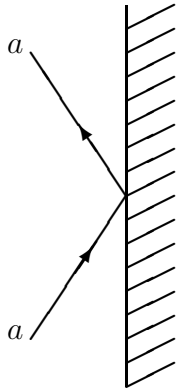
\begin{figure}[h]
\setlength{\unitlength}{0.0125in}
\begin{picture}(40,0)(60,470)
\thicklines \put(330,460){\line( 0,1){160}}

\put(330,460){\line(2,1){20}} \put(330,470){\line(2,1){20}} \put(330,480){\line(2,1){20}}
\put(330,490){\line(2,1){20}} \put(330,500){\line(2,1){20}} \put(330,510){\line(2,1){20}}
\put(330,520){\line(2,1){20}} \put(330,530){\line(2,1){20}} \put(330,540){\line(2,1){20}}
\put(330,550){\line(2,1){20}} \put(330,560){\line(2,1){20}} \put(330,570){\line(2,1){20}}
\put(330,580){\line(2,1){20}} \put(330,590){\line(2,1){20}} \put(330,600){\line(2,1){20}}
\put(330,610){\line(2,1){20}}

\put(330,540){\line(-2,-3){20}} \put(290,480){\vector(2,3){26}} \put(280,480){$a$}

\put(330,540){\vector(-2,3){20}} \put(310,570){\line(-2,3){20}} \put(280,600){$a$}

\end{picture}
 \caption{Scattering off the boundary}
 \end{figure}

With this definition we restrict our attention to theories with distinguishable particles,
otherwise the reflection factor $K_{a}$ should be a matrix to allow for a mixing of states.

\vspace{5cm}

\begin{figure}[h]
\setlength{\unitlength}{0.0125in}
\begin{picture}(40,0)(60,470)
\thicklines \put(250,460){\line( 0,1){160}}

\put(250,460){\line(2,1){20}} \put(250,470){\line(2,1){20}} \put(250,480){\line(2,1){20}}
\put(250,490){\line(2,1){20}} \put(250,500){\line(2,1){20}} \put(250,510){\line(2,1){20}}
\put(250,520){\line(2,1){20}} \put(250,530){\line(2,1){20}} \put(250,540){\line(2,1){20}}
\put(250,550){\line(2,1){20}} \put(250,560){\line(2,1){20}} \put(250,570){\line(2,1){20}}
\put(250,580){\line(2,1){20}} \put(250,590){\line(2,1){20}} \put(250,600){\line(2,1){20}}
\put(250,610){\line(2,1){20}}

\put(250,540){\line(-2,-5){30}}  \put(210,465){$a$} \put(250,540){\line(-2,5){30}}  \put(210,615){$a$}
\put(250,540){\line(-3,-2){70}}  \put(170,495){$b$} \put(250,540){\line(-3,2){70}}  \put(170,580){$b$}

\put(310,540){$=$}

\thicklines \put(410,460){\line( 0,1){160}}

\put(410,460){\line(2,1){20}} \put(410,470){\line(2,1){20}} \put(410,480){\line(2,1){20}}
\put(410,490){\line(2,1){20}} \put(410,500){\line(2,1){20}} \put(410,510){\line(2,1){20}}
\put(410,520){\line(2,1){20}} \put(410,530){\line(2,1){20}} \put(410,540){\line(2,1){20}}
\put(410,550){\line(2,1){20}} \put(410,560){\line(2,1){20}} \put(410,570){\line(2,1){20}}
\put(410,580){\line(2,1){20}} \put(410,590){\line(2,1){20}} \put(410,600){\line(2,1){20}}
\put(410,610){\line(2,1){20}}

\put(410,560){\line(-2,-5){40}}  \put(360,460){$a$} \put(410,560){\line(-2,5){24}}  \put(375,620){$a$}
\put(410,520){\line(-3,-2){70}}  \put(330,475){$b$} \put(410,520){\line(-3,2){70}}  \put(330,560){$b$}

\end{picture}
 \caption{Boundary factorization}
 \end{figure}
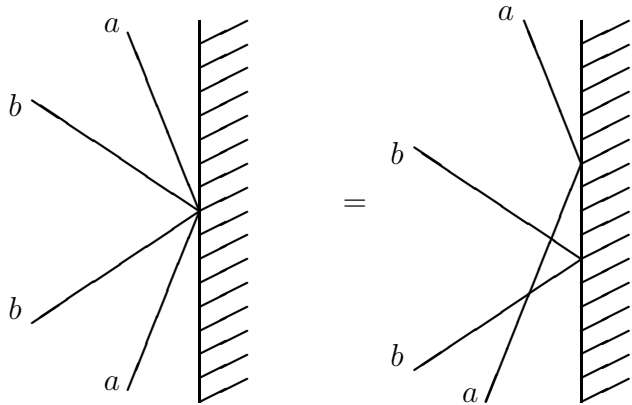

\vspace{0.5cm}

Factorization means now that the reflection matrix can always be decomposed into products of two-particle bulk
scattering amplitudes and single-particle boundary reflection amplitudes.

In every integrable boundary field theory we expect to find a certain number of different reflection matrices,
corresponding to the various boundary conditions compatible with integrability.

In the bulk case, the two-particle scattering amplitudes can be exactly calculated (up to the so-called \lq\lq CDD
ambiguity\rq\rq) as solutions of a number of constraints, which are expressed by unitarity, crossing, Yang-Baxter
and bootstrap equations. In the presence of a boundary, the reflection amplitudes are constrained by analogous
equations, which relate them to the bulk $S$-matrix. In this way, starting from a theory whose bulk scattering
matrix is known, one can investigate which are the possible reflection amplitudes.

\vspace{0.5cm}

In the \lq\lq perturbed conformal field theory\rq\rq \, approach the integrable boundary conditions must be
associated in the ultraviolet limit to some of the possible conformal ones, which, in the case of Virasoro
minimal models, are in one-to-one correspondence with the primary fields of the examined CFT (see
eq.(\ref{bstates})). Hence, if the examined theory is a perturbed minimal model, we know explicitly from the
primary fields content of the corresponding CFT which are the conformal boundary conditions that we have to
recover in the UV limit.

On the contrary, the affine Toda field theory related to an algebra $g$ of rank $r$ has in general a Lagrangian
description given by (\ref{todalagr}). We can now define a boundary field theory with Lagrangian
\begin{equation}\label{todabound}
{\cal L}_{B}=\theta(-x){\cal L}-\delta(x){\cal B}(\phi)
\end{equation}
and investigate which reflection amplitudes are compatible with the bulk $S$-matrix. Restrictions on the boundary
interaction ${\cal B}(\phi)$ due to the integrability requirement have been discussed in \cite{durb},\cite{durb2}.
Also this theory can be seen as a perturbation of a certain CFT, which has however a spectrum of primary operators
generally much richer than in minimal models. Furthermore, this CFT possesses a larger symmetry than just
conformal invariance, and in this case the structure of the boundary states for a boundary condition that is only
conformally invariant is not yet known.

\newpage

The main equations for the reflection amplitudes have been derived in \cite{ghoszam} and discussed in
\cite{fring0}-\cite{sas},\cite{corr}. The unitarity condition
\begin{equation}\label{unit}
K_{a}\left(\theta\right)K_{a}\left(-\theta\right)=1
\end{equation}
is a straightforward generalization of the bulk one, while the boundary crossing equation
\begin{equation}\label{cross}
K_{a}\left(\theta\right)K_{\bar{a}}\left(\theta+i\pi\right)=S_{aa}\left(2\theta\right)
\end{equation}
required more attention, and will be of great importance in our considerations.

\vspace{3cm}

\begin{figure}[h]
\setlength{\unitlength}{0.0125in}
\begin{picture}(40,90)(60,420)

\thicklines \put(200,420){\line( 0,1){150}} \put(210,420){\line( 0,1){150}} \put(200,420){\line( 1,0){10}}
\put(200,570){\line( 1,0){10}}

\put(200,420){\line( 1,2){10}} \put(200,430){\line( 1,2){10}} \put(200,440){\line( 1,2){10}} \put(200,450){\line(
1,2){10}} \put(200,460){\line( 1,2){10}} \put(200,470){\line( 1,2){10}} \put(200,480){\line( 1,2){10}}
\put(200,490){\line( 1,2){10}} \put(200,500){\line( 1,2){10}} \put(200,510){\line( 1,2){10}} \put(200,520){\line(
1,2){10}} \put(200,530){\line( 1,2){10}} \put(200,540){\line( 1,2){10}} \put(200,550){\line( 1,2){10}}

\put(200,500){\line(-2,-3){20}} \put(150,425){\vector(2,3){30}}

\put(140,420){$a$} \put(192,475){$ \theta$}

\put(200,500){\vector(-2,3){30}} \put(170,545){\line(-2,3){20}}

\put(210,500){\line(2,-3){20}} \put(260,425){\vector(-2,3){30}}

\put(210,500){\vector(2,3){30}} \put(240,545){\line(2,3){20}}

\put(450,420){\vector(-2,3){100}} \put(350,420){\vector(2,3){100}}

\put(395,470){$ 2\theta$} \put(213,515){$ \theta + i \pi$} \put(455,415){$a$} \put(340,415){$a$}
\put(310,505){$=$}
\end{picture}
\caption{The boundary crossing-unitarity relation}
 \end{figure}
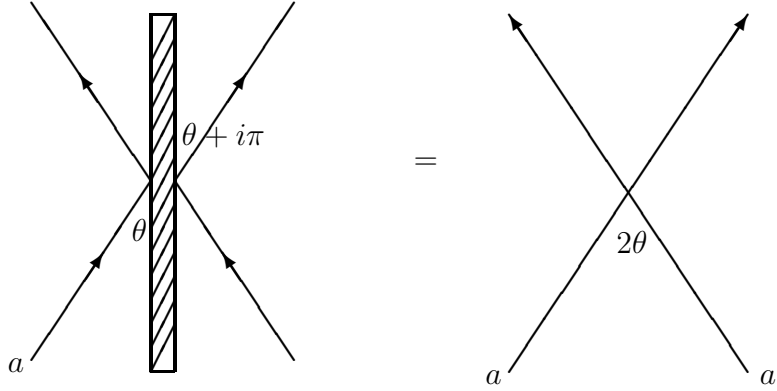\label{figcross}

\vspace{1cm}

In the case of distinguishable particles the Yang-Baxter equations
\begin{equation}\label{YB}
K_{a}\left(\theta_{a}\right)S_{ab}\left(\theta_{b}+\theta_{a}\right)K_{b}\left(\theta_{b}\right)S_{ab}\left(\theta_{b}-\theta_{a}\right)=
S_{ab}\left(\theta_{b}-\theta_{a}\right)K_{b}\left(\theta_{b}\right)S_{ab}\left(\theta_{b}+\theta_{a}\right)K_{a}\left(\theta_{a}\right)
\end{equation}
are trivially satisfied.

\vspace{4cm}

\begin{figure}[h]
\setlength{\unitlength}{0.0125in}
\begin{picture}(40,0)(60,470)
\thicklines \put(120,480){\line( 1,0){170}} \put(340,480){\line(1,0){170}} \put(120,470){\line(2,1){20}}
\put(130,470){\line(2,1){20}} \put(140,470){\line(2,1){20}} \put(150,470){\line(2,1){20}}
\put(160,470){\line(2,1){20}} \put(170,470){\line(2,1){20}} \put(180,470){\line(2,1){20}}
\put(190,470){\line(2,1){20}} \put(200,470){\line(2,1){20}} \put(210,470){\line(2,1){20}}
\put(220,470){\line(2,1){20}} \put(230,470){\line(2,1){20}} \put(240,470){\line(2,1){20}}
\put(250,470){\line(2,1){20}} \put(260,470){\line(2,1){20}} \put(270,470){\line(2,1){20}}
\put(340,470){\line(2,1){20}} \put(350,470){\line(2,1){20}} \put(360,470){\line(2,1){20}}
\put(370,470){\line(2,1){20}} \put(380,470){\line(2,1){20}} \put(390,470){\line(2,1){20}}
\put(400,470){\line(2,1){20}} \put(410,470){\line(2,1){20}} \put(420,470){\line(2,1){20}}
\put(430,470){\line(2,1){20}} \put(440,470){\line(2,1){20}} \put(450,470){\line(2,1){20}}
\put(460,470){\line(2,1){20}} \put(470,470){\line(2,1){20}} \put(480,470){\line(2,1){20}}
\put(490,470){\line(2,1){20}}
\put(310,510){$=$} \put(220,480){\vector(1,3){25}} \put(220,480){\line(-1,3){25}} \put(410,480){\vector(1,3){25}}
\put(410,480){\line(-1,3){25}} \put(170,480){\vector(2,1){120}} \put(170,480){\line(-2,1){50}}
\put(460,480){\vector(2,1){50}} \put(460,480){\line(-2,1){120}} \put(125,505){$b$} \put(136,484){$\theta_b$}
\put(187,545){$a$} \put(390,545){$a$} \put(350,538){$b$}
\end{picture}
 \caption{The boundary Yang-Baxter equation}
 \end{figure}
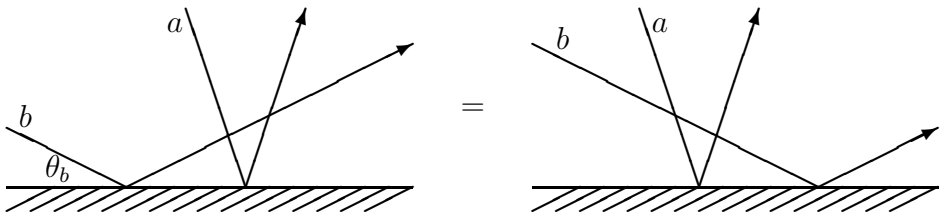

\vspace{0.5cm}

On the contrary, a strong constraint is given by the bootstrap equations
\begin{equation}\label{boot}
K_{c}\left(\theta\right)= K_{a}(\theta-i \bar{u}_{ac}^{b})S_{ab}(2\theta-i \bar{u}_{ac}^{b}+i
\bar{u}_{bc}^{a})K_{b}(\theta+i \bar{u}_{bc}^{a})\,,
\end{equation}
with $\bar{u}_{ab}^{c}\equiv\pi-u_{ab}^{c}$, where $u_{ab}^{c}$ are the bulk \lq\lq fusing angles\rq\rq.

\vspace{4.5cm}

\begin{figure}[h]
\setlength{\unitlength}{0.0125in}
\begin{picture}(40,0)(60,470)
\thicklines \put(120,480){\line( 1,0){170}} \put(340,480){\line(1,0){170}} \put(120,470){\line(2,1){20}}
\put(130,470){\line(2,1){20}} \put(140,470){\line(2,1){20}} \put(150,470){\line(2,1){20}}
\put(160,470){\line(2,1){20}} \put(170,470){\line(2,1){20}} \put(180,470){\line(2,1){20}}
\put(190,470){\line(2,1){20}} \put(200,470){\line(2,1){20}} \put(210,470){\line(2,1){20}}
\put(220,470){\line(2,1){20}} \put(230,470){\line(2,1){20}} \put(240,470){\line(2,1){20}}
\put(250,470){\line(2,1){20}} \put(260,470){\line(2,1){20}} \put(270,470){\line(2,1){20}}
\put(340,470){\line(2,1){20}} \put(350,470){\line(2,1){20}} \put(360,470){\line(2,1){20}}
\put(370,470){\line(2,1){20}} \put(380,470){\line(2,1){20}} \put(390,470){\line(2,1){20}}
\put(400,470){\line(2,1){20}} \put(410,470){\line(2,1){20}} \put(420,470){\line(2,1){20}}
\put(430,470){\line(2,1){20}} \put(440,470){\line(2,1){20}} \put(450,470){\line(2,1){20}}
\put(460,470){\line(2,1){20}} \put(470,470){\line(2,1){20}} \put(480,470){\line(2,1){20}}
\put(490,470){\line(2,1){20}}
\put(310,510){$=$} \put(200,480){\line(2,3){40}} \put(240,540){\line(3,1){40}} \put(240,540){\line(-1,3){10}}

\put(200,480){\line(-2,3){40}} \put(160,540){\line(-3,1){40}} \put(160,540){\line(1,3){10}} \put(115,545){$a$}
\put(285,545){$a$} \put(175,570){$b$} \put(220,570){$b$} \put(165,515){$c$} \put(230,515){$c$}

\put(400,480){\line(-5,3){65}} \put(400,480){\line(5,3){50}} \put(440,480){\line(-1,3){30}}
\put(440,480){\line(1,3){10}}

\put(450,510){\line(5,6){30}} \put(480,546){\line(5,3){40}} \put(480,546){\line(1,3){15}}

\put(335,508){$a$} \put(400,565){$b$} \put(525,565){$a$} \put(485,585){$b$} \put(470,525){$c$}

\end{picture}
 \caption{The boundary bootstrap equation}
 \end{figure}
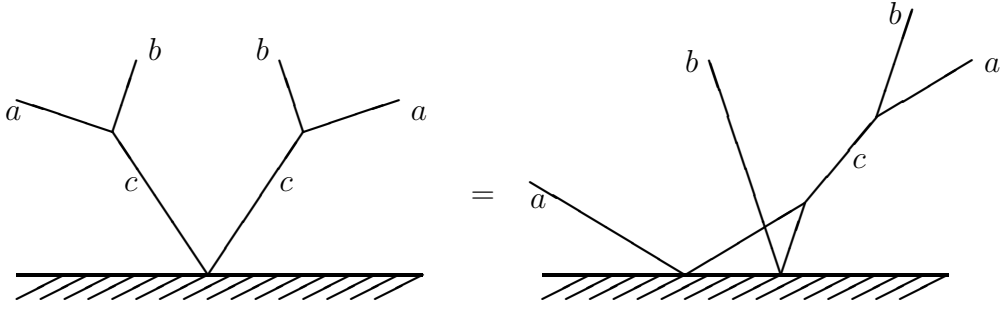

\vspace{0.5cm}

As noticed in \cite{ghoszam}, the boundary can exist in several stable states, and the presence of poles in
$K_{a}(\theta)$ indicates the possibility for particle $a$ to excite it. For reflection amplitudes, the physical
strip is the region $0\leq \rm{Im}\,\theta\leq\frac{\pi}{2}$ of the complex $\theta$-plane.

\vspace{5cm}

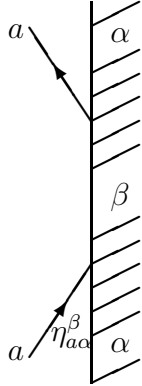
\begin{figure}[h]
\setlength{\unitlength}{0.0125in}
\begin{picture}(40,0)(60,470)
\thicklines \put(330,460){\line( 0,1){160}}

\put(330,460){\line(2,1){20}}

\put(330,480){\line(2,1){20}} \put(330,490){\line(2,1){20}} \put(330,500){\line(2,1){20}}
\put(330,510){\line(2,1){20}} \put(330,520){\line(2,1){20}}

\put(330,550){\line(2,1){20}} \put(330,560){\line(2,1){20}} \put(330,570){\line(2,1){20}}
\put(330,580){\line(2,1){20}} \put(330,590){\line(2,1){20}}

\put(330,610){\line(2,1){20}}

\put(338,473){$\alpha $} \put(338,535){$\beta $} \put(338,603){$\alpha $}
\put(330,510){\line(-2,-3){10}} \put(304,471){\vector(2,3){16}} \put(295,605){$a$}

\put(313,478){$\eta_{a \alpha}^{\beta}$}

\put(330,570){\vector(-2,3){16}} \put(314,594){\line(-2,3){10}} \put(295,470){$a$}

\end{picture}
 \caption{Boundary excitation}\label{figexcit}
 \end{figure}

\vspace{0.5cm}

We will call $\eta_{a\alpha}^{\beta}$ the \lq\lq fusing angle\rq\rq \, related to an odd-order pole with positive
residue at $\theta_{a}=i\eta_{a\alpha}^{\beta}$ in the reflection amplitude $K_{a}^{\alpha}$ of particle $a$ on
the boundary state $\alpha$, creating the state $\beta$. The corresponding bootstrap equation for the boundary
bound states is
\begin{equation}\label{boundst}
K_{b}^{\beta}\left(\theta\right)=S_{ab}\left(\theta+i\eta_{a\alpha}^{\beta}\right)K_{b}^{\alpha}\left(\theta\right)S_{ab}\left(\theta-i\eta_{a\alpha}^{\beta}\right),
\end{equation}
and the energies of the two states $|\alpha\rangle$ and $|\beta\rangle$ are related by
\begin{equation}\label{energy}
E_{\beta}=E_{\alpha}+m_{a}\cos\left(\eta_{a\alpha}^{\beta}\right).
\end{equation}

\vspace{4.5cm}

\begin{figure}[h]
\setlength{\unitlength}{0.0125in}
\begin{picture}(40,0)(60,470)
\thicklines \put(120,510){\line( 1,0){170}} \put(340,510){\line(1,0){170}} \put(120,500){\line(2,1){20}}
\put(150,500){\line(2,1){20}} \put(160,500){\line(2,1){20}} \put(170,500){\line(2,1){20}}
\put(180,500){\line(2,1){20}} \put(190,500){\line(2,1){20}} \put(220,500){\line(2,1){20}}
\put(230,500){\line(2,1){20}} \put(240,500){\line(2,1){20}} \put(250,500){\line(2,1){20}}
\put(260,500){\line(2,1){20}} \put(270,500){\line(2,1){20}} \put(340,500){\line(2,1){20}}
\put(350,500){\line(2,1){20}} \put(380,500){\line(2,1){20}} \put(390,500){\line(2,1){20}}
\put(400,500){\line(2,1){20}} \put(410,500){\line(2,1){20}} \put(420,500){\line(2,1){20}}
\put(430,500){\line(2,1){20}} \put(440,500){\line(2,1){20}} \put(450,500){\line(2,1){20}}
\put(480,500){\line(2,1){20}} \put(490,500){\line(2,1){20}}
\put(310,540){$=$} \put(220,510){\vector(1,3){25}} \put(220,510){\line(-1,3){25}} \put(410,510){\vector(1,3){25}}
\put(410,510){\line(-1,3){25}} \put(170,510){\line(-2,1){50}} \put(120,535){\vector(2,-1){20}}
\put(460,510){\line(-2,1){120}} \put(340,570){\vector(2,-1){25}} \put(125,535){$a$} \put(120,516){$\eta_{a
\alpha}^{\beta}$} \put(187,575){$b$} \put(390,575){$b$} \put(350,568){$a$} \put(138,501){$\alpha $}
\put(210,500){$\beta $} \put(368,501){$\alpha $} \put(470,500){$\beta $}
\end{picture}
 \caption{The boundary bound state bootstrap equation}
 \end{figure}
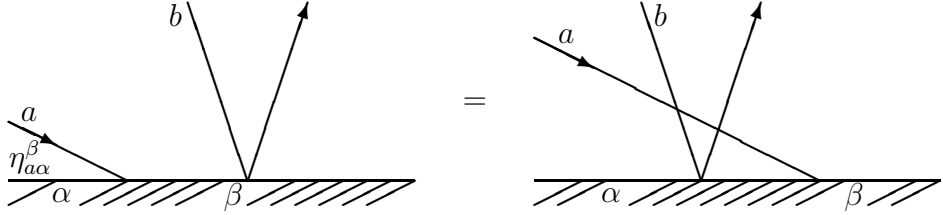

\vspace{1.5cm}

\section{Boundary affine Toda field theories}

Fring and Koberle (\cite{fring1}) have demonstrated that in affine Toda field theories, where the $S$-matrix
elements are always of the form (\ref{todablocks}), the boundary crossing equation (\ref{cross}) has factorized
solutions of the form
\begin{equation}\label{K}
K_{a}(\theta)=\prod_{x}{\cal K}_{x}(\theta),\qquad {\cal
K}_{x}(\theta)=\frac{s_{\frac{1-x-h}{2h}}(\theta)s_{\frac{-1-x-h}{2h}}(\theta)}{s_{\frac{1-x-B-h}{2h}}(\theta)s_{\frac{-1-x+B-h}{2h}}(\theta)}
\end{equation}
where the blocks ${\cal K}_{x}(\theta)$ are in one-to-one correspondence to the blocks $\{x\}_{2\theta}$ in
$S_{a\bar{a}}$ up to a shift of $2h$ in $x$. In order to determine which of the blocks are shifted by $2h$ and
which are not, one has to write a bootstrap equation (\ref{boot}) involving only one particular $K_{a}(\theta)$,
find its most general solution, and then compute the other $K_{b}(\theta)$ consistently.

\vspace{0.5cm}

Given a solution $K_{a}(\theta)$ of the equations (\ref{unit})-(\ref{boot}), a function of the form
\begin{equation}
K_{a}(\theta)\psi_{a}(\theta)
\end{equation}
is again a solution if $\psi_{a}$ satisfies the homogeneous equations
\begin{equation}\label{CDD1}
\psi_{\bar{a}}(\theta+i\frac{\pi}{2})\psi_{a}(\theta-i\frac{\pi}{2})=1,
\end{equation}
\begin{equation}\label{CDD2}
\psi_{c}(\theta)=\psi_{a}(\theta-i\bar{u}_{ac}^{b})\psi_{b}(\theta+i\bar{u}_{bc}^{a}).
\end{equation}
Some possible choices of these functions (which are CDD-factors and have been analyzed in \cite{sas}) are

\begin{enumerate}

\item {$\psi_{a}(\theta)=\frac{K_{a}(\theta+i\pi)}{K_{a}(\theta)}$}

Every block $\psi_{x}(\theta)=\frac{{\cal K}_{x}(\theta+i\pi)}{{\cal K}_{x}(\theta)}=\frac{{\cal
K}_{x+2h}(\theta)}{{\cal K}_{x}(\theta)}$ satisfies eq. (\ref{CDD1}) and (\ref{CDD2}), so that the new solution
can be obtained from the first one by shifting all the $x$'s by $2h$.

\item {$\psi_{a}(\theta)=\left(S_{ab}(\theta)\right)^{\pm 1}$, $\forall b$  }

\item {$\psi_{a}(\theta)=\prod_{b}S_{ab}(\theta)$ }

\end{enumerate}
What we expect is that these CDD-factors relate different sets of reflection amplitudes corresponding to the
various boundary conditions compatible with integrability.

\vspace{0.5cm}

In \cite{io}, for the theories in exam we have analyzed the two solutions of the form (\ref{K}) $K_{a}(\theta)$
and $K_{a}(\theta+i\pi)$, related by the first kind of CDD-ambiguity mentioned, calling \lq\lq minimal\rq\rq \,
the one with the minimum number of poles in the physical strip.

The blocks ${\cal K}_{x}(\theta)$ have poles at
\begin{equation}\label{poles}
\theta_{\pm}=\frac{\pm 1-x-h}{2h}i\pi \;(\textrm{mod} \, 2\pi i)\qquad \textrm{and} \qquad
\theta_{\pm}^{B}=\frac{\pm B\mp 1 +x+h}{2h}i\pi \;(\textrm{mod} \, 2\pi i)
\end{equation}
and zeroes at
\begin{equation}\label{zeroes}
^{0}\theta_{\pm}=\frac{\pm 1+x+h}{2h}i\pi \;(\textrm{mod} \, 2\pi i)\qquad \textrm{and} \qquad
^{0}\theta_{\pm}^{B}=\frac{\pm 1\mp B -x-h}{2h}i\pi \;(\textrm{mod} \, 2\pi i).
\end{equation}
Only shifted blocks can give $\theta_{\pm}$ poles and $^{0}\theta_{\pm}^{B}$ zeroes in the physical strip.
Contrary to the case of the bulk $S$-matrix, the coupling constant dependent $\theta_{\pm}^{B}$ poles of the
unshifted blocks can move inside the strip $0\leq \rm{Im}\,\theta\leq\pi$, but they are confined inside the
interval $\frac{\pi}{2}\leq \rm{Im}\,\theta \leq\pi$ (they can reach the value $i\frac{\pi}{2}$ only in the
trivial cases $B=0,2$); the $^{0}\theta_{\pm}$ zeroes of unshifted blocks are also located in $\frac{\pi}{2}\leq
\rm{Im}\,\theta \leq\pi$.

\vspace{0.5cm}

Fring and Koberle have demonstrated in \cite{fring2} that every excited state reflection amplitude obtained from
(\ref{K}) iterating eq.(\ref{boundst}) can be expressed in terms of the ground state one as
\begin{equation}\label{colour}
K_{a}^{\mu}\left(\theta\right)=\left(\prod_{b}S_{ab}\left(\theta\right)\right)K_{a}^{0}\left(\theta\right),
\end{equation}
and the corresponding energy levels are related by
\begin{equation}\label{level}
E_{\mu}=E_{0}+\frac{1}{2}\sum_{b}m_{b}.
\end{equation}
The important feature of expression (\ref{colour}) is that all the $b$'s are of the same colour with respect to
the bicolouration of the Dynkin diagram of $g$. Bicolouration consists in assigning a colour (black or white) to
every node in the Dynkin diagram, such that two nodes linked to each other are differently coloured (for the
$E_{n}$ case, see appendix \ref{Smatrlist}).

\subsection{$E_{6}$ affine Toda field theory}

A complete analysis of the \lq\lq minimal\rq\rq \, solution was performed in \cite{fring2}. The bootstrap closes
on eight boundary bound states, and there are six different energy levels, two of which are degenerate.

Labeling the particles with the conventions of appendix \ref{Smatrlist} and omitting the $\theta$-dependence, the
\lq\lq minimal\rq\rq \, set of reflection amplitudes is:

\begin{eqnarray}
\label{min1}K_{1}&=& {\cal K}_{5}{\cal K}_{35}  \\
\label{min2}K_{2}&=& {\cal K}_{1}{\cal K}_{7}{\cal K}_{11} {\cal K}_{29} \\
\label{min3}K_{3}&=& {\cal K}_{3}{\cal K}_{5}{\cal K}_{7}{\cal K}_{11}{\cal K}_{29}{\cal K}_{33} \\
\label{min4}K_{4}&=& {\cal K}_{1}{\cal K}_{3}{\cal K}^{2}_{5}{\cal K}_{7}{\cal K}^{2}_{9}{\cal K}_{27}{\cal
K}_{29} {\cal K}^{2}_{31}{\cal K}_{35}
\end{eqnarray}
with $K_{6}=K_{1}$ and $K_{5}=K_{3}$.

Fixing $E_{0}=0$, the energy levels are:
\begin{center}
\begin{tabular}{cclc|}
$E_{\alpha}$ &=& $0.796 m =\frac{m_{2}}{2}$ \\
$E_{\beta}$ &=& $1.087 m =\frac{m_{3}}{2}$ \\
$E_{\gamma}$ &=& $1.884 m =\frac{m_{2}+m_{3}}{2}$ \\
$E_{\delta}$ &=& $2.175 m =\frac{m_{3}+m_{5}}{2}$ \\
$E_{\varepsilon}$ &=& $2.971 m =\frac{m_{2}+m_{3}+m_{5}}{2}$ \\
\end{tabular}
\end{center}
where the mass parameter $m$ is defined by $m_{1}^{2}=\left(3-\sqrt{3}\right)m^{2}$. Levels $\beta$ and $\gamma$
are degenerate, due to the equality $m_{3}=m_{5}$; they correspond respectively to the states
$\beta_{1},\beta_{2}$ and $\gamma_{1},\gamma_{2}$, with
\begin{center}
\begin{tabular}{cclc|cclc|}
$K_{b}^{\beta_{1}}(\theta)$ &=& $S_{b3}(\theta)K_{b}^{0}(\theta)$ && $K_{b}^{\gamma_{1}}(\theta)$ &=& $S_{b2}(\theta)S_{b5}(\theta)K_{b}^{0}(\theta)$ \\
$K_{b}^{\beta_{2}}(\theta)$ &=& $S_{b5}(\theta)K_{b}^{0}(\theta)$ && $K_{b}^{\gamma_{2}}(\theta)$ &=& $S_{b2}(\theta)S_{b3}(\theta)K_{b}^{0}(\theta)$ \\
\end{tabular}
\end{center}

\vspace{1cm}

We list the \lq\lq fusing angles\rq\rq \, following the conventions of \cite{fring2}:

\begin{center}
\begin{tabular}{|c|c|c|c|c|c|c|c|c|}\hline
 $ a\setminus\mu$  &  $ 0$  &  $ \alpha$ & $ \beta_{1}$ & $ \beta_{2}$ &  $ \gamma_{1}$ & $\gamma_{2}$ & $\delta$ &$\varepsilon$  \\ \hline
 1 & $1^{\beta_{1}}$ & $1^{\gamma_{2}}5^{\beta_{1}}$& $3^{\gamma_{1}}$& $1^{\delta}$& $1^{\varepsilon}5_{3}^{\delta}$& & &  \\ \hline
 2 & $4^{\alpha}$ & $2^{\delta}6^{\alpha}$ & $4_{3}^{\gamma_{2}}6^{\beta_{1}}$& $4_{3}^{\gamma_{1}}6^{\beta_{2}}$& $6_{3}^{\gamma_{1}}$& $6_{3}^{\gamma_{2}}$& $4_{5}^{\varepsilon}6_{3}^{\delta}$& $6_{5}^{\varepsilon}$\\ \hline
 3 & $2^{\gamma_{1}}4^{\beta_{2}}$ & $4_{3}^{\gamma_{1}}$ & $2_{3}^{\varepsilon}4_{3}^{\delta}$ & $6_{3}^{\beta_{1}}$ & $6_{5}^{\gamma_{2}}$ & $4_{5}^{\varepsilon}$ & & \\ \hline
 4 & $1^{\varepsilon}3_{3}^{\delta}5^{\alpha}$& $3_{5}^{\varepsilon}$ & $5_{5}^{\gamma_{2}}$ & $5_{5}^{\gamma_{1}}$& & & $5_{9}^{\varepsilon}$& \\ \hline
 5 & $2^{\gamma_{2}}4^{\beta_{1}}$ & $4_{3}^{\gamma_{2}}$ & $6_{3}^{\beta_{2}}$ & $2_{3}^{\varepsilon}4_{3}^{\delta}$   & $4_{5}^{\varepsilon}$ & $6_{5}^{\gamma_{1}}$& & \\ \hline
 6 & $1^{\beta_{2}}$ & $1^{\gamma_{1}}5^{\beta_{2}}$& $1^{\delta}$& $3^{\gamma_{2}}$& & $1^{\varepsilon}5_{3}^{\delta}$& &   \\ \hline
\end{tabular}
\end{center}

\vspace{0.5cm}

Each entry in the table indicates a fusing angle as a multiple of $\frac{i\pi}{12}$; the left column refers to the
particle type which scatters off the boundary in the state indicated in the first row. The superscript refers to
the state the boundary is changing into, and the subscript refers to the order of the pole, if multiple. As in
all the other cases we will examine, the residues have always the same sign as $B$ varies in $[0,2]$, they vanish
at the extremes of the interval and sometimes also in $B=1$.

\vspace{0.5cm}

The second solution gives the same number of boundary bound states, with energies
\begin{center}
\begin{tabular}{cclc|}
$E_{\alpha}$ &=& $0.563 m =\frac{m_{1}}{2}$ \\
$E_{\beta}$ &=& $1.126 m =\frac{m_{1}+m_{6}}{2}$ \\
$E_{\gamma}$ &=& $1.538 m =\frac{m_{4}}{2}$ \\
$E_{\delta}$ &=& $2.101 m =\frac{m_{1}+m_{4}}{2}$ \\
$E_{\varepsilon}$ &=& $2.664 m =\frac{m_{1}+m_{4}+m_{6}}{2}$ \\
\end{tabular}
\end{center}
Levels $\alpha$ and $\delta$ are degenerate, with
\begin{center}
\begin{tabular}{cclc|cclc|}
$K_{b}^{\alpha_{1}}(\theta)$ &=& $S_{b6}(\theta)K_{b}^{0}(\theta)$ && $K_{b}^{\delta_{1}}(\theta)$ &=& $S_{b4}(\theta)S_{b1}(\theta)K_{b}^{0}(\theta)$ \\
$K_{b}^{\alpha_{2}}(\theta)$ &=& $S_{b1}(\theta)K_{b}^{0}(\theta)$ && $K_{b}^{\delta_{2}}(\theta)$ &=& $S_{b4}(\theta)S_{b6}(\theta)K_{b}^{0}(\theta)$ \\
\end{tabular}
\end{center}

The \lq\lq fusing angles\rq\rq \, are:

\begin{center}
\begin{tabular}{|c|c|c|c|c|c|c|c|c|}\hline
 $ a\setminus\mu$  &  $ 0$  &  $ \alpha_{1}$ & $ \alpha_{2}$ & $ \beta$ &  $ \gamma$ & $\delta_{1}$ & $\delta_{2}$ &$\varepsilon$  \\ \hline
 1 & $4^{\alpha_{1}}$ & $6^{\alpha_{2}}$& $2^{\gamma}4^{\beta}$& $2^{\delta_{2}}$& $4_{3}^{\delta_{2}}$& $4_{3}^{\varepsilon}$& $6_{3}^{\delta_{1}}$&  \\ \hline
 2 & $1^{\gamma}3^{\beta}6^{0}$ & $1^{\delta_{2}}6^{\alpha_{1}}$ & $1^{\delta_{1}}6^{\alpha_{2}}$& $1^{\varepsilon}5_{3}^{\gamma}6^{\beta}$& $3_{3}^{\varepsilon}6^{\gamma}$& $6^{\delta_{1}}$& $6^{\delta_{2}}$& $6^{\varepsilon}$\\ \hline
 3 & $1^{\delta_{1}}5^{\alpha_{2}}$ & $1^{\varepsilon}5_{3}^{\beta}$ & $3_{3}^{\delta_{2}}$ & & $5_{5}^{\delta_{1}}$ & &$5_{7}^{\varepsilon}$ & \\ \hline
 4 & $2_{3}^{\varepsilon}4_{3}^{\gamma}6^{0}$& $4_{5}^{\delta_{2}}6_{3}^{\alpha_{1}}$ & $4_{5}^{\delta_{1}}6_{3}^{\alpha_{2}}$ & $4_{7}^{\varepsilon}6_{5}^{\beta}$& $6_{7}^{\gamma}$& $6_{9}^{\delta_{1}}$& $6_{9}^{\delta_{2}}$& $6_{11}^{\varepsilon}$\\ \hline
 5 & $1^{\delta_{2}}5^{\alpha_{1}}$ & $3_{3}^{\delta_{1}}$ & $1^{\varepsilon}5_{3}^{\beta}$ & & $5_{5}^{\delta_{2}}$ & $5_{7}^{\varepsilon}$ & & \\ \hline
 6 & $4^{\alpha_{2}}$ & $2^{\gamma}4^{\beta}$& $6^{\alpha_{1}}$& $2^{\delta_{1}}$& $4_{3}^{\delta_{1}}$& $6_{3}^{\delta_{2}}$&$4_{3}^{\varepsilon}$ &  \\ \hline
\end{tabular}
\end{center}

\vspace{0.5cm}

The two solutions have different behaviours with respect to the $Z_{3}$ symmetry of the extended Dynkin diagram.
In fact, the boundary bound states obtained from the \lq\lq minimal\rq\rq \, solution also enjoy this symmetry,
while the ones obtained from the second solution don't. This is an indication that the \lq\lq minimal\rq\rq \,
solution should correspond to the free boundary condition or to another boundary condition which preserves the
$Z_{3}$ symmetry, while in the second case an operator which breaks this symmetry lives on the boundary.

\subsection{$E_{7}$ affine Toda field theory}

Labeling the particles with the conventions of appendix \ref{Smatrlist} and omitting the $\theta$-dependence, the
\lq\lq minimal\rq\rq \, set of reflection amplitudes is:

\begin{center}
\begin{tabular}{cclc|}
$K_{1}$&=& ${\cal K}_{1}{\cal K}_{9}{\cal K}_{17}$  \\
$K_{2}$&=& ${\cal K}_{1}{\cal K}_{7}{\cal K}_{11} {\cal K}_{53} $\\
$K_{3}$&=& ${\cal K}_{1}{\cal K}_{5}{\cal K}_{7}{\cal K}_{9}{\cal K}_{47}{\cal K}_{13} {\cal K}_{17}$ \\
$K_{4}$&=& ${\cal K}_{1}{\cal K}_{3}{\cal K}_{7}{\cal K}_{9}{\cal K}_{45}{\cal K}_{11} {\cal K}_{15}{\cal K}_{53}$ \\
$K_{5}$&=& ${\cal K}_{1}{\cal K}_{3}{\cal K}_{5}{\cal K}_{7}{\cal K}_{43}{\cal K}_{9}^{2}{\cal K}_{11}{\cal K}_{47} {\cal K}_{13}{\cal K}_{51}{\cal K}_{17}$ \\
$K_{6}$&=& ${\cal K}_{1}{\cal K}_{3}{\cal K}_{5}^{2}{\cal K}_{7}{\cal K}_{43}{\cal K}_{9}^{2}{\cal K}_{45}{\cal K}_{11}{\cal K}_{47} {\cal K}_{13}^{2}{\cal K}_{51}{\cal K}_{17}$ \\
$K_{7}$&=& ${\cal K}_{1}{\cal K}_{3}^{2}{\cal K}_{5}^{2}{\cal K}_{41}{\cal K}_{7}^{3}{\cal K}_{43}{\cal
K}_{9}^{2}{\cal K}_{45}^{2}{\cal K}_{11}^{3}{\cal K}_{47} {\cal K}_{13}{\cal K}_{49}^{2}{\cal K}_{15}^{2}{\cal
K}_{53}$ \\
\end{tabular}
\end{center}

Fixing $E_{0}=0$ and defining $M$ the mass of the lightest particle, the corresponding energy levels are:

\begin{center}
\begin{tabular}{cclc|cclc|}
$E_{\alpha}$ &=& $1.266 M =\frac{m_{5}}{2}$ && $E_{\varepsilon}$ &=&2.706 $M =\frac{m_{1}+m_{3}+m_{5}}{2}$ \\
$E_{\beta}$ &=& $1.440 M =\frac{m_{1}+m_{3}}{2}$ && $E_{\sigma}$ &=& $3.206 M =\frac{m_{1}+m_{5}+m_{6}}{2}$ \\
$E_{\gamma}$ &=& $1.940 M =\frac{m_{1}+m_{6}}{2}$ && $E_{\tau}$ &=& $3.645 M =\frac{m_{3}+m_{5}+m_{6}}{2}$ \\
$E_{\delta}$ &=& $2.380 M =\frac{m_{3}+m_{6}}{2}$ && \\
\end{tabular}
\end{center}

\vspace{0.5cm}

We list now the \lq\lq fusing angles\rq\rq \, as multiples of $\frac{i\pi}{18}$:

\begin{center}
\begin{tabular}{|c|c|c|c|c|c|c|c|c|}\hline
 $ a\setminus\mu$  &  $ 0^{+}$  &  $ \alpha^{+}$ &  $ \beta^{-}$ &  $ \gamma^{+}$& $\delta^{+}$ & $\varepsilon^{-}$ & $\sigma^{+}$ & $\tau^{+}$\\ \hline
 $1^{-}$ &   &  $8^{\beta}$ &  $2^{\delta}6^{\gamma}$ &  $4^{\varepsilon}$ &  & $2^{\tau}6_{3}^{\sigma}$ &  & \\ \hline
  $2^{+}$ & $1^{\alpha}$  & $3^{\delta}9^{\alpha}$ & $1^{\varepsilon}9^{\beta}$ & $1^{\sigma}7_{3}^{\delta}9^{\gamma}$ & $1^{\tau}5_{3}^{\sigma}9_{3}^{\delta}$ & $9_{3}^{\varepsilon}$ & $7_{5}^{\tau}9_{3}^{\sigma}$ & $9_{5}^{\tau}$\\ \hline
  $3^{-}$ & $4^{\beta}$   & $4_{3}^{\varepsilon}$ & $2^{\sigma}6_{3}^{\delta}$ & & $8_{5}^{\varepsilon}$ & $6_{5}^{\tau}$ &  & \\ \hline
  $4^{+}$ & $1^{\gamma}5^{\alpha}$ &  $1^{\sigma}7_{3}^{\gamma}9^{\alpha}$ & $5_{3}^{\varepsilon}9_{3}^{\beta}$ & $3_{3}^{\tau}5_{3}^{\sigma}9_{5}^{\gamma}$ & $5_{5}^{\tau}9^{\delta}$ & $9_{5}^{\varepsilon}$ & $9_{7}^{\sigma}$ & $9_{7}^{\tau}$\\ \hline
  $5^{+}$ & $2^{\delta}4^{\gamma}6^{\alpha}$ &  $2_{3}^{\tau}4_{3}^{\sigma}$ & $6_{5}^{\varepsilon}$ & $6_{7}^{\sigma}8_{5}^{\delta}$ & $6_{7}^{\tau}$ &  & $8_{9}^{\tau}$ & \\ \hline
  $6^{-}$ & $2^{\varepsilon}6^{\beta}$    & $6_{5}^{\varepsilon}$ & $4_{5}^{\tau}8_{5}^{\gamma}$ &  & & $8_{9}^{\sigma}$ &  & \\ \hline
  $7^{+}$ & $1^{\tau}3_{3}^{\sigma}5_{3}^{\delta}7^{\alpha}$  & $5_{7}^{\tau}9_{5}^{\alpha}$ & $7_{7}^{\varepsilon}9_{5}^{\beta}$ & $7_{9}^{\sigma}9_{7}^{\gamma}$ & $7_{11}^{\tau}9_{9}^{\delta}$ & $9_{11}^{\varepsilon}$ & $9_{13}^{\sigma}$ & $9_{15}^{\tau}$\\ \hline

  \end{tabular}
\end{center}

\vspace{0.5cm}

The signs refer to the $Z_{2}$ symmetry of the extended Dynkin diagram, choosing the convention in which the
boundary ground state is even. All the poles in the reflection amplitudes are consistent with the change of parity
induced in the boundary by the particles which create the bound states, so that this solution corresponds to a
boundary condition which preserves parity.

\vspace {0.5cm}

Starting from the second solution, we obtain the same number of boundary bound states, and energy levels:

\begin{center}
\begin{tabular}{cclc|cclc|}
$E_{\alpha}$ &=& $0.643 M =\frac{m_{2}}{2}$ && $E_{\varepsilon}$ &=& $2.494 M =\frac{m_{2}+m_{7}}{2}$ \\
$E_{\beta}$ &=& $0.985 M =\frac{m_{4}}{2}$ && $E_{\sigma}$ &=& $2.836 M =\frac{m_{4}+m_{7}}{2}$ \\
$E_{\gamma}$ &=& $1.627 M =\frac{m_{2}+m_{4}}{2}$ && $E_{\tau}$ &=& $3.478 M =\frac{m_{2}+m_{4}+m_{7}}{2}$ \\
$E_{\delta}$ &=& $1.851 M =\frac{m_{7}}{2}$ && \\
\end{tabular}
\end{center}

\vspace{0.5cm}

The \lq\lq fusing angles\rq\rq \, are:

\begin{center}
\begin{tabular}{|c|c|c|c|c|c|c|c|c|}\hline
 $ a\setminus\mu$  &  $ 0$  &  $ \alpha$ &  $ \beta$ &  $ \gamma$& $\delta$ & $\varepsilon$ & $\sigma$ & $\tau$ \\ \hline
 $1^{-}$ & $1^{\beta}5^{\alpha}9^{0}$  &  $1^{\gamma}7^{\beta}9^{\alpha}$ &  $3^{\delta}5^{\gamma}9_{3}^{\beta}$ &  $3^{\varepsilon}9_{3}^{\gamma}$ & $1^{\sigma}5_{3}^{\varepsilon}9_{3}^{\delta}$ & $1^{\tau}7_{3}^{\sigma}9_{3}^{\varepsilon}$ & $5_{3}^{\tau}9_{5}^{\sigma}$ & $9_{5}^{\tau}$\\ \hline
 $2^{+}$ & $4^{\beta}6^{\alpha}9^{0}$  & $2^{\delta}4^{\gamma}9^{\alpha}$ & $6_{3}^{\gamma}9^{\beta}$ & $2^{\sigma}8_{3}^{\delta}9^{\gamma}$ &  $4_{3}^{\sigma}6_{3}^{\varepsilon}9^{\delta}$ & $4_{3}^{\tau}9^{\varepsilon}$ & $6_{5}^{\tau}9^{\sigma}$ & $9^{\tau}$\\ \hline
 $3^{-}$ & $1^{\delta}3^{\gamma}7^{\alpha}9^{0}$   & $1^{\varepsilon}5^{\delta}9_{3}^{\alpha}$ & $1^{\sigma}7_{3}^{\gamma}9_{3}^{\beta}$ & $1^{\tau}5_{5}^{\sigma}9_{5}^{\gamma}$& $3_{3}^{\tau}7_{5}^{\varepsilon}9_{5}^{\delta}$ & $9_{7}^{\varepsilon}$ & $7_{7}^{\tau}9_{7}^{\sigma}$ & $9_{9}^{\tau}$\\ \hline
 $4^{+}$ & $2^{\delta}6^{\beta}9^{0}$ &  $2^{\varepsilon}6_{3}^{\gamma}8_{3}^{\beta}9^{\alpha}$ & $2_{3}^{\sigma}4_{3}^{\varepsilon}9^{\beta}$ & $2_{3}^{\tau}9^{\gamma}$ & $6_{5}^{\sigma}9^{\delta}$ & $6_{7}^{\tau}8_{7}^{\sigma}9^{\varepsilon}$ & $9^{\sigma}$ & $9^{\tau}$\\ \hline
 $5^{+}$ & $1^{\varepsilon}5_{3}^{\gamma}9^{0}$ &  $3_{3}^{\sigma}9_{3}^{\alpha}$ & $1^{\tau}7_{5}^{\delta}9_{3}^{\beta}$ & $7_{7}^{\varepsilon}9_{5}^{\gamma}$ & $9_{7}^{\delta}$ & $9_{9}^{\varepsilon}$ & $9_{9}^{\sigma}$ & $9_{11}^{\tau}$\\ \hline
 $6^{-}$ & $1^{\sigma}3_{3}^{\varepsilon}5_{3}^{\delta}7_{3}^{\beta}9^{0}$ & $1^{\tau}5_{5}^{\varepsilon}7_{5}^{\gamma}9_{3}^{\alpha}$ & $3_{5}^{\tau}5_{5}^{\sigma}9_{5}^{\beta}$ & $5_{7}^{\tau}9_{7}^{\gamma}$ & $7_{9}^{\sigma}9_{7}^{\delta}$ & $7_{11}^{\tau}9_{9}^{\varepsilon}$ & $9_{11}^{\sigma}$ & $9_{13}^{\tau}$\\ \hline
 $7^{+}$ & $2_{3}^{\tau}4_{5}^{\sigma}6_{5}^{\delta}8_{3}^{\alpha}9^{0}$  & $4_{7}^{\tau}6_{7}^{\varepsilon}9^{\alpha}$ & $6_{9}^{\sigma}8_{7}^{\gamma}9^{\beta}$ & $6_{11}^{\tau}9^{\gamma}$ & $8_{11}^{\varepsilon}9^{\delta}$ & $9^{\varepsilon}$ & $8_{15}^{\tau}9^{\sigma}$ & $9^{\tau}$\\ \hline

  \end{tabular}
\end{center}

\vspace{0.5cm}

In this case the boundary bound states don't have a definite parity, and this again signals the presence on the
boundary of an operator which breaks this symmetry.

\subsection{$E_{8}$ affine Toda field theory}

Labeling the particles with the conventions of appendix \ref{Smatrlist} and omitting the $\theta$-dependence, the
\lq\lq minimal\rq\rq \, set of reflection amplitudes is

\begin{center}
\begin{tabular}{l}
$K_{1}={\cal K}_{1}{\cal K}_{11}{\cal K}_{19}{\cal K}_{89}$  \\
$K_{2}={\cal K}_{1}{\cal K}_{7}{\cal K}_{11} {\cal K}_{13}{\cal K}_{77}{\cal K}_{19}{\cal K}_{23}{\cal K}_{89} $\\
$K_{3}={\cal K}_{1}{\cal K}_{3}{\cal K}_{9}{\cal K}_{11}{\cal K}_{71}{\cal K}_{13} {\cal K}_{17}{\cal K}_{19}
                     {\cal K}_{79}{\cal K}_{21}{\cal K}_{87}{\cal K}_{29}$ \\
$K_{4}={\cal K}_{1}{\cal K}_{5}{\cal K}_{7}{\cal K}_{9}{\cal K}_{11}{\cal K}_{71}{\cal K}_{13} {\cal K}_{15}
                     {\cal K}_{75}{\cal K}_{17}{\cal K}_{19}{\cal K}_{79}{\cal K}_{21}{\cal K}_{83}{\cal K}_{25}{\cal K}_{29}$ \\
$K_{5}={\cal K}_{1}{\cal K}_{3}{\cal K}_{5}{\cal K}_{7}{\cal K}_{9}{\cal K}_{69}{\cal K}_{11}^{2}{\cal K}_{71}
                     {\cal K}_{13}{\cal K}_{73}{\cal K}_{15}^{2}{\cal K}_{17}{\cal K}_{77}{\cal K}_{19}^{2}{\cal K}_{79}
                     {\cal K}_{21}{\cal K}_{81}$\\
                   $ \quad \;\;\; \times{\cal K}_{23}{\cal K}_{85}{\cal K}_{27}{\cal K}_{89} $ \\
\end{tabular}

\begin{tabular}{l}
$K_{6}={\cal K}_{1}{\cal K}_{3}{\cal K}_{5}{\cal K}_{7}{\cal K}_{67}{\cal K}_{9}^{2}{\cal K}_{11}^{2}
                     {\cal K}_{71} {\cal K}_{13}^{2}{\cal K}_{73}{\cal K}_{15}{\cal K}_{75}{\cal K}_{17}^{2}{\cal K}_{77}
                     {\cal K}_{19}{\cal K}_{79}^{2}{\cal K}_{21}^{2}$\\

                      $ \quad \;\;\; \times{\cal K}_{23}{\cal K}_{83}{\cal K}_{25}{\cal K}_{87}{\cal K}_{29}$ \\
$K_{7}={\cal K}_{1}{\cal K}_{3}{\cal K}_{5}^{2}{\cal K}_{7}^{2}{\cal K}_{67}{\cal K}_{9}^{2}{\cal K}_{69}
                     {\cal K}_{11}^{2}
                     {\cal K}_{71}^{2} {\cal K}_{13}^{3}{\cal K}_{73}{\cal K}_{15}^{2}{\cal K}_{75}^{2}{\cal K}_{17}^{3}
                     {\cal K}_{77}
                     {\cal K}_{19}^{2}{\cal K}_{79}^{2}$\\

                     $ \quad \;\;\; \times {\cal K}_{21}^{2}{\cal K}_{81}{\cal K}_{23}{\cal K}_{83}^{2}
                     {\cal K}_{25}^{2}{\cal K}_{87}
                     {\cal K}_{29}$ \\
$K_{8}={\cal K}_{1}{\cal K}_{3}^{2}{\cal K}_{5}^{2}{\cal K}_{65}{\cal K}_{7}^{3}{\cal K}_{67}{\cal
K}_{9}^{3}{\cal K}_{69}^{2}{\cal K}_{11}^{4}
                     {\cal K}_{71}^{2} {\cal K}_{13}^{3}{\cal K}_{73}^{3}{\cal K}_{15}^{4}{\cal K}_{75}^{2}{\cal K}_{17}^{3}
                     {\cal K}_{77}^{3}
                     {\cal K}_{19}^{4}$\\

                     $ \quad \;\;\; \times {\cal K}_{79}^{2}{\cal K}_{21}^{2}{\cal K}_{81}^{3}{\cal K}_{23}^{3}{\cal K}_{83}
                     {\cal K}_{25}{\cal K}_{85}^{2}{\cal K}_{27}^{2}{\cal K}_{89}$ \\
\end{tabular}
\end{center}

\vspace{1cm}

Fixing $E_{0}=0$ and defining $M$ the mass of the lightest particle, the corresponding energy levels are:

\begin{center}
\begin{tabular}{cclc|cclc}
$E_{\alpha}$ &=& $0.9945 M =\frac{m_{3}}{2}$ && $E_{\nu}$ &=& $3.1480 M =\frac{m_{4}+m_{7}}{2}$ \\
$E_{\beta}$ &=& $1.2024 M =\frac{m_{4}}{2}$ && $E_{\rho}$ &=& $3.5547 M =\frac{m_{6}+m_{7}}{2}$ \\
$E_{\gamma}$ &=& $1.6092 M =\frac{m_{6}}{2}$ && $E_{\sigma}$ &=& $3.8061 M =\frac{m_{3}+m_{4}+m_{6}}{2}$ \\
$E_{\delta}$ &=& $1.9456 M =\frac{m_{7}}{2}$ && $E_{\tau}$ &=& $4.1425 M =\frac{m_{3}+m_{4}+m_{7}}{2}$ \\
$E_{\varepsilon}$ &=& $2.1969 M =\frac{m_{3}+m_{4}}{2}$ && $E_{\psi}$ &=& $4.5493 M =\frac{m_{3}+m_{6}+m_{7}}{2}$ \\
$E_{\kappa}$ &=& $2.6037 M =\frac{m_{3}+m_{6}}{2}$ && $E_{\omega}$ &=& $4.7572 M =\frac{m_{4}+m_{6}+m_{7}}{2}$ \\
$E_{\lambda}$ &=& $2.8116 M =\frac{m_{4}+m_{6}}{2}$ && $E_{\phi}$ &=& $5.7517 M =\frac{m_{3}+m_{4}+m_{6}+m_{7}}{2}$ \\
$E_{\mu}$ &=& $2.9401 M =\frac{m_{3}+m_{7}}{2}$ && \\
\end{tabular}
\end{center}

\vspace{1cm}

We list now the \lq\lq fusing angles\rq\rq \, as multiples of $\frac{i\pi}{30}$:

\begin{center}

\begin{tabular}{|c|c|c|c|c|}\hline
 $ a\setminus\mu$  &  $ 0$  &  $ \alpha$ &  $ \beta$ &  $ \gamma$   \\ \hline
1 & $1^{\alpha}$ & $3^{\delta}13^{\beta}$ & $1^{\varepsilon}7^{\delta}11^{\gamma}15^{\beta}$&
$1^{\kappa}9^{\varepsilon}15^{\gamma}$ \\ \hline

2 & $1^{\gamma}7^{\beta}$& $1^{\kappa}7_{3}^{\varepsilon}9^{\delta}15^{\alpha}$ &
$1^{\lambda}5^{\kappa}15^{\beta}$ & $3^{\nu}7_{3}^{\lambda}13_{3}^{\delta}15^{\gamma}$   \\ \hline

3 & $2^{\delta}6^{\gamma}10^{\alpha}$ & $2_{3}^{\mu}4^{\lambda}6^{\kappa}12_{3}^{\gamma}14^{\beta}$ &
$2^{\nu}6_{3}^{\lambda}10_{3}^{\varepsilon}$ & $2^{\rho}8_{3}^{\mu}10_{3}^{\kappa}$    \\
\hline

4 & $4^{\varepsilon}6^{\delta}8^{\gamma}10^{\beta}$ & $6_{3}^{\mu}8_{3}^{\kappa}10_{3}^{\varepsilon}$ &
$2^{\rho}6_{3}^{\nu}8_{3}^{\lambda}12_{3}^{\delta}$ & $4_{3}^{\sigma}6_{3}^{\rho}10_{5}^{\lambda}$  \\
\hline

5 & $1^{\mu}3^{\lambda}7^{\varepsilon}11^{\beta}$&
$3_{3}^{\sigma}5_{3}^{\rho}11_{5}^{\varepsilon}13_{3}^{\gamma}15^{\alpha}$ & $1^{\tau}9_{5}^{\mu}15_{3}^{\beta}$ &
$1^{\psi}7_{5}^{\sigma}11_{5}^{\lambda}15_{5}^{\gamma}$     \\
\hline

6 & $2^{\nu}4^{\mu}6_{3}^{\kappa}10^{\gamma}12^{\alpha}$ & $2^{\tau}8_{5}^{\nu}10_{3}^{\kappa}$ &
$4_{3}^{\tau}6_{5}^{\sigma}10_{5}^{\lambda}12_{5}^{\varepsilon}$ &
$2_{3}^{\omega}4_{3}^{\psi}12_{7}^{\kappa}14_{5}^{\delta}$
\\ \hline

7 & $2^{\sigma}4_{3}^{\rho}6_{3}^{\nu}8_{3}^{\kappa}10_{3}^{\delta}12^{\beta}$&
$4_{5}^{\psi}6_{5}^{\tau}10_{7}^{\mu}12_{5}^{\varepsilon}$ &
$4_{5}^{\omega}8_{7}^{\sigma}10_{7}^{\nu}14_{5}^{\gamma}$ &
$6_{7}^{\omega}10_{9}^{\rho}12_{7}^{\lambda}$    \\
\hline

8 & $1^{\omega}3_{3}^{\psi}5_{5}^{\tau}7_{5}^{\rho}9_{5}^{\lambda}11_{3}^{\delta}13^{\alpha}$ &
$1^{\phi}7_{9}^{\psi}9_{9}^{\sigma}11_{7}^{\mu}15_{5}^{\alpha}$ &
$3^{\phi}7_{9}^{\omega}11_{9}^{\nu}13_{7}^{\varepsilon}15_{5}^{\beta}$ &
$5_{9}^{\phi}11_{11}^{\rho}13_{9}^{\kappa}15_{7}^{\gamma}$    \\
\hline
  \end{tabular}

\begin{tabular}{|c|c|c|c|c|c|}\hline
 $ a\setminus\mu$    & $\delta$ &$\varepsilon$  & $\kappa$ & $\lambda$ & $\mu$ \\ \hline
1 & $1^{\mu}5^{\lambda}15^{\delta}$ & $3^{\nu}7^{\mu}11_{3}^{\kappa}15^{\varepsilon}$ & $3^{\rho}13_{3}^{\lambda}15^{\kappa}$& $1^{\sigma}7_{3}^{\rho}15_{3}^{\lambda}$ & $5^{\sigma}13_{3}^{\nu}15^{\mu}$\\
\hline

2 &
$1^{\rho}7_{3}^{\nu}11_{3}^{\kappa}15_{3}^{\delta}$ &$1^{\sigma}9_{3}^{\nu}15_{3}^{\varepsilon}$& $3^{\tau}7_{5}^{\sigma}9_{3}^{\rho}15_{3}^{\kappa}$ & $13_{5}^{\nu}15_{3}^{\lambda}$ &$1^{\psi}7_{5}^{\tau}15_{5}^{\mu}$  \\
\hline

3 & $6_{3}^{\rho}10_{5}^{\mu}$ & $2_{3}^{\tau}6_{3}^{\sigma}12_{5}^{\lambda}$ & $2_{3}^{\psi}14_{5}^{\lambda}$ & $2^{\omega}8_{5}^{\tau}10_{5}^{\sigma}$ & $4_{3}^{\omega}6_{3}^{\psi}12_{7}^{\rho}14_{5}^{\nu}$  \\
\hline

4 &
$4_{3}^{\tau}8_{5}^{\rho}10_{5}^{\nu}14_{5}^{\varepsilon}$  & $2^{\psi}6_{5}^{\tau}8_{5}^{\sigma}12_{5}^{\mu}$& $6_{5}^{\psi}10_{7}^{\sigma}$& $6_{5}^{\omega}12_{7}^{\rho}$ & $8_{7}^{\psi}10_{7}^{\tau}$ \\
\hline

5 & $3_{3}^{\omega}7_{5}^{\tau}11_{7}^{\nu}15_{5}^{\delta}$ & $5_{5}^{\omega}13_{7}^{\lambda}15_{5}^{\varepsilon}$& $11_{9}^{\sigma}15_{7}^{\kappa}$ & $1^{\phi}9_{9}^{\psi}15_{9}^{\lambda}$ & $3_{5}^{\phi}11_{11}^{\tau}13_{9}^{\rho}15_{7}^{\mu}$   \\
\hline

6 & $6_{7}^{\psi}10_{7}^{\rho}12_{7}^{\mu}$ & $10_{7}^{\sigma}$& $2_{3}^{\phi}8_{9}^{\omega}14_{9}^{\mu}$&
$4_{5}^{\phi}12_{11}^{\sigma}14_{9}^{\nu}$ & $10_{9}^{\psi}$
\\ \hline

7  & $2_{3}^{\phi}8_{9}^{\psi}12_{9}^{\nu}$ & $4_{7}^{\phi}10_{11}^{\tau}14_{9}^{\kappa}$ & $6_{9}^{\phi}10_{13}^{\psi}12_{11}^{\sigma}$ & $10_{13}^{\omega}$ & $12_{13}^{\tau}$  \\
\hline

8 & $9_{13}^{\omega}13_{11}^{\mu}15_{9}^{\delta}$ & $7_{13}^{\phi}11_{13}^{\tau}15_{11}^{\varepsilon}$& $11_{15}^{\psi}15_{13}^{\kappa}$ & $11_{17}^{\omega}13_{15}^{\sigma}15_{13}^{\lambda}$ & $9_{17}^{\phi}15_{15}^{\mu}$   \\
\hline
  \end{tabular}

\begin{tabular}{|c|c|c|c|c|c|c|c|}\hline
 $ a\setminus\mu$ & $\nu$ & $\rho$ & $\sigma$ & $\tau$ & $\psi$ & $\omega$ & $\phi$\\ \hline
1 & $1^{\tau}11_{3}^{\rho}15_{3}^{\nu}$ & $1^{\psi}9_{3}^{\tau}15_{3}^{\rho}$& $3^{\omega}7_{3}^{\psi}15_{3}^{\sigma}$ & $11_{5}^{\psi}15_{3}^{\tau}$ & $13_{5}^{\omega}15_{3}^{\psi}$ & $1^{\phi}15_{5}^{\omega}$ & $15_{5}^{\phi}$\\
\hline

2 & $1^{\omega}5_{3}^{\psi}11_{5}^{\sigma}15_{5}^{\nu}$ & $7_{5}^{\omega}15_{5}^{\rho}$& $9_{5}^{\omega}13_{7}^{\tau}15_{5}^{\sigma}$ & $1^{\phi}15_{7}^{\tau}$ & $7_{7}^{\phi}15_{7}^{\psi}$ & $15_{7}^{\omega}$ & $15_{9}^{\phi}$\\
\hline

3 &  $6_{5}^{\omega}10_{7}^{\tau}$ & $10_{7}^{\psi}$ & $2_{3}^{\phi}$ & $6_{5}^{\phi}12_{9}^{\omega}$ &  $14_{9}^{\omega}$ & $10_{9}^{\phi}$ &\\
\hline

4 &  $8_{7}^{\omega}$ & $4_{5}^{\phi}10_{9}^{\omega}14_{9}^{\sigma}$& $6_{7}^{\phi}12_{9}^{\psi}$& $8_{9}^{\phi}$ & $10_{9}^{\phi}$ & & \\
\hline

5 &  $15_{9}^{\nu}$ & $7_{9}^{\phi}11_{11}^{\omega}15_{11}^{\rho}$& $15_{11}^{\sigma}$& $13_{13}^{\omega}15_{11}^{\tau}$ & $11_{15}^{\phi}15_{13}^{\psi}$ & $15_{15}^{\omega}$ & $15_{17}^{\phi}$ \\
\hline

6 & $6_{9}^{\phi}10_{11}^{\omega}12_{11}^{\tau}$ & $12_{13}^{\psi}$ & $14_{13}^{\tau}$ & $10_{13}^{\phi}$ &&
$12_{17}^{\phi}$ &
\\ \hline

7 & $8_{13}^{\phi}14_{13}^{\rho}$  & $12_{15}^{\omega}$& $10_{17}^{\phi}$ & $14_{17}^{\psi}$ &  $12_{19}^{\phi}$ & &   \\
\hline

8 & $13_{17}^{\tau}15_{15}^{\nu}$ & $13_{19}^{\psi}15_{17}^{\rho}$ & $11_{21}^{\phi}15_{19}^{\sigma}$ & $15_{21}^{\tau}$ &  $15_{23}^{\psi}$ & $13_{25}^{\phi}15_{23}^{\omega}$& $15_{29}^{\phi}$ \\
\hline
  \end{tabular}

\end{center}

\vspace{1.5cm}

Starting from the second solution, we obtain the same number of boundary bound states, and energy levels:

\begin{center}
\begin{tabular}{cclc|cclc}
$E_{\alpha}$ &=& $0.5000 M =\frac{m_{1}}{2}$ && $E_{\nu}$ &=& $2.8917 M =\frac{m_{1}+m_{8}}{2} $ \\
$E_{\beta}$ &=& $0.8090 M =\frac{m_{2}}{2}$ && $E_{\rho}$ &=& $3.2007 M =\frac{m_{2}+m_{8}}{2}$ \\
$E_{\gamma}$ &=& $1.3090 M =\frac{m_{1}+m_{2}}{2}$ && $E_{\sigma}$ &=& $3.7007 M =\frac{m_{1}+m_{2}+m_{8}}{2}$ \\
$E_{\delta}$ &=& $1.4781 M =\frac{m_{5}}{2}$ && $E_{\tau}$ &=& $3.8698 M =\frac{m_{5}+m_{8}}{2}$ \\
$E_{\varepsilon}$ &=& $1.9781 M =\frac{m_{1}+m_{5}}{2}$ && $E_{\psi}$ &=& $4.3698 M =\frac{m_{1}+m_{5}+m_{8}}{2}$ \\
$E_{\kappa}$ &=& $2.2872 M =\frac{m_{2}+m_{5}}{2}$ && $E_{\omega}$ &=& $4.6789 M =\frac{m_{2}+m_{5}+m_{8}}{2}$ \\
$E_{\lambda}$ &=& $2.3917 M =\frac{m_{8}}{2}$ && $E_{\phi}$ &=& $5.1789 M =\frac{m_{1}+m_{2}+m_{5}+m_{8}}{2}$ \\
$E_{\mu}$ &=& $2.7872 M =\frac{m_{1}+m_{2}+m_{5}}{2}$ && \\
\end{tabular}
\end{center}

\vspace{1.5cm}

The \lq\lq fusing angles\rq\rq \, are:

\begin{center}

\begin{tabular}{|c|c|c|c|}\hline
 $ a\setminus\mu$    &$0$  & $\alpha$ & $\beta$ \\ \hline
1 & $6^{\beta}10^{\alpha}15^{0}$& $2^{\delta}6^{\gamma}12^{\beta}15^{\alpha}$ & $8^{\delta}10^{\gamma}15^{\beta}$ \\
\hline

2 & $4^{\delta}6^{\gamma}10^{\beta}12^{\alpha}15^{0}$ & $4^{\varepsilon}10^{\gamma}15^{\alpha}$ & $2^{\lambda}4^{\kappa}12_{3}^{\gamma}15^{\beta}$ \\
\hline

3  & $1^{\varepsilon}7^{\delta}11^{\beta}15^{0}$ & $3^{\lambda}7^{\varepsilon}11_{3}^{\gamma}15^{\alpha}$ & $1^{\mu}7_{3}^{\kappa}9_{3}^{\varepsilon}15_{3}^{\beta}$ \\
\hline

4 & $1^{\lambda}3^{\kappa}13^{\alpha}15^{0}$ & $1^{\nu}3^{\mu}7_{3}^{\kappa}11_{3}^{\delta}15_{3}^{\alpha}$ & $1^{\rho}5_{3}^{\nu}13_{3}^{\gamma}15_{3}^{\beta}$ \\
\hline

5 & $2^{\nu}6_{3}^{\lambda}8_{3}^{\varepsilon}10_{3}^{\delta}15^{0}$& $4_{3}^{\rho}6_{3}^{\nu}10_{5}^{\varepsilon}14_{3}^{\beta}15^{\alpha}$ & $2^{\sigma}6_{5}^{\rho}8_{5}^{\mu}10_{5}^{\kappa}15^{\beta}$ \\
\hline

6  &$1^{\rho}5_{3}^{\mu}7_{3}^{\lambda}11_{3}^{\gamma}15^{0}$&
$1^{\sigma}7_{5}^{\nu}9_{5}^{\lambda}15_{3}^{\alpha}$ & $3_{3}^{\tau}7_{5}^{\rho}13_{5}^{\delta}15_{3}^{\beta}$ \\
\hline

7   & $1^{\tau}3_{3}^{\sigma}7_{5}^{\nu}9_{5}^{\kappa}13_{3}^{\beta}15^{0}$& $1^{\psi}5_{5}^{\tau}9_{7}^{\mu}13_{5}^{\gamma}15_{3}^{\alpha}$ & $1^{\omega}7_{7}^{\sigma}11_{7}^{\lambda}15_{5}^{\beta}$  \\
\hline

8  & $2_{3}^{\omega}4_{5}^{\psi}6_{7}^{\tau}8_{7}^{\rho}10_{7}^{\lambda}12_{5}^{\delta}14_{3}^{\alpha}15^{0}$ & $2_{3}^{\phi}6_{9}^{\psi}8_{9}^{\sigma}10_{9}^{\nu}12_{7}^{\varepsilon}15^{\alpha}$ & $4_{7}^{\phi}6_{9}^{\omega}10_{11}^{\rho}12_{9}^{\kappa}14_{7}^{\gamma}15^{\beta}$  \\
\hline
  \end{tabular}

\begin{tabular}{|c|c|c|c|c|}\hline
 $ a\setminus\mu$   & $\gamma$ &$\delta$  & $\varepsilon$ & $\kappa$  \\ \hline
1 & $2^{\kappa}8^{\varepsilon}15^{\gamma}$ & $4^{\lambda}6^{\kappa}10_{3}^{\varepsilon}15^{\delta}$ & $4^{\nu}6^{\mu}12_{3}^{\kappa}15^{\varepsilon}$ & $4^{\rho}10_{3}^{\mu}14_{3}^{\lambda}15^{\kappa}$  \\
\hline

2 & $2^{\nu}4^{\mu}8_{3}^{\lambda}14_{3}^{\delta}15^{\gamma}$ & $6_{3}^{\mu}10_{3}^{\kappa}12_{3}^{\varepsilon}15^{\delta}$ & $10_{3}^{\mu}15^{\varepsilon}$ & $2^{\tau}12_{5}^{\mu}15^{\kappa}$  \\
\hline

3 & $3^{\rho}7_{3}^{\mu}15_{3}^{\gamma}$ & $5_{3}^{\rho}11_{5}^{\kappa}15_{3}^{\delta}$ & $3_{3}^{\tau}5_{3}^{\sigma}11_{7}^{\mu}13_{5}^{\lambda}15_{3}^{\varepsilon}$ & $15_{5}^{\kappa}$  \\
\hline

4 & $1^{\sigma}11_{5}^{\kappa}15_{5}^{\gamma}$ & $1^{\tau}9_{5}^{\nu}13_{5}^{\varepsilon}15_{5}^{\delta}$ & $1^{\psi}15_{7}^{\varepsilon}$ & $1^{\omega}5_{5}^{\psi}9_{7}^{\sigma}13_{7}^{\mu}15_{7}^{\kappa}$  \\
\hline

5 & $6_{5}^{\sigma}10_{7}^{\mu}15^{\gamma}$ & $2_{3}^{\psi}6_{5}^{\tau}12_{7}^{\lambda}15^{\delta}$ & $4_{5}^{\omega}6_{5}^{\psi}12_{9}^{\nu}14_{7}^{\kappa}15^{\varepsilon}$ & $2_{3}^{\phi}6_{7}^{\omega}12_{9}^{\rho}15^{\kappa}$  \\
\hline

6 & $3_{3}^{\psi}7_{7}^{\sigma}9_{7}^{\rho}13_{7}^{\varepsilon}15_{5}^{\gamma}$ &
$1^{\omega}7_{7}^{\tau}11_{7}^{\mu}15_{7}^{\delta}$ & $1^{\phi}7_{9}^{\psi}9_{9}^{\tau}15_{9}^{\varepsilon}$ &
$7_{9}^{\omega}15_{9}^{\kappa}$
\\ \hline

7 & $1^{\phi}5_{7}^{\omega}11_{9}^{\nu}15_{7}^{\gamma}$ & $3_{5}^{\phi}7_{9}^{\psi}13_{9}^{\kappa}15_{7}^{\delta}$ & $13_{11}^{\mu}15_{9}^{\varepsilon}$ & $7_{11}^{\phi}11_{13}^{\tau}15_{11}^{\kappa}$  \\
\hline

8 & $6_{11}^{\phi}10_{13}^{\sigma}12_{11}^{\mu}15^{\gamma}$ & $8_{13}^{\omega}10_{13}^{\tau}14_{11}^{\varepsilon}15^{\delta}$ & $8_{15}^{\phi}10_{15}^{\psi}15^{\varepsilon}$ & $10_{17}^{\omega}14_{15}^{\mu}15^{\kappa}$ \\
\hline

\end{tabular}

\begin{tabular}{|c|c|c|c|c|}\hline
 $ a\setminus\mu$  & $\lambda$ & \hspace{1cm} $\mu$ \hspace{1cm} & \hspace{1cm} $\nu$ \hspace{1cm} & \hspace{1cm} $\rho$ \hspace{1cm}   \\ \hline
1 & $6_{3}^{\rho}10_{3}^{\nu}15^{\lambda}$ & $4^{\sigma}14_{3}^{\nu}15^{\mu}$ & $2^{\tau}6_{3}^{\sigma}12_{3}^{\rho}15^{\nu}$ & $8_{3}^{\tau}10_{3}^{\sigma}15^{\rho}$  \\
\hline

2 & $4_{3}^{\tau}6_{3}^{\sigma}10_{5}^{\rho}12_{5}^{\nu}15^{\lambda}$ & $2^{\psi}8_{5}^{\tau}15^{\mu}$ & $4_{3}^{\psi}10_{5}^{\sigma}15^{\nu}$ & $4_{3}^{\omega}12_{7}^{\sigma}15^{\rho}$   \\
\hline

3 & $1^{\psi}7_{5}^{\tau}11_{5}^{\rho}15_{7}^{\lambda}$ & $3_{3}^{\omega}13_{7}^{\rho}15_{5}^{\mu}$ & $7_{5}^{\psi}11_{7}^{\sigma}15_{7}^{\nu}$ & $1^{\phi}7_{7}^{\omega}9_{7}^{\psi}15_{9}^{\rho}$  \\
\hline

4 & $3_{3}^{\omega}13_{7}^{\nu}15_{7}^{\lambda}$ & $1^{\phi}15_{9}^{\mu}$ & $3_{3}^{\phi}7_{7}^{\omega}11_{9}^{\tau}15_{9}^{\nu}$ & $13_{9}^{\sigma}15_{9}^{\rho}$  \\
\hline

5 & $8_{9}^{\psi}10_{9}^{\tau}15^{\lambda}$ & $6_{7}^{\phi}12_{11}^{\sigma}15^{\mu}$ & $10_{11}^{\psi}14_{11}^{\rho}15^{\nu}$ & $8_{11}^{\phi}10_{11}^{\omega}15^{\rho}$  \\
\hline

6 & $5_{7}^{\phi}11_{11}^{\sigma}15_{9}^{\lambda}$ &
$7_{11}^{\phi}9_{11}^{\omega}15_{11}^{\mu}$ & $15_{11}^{\nu}$ & $13_{13}^{\tau}15_{11}^{\rho}$   \\
\hline

7 & $9_{13}^{\omega}13_{13}^{\rho}15_{11}^{\lambda}$ & $11_{15}^{\psi}15_{13}^{\mu}$ & $9_{15}^{\phi}13_{15}^{\sigma}15_{13}^{\nu}$ & $15_{15}^{\rho}$ \\
\hline

8 & $12_{17}^{\tau}14_{15}^{\nu}15^{\lambda}$ & $10_{19}^{\phi}15^{\mu}$ & $12_{19}^{\psi}15^{\nu}$ & $12_{21}^{\omega}15^{\rho}$  \\
\hline

\end{tabular}

\begin{tabular}{|c|c|c|c|c|c|}\hline
 $ a\setminus\mu$ & \hspace{1cm} $\sigma$ \hspace{1cm} & \hspace{5mm} $\tau$ \hspace{5mm} & \hspace{5mm} $\psi$ \hspace{5mm} & \hspace{5mm} $\omega$ \hspace{5mm} & \hspace{0.5cm} $\phi$ \hspace{0.5cm} \\ \hline
1 & $2^{\omega}8_{3}^{\psi}15^{\sigma}$ & $6_{3}^{\omega}10_{5}^{\psi}15^{\tau}$ & $6_{3}^{\phi}12_{5}^{\omega}15^{\psi}$ & $10_{5}^{\phi}15^{\omega}$ & $15^{\phi}$ \\
\hline

2 & $4_{3}^{\phi}14_{7}^{\tau}15^{\sigma}$ & $6_{5}^{\phi}10_{7}^{\omega}12_{7}^{\psi}15^{\tau}$ & $10_{7}^{\phi}15^{\psi}$ & $12_{9}^{\phi}15^{\omega}$ &$15^{\phi}$  \\
\hline

3  & $7_{7}^{\phi}15_{9}^{\sigma}$ & $11_{9}^{\omega}15_{9}^{\tau}$ & $11_{11}^{\phi}15_{9}^{\psi}$ & $15_{11}^{\omega}$ & $15_{11}^{\phi}$ \\
\hline

4 & $11_{11}^{\omega}15_{11}^{\sigma}$ & $13_{11}^{\psi}15_{11}^{\tau}$ & $15_{13}^{\psi}$ & $13_{13}^{\phi}15_{13}^{\omega}$ & $15_{15}^{\phi}$ \\
\hline

5  & $10_{13}^{\phi}15^{\sigma}$ & $15^{\tau}$ & $14_{15}^{\omega}15^{\psi}$ &$15^{\omega}$ & $15^{\phi}$ \\
\hline

6  & $13_{15}^{\psi}15_{13}^{\sigma}$ & $11_{15}^{\phi}15_{15}^{\tau}$ & $15_{17}^{\psi}$ & $15_{17}^{\omega}$ & $15_{19}^{\phi}$ \\
\hline

7 & $15_{17}^{\sigma}$ & $13_{19}^{\omega}15_{17}^{\tau}$ & $13_{21}^{\phi}15_{19}^{\psi}$ & $15_{21}^{\omega}$ & $15_{23}^{\phi}$  \\
\hline

8  & $12_{23}^{\phi}15^{\sigma}$ & $14_{23}^{\psi}15^{\tau}$ &$15^{\psi}$ & $14_{27}^{\phi}15^{\omega}$ & $15^{\phi}$\\
\hline

\end{tabular}
\end{center}

\vspace{1cm}

\section{Pole interpretation}

In the procedure of applying the bootstrap equation (\ref{boundst}) we have considered all the odd order poles
with positive residue. This property, however, is necessary but not sufficient for the creation of a boundary
bound state. In fact, Dorey, Tateo and Watts (\cite{tateo}) have proposed two kinds of mechanisms which can
describe some of the poles without involving new boundary bound states.

The first one is a \lq $u$-channel\rq \, mechanism, and it is invoked when an excited state reflection factor
$K_{a}^{\beta}$ has a pole at the same place $\theta_{a}=i\eta_{a\alpha}^{\beta}$ as the pole in $K_{a}^{\alpha}$
which generated $\beta$.

\vspace{5.5cm}

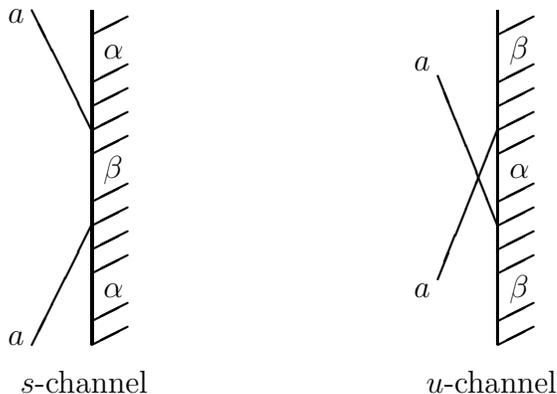
\begin{figure}[h]
\setlength{\unitlength}{0.0125in}
\begin{picture}(40,0)(60,470)

\put(180,470){$s$-channel}
\thicklines \put(210,490){\line(0,1){140}}

\put(210,490){\line(2,1){15}} \put(210,500){\line(2,1){15}} \put(210,520){\line(2,1){15}}
\put(210,530){\line(2,1){15}} \put(210,540){\line(2,1){15}} \put(210,550){\line(2,1){15}}
 \put(210,570){\line(2,1){15}} \put(210,580){\line(2,1){15}}
\put(210,590){\line(2,1){15}} \put(210,600){\line(2,1){15}} \put(210,620){\line(2,1){15}}

\put(210,580){\line(-1,2){25}} \put(210,540){\line(-1,-2){25}}

\put(215,510){$\alpha$}\put(215,560){$\beta$}\put(215,610){$\alpha$}

\put(175,490){$a$} \put(175,625){$a$}

\put(350,470){$u$-channel}
\thicklines \put(380,490){\line(0,1){140}}

\put(380,490){\line(2,1){15}} \put(380,500){\line(2,1){15}} \put(380,520){\line(2,1){15}}
\put(380,530){\line(2,1){15}} \put(380,540){\line(2,1){15}} \put(380,550){\line(2,1){15}}
 \put(380,570){\line(2,1){15}} \put(380,580){\line(2,1){15}}
\put(380,590){\line(2,1){15}} \put(380,600){\line(2,1){15}} \put(380,620){\line(2,1){15}}

\put(380,580){\line(-2,-5){25}} \put(380,540){\line(-2,5){25}}

\put(385,510){$\beta$}\put(385,560){$\alpha$}\put(385,610){$\beta$}

\put(345,510){$a$} \put(345,605){$a$}
\end{picture}
 \caption{First mechanism}
 \end{figure}

The pole in exam doesn't excite the boundary to a new bound state, but simply corresponds to going back from
$\beta$ to $\alpha$. This rule, reasonable but non properly founded, has a clear explanation in the case of a
theory with defect (\cite{defect}), where not just reflection but also transmission is allowed. In that case it
is shown with an explicit example how a pole at $\theta=i\eta$ with this property can be neglected, because,
although it has positive residue in the reflection amplitude, its residue in the transmission one is negative. At
the same time, both amplitudes have a positive residue pole at $\theta=i(\pi-\eta)$, which exactly corresponds to
going back to the original boundary state. Unfortunately, integrable defect theories seem to apply only to
quasi-free systems.

The other possibility is a boundary generalization of the Coleman-Thun mechanism: in some cases a pole can be
described by on-shell diagrams, different from the one in Figure \ref{figexcit}, which correspond to multiple
rescattering processes. These methods have also been applied in \cite{delius} and \cite{dorey}, with the
hypothesis that, if an alternative diagram can be drawn, then the pole in exam doesn't correspond to the creation
of a boundary bound state.

In general, the order of a certain diagram is given by $P-2L$, where $P$ and $L$ are respectively the number of
propagators and loops. When dealing with a boundary, however, vertex factors can also be given by reflection
amplitudes. If these amplitudes have poles (or zeroes) at the rapidities dictated by the on-shell condition, then
their effect will be to raise (or lower) the order of the diagram.

As we have seen, the reflection amplitudes of the form (\ref{K}) never have coupling-independent zeroes in the
physical strip, hence the order of a given diagram could only be raised by their insertions. In this case, the
only two diagrams which can describe a first order pole are:

\vspace{5cm}

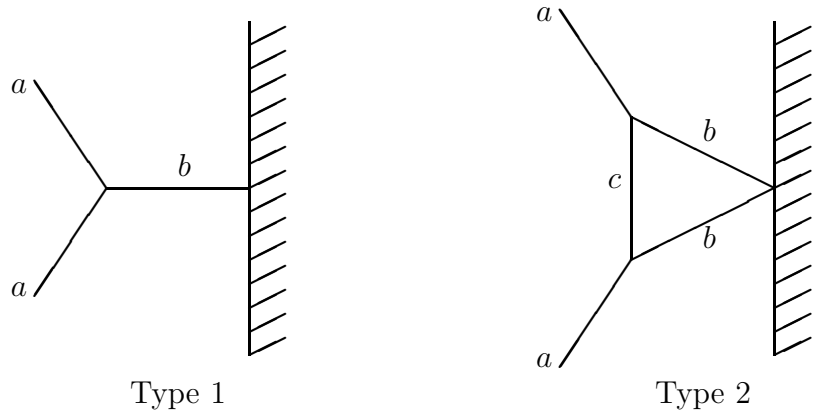
\begin{figure}[h]
\setlength{\unitlength}{0.0125in}
\begin{picture}(40,0)(60,470)

\put(190,470){Type 1}
\thicklines \put(240,490){\line(0,1){140}}

\put(240,490){\line(2,1){15}} \put(240,500){\line(2,1){15}} \put(240,510){\line(2,1){15}}
\put(240,520){\line(2,1){15}} \put(240,530){\line(2,1){15}} \put(240,540){\line(2,1){15}}
\put(240,550){\line(2,1){15}} \put(240,560){\line(2,1){15}} \put(240,570){\line(2,1){15}}
\put(240,580){\line(2,1){15}} \put(240,590){\line(2,1){15}} \put(240,600){\line(2,1){15}}
\put(240,610){\line(2,1){15}} \put(240,620){\line(2,1){15}}

\put(240,560){\line(-1,0){60}} \put(180,560){\line(-2,3){30}} \put(180,560){\line(-2,-3){30}}

\put(210,565){$b$} \put(140,515){$a$} \put(140,600){$a$}

\put(410,470){Type 2}
\thicklines \put(460,490){\line(0,1){140}}

\put(460,490){\line(2,1){15}} \put(460,500){\line(2,1){15}} \put(460,510){\line(2,1){15}}
\put(460,520){\line(2,1){15}} \put(460,530){\line(2,1){15}} \put(460,540){\line(2,1){15}}
\put(460,550){\line(2,1){15}} \put(460,560){\line(2,1){15}} \put(460,570){\line(2,1){15}}
\put(460,580){\line(2,1){15}} \put(460,590){\line(2,1){15}} \put(460,600){\line(2,1){15}}
\put(460,610){\line(2,1){15}} \put(460,620){\line(2,1){15}}

\put(460,560){\line(-2,1){60}} \put(460,560){\line(-2,-1){60}}

\put(400,530){\line(0,1){60}}

\put(400,590){\line(-2,3){30}} \put(400,530){\line(-2,-3){30}}

\put(390,560){$c$} \put(360,485){$a$} \put(360,630){$a$} \put(430,580){$b$} \put(430,535){$b$}

\end{picture}
 \caption{First order diagrams}
 \end{figure}

If we call $\eta_{a}$ the pole we are interested in, we can see that it is possible to draw Type 1 diagram
(already introduced in \cite{ghoszam}) if there is a particle $b$ such that $i u_{ab}^{a}=\eta_{a}+i
\frac{\pi}{2}$ and such that the corresponding reflection amplitude $K_{b}(\theta)$ has an odd order pole with
positive residue at $\theta=i\frac{\pi}{2}$. Type 2 diagram can be drawn if there are two particles $b$ and $c$
such that $u_{bc}^{a}<\frac{\pi}{2}$ and $\eta_{a}=i \left(u_{bc}^{a}+u_{ab}^{c}-\pi\right)$; the amplitude
$K_{b}(\theta)$ has to be evaluated at $\eta_{b}=i u_{bc}^{a}$.

A second order diagram, which will describe third or higher order poles of $K_{a}^{\alpha}$, is:

\vspace{4.2cm}

\begin{figure}[h]
\setlength{\unitlength}{0.0125in}
\begin{picture}(40,0)(60,470)
\thicklines \put(330,460){\line(0,1){160}}

\put(330,460){\line(2,1){15}} \put(330,480){\line(2,1){15}} \put(330,490){\line(2,1){15}}
\put(330,500){\line(2,1){15}} \put(330,510){\line(2,1){15}} \put(330,520){\line(2,1){15}}
\put(330,550){\line(2,1){15}} \put(330,560){\line(2,1){15}}

\put(330,570){\line(2,1){15}} \put(330,580){\line(2,1){15}} \put(330,590){\line(2,1){15}}
\put(330,610){\line(2,1){15}}

\put(330,540){\line(-2,5){14}} \put(330,540){\line(-2,-5){14}}

\put(316,575){\line(2,3){14}} \put(316,505){\line(2,-3){14}}

\put(316,575){\line(-3,1){40}} \put(316,505){\line(-3,-1){40}}

\put(335,535){$\beta$} \put(335,470){$\alpha$} \put(335,600){$\alpha$}

\put(265,588){$a$} \put(265,490){$a$} \put(312,555){$b$} \put(313,520){$b$} \put(318,590){$c$} \put(318,483){$c$}
\end{picture}
 \caption{Type 3 (second order)}
 \end{figure}
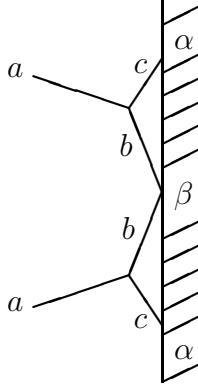

This diagram can be drawn if there are $b$ and $c$ such that $K_{c}^{\beta}$ has a pole at $\eta_{c}=i
u_{ac}^{b}-\eta_{a}$ creating the boundary bound state $\alpha$ or, if \,
$\rm{Im}\left(\eta_{c}\right)>\frac{\pi}{2}$, such that  $K_{c}^{\alpha}$ has a pole at $i\pi-\eta_{c}$ creating
$\beta$. The amplitude $K_{b}^{\beta}$ has to be evaluated at $\eta_{b}=i u_{ab}^{c}+\eta_{a}-i\pi$.

A third order diagram, with 13 propagators and 5 loops, is:

\vspace{7cm}

\begin{figure}[h]
\setlength{\unitlength}{0.0125in}
\begin{picture}(40,0)(60,470)
\thicklines \put(350,460){\line(0,1){220}}

\put(350,460){\line(2,1){15}} \put(350,480){\line(2,1){15}} \put(350,490){\line(2,1){15}}
\put(350,500){\line(2,1){15}} \put(350,510){\line(2,1){15}} \put(350,520){\line(2,1){15}}
\put(350,530){\line(2,1){15}} \put(350,540){\line(2,1){15}} \put(350,550){\line(2,1){15}}
\put(350,580){\line(2,1){15}}

\put(350,590){\line(2,1){15}} \put(350,600){\line(2,1){15}} \put(350,610){\line(2,1){15}}
\put(350,620){\line(2,1){15}} \put(350,630){\line(2,1){15}} \put(350,640){\line(2,1){15}}
\put(350,650){\line(2,1){15}} \put(350,670){\line(2,1){15}}
\put(355,565){$\beta$} \put(355,470){$\alpha$} \put(355,660){$\alpha$}

\put(350,550){\line(-2,3){40}} \put(350,590){\line(-2,-3){40}}

\put(310,610){\line(3,1){21}} \put(310,530){\line(3,-1){21}}

\put(331,617){\line(2,-3){19}} \put(331,617){\line(1,2){19}}

\put(331,523){\line(2,3){19}} \put(331,523){\line(1,-2){19}}

\put(310,610){\line(-3,1){30}} \put(310,530){\line(-3,-1){30}}

\put(270,620){$a$} \put(270,520){$a$}

\put(315,583){$b$} \put(318,617){$c$} \put(315,550){$d$} \put(333,635){$f$} \put(333,490){$f$}

\put(318,517){$e$}

\put(335,540){$b$} \put(335,593){$d$}
\end{picture}
 \caption{Type 4 (third order)}
 \end{figure}
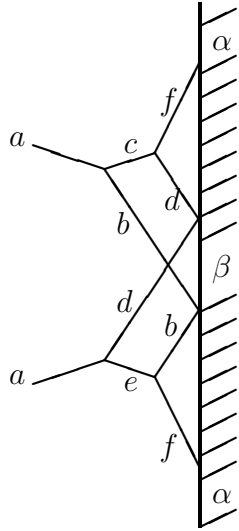

This diagram can be drawn if there are all the opportune bulk fusing angles, and one has to evaluate the two
amplitudes $K_{b}^{\beta}$ and $K_{d}^{\beta}$ at the rapidities dictated by the on-shell condition.

\vspace{0.5cm}

We will now investigate if some of our poles can be described by these mechanisms. In none of the examined
reflection amplitudes \lq$u$-channel\rq \, diagrams can explain any pole. However, many generalized Coleman-Thun
diagrams can be drawn, with interesting consequences.

\vspace{0.5cm}

\subsection{Analysis of the $E_{7}$ pole structure}

\subsubsection{\lq\lq Minimal\rq\rq \, solution}

We start considering the reflection matrix in the ground state. Type 1 diagram can never be drawn (none of the
seven particles couples to the boundary at $\theta=i\frac{\pi}{2}$). If $a=1,2$, neither can type 2, lacking $b$
and $c$ such that $u_{bc}^{a}<\frac{\pi}{2}$.

However, many poles of the remaining amplitudes can be explained by this diagram; we list all the possible
corresponding choices of $b$ and $c$ in the following table:

\begin{center}
\begin{tabular}{|l|l|l|l|l|l|}\hline
\hspace{1mm} $a$  & \hspace{2mm} 3 & \hspace{2mm} 4 & \hspace{9mm} 5 & \hspace{9mm} 6 & \hspace{1.5cm} 7 \\ \hline
\hspace{1mm} $\eta_{a}$ & \hspace{2mm} $4^{\beta}$ & \hspace{2mm} $1^{\gamma}$ &\hspace{2mm} $2^{\delta}$
\hspace{6mm} $4^{\gamma}$ &\hspace{2mm} $2^{\varepsilon}$ \hspace{6mm} $6^{\beta}$ &\hspace{2mm} $1^{\tau}$
\hspace{6mm} $3_{3}^{\sigma}$ \hspace{6mm} $5_{3}^{\delta}$ \\ \hline
 $(b,c)$ & $(2,1)$& $(1,1)$ & $(1,3)$ \hspace{1mm} $(3,1)$ & $(1,5)$ \hspace{1mm} $(5,1)$ & $(1,6)$ \hspace{1mm} $(6,1)$ \hspace{1mm} $(6,3)$ \\
         &        &         &                              & $(4,1)$                      & $(2,5)$ \hspace{1mm} $(3,6)$ \\
         &        &         &                              & $(2,3)$                      &                              \\ \hline
  \end{tabular}
\end{center}

\vspace{0.5cm}

The two triple poles of $K_{7}^{0}$ are described by this diagram because, in the case $(b,c)=(6,1)$, $K_{6}^{0}$
has a double pole at $\eta_{6}=i u_{16}^{7}=i \frac{4}{18}\pi$, and in the other two cases the fusing angle
$u_{36}^{7}$ corresponds to a triple pole of the $S$-matrix, so that the correct diagram to be drawn is:

\vspace{5.5cm}

\begin{figure}[h]
\setlength{\unitlength}{0.0125in}
\begin{picture}(40,0)(60,470)
\thicklines \put(350,490){\line(0,1){140}}

\put(350,490){\line(2,1){15}} \put(350,500){\line(2,1){15}} \put(350,510){\line(2,1){15}}
\put(350,520){\line(2,1){15}} \put(350,530){\line(2,1){15}} \put(350,540){\line(2,1){15}}
\put(350,550){\line(2,1){15}} \put(350,560){\line(2,1){15}} \put(350,570){\line(2,1){15}}
\put(350,580){\line(2,1){15}} \put(350,590){\line(2,1){15}} \put(350,600){\line(2,1){15}}
\put(350,610){\line(2,1){15}} \put(350,620){\line(2,1){15}}

\put(350,560){\line(-3,2){39}} \put(350,560){\line(-3,-2){39}} \put(311,586){\line(-3,1){21}}
\put(311,586){\line(-3,-2){21}} \put(311,534){\line(-3,-1){21}} \put(311,534){\line(-3,2){21}}

\put(290,527){\line(0,1){66}}

\put(290,593){\line(-2,3){30}} \put(290,527){\line(-2,-3){30}}

\put(280,560){$c$} \put(250,485){$a$} \put(250,630){$a$} \put(325,585){$b$} \put(325,530){$b$}

\end{picture}
\caption{Third order diagram}
 \end{figure}
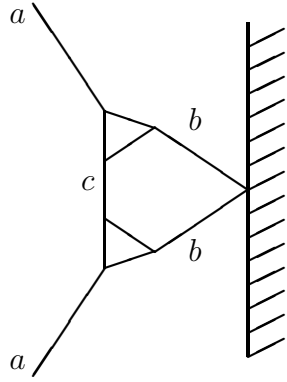

In this way, excluding the excitation diagrams, the only boundary bound state that we can get from the ground
state is $\alpha$. If we repeat the above procedure for the $K_{b}^{\alpha}$ amplitudes, we can exclude the
creation of other boundary bound states. Here we list which poles are explained by the three kinds of diagram
seen:

\begin{center}
\begin{tabular}{|l|l|l|l|l|l|l|l|}\hline
\hspace{1mm} $a$  & \hspace{1mm} 1                 & \hspace{1mm} 2           & \hspace{2mm} 3 & \hspace{9mm}
4                                         & \hspace{9mm} 5                                           &
\hspace{2mm} 6        & \hspace{2mm} 7 \\ \hline \hspace{1mm} $\eta_{a}$ & \hspace{1mm} $8^{\beta}$ &\hspace{1mm}
$3^{\delta}$ &\hspace{2mm} $4_{3}^{\varepsilon}$ &\hspace{2mm} $1^{\sigma}$ \hspace{6mm} $7_{3}^{\gamma}$
&\hspace{2mm} $2_{3}^{\tau}$ \hspace{6mm} $4_{3}^{\sigma}$ & \hspace{2mm} $6_{5}^{\varepsilon}$ & \hspace{2mm}
$5_{7}^{\tau}$ \\ \hline Type                    & \hspace{1mm} 1            & \hspace{1mm} 1 &\hspace{2mm}
2                     & \hspace{2mm} 2           \hspace{8mm} 3                &\hspace{2mm} 2 \hspace{8mm}
2                             & \hspace{2mm} 2        & \hspace{2mm} 3 \\ \hline
 $(b,c)$ & $(4)$                       & $(2)$       & $(2,1)$                           & $(1,1)$                  \hspace{1mm} $(4,4)$          & $(1,3)$ \hspace{1mm} $(3,1)$                             & $(5,1)$               & $(7,5)$        \\ \hline

\end{tabular}
\end{center}

\vspace{0.5cm}

All the type 3 diagrams mentioned have $\alpha$ as external boundary state, and the ground state as intermediate
one. Poles of high order can be described by many different diagrams, but we have listed just one of the possible
choices. As an example, the  triple pole at $\eta_{5}=i\frac{2}{18}\pi$ admits also a description in terms of
type 3 diagram with $(b,c)=(7,2)$. The order seven for the pole at $\eta_{7}=i\frac{5}{18}\pi$ is due to the facts
that $K_{7}^{0}$ has a simple pole at $\eta_{7}=i\frac{1}{18}\pi$, and $u_{57}^{5}$ corresponds to a fifth-order
pole of the $S$-matrix.

\subsubsection{Second solution}

As we will see, in this case the above mechanisms cannot explain a number of poles sufficient to exclude the
existence of some boundary bound states.

Let's start from the ground state. Type 1 diagram cannot be drawn, because there are not the appropriate bulk
fusing angles. If $a=1,2$, we already know that neither can type 2, but for the remaining particles it explains
the following poles:

\begin{center}
\begin{tabular}{|l|l|l|l|l|l|}\hline
\hspace{1mm} $a$        & \hspace{2mm} 3            & \hspace{9mm}
5                                               & \hspace{9mm} 6                                              &
\hspace{9mm} 7 \\ \hline \hspace{1mm} $\eta_{a}$ & \hspace{2mm} $3^{\gamma}$ & \hspace{2mm} $1^{\varepsilon}$
\hspace{6mm} $5_{3}^{\gamma}$ &\hspace{2mm} $1^{\sigma}$ \hspace{6mm} $3_{3}^{\varepsilon}$ &\hspace{2mm}
$2_{3}^{\tau}$ \hspace{6mm} $4_{5}^{\sigma}$  \\ \hline
 $(b,c)$                & $(1,2)$                   & $(2,2)$ \hspace{1mm} $(4,2)$                                 & $(1,4)$ \hspace{1mm} $(3,2)$                                & $(4,4)$ \hspace{1mm} $(5,4)$  \\ \hline

\end{tabular}
\end{center}

\vspace{0.5cm}

In this way, from the ground state we get the excited stated $\alpha$, $\beta$, $\delta$. It is now easy to see
that in the corresponding amplitudes there are simple poles which cannot be explained with alternative diagrams,
and which excite the boundary to all the other bound states $\gamma$, $\varepsilon$, $\sigma$, $\tau$. These are

\begin{center}
\begin{tabular}{|c|c|c|c|c|}\hline
 $K_{a}^{\mu}$ & $K_{1}^{\alpha}$ & $K_{3}^{\alpha}$  & $K_{1}^{\delta}$ & $K_{1}^{\varepsilon}$\\ \hline
 $\eta_{a}$    & $1^{\gamma}$     & $1^{\varepsilon}$ & $1^{\sigma}$     & $1^{\tau}$\\ \hline

\end{tabular}
\end{center}

\vspace{0.5cm}

\subsection{The same analysis for $E_{6}$ and $E_{8}$}

In these cases, none of the four solutions admits a reduction of the boundary bound states number by means of
generalized Coleman-Thun diagrams.

In the $E_{6}$ case, this can be easily seen if we notice that particles $1$, $2$ and $6$, for which Type 2
diagram is not allowed, have simple poles at rapidities which forbid also Type 1 diagram, and are able to generate
the whole set of boundary bound states.

An analogous mechanism works in the $E_{8}$ case, because Type 2 diagram, when applied to light particles,
describes a narrow range of possible poles.

\vspace{1cm}

\section{Perturbed Minimal Models}

It is now interesting to see if the results obtained for affine Toda field theories can be extended to minimal
models perturbations.

As we have seen, the minimal parts of the Toda $S$-matrices correspond to the scattering amplitudes of certain
perturbed conformal field theories. The $E_{6}$, $E_{7}$ and $E_{8}$ Toda theories are related respectively to
the thermal perturbation of the tricritical 3-state Potts model, the thermal perturbation of the tricritical
Ising model and the magnetic perturbation of the Ising model.

In the bulk theory, \lq\lq dressing\rq\rq \, a minimal $S$-matrix with coupling constant-dependent CDD-factors
doesn't induce any change in the bound states spectrum. These factors, in fact, don't introduce new poles in the
physical strip, and don't alter the sign of the existing ones' residues.

Starting from a Toda $S$-matrix of the form (\ref{todablocks}), where $B$ is real and varies in $[0,2]$, we can
recover its minimal part performing the so-called \lq\lq roaming limit\rq\rq, which consists in taking $B=1+iC$
and sending the real quantity $C$ to infinity.

In general, the residue of a given pole is a real function of $C$ which preserves the same sign it had for real
$B$; its limit as $C$ tends to infinity is the value dictated by the minimal $S$-matrix.

Reflection matrices of the form (\ref{K}) are manifestly factorized in a minimal part, which satisfies equations
(\ref{unit})-(\ref{boot}) with the minimal $S$-matrix, and a set of coupling constant-dependent factors, which
admit a \lq\lq roaming limit\rq\rq \, of the same form as for the $\{x\}_{\theta}$ blocks. We already know from
(\ref{poles}) that these factors don't introduce new poles in the physical strip $0\leq
Im\theta\leq\frac{\pi}{2}$.

As we have seen in eq.(\ref{colour}), a general reflection amplitude is a product of the two kinds of blocks
$\{x\}_{\theta}$ and ${\cal K}_{y}(\theta)$. Let us assume that ${\cal K}_{y}(\theta)$ has a pole at
$\theta_{0}=i\frac{\pi}{h}\eta$; the CDD-factors of the block $\{x\}_{\theta}$, evaluated at that rapidity, give:
\begin{equation}
\frac{1}{s_{\frac{x+1-B}{h}}(\theta_{0})s_{\frac{x-1+B}{h}}(\theta_{0})}=\frac{\sin\left[\frac{\pi}{2h}(\eta-x-1+B)\right]\sin\left[\frac{\pi}{2h}(\eta-x+1-B)\right]}{\sin\left[\frac{\pi}{2h}(\eta+x+1-B)\right]\sin\left[\frac{\pi}{2h}(\eta+x-1+B)\right]}.
\end{equation}
If $\eta\neq x$, this quantity is always positive, while if $\eta=x$ we have
\begin{equation}\label{neg}
\frac{1}{s_{\frac{x+1-B}{h}}(\theta_{0})s_{\frac{x-1+B}{h}}(\theta_{0})}=-\frac{\sin^{2}\left[\frac{\pi}{2h}(1-B)\right]}{\sin\left[\frac{\pi}{2h}(2x+1-B)\right]\sin\left[\frac{\pi}{2h}(2x-1+B)\right]},
\end{equation}
which is always negative as $B$ varies in $[0,2]$, and vanishes in $B=1$. If we now parameterize $B=1+iC$, as a
function of $C$ we have
\begin{equation}\label{neg}
\frac{1}{s_{\frac{x+1-B}{h}}(\theta_{0})s_{\frac{x-1+B}{h}}(\theta_{0})}=\frac{\cos\left[\frac{\pi}{h}(\eta-x)\right]-\cosh\left(\frac{\pi}{h}C\right)}{\cos\left[\frac{\pi}{h}(\eta+x)\right]-\cosh\left(\frac{\pi}{h}C\right)},
\end{equation}
which is a positive quantity for every $C$, and if $\eta=x$ vanishes in $C=0$. This means that, in the presence of
a block $\{\eta\}$, the corresponding pole can have different signs whether we are considering the minimal or the
Toda reflection amplitude.

This phenomenon is not present in the bulk theory, because $S$-matrix poles are always located at positions
shifted by $\pm 1$ with respect to the blocks parameters $x$.

An analogous behaviour is produced by the CDD-factors of a block ${\cal K}_{y}(\theta)$, evaluated at
$\theta_{0}=i\frac{\pi}{h}\eta$ with $\eta=\frac{3h-y}{2}$. This, however, never happens for the $E_{n}$ series
elements, because all the parameters $y$ are odd, but the poles are always located at entire multiples of
$i\frac{\pi}{h}$.

From this analysis we can conclude that a reflection amplitude pole located at $\theta_{0}=i\frac{\pi}{h}\eta$
will have a residue with different sign in the minimal and in the Toda theory if this reflection amplitude has an
odd number of $\{\eta\}$ blocks.

If we think to the bulk situation this new possibility seems problematic. We have to remember, however, that the
two integrable field theories defined by a Toda Lagrangian and a minimal model perturbation correspond in the UV
limit to completely different conformal field theories. Hence, although they share the minimal part of the
$S$-matrix, they could be governed by very different integrable boundary conditions, and this might become
manifest in distinct bound states structures of the corresponding reflection amplitudes.

We will now study this phenomenon in the three examined theories.

\subsection{The Tricritical Ising Model}

Analyzing the \lq\lq minimal\rq\rq \, solution, we find many situations of the type described above, always
corresponding to poles with negative residue in the Toda theory, and positive in the minimal one. We list the
additional \lq\lq fusing angles\rq\rq \, with the usual conventions:

\begin{center}
\begin{tabular}{|c|c|c|c|c|c|c|c|c|}\hline
 $ a\setminus\mu$  &  $ 0$  &  $ \alpha$ &  $ \beta$ &  $ \gamma$& $\delta$ & $\varepsilon$ & $\sigma$ & $\tau$\\ \hline
  $1$ &   &   &   &   &  &  &  & \\ \hline
  $2$ &   &   &   &   &  &  &  & \\ \hline
  $3$ &   & $3$ &  & $3$ & $3$ & $3$ &  & \\ \hline
  $4$ &   &  $4$ & $4$  & $4$  &  &  &  & $4$ \\ \hline
  $5$ &   &  $1,\:3,\:5$  & $3$ & $3,\:5$ & $5$ & $1,\:5$ & $1$ & $1,\:3$ \\ \hline
  $6$ &   & $3$ &  & $1$ & $1$ & $3$ & $1,\:3$ & $1,\:3$\\ \hline
  $7$ &   &  &  & $4_{3}$ & $4_{3}$ &  & $4_{3}$ & $4_{3}$\\ \hline

  \end{tabular}
\end{center}

\vspace{0.5cm}

This poles create twenty new different boundary bound states, with reflection amplitudes of the form
\begin{equation}\label{noncolour}
K_{a}^{\mu}\left(\theta\right)=\left(\prod_{b}S_{ab}^{\pm
1}\left(\theta\right)\right)K_{a}^{0}\left(\theta\right),
\end{equation}
where the $b$'s can now have different colours with respect of the bicolouration of the Dynkin diagram, and the
corresponding energy levels are related by
\begin{equation}\label{level}
E_{\mu}=E_{0}+\frac{1}{2}\sum_{b}\left(\pm m_{b}\right).
\end{equation}

If we analyze the reflection amplitudes on this new boundary states, we can see that their poles generate a
cascade of other bound states, with the possibility to have
\begin{equation}\label{noncolour}
K_{a}^{\mu}\left(\theta\right)=\left(\prod_{b}S_{ab}^{\pm
n}\left(\theta\right)\right)K_{a}^{0}\left(\theta\right), \qquad E_{\mu}=E_{0}+\frac{1}{2}\sum_{b}\left(\pm
n\,m_{b}\right),
\end{equation}
with $n=1,2,...$\,.

This seems an indication that in this case the bootstrap doesn't close on a finite number of boundary states.

For the states examined we have checked that the mentioned \lq$u$-channel\rq \, mechanism cannot be applied to any
new pole. However, if we consider the generalized Coleman-Thun diagrams we can explain all the new poles
introduced on the eight Toda states by the \lq\lq roaming\rq\rq \, limit. This seems an amazing coincidence,
because we can always use Type 2 diagram, and exactly particles $1$ and $2$, for which this diagram can't be
drawn, don't produce any new pole. In this way we are also left with the two possibilities of a bootstrap closing
on two or eight boundary bound states.

Essentially the same situation arises with the second solution: the \lq\lq roaming\rq\rq \, limit introduces many
new positive residue poles in a similar way, and again Coleman-Thun diagrams can describe all of them. This time
also particles $1$ and $2$ generate new states, but we can use both Type 1 and Type 2 diagrams.

\subsection{The same analysis for $E_{6}$ and $E_{8}$}

Also in the $E_{8}$ case the \lq\lq roaming\rq\rq \, limit produces many additional positive residue poles, which
seem to indicate a non-closing bootstrap, but again we have an almost incredible coincidence between the new
poles and the ones we can explain with Coleman-Thun diagrams, so that for both solutions these mechanisms allow
the bootstrap to close on sixteen states.

\vspace{0.5cm}

In the $E_{6}$ case the situation is more delicate. As before, the \lq\lq roaming\rq\rq \, limit introduces many
new poles, and again the bootstrap presumably doesn't close. The problem is that now the Coleman-Thun diagrams
can describe almost all these poles, but in both solutions four of them remain unexplained. For the \lq\lq
minimal\rq\rq \, solution these are simple poles located at $\theta=i\frac{3}{12}\pi$ in $K_{2}^{\mu}$ with
$\mu=\beta_{1},\beta_{2},\gamma_{1},\gamma_{2}$, and they generate a cascade of states. In the second solution
case these poles, located at $\theta=i\frac{3}{12}\pi$ in $K_{4}^{\mu}$ with
$\mu=\alpha_{1},\alpha_{2},\delta_{1},\delta_{2}$, are triple, so that it is more difficult to conclude that they
don't admit alternative diagrams. However, we have tried to explain them with all kinds of diagram mentioned
(including Type 4), but we haven't succeeded.

This seems to indicate that both the solutions analyzed don't have a physical meaning for the minimal model. At
this point, in order to confirm the whole construction, we need to find at least one physical reflection matrix,
using the CDD-ambiguity mentioned.

\vspace{1cm}

\section{CDD-Ambiguity: the $E_{6}$ case}

We start considering the \lq\lq minimal\rq\rq \, reflection matrix (\ref{min1})-(\ref{min4}). The problem with the
poles $\theta=i\frac{3}{12}\pi$ in $K_{2}^{\mu}$ ($\mu=\beta_{1},\beta_{2},\gamma_{1},\gamma_{2}$) is that they
generate states characterized by a mixing of the Dynkin diagram colourations, because
\begin{equation}
S_{b2}\left(\theta+i\frac{3}{12}\pi\right)S_{b2}\left(\theta-i\frac{3}{12}\pi\right)=S_{b1}\left(\theta\right)S_{b6}\left(\theta\right).
\end{equation}
Their appearance is due to the fact that every minimal block $\{x\}$ evaluated at $\theta=i\frac{x}{h}\pi$ is a
negative quantity, hence it changes the residue sign as it enters the reflection amplitude expression at some
excited state. This doesn't happen in the Toda theory because another compensating negative sign arises from the
coupling-dependent factors. It is then clear that, adding a $S$-matrix factor to the initial reflection amplitude,
we will hardly overcome this problem, because we will get a more complicated pole structure and presumably we
will recover most of the previous boundary bound states.

The best strategy seems then to divide the reflection amplitude by an opportune $S$-matrix element. The Dynkin
colour which characterize the boundary bootstrap for the \lq\lq minimal\rq\rq \, solution is the one of particles
$2$, $3$ and $5$, and we will start with the simplest choice, i.e. the self-conjugate particle $2$. We then
define:
\begin{equation}\label{E6CDD}
\widetilde{K}_{a}^{0}(\theta)=S_{a2}^{-1}(\theta)K_{a}^{0}(\theta)
\end{equation}

The \lq\lq fusing angles\rq\rq \, are

\begin{center}
\begin{tabular}{|c|c|c|c|}\hline
 $ a\setminus\mu$  &  $ 0$  &  $\delta$ \\ \hline
  $1$ & $1^{\beta_{1}}$  & $(3)_{1}$   \\ \hline
  $2$ &   & $(2)_{1}\,6^{\delta}$   \\ \hline
  $3$ & $(3)_{2}$  &  $(1)_{2}(2_{3})_{2}(3)_{2}$  \\ \hline
  $4$ & $3^{\delta}$  &    \\ \hline
  $5$ & $(3)_{2}$  &  $(1)_{2}(2_{3})_{2}(3)_{2}$  \\ \hline
  $6$ & $1^{\beta_{2}}$  & $(3)_{1}$   \\ \hline
    \end{tabular}
\end{center}

\vspace{0.5cm}

where the brackets mean that the corresponding pole can be explained by Coleman-Thun diagrams, of the type
indicated by the external subscript. We have kept the same letters for the boundary states, because the relation
between the excited state reflection amplitudes and the ground state ones is the same as in the case of the
initial solution (\ref{min1})-(\ref{min4}).

The poles at $1^{\beta_{1,2}}$ in $K_{1,6}^{0}$ deserve a separate discussion. They would create again states on
which the reflection amplitude of particle 2 has a change of sign from negative to positive in the residue of the
pole at $i\frac{3}{12}\pi$, due to the appearance of the block $\{3\}$. We already know that Type 1 diagram
cannot be used, and there are not $b$ and $c$ such that $u_{bc}^{2}<\frac{\pi}{2}$. However, we know that if a
scattering amplitude $S_{bc}(\theta)$ has a pole at $\theta=i u_{bc}^{a}$, then the amplitude
$S_{\bar{b}c}(\theta)$ will have a corresponding \lq\lq crossed-channel\rq\rq \, pole at
$\theta=i(\pi-u_{bc}^{a})$. In this way, we could try to draw a \lq\lq crossed\rq\rq \, version of Type 2
diagram, represented by:

\vspace{6cm}

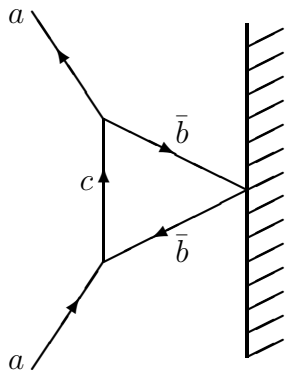
\begin{figure}[h]
\setlength{\unitlength}{0.0125in}
\begin{picture}(40,0)(60,470)

\thicklines \put(360,490){\line(0,1){140}}

\put(360,490){\line(2,1){15}} \put(360,500){\line(2,1){15}} \put(360,510){\line(2,1){15}}
\put(360,520){\line(2,1){15}} \put(360,530){\line(2,1){15}} \put(360,540){\line(2,1){15}}
\put(360,550){\line(2,1){15}} \put(360,560){\line(2,1){15}} \put(360,570){\line(2,1){15}}
\put(360,580){\line(2,1){15}} \put(360,590){\line(2,1){15}} \put(360,600){\line(2,1){15}}
\put(360,610){\line(2,1){15}} \put(360,620){\line(2,1){15}}

\put(360,560){\line(-2,1){30}} \put(300,590){\vector(2,-1){30}}

\put(360,560){\vector(-2,-1){40}} \put(320,540){\line(-2,-1){20}}

\put(300,530){\vector(0,1){40}} \put(300,570){\line(0,1){20}}

\put(300,590){\vector(-2,3){20}} \put(280,620){\line(-2,3){10}}

\put(300,530){\line(-2,-3){10}} \put(270,485){\vector(2,3){20}}

\put(290,560){$c$} \put(260,485){$a$} \put(260,630){$a$} \put(330,580){$\bar{b}$} \put(330,530){$\bar{b}$}

\end{picture}
 \caption{Crossed version of Type 2 diagram}
 \end{figure}

We will indicate the \lq\lq modified fusing angles\rq\rq \, of this diagram by a tilde; these are related to the
direct-channel ones in the following way:
\begin{eqnarray*}
\tilde{u}_{bc}^{a}&=&\pi-u_{bc}^{a}   \\
\tilde{u}_{ac}^{b}&=&u_{ac}^{b}   \\
\tilde{u}_{ab}^{c}&=&u_{ab}^{c}+(u_{bc}^{a}-\tilde{u}_{bc}^{a})=u_{ab}^{c}+(2u_{bc}^{a}-\pi)   \\
\tilde{\eta}_{a}&=&i(\tilde{u}_{bc}^{a}+\tilde{u}_{ab}^{c}-\pi)=i(u_{bc}^{a}+u_{ab}^{c}-\pi)   \\
\end{eqnarray*}

The boundary crossing equation (\ref{cross}) implies that the antiparticle reflection amplitude to be evaluated
at $\tilde{\eta}_{b}=\tilde{u}_{bc}^{a}$ is
\begin{equation}\label{crampl}
K_{\bar{b}}(\theta+i\pi)=\frac{S_{bb}(2\theta)}{K_{b}(\theta)}.
\end{equation}

This diagram can effectively explain the two poles in exam, with $(a,b,c)=(1,6,3)$ and $(a,b,c)=(6,1,5)$.

The same idea can be applied to Type 3 diagram, when the version to be used is the one with \,
$\rm{Im}\left(\eta_{c}\right)>\frac{\pi}{2}$, which again requires $b$ and $c$ such that
$u_{bc}^{a}<\frac{\pi}{2}$. The \lq\lq modified fusing angles\rq\rq \, are defined as before, and the process is
represented by:

\vspace{5cm}

\begin{figure}[h]
\setlength{\unitlength}{0.0125in}
\begin{picture}(40,0)(60,470)
\thicklines \put(350,460){\line(0,1){160}}

\put(350,460){\line(2,1){15}} \put(350,480){\line(2,1){15}} \put(350,490){\line(2,1){15}}
\put(350,500){\line(2,1){15}} \put(350,510){\line(2,1){15}} \put(350,520){\line(2,1){15}}
\put(350,550){\line(2,1){15}} \put(350,560){\line(2,1){15}}

\put(350,570){\line(2,1){15}} \put(350,580){\line(2,1){15}} \put(350,590){\line(2,1){15}}
\put(350,610){\line(2,1){15}}

\put(350,540){\vector(-1,-2){15}} \put(335,510){\line(-1,-2){10}}

\put(325,590){\vector(1,-2){15}} \put(350,540){\line(-1,2){10}}

\put(350,582){\vector(-3,1){13}} \put(325,590){\line(3,-1){12}}

\put(325,490){\vector(3,1){17}}  \put(350,498){\line(-3,-1){12}}

\put(325,590){\vector(-3,2){18}}  \put(295,610){\line(3,-2){12}}

\put(293,468){\vector(3,2){20}}  \put(310,480){\line(3,2){14}}

\put(355,535){$\beta$} \put(355,470){$\alpha$} \put(355,600){$\alpha$}

\put(285,608){$a$} \put(285,470){$a$} \put(328,555){$\bar{b}$} \put(330,520){$\bar{b}$} \put(338,590){$c$}
\put(338,483){$c$}

\end{picture}
\caption{Crossed version of Type 3 diagram}
 \end{figure}
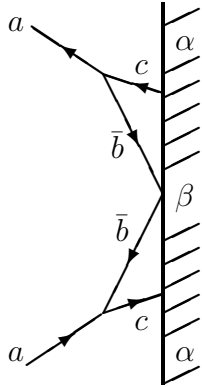

Also this crossed diagram, although of second order, has an immediate application to the two simple poles
mentioned, because now the reflection amplitudes can have zeroes in the physical strip, due to the $S_{a2}^{-1}$
term. If we choose $(a,b,c)=(1,5,4)$ (and $(a,b,c)=(6,3,4)$), $\alpha=0$ and $\beta=\delta$, we can describe the
poles at $\tilde{\eta}_{a}=i\frac{1}{12}\pi$, because particle 4 couples to the ground state at
$i\pi-\tilde{\eta}_{c}=i\frac{3}{12}\pi$ and
$\frac{S_{55}\left(2\theta\right)}{K_{5}^{\delta}\left(\theta\right)}=\frac{S_{33}\left(2\theta\right)}{K_{3}^{\delta}\left(\theta\right)}$
has a simple zero in $\tilde{\eta}_{b}=i\frac{2}{12}\pi$.

In this way, we have a bootstrap closing on the two states $0$ and $\delta$: the fact that we have skipped states
$\beta_{1,2}$ and $\gamma_{1,2}$ let us conclude that the same situation arises in the Toda theory and in the
minimal model.

\vspace{1cm}

\section{Consequences of \lq\lq crossed diagrams\rq\rq}

It is now interesting to see what happens if we extend the use of \lq\lq crossed diagrams\rq\rq, fundamental in
the last discussion, also to the other sets of reflection amplitudes examined.

First of all we summarize the various possibilities given by the application of the direct-channel diagrams,
listing in the following table the number of states on which the boundary bootstrap closes (or presumably doesn't,
if we indicate \lq\lq$\infty$\rq\rq) in the various systems analyzed:

\begin{center}
\begin{tabular}{|c|c c|c c|c c|}\hline
&  $ E_{6}^{(1)}$  & $ E_{6}^{(2)}$ & $ E_{7}^{(1)}$ & $ E_{7}^{(2)}$ & $ E_{8}^{(1)}$ & $ E_{8}^{(2)}$ \\ \hline
Toda Field Theory  & 8 & 8 & 8 & 8 & 16 & 16 \\
Toda\,+\,Coleman-Thun  & 8 & 8 & 2 & 8 & 16 & 16 \\ \hline
Minimal Model (MM) & $\infty$ & $\infty$ & $\infty$ & $\infty$ & $\infty$ & $\infty$ \\
MM\,+\,Coleman-Thun & $\infty$ & $\infty$ & 2/8 & 8 & 16 & 16 \\ \hline

  \end{tabular}
\end{center}

\vspace{0.5cm}

The first column refers to the kind of theory: the voices \lq\lq Toda Field Theory\rq\rq \, and \lq\lq Minimal
Model \rq\rq \, discriminate between \lq\lq dressed\rq\rq \, and minimal scattering matrices, with a bootstrap
carried on all odd-order poles with positive residue, while with \lq\lq +\,Coleman-Thun\rq\rq \, we mean the
exclusion of boundary bound state creations if alternative diagrams can be drawn. The first row indicates the six
solutions examined (two for each algebra), with $(1)$ and $(2)$ referring respectively to the \lq\lq
minimal\rq\rq \, solution and to the one shifted by $i\pi$.

\vspace{0.5cm}

Obviously the new diagrams increase the number of explicable poles, most of all for light particles, but the
principal novelty is that their order can also be lowered, because in general the \lq\lq crossed\rq\rq \,
reflection amplitudes (\ref{crampl}) can have zeroes in the physical strip, even if this is impossible for the
\lq\lq direct-channel\rq\rq \, ones. However, this is true only if we are on excited states, because on the ground
state (\ref{crampl}) exactly corresponds to going from the \lq\lq minimal\rq\rq \, solution to the shifted one or
vice versa.

This implies that if a simple pole of $K_{b}^{0}$ cannot be explained by Coleman-Thun (normal or crossed) Type 1
or Type 2 diagrams, then we can conclude that it creates a boundary bound state, but this argument is not valid
on excited states. On these states, in fact, even if we don't find an opportune diagram to describe a certain pole
(simple or multiple), we cannot conclude that this pole corresponds to a boundary excitation, because we cannot
check all possible order diagrams with the opportune zeroes insertions.

\vspace{0.5cm}

We will now describe the combined effect of \lq\lq normal\rq\rq \, and \lq\lq crossed\rq\rq \, diagrams on the
various amplitudes considered.

\vspace{0.5cm}

Let's start with the $E_{6}$ \lq\lq minimal\rq\rq \, solution. In the Toda case we are able to reduce the number
of boundary states from 8 to 4, getting a bootstrap which closes on $0,\alpha,\beta_{1},\beta_{2}$. This doesn't
solve the minimal model problem, because it seems again that the poles at $\theta=i\frac{3}{12}\pi$ in
$K_{2}^{\beta_{1,2}}$ generate an infinite cascade of boundary states; we couldn't find opportune crossed
diagrams to describe this simple poles, but as we have explained we could not investigate all the possibilities.

As it regards the second solution, from the ground state we are sure to obtain states $\beta$ and $\gamma$,
skipping $\alpha_{1,2}$, but we cannot exclude the subsequent creation of $\delta_{1,2}$ and $\varepsilon$. Hence
for the Toda theory we can conclude that the bootstrap closes on a number of states between 3 and 6, while we
have not a definite answer for the minimal model.

\vspace{0.5cm}

For the other two algebras we already know that the use of standard Coleman-Thun diagrams gives the same boundary
states spectrum in the Toda theory and in the perturbed minimal model.

The bootstrap generated by the $E_{7}$ \lq\lq minimal\rq\rq \, solution remains unchanged on the two states $0$
and $\alpha$. With the second solution we certainly have the $\delta$ creation from the ground state (without
$\alpha$, $\beta$ and $\gamma$), but we cannot decide what happens with the following states $\varepsilon$,
$\sigma$ and $\tau$; the bootstrap will then close on a number of states between 2 and 5.

The $E_{8}$ case, finally, is very similar. The \lq\lq minimal\rq\rq \, solution generates states
$\alpha,\beta,\gamma,\delta,\varepsilon$ from the ground state, hence the bootstrap will close on a number of
states between 6 and all the 16. With the second solution, instead, we are sure to get states $\delta$ and
$\lambda$ avoiding $\alpha,\beta$ and $\gamma$, so that the uncertainty is between 3 and 13 states.

\vspace{0.5cm}

We summarize these results in the following table, with the same conventions as described before:

\begin{center}
\begin{tabular}{|c|c c c|c c|c c|}\hline
&  $ E_{6}^{(1)}$  & $ E_{6}^{(2)}$ & $ E_{6}^{(3)}$ & $ E_{7}^{(1)}$ & $ E_{7}^{(2)}$ & $ E_{8}^{(1)}$ & $
E_{8}^{(2)}$ \\ \hline
Toda  & 4 & 3-6 & 2 & 2 & 2-5 & 6-16 & 3-13 \\
MM & 4-$\infty$ & 3-$\infty$ & 2 & 2 & 2-5 & 6-16 & 3-13 \\ \hline
  \end{tabular}
\end{center}

\vspace{0.5cm}

Every entry of the form \lq\lq $n-m$\rq\rq \, means that we cannot decide on how many states the bootstrap
closes, but this number should lie between $n$ and $m$. The additional solution indicated by $E_{6}^{(3)}$ is the
one with ground state amplitudes (\ref{E6CDD}).

\vspace{1cm}

\section{Discussion}

We have performed a detailed analysis of the boundary states structures arising from the reflection amplitudes
found by Fring and Koberle, showing how generalized Coleman-Thun mechanisms can have interesting consequences
compared with a blind iteration of the bootstrap on all odd-order poles with positive residue. However, these
on-shell methods are not sufficient to outline a clear and definitive picture of the phenomenon. We are in fact
left with various kinds of problems.

The first one is to understand if the possibility of drawing generalized Coleman-Thun diagrams really excludes the
creation of a boundary bound state. This seems reasonable in cases where the bootstrap closes only under this
assumption, as for perturbed minimal models, but we have seen that, for the \lq\lq minimal\rq\rq \, solution in
the $E_{7}$ Toda theory, the alternative is between two closing bootstraps, one on eight states, and the other on
two. Another eventuality is that the same ground state reflection amplitudes correspond to distinct boundary
conditions, whose different physical properties become manifest in the interpretation of certain poles. This
phenomenon was recognized in \cite{tateo} in the case of the scaling Lee-Yang model, knowing independently the
different spectra from a boundary generalization of the truncated conformal space approach.

The direct strategy to face this problem would be to calculate the residues of the various diagrams and compare
them with the actual residue of the corresponding pole in the reflection amplitude, but it is not known how to
treat this perturbative calculations in the presence of a boundary.

Another delicate point is the use of crossed diagrams, which are so important in the $E_{6}$ case; again it would
be necessary to calculate their contribute to the residues. Furthermore, the possibility of inserting in these
diagrams crossed reflection amplitudes with zeroes in the physical strip makes it very difficult to conclude
something about many poles, and alternative methods are essential to check the existence of the related boundary
states.

Finally, in this context we have no way to understand which is the boundary condition related to a certain
reflection matrix, and the best we can do is just to notice whether an eventual symmetry of the systems is
preserved or not by the corresponding excited boundary states structure. This is a particularly delicate problem,
especially in the light of the difference between the bound states spectra displayed by the minimal and the Toda
reflection amplitudes. It could be, in fact, that the two basic solutions analyzed in the $E_{6}$ case are
related to boundary conditions that in the UV limit correspond to conformal boundary conditions present in the CFT
associated to the Toda Lagrangian but not in the tricritical 3-state Potts model.

\vspace{0.5cm}

To solve these problems, we intend to proceed in the future work performing indirect checks on certain properties
of the theory, using thermodynamic Bethe ansatz (TBA) equations (\cite{TBA1},\cite{TBA2}) and analyzing one-point
function behaviours (\cite{FF}). The basic idea is to perform the UV limit in order to calculate the so-called
ground state degeneracy $g$ given by a certain set of reflection amplitudes. This quantity, introduced by Affleck
and Ludwig in \cite{affllud} and related to a conformal boundary state $|A\rangle$, is defined as
\begin{equation}\label{g}
g_{A}=\langle 0|A\rangle,
\end{equation}
where $|0\rangle$ is the ground state of the bulk Hamiltonian. If the state $|A\rangle$ is of the form
(\ref{bstates}), $g$ has the generally noninteger value
\begin{equation}\label{valg}
g_{A}=\frac{S_{A0}}{\sqrt{S_{00}}}.
\end{equation}
The various possible choices of excited boundary states should produce different values of $g$, which in the case
of minimal models can be compared to the ones given by the primary operators present in the CFT using
eq.(\ref{valg}). In this way it should be possible to identify the physical meaning of the reflection matrices,
associating them to specific boundary conditions, and to deduce which of the initial boundary bound states are
really involved in the bootstrap.

\newpage

\appendix

\chapter{Conformal correlators with boundary}
\label{confcalc}

\section{Ising model}

\subsubsection{Boundary fields}
\begin{eqnarray*}
&&|\tilde{1}\rangle=\frac{1}{\sqrt{2}}|1\rangle+\frac{1}{\sqrt{2}}|\varepsilon\rangle+\frac{1}{\sqrt[4]{2}}|\sigma\rangle\\
&&|\tilde{\varepsilon}\rangle=\frac{1}{\sqrt{2}}|1\rangle+\frac{1}{\sqrt{2}}|\varepsilon\rangle-\frac{1}{\sqrt[4]{2}}|\sigma\rangle\\
&&|\tilde{\sigma}\rangle=|1\rangle-|\varepsilon\rangle\\
\end{eqnarray*}

\subsubsection{Bulk structure constants}
\begin{displaymath}
C_{\sigma\sigma}^{\varepsilon}=C_{\sigma\varepsilon}^{\sigma}=\frac{1}{2}
\end{displaymath}

\subsubsection{Correlator $\langle\sigma\sigma\rangle_{\alpha}$}
\begin{center}
\begin{tabular}{|c|c|c|c|c|}\hline
$\alpha$ &  $A$  & $ B$ & $ C_{\sigma,\psi_{1,1}}^{\alpha}$ & $ C_{\sigma,\psi_{1,3}}^{\alpha}$  \\ \hline

$\tilde{1}, \tilde{\varepsilon}$ & $\frac{\sqrt{\pi}}{\Gamma^{2}\left(\frac{1}{4}\right)}$ &
$\frac{\sqrt{\pi}}{\Gamma^{2}\left(\frac{1}{4}\right)}$& $2^{\frac{1}{4}}$ & $0$ \\
$\tilde{\sigma}$ & $-\frac{\sqrt{\pi}}{\Gamma^{2}\left(\frac{1}{4}\right)}$ &
$\frac{\sqrt{\pi}}{\Gamma^{2}\left(\frac{1}{4}\right)}$&  $0$ & $2^{-\frac{1}{4}}$ \\ \hline

\end{tabular}
\end{center}

\subsubsection{Correlator $\langle\varepsilon\sigma\rangle_{\alpha}$}
\begin{center}
\begin{tabular}{|c|c|c|c|}\hline
$\alpha$ &  $A$  & $ B$ & $ C_{\varepsilon\,\psi_{1,1}}^{\alpha}$  \\
\hline

$\tilde{1}$ & $0$ &
$\frac{2^{\frac{3}{4}}\sqrt{\pi}}{\Gamma^{2}\left(\frac{1}{4}\right)}$& $1$  \\
$\tilde{\varepsilon}$ & $0$ &
$-\frac{2^{\frac{3}{4}}\sqrt{\pi}}{\Gamma^{2}\left(\frac{1}{4}\right)}$& $-1$  \\
$\tilde{\sigma}$ & $0$ & $0$&  $0$  \\ \hline

\end{tabular}
\end{center}

\vspace{0.5cm}

In the case of the correlator $\langle\sigma\sigma\rangle_{\alpha}$, with $\phi _{n,m}=\phi _{1,2}=\sigma$, there
are both channels $(m+1)$ and $(m-1)$, hence it is possible to determine $A$ and $B$ independently. Furthermore,
due to the corresponding bulk OPE's, there are both channels also in the expansion with bulk-boundary structure
constants.

However, for the correlator $\langle\varepsilon\sigma\rangle_{\alpha}$, with $\phi _{n,m}=\phi
_{1,3}=\varepsilon$, there is only the $(m-1)$ channel, hence we can determine just $B$, and we have to deduce the
value of $A$ from the result for $A+B$. As expected, we always find $A=0$, as a consequence of the truncation of
the OPE $\; \sigma\times\varepsilon=\sigma$, related to the fact that one of the two independent solutions of the
differential equation for the correlator decouples. As a check, we can consider the same operator in the
representation $\phi _{2,1}$. Now we find just the $(m+1)$ channel, and we have the same situation with the roles
of $A$ e $B$ interchanged. This interchange is a general fact, consequence of the substitution $\alpha
_{nm}\rightarrow 2\alpha _{0}-\alpha _{nm}$, with $\alpha _{0}=(\alpha _{+}+\alpha _{-})/2$. In fact, making this
replacement, nothing changes in the correlator if we consider the limit $\eta \rightarrow 1$, and so we obtain
the same value for $A+B$ and the same bulk-boundary structure constants. On the contrary, in the limit $ \eta
\rightarrow 0$, in the prefactor emerges a multiplicative term proportional to $\eta $, with exponent $4\alpha
_{0}\alpha _{1,2}-4\alpha _{n,m}\alpha _{1,2}$, which gives exactly the behaviour
$|z_{1}-z_{2}|^{-2\left(h_{n,m+1}-h_{n,m-1} \right)}$, needed to interchange the two channels and hence the
values of $A$ and $B$. As a consequence of the corresponding bulk OPE, in the expansion with bulk-boundary
structure constants we have just the identity channel, hence we can fix only $ C_{(n,m)\psi _{1,1}}^{\alpha}$.

\newpage

\section{Tricritical Ising model}

\subsubsection{Boundary fields}

\begin{eqnarray*}
|\tilde{1}\rangle &=&\frac{\left[\frac{1}{5}\left(5-\sqrt{5} \right)
\right]^{\frac{1}{4}}}{2^{\frac{3}{4}}}\left(| 1\rangle +| \varepsilon ''\rangle
\right)+\frac{\left[\frac{1}{5}\left(5+\sqrt{5} \right)
\right]^{\frac{1}{4}}}{2^{\frac{3}{4}}}\left(| \varepsilon \rangle +\mid t\rangle \right)+ \\
&& +\frac{\left[\frac{1}{5}\left(5+\sqrt{5} \right) \right]^{\frac{1}{4}}}{\sqrt{2}}|\sigma \rangle
+\frac{\left[\frac{1}{5}\left(5-\sqrt{5} \right)
\right]^{\frac{1}{4}}}{\sqrt{2}}| \sigma' \rangle \\
|\tilde{\varepsilon''}\rangle &=& \frac{\left[\frac{1}{5}\left(5-\sqrt{5} \right)
\right]^{\frac{1}{4}}}{2^{\frac{3}{4}}}\left(| 1\rangle +| \varepsilon ''\rangle
\right)+\frac{\left[\frac{1}{5}\left(5+\sqrt{5} \right) \right]^{\frac{1}{4}}}{2^{\frac{3}{4}}}\left(|
\varepsilon \rangle +| t\rangle  \right)+ \\
&&-\frac{\left[\frac{1}{5}\left(5+\sqrt{5} \right) \right]^{\frac{1}{4}}}{\sqrt{2}}| \sigma \rangle
-\frac{\left[\frac{1}{5}\left(5-\sqrt{5} \right) \right]^{\frac{1}{4}}}{\sqrt{2}}| \sigma' \rangle
 \\
|\tilde{\varepsilon}\rangle &=&\frac{\sqrt{5+\sqrt{5}} }{2^{\frac{3}{4}}\left[5\left(5-\sqrt{5} \right)
\right]^{\frac{1}{4}}}\left(| 1\rangle +| \varepsilon ''\rangle  \right)-\frac{\sqrt{5-\sqrt{5}}
}{2^{\frac{3}{4}}\left[5\left(5+\sqrt{5} \right) \right]^{\frac{1}{4}}}\left(| \varepsilon \rangle +| t\rangle
\right) + \\
&&+\frac{\sqrt{5-\sqrt{5}} }{\sqrt{2}\left[5\left(5+\sqrt{5} \right) \right]^{\frac{1}{4}}}| \sigma
\rangle-\frac{\sqrt{5+\sqrt{5}} }{\sqrt{2}\left[5\left(5-\sqrt{5} \right) \right]^{\frac{1}{4}}}| \sigma' \rangle
 \\
| \tilde{t}\rangle &=&\frac{\sqrt{5+\sqrt{5}} }{2^{\frac{3}{4}}\left[5\left(5-\sqrt{5} \right)
\right]^{\frac{1}{4}}}\left(| 1\rangle +|\varepsilon ''\rangle  \right)-\frac{\sqrt{5-\sqrt{5}}
}{2^{\frac{3}{4}}\left[5\left(5+\sqrt{5} \right) \right]^{\frac{1}{4}}}\left(| \varepsilon \rangle +| t\rangle
\right)+ \\
&&-\frac{\sqrt{5-\sqrt{5}} }{\sqrt{2}\left[5\left(5+\sqrt{5} \right) \right]^{\frac{1}{4}}}| \sigma
\rangle+\frac{\sqrt{5+\sqrt{5}} }{\sqrt{2}\left[5\left(5-\sqrt{5} \right) \right]^{\frac{1}{4}}}| \sigma' \rangle
 \\
| \tilde{\sigma }\rangle &=& \frac{\sqrt{5+\sqrt{5}}}{\left[10\left(5-\sqrt{5} \right)
\right]^{\frac{1}{4}}}\left(| 1\rangle -| \varepsilon ''\rangle
\right)+\frac{\sqrt{5-\sqrt{5}}}{\left[10\left(5+\sqrt{5} \right) \right]^{\frac{1}{4}}}\left(| \varepsilon
\rangle -| t\rangle  \right)
 \\
|\tilde{\sigma' }\rangle &=&\frac{\left(5-\sqrt{5} \right)^{\frac{1}{4}}}{10^{\frac{1}{4}}} \left(| 1\rangle -|
\varepsilon ''\rangle \right)-\frac{\left(5+\sqrt{5} \right)^{\frac{1}{4}}}{10^{\frac{1}{4}}} \left(| \varepsilon
\rangle -| t\rangle \right)
\end{eqnarray*}

\subsubsection{Bulk structure constants (\cite{TIMscaling})}
\begin{center}
\begin{tabular}{|l|l|l|l|l|}\hline
$ C_{nm,12}^{n,m-1}$  &  $ C_{nm,12}^{n,m+1}$ &&  $ C_{nm,21}^{n-1,m}$  &  $ C_{nm,21}^{n+1,m}$  \\ \hline

$ C_{\varepsilon,\varepsilon}^{1}=1$ & $ C_{\varepsilon,\varepsilon}^{t}=\frac{2}{3}\sqrt{\frac{\Gamma
\left(\frac{4}{5} \right)\Gamma
^{3}\left(\frac{2}{5} \right)}{\Gamma \left(\frac{1}{5} \right)\Gamma ^{3}\left(\frac{3}{5} \right)}}$ &&  & $ C_{\varepsilon,\sigma'}^{\sigma}=\frac{1}{2}$ \\

$ C_{t,\varepsilon}^{\varepsilon}=\frac{2}{3}\sqrt{\frac{\Gamma \left(\frac{4}{5} \right)\Gamma
^{3}\left(\frac{2}{5} \right)}{\Gamma \left(\frac{1}{5} \right)\Gamma ^{3}\left(\frac{3}{5} \right)}}$   &  $
C_{t,\varepsilon}^{\varepsilon''}=\frac{3}{7}$  &&     &  $ C_{t,\sigma'}^{\sigma}=\frac{3}{4}$    \\

$ C_{\varepsilon'',\varepsilon}^{t}=\frac{3}{7}$  &  &&    &  $ C_{\varepsilon'',\sigma'}^{\sigma'}=\frac{7}{8}$   \\

$ C_{\sigma,\varepsilon}^{\sigma}=\sqrt{\frac{\Gamma \left(\frac{4}{5} \right)\Gamma ^{3}\left(\frac{2}{5}
\right)}{\Gamma \left(\frac{1}{5} \right)\Gamma ^{3}\left(\frac{3}{5} \right)}}$  &  $
C_{\sigma,\varepsilon}^{\sigma'}=\frac{1}{2}$  && $ C_{\sigma,\sigma'}^{t}=\frac{3}{4}$  &  $ C_{\sigma,\sigma'}^{\varepsilon}=\frac{1}{2}$    \\

$ C_{\sigma',\varepsilon}^{\sigma}=\frac{1}{2}$ &  && $ C_{\sigma',\sigma'}^{\varepsilon''}=\frac{7}{8}$ &  $ C_{\sigma',\sigma'}^{1}=1$    \\
\hline
\end{tabular}
\end{center}

\vspace{1cm}

In the cases of correlators for which only one of the two channels is present, we can repeat the same
considerations seen for the Ising model.

\subsubsection{Correlator $\langle\varepsilon\varepsilon\rangle_{\alpha}$}
\begin{center}
\begin{tabular}{|c|c|c|c|c|}\hline
$\alpha$ &  $A$  & $ B$ & $ C_{\varepsilon,\psi_{1,1}}^{\alpha}$ & $ C_{\varepsilon,\psi_{1,3}}^{\alpha}$  \\
\hline

$\tilde{1}, \tilde{\varepsilon''}$ & $\sqrt{\frac{5+\sqrt{5}}{5-\sqrt{5}}}\frac{\Gamma \left(\frac{2}{5}
\right)}{\Gamma ^{2}\left(\frac{1}{5} \right)}$ &
$\frac{\Gamma \left(\frac{2}{5} \right)}{\Gamma ^{2}\left(\frac{1}{5} \right)}$& $\left(\frac{5+\sqrt{5}}{5-\sqrt{5}} \right)^{\frac{1}{4}}$ & $0$ \\

$\tilde{\sigma'}$ & $\sqrt{\frac{5+\sqrt{5}}{5-\sqrt{5}}}\frac{\Gamma \left(\frac{2}{5} \right)}{\Gamma
^{2}\left(\frac{1}{5} \right)}$ & $\frac{\Gamma \left(\frac{2}{5} \right)}{\Gamma ^{2}\left(\frac{1}{5}
\right)}$&  $-\left(\frac{5+\sqrt{5}}{5-\sqrt{5}} \right)^{\frac{1}{4}}$ & $0$ \\

$\tilde{\varepsilon}, \tilde{t}$ & $-\sqrt{\frac{5-\sqrt{5}}{5+\sqrt{5}}}\frac{\Gamma \left(\frac{2}{5}
\right)}{\Gamma ^{2}\left(\frac{1}{5} \right)}$ & $\frac{\Gamma \left(\frac{2}{5} \right)}{\Gamma ^{2}\left(\frac{1}{5} \right)}$&  $-\left(\frac{5-\sqrt{5}}{5+\sqrt{5}} \right)^{\frac{3}{4}}$ & $\sqrt{\frac{10}{5+\sqrt{5}}}\sqrt{\frac{\Gamma \left(\frac{2}{5} \right)\Gamma \left(\frac{7}{5} \right)}{\Gamma \left(\frac{1}{5} \right)\Gamma \left(\frac{8}{5} \right)}}$ \\

$\tilde{\sigma}$ & $-\sqrt{\frac{5-\sqrt{5}}{5+\sqrt{5}}}\frac{\Gamma \left(\frac{2}{5} \right)}{\Gamma
^{2}\left(\frac{1}{5} \right)}$ & $\frac{\Gamma \left(\frac{2}{5} \right)}{\Gamma ^{2}\left(\frac{1}{5}
\right)}$&  $\left(\frac{5-\sqrt{5}}{5+\sqrt{5}} \right)^{\frac{3}{4}}$ &
$\sqrt{\frac{10}{5+\sqrt{5}}}\sqrt{\frac{\Gamma \left(\frac{2}{5} \right)\Gamma \left(\frac{7}{5} \right)}{\Gamma
\left(\frac{1}{5} \right)\Gamma \left(\frac{8}{5} \right)}}$ \\ \hline
\end{tabular}
\end{center}

\subsubsection{Correlator $\langle t\varepsilon\rangle_{\alpha}$}
\begin{center}
\begin{tabular}{|c|c|c|c|c|}\hline
$\alpha$ &  $A$  & $ B$ & $ C_{t,\psi_{1,1}}^{\alpha}$ & $ C_{t,\psi_{1,3}}^{\alpha}$  \\
\hline

$\tilde{1}, \tilde{\varepsilon''}$ & $\frac{\Gamma \left(\frac{2}{5} \right)}{\Gamma ^{2}\left(\frac{1}{5}
\right)}$ &
$\sqrt{\frac{5+\sqrt{5}}{5-\sqrt{5}}}\frac{\Gamma \left(\frac{2}{5} \right)}{\Gamma ^{2}\left(\frac{1}{5} \right)}$& $\left(\frac{5+\sqrt{5}}{5-\sqrt{5}} \right)^{\frac{1}{4}}$ & $0$ \\

$\tilde{\sigma'}$ & $-\frac{\Gamma \left(\frac{2}{5} \right)}{\Gamma ^{2}\left(\frac{1}{5} \right)}$ & $-\sqrt{\frac{5+\sqrt{5}}{5-\sqrt{5}}}\frac{\Gamma \left(\frac{2}{5} \right)}{\Gamma ^{2}\left(\frac{1}{5} \right)}$&  $\left(\frac{5+\sqrt{5}}{5-\sqrt{5}} \right)^{\frac{1}{4}}$ & $0$ \\

$\tilde{\varepsilon}, \tilde{t}$ & $\frac{\Gamma \left(\frac{2}{5} \right)}{\Gamma ^{2}\left(\frac{1}{5} \right)}$ & $-\sqrt{\frac{5-\sqrt{5}}{5+\sqrt{5}}}\frac{\Gamma \left(\frac{2}{5} \right)}{\Gamma ^{2}\left(\frac{1}{5} \right)}$&  $-\left(\frac{5-\sqrt{5}}{5+\sqrt{5}} \right)^{\frac{3}{4}}$ & $-\sqrt{\frac{10}{5+\sqrt{5}}}\sqrt{\frac{\Gamma \left(\frac{2}{5} \right)\Gamma \left(\frac{7}{5} \right)}{\Gamma \left(\frac{1}{5} \right)\Gamma \left(\frac{8}{5} \right)}}$ \\

$\tilde{\sigma}$ & $-\frac{\Gamma \left(\frac{2}{5} \right)}{\Gamma ^{2}\left(\frac{1}{5} \right)}$ &
$\sqrt{\frac{5-\sqrt{5}}{5+\sqrt{5}}}\frac{\Gamma \left(\frac{2}{5} \right)}{\Gamma ^{2}\left(\frac{1}{5}
\right)}$& $-\left(\frac{5-\sqrt{5}}{5+\sqrt{5}} \right)^{\frac{3}{4}}$ &
$\sqrt{\frac{10}{5+\sqrt{5}}}\sqrt{\frac{\Gamma \left(\frac{2}{5} \right)\Gamma \left(\frac{7}{5} \right)}{\Gamma
\left(\frac{1}{5} \right)\Gamma \left(\frac{8}{5} \right)}}$
\\ \hline
\end{tabular}
\end{center}

In the case $ \alpha=\tilde{1},\tilde{\varepsilon''},\tilde{\sigma'}$, it was not possible to deduce the value of
$C_{ t, \psi _{1,3}}^{\alpha}$ directly from the product $\left[C_{ \varepsilon, \psi _{1,3}}^{\alpha}
\right]\left[C_{ t, \psi _{1,3}}^{\alpha} \right]$, because this last quantity is equal to zero, as $ C_{
\varepsilon,\psi _{1,3}}^{\alpha}$. However, since the vanishing of this structure constant is a special feature
of the model ${\cal M}(5,4)$, we have computed the ratio $ \frac{C_{t,\psi _{1,3}}^{\alpha}C_{\varepsilon,\psi
_{1,3}}^{\alpha}}{C_{\varepsilon,\psi _{1,3}}^{\alpha}}$ keeping $p$ arbitrary, and then we have taken the limit
$ p\rightarrow 5$. The same procedure has been used in all the other analogous situations.

\subsubsection{Correlator $\langle \sigma\varepsilon\rangle_{\alpha}$}
\begin{center}
\begin{tabular}{|c|c|c|c|c|}\hline
$\alpha$ &  $A$  & $B$ & $ C_{\sigma,\psi_{1,1}}^{\alpha}$ & $ C_{\sigma,\psi_{1,3}}^{\alpha}$  \\
\hline

$\tilde{1}$ & $\frac{\Gamma \left(\frac{7}{5} \right)}{2^{\frac{3}{4}}\Gamma \left(\frac{1}{5} \right)\Gamma \left(\frac{6}{5} \right)}$ & $\sqrt{\frac{5+\sqrt{5}}{5-\sqrt{5}}}\frac{\Gamma \left(\frac{7}{5} \right)}{2^{\frac{3}{4}}\Gamma \left(\frac{1}{5} \right)\Gamma \left(\frac{6}{5} \right)}$& $2^{\frac{1}{4}}\left(\frac{5+\sqrt{5}}{5-\sqrt{5}} \right)^{\frac{1}{4}}$ & $0$ \\

$\tilde{\varepsilon''}$ & $-\frac{\Gamma \left(\frac{7}{5} \right)}{2^{\frac{3}{4}}\Gamma \left(\frac{1}{5} \right)\Gamma \left(\frac{6}{5} \right)}$ & $-\sqrt{\frac{5+\sqrt{5}}{5-\sqrt{5}}}\frac{\Gamma \left(\frac{7}{5} \right)}{2^{\frac{3}{4}}\Gamma \left(\frac{1}{5} \right)\Gamma \left(\frac{6}{5} \right)}$& $-2^{\frac{1}{4}}\left(\frac{5+\sqrt{5}}{5-\sqrt{5}} \right)^{\frac{1}{4}}$ & $0$ \\

$\tilde{\varepsilon}$ & $-\frac{\Gamma \left(\frac{7}{5} \right)}{2^{\frac{3}{4}}\Gamma \left(\frac{1}{5} \right)\Gamma \left(\frac{6}{5} \right)}$ & $\sqrt{\frac{5-\sqrt{5}}{5+\sqrt{5}}}\frac{\Gamma \left(\frac{7}{5} \right)}{2^{\frac{3}{4}}\Gamma \left(\frac{1}{5} \right)\Gamma \left(\frac{6}{5} \right)}$& $2^{\frac{1}{4}}\left(\frac{5-\sqrt{5}}{5+\sqrt{5}} \right)^{\frac{3}{4}}$ & $\frac{1}{2^{\frac{3}{4}}}\sqrt{\frac{10}{5+\sqrt{5}}}\sqrt{\frac{\Gamma \left(\frac{2}{5} \right)\Gamma \left(\frac{7}{5} \right)}{\Gamma \left(\frac{1}{5} \right)\Gamma \left(\frac{8}{5} \right)}}$ \\

$\tilde{t}$ & $\frac{\Gamma \left(\frac{7}{5} \right)}{2^{\frac{3}{4}}\Gamma \left(\frac{1}{5} \right)\Gamma \left(\frac{6}{5} \right)}$ & $-\sqrt{\frac{5-\sqrt{5}}{5+\sqrt{5}}}\frac{\Gamma \left(\frac{7}{5} \right)}{2^{\frac{3}{4}}\Gamma \left(\frac{1}{5} \right)\Gamma \left(\frac{6}{5} \right)}$& $-2^{\frac{1}{4}}\left(\frac{5-\sqrt{5}}{5+\sqrt{5}} \right)^{\frac{3}{4}}$ & $-\frac{1}{2^{\frac{3}{4}}}\sqrt{\frac{10}{5+\sqrt{5}}}\sqrt{\frac{\Gamma \left(\frac{2}{5} \right)\Gamma \left(\frac{7}{5} \right)}{\Gamma \left(\frac{1}{5} \right)\Gamma \left(\frac{8}{5} \right)}}$ \\

$\tilde{\sigma}, \tilde{\sigma'}$ & $0$ & $0$& $0$ & $0$ \\ \hline
\end{tabular}
\end{center}

\subsubsection{Correlator $\langle \varepsilon''\varepsilon\rangle_{\alpha}$}
\begin{center}
\begin{tabular}{|c|c|c|c|}\hline
$\alpha$ &  $A$  & $ B$ & $ C_{\varepsilon'',\psi_{1,1}}^{\alpha}$  \\
\hline

$\tilde{1}, \tilde{\varepsilon''}$ & $0$ & $\left(\frac{5+\sqrt{5}}{5-\sqrt{5}} \right)^{\frac{3}{4}}\frac{\Gamma \left(\frac{2}{5} \right)}{\Gamma ^{2}\left(\frac{1}{5} \right)}$ & $1$  \\

$\tilde{\sigma'}$ &  $0$ & $\left(\frac{5+\sqrt{5}}{5-\sqrt{5}} \right)^{\frac{3}{4}}\frac{\Gamma \left(\frac{2}{5} \right)}{\Gamma ^{2}\left(\frac{1}{5} \right)}$ & $-1$ \\

$\tilde{\varepsilon}, \tilde{t}$ & $0$ & $-\left(\frac{5-\sqrt{5}}{5+\sqrt{5}} \right)^{\frac{1}{4}}\frac{\Gamma \left(\frac{2}{5} \right)}{\Gamma ^{2}\left(\frac{1}{5} \right)}$& $1$  \\

$\tilde{\sigma}$ &  $0$ & $-\left(\frac{5-\sqrt{5}}{5+\sqrt{5}} \right)^{\frac{1}{4}}\frac{\Gamma
\left(\frac{2}{5} \right)}{\Gamma ^{2}\left(\frac{1}{5} \right)}$ & $-1$
\\ \hline
\end{tabular}
\end{center}

\subsubsection{Correlator $\langle \sigma'\varepsilon\rangle_{\alpha}$}
\begin{center}
\begin{tabular}{|c|c|c|c|}\hline
$\alpha$ &  $A$  & $B$ & $ C_{\sigma',\psi_{1,1}}^{\alpha}$   \\
\hline

$\tilde{1}$ & $0$ & $2^{\frac{1}{4}}\left(\frac{5+\sqrt{5}}{5-\sqrt{5}} \right)^{\frac{3}{4}}\frac{\Gamma \left(\frac{2}{5} \right)}{\Gamma ^{2}\left(\frac{1}{5} \right)}$ & $2^{\frac{1}{4}}$  \\

$\tilde{\varepsilon''}$ & $0$ & $-2^{\frac{1}{4}}\left(\frac{5+\sqrt{5}}{5-\sqrt{5}} \right)^{\frac{3}{4}}\frac{\Gamma \left(\frac{2}{5} \right)}{\Gamma ^{2}\left(\frac{1}{5} \right)}$ & $-2^{\frac{1}{4}}$  \\

$\tilde{\varepsilon}$ & $0$ & $2^{\frac{1}{4}}\left(\frac{5-\sqrt{5}}{5+\sqrt{5}} \right)^{\frac{1}{4}}\frac{\Gamma \left(\frac{2}{5} \right)}{\Gamma ^{2}\left(\frac{1}{5} \right)}$ & $-2^{\frac{1}{4}}$  \\

$\tilde{t}$ &  $0$ & $-2^{\frac{1}{4}}\left(\frac{5-\sqrt{5}}{5+\sqrt{5}} \right)^{\frac{1}{4}}\frac{\Gamma \left(\frac{2}{5} \right)}{\Gamma ^{2}\left(\frac{1}{5} \right)}$ & $2^{\frac{1}{4}}$  \\

$\tilde{\sigma}, \tilde{\sigma'}$ & $0$ & $0$& $0$  \\ \hline
\end{tabular}
\end{center}

\subsubsection{Correlator $\langle \sigma'\sigma'\rangle_{\alpha}$}
\begin{center}
\begin{tabular}{|c|c|c|c|c|}\hline
$\alpha$ &  $A$  & $B$ & $ C_{\sigma',\psi_{1,1}}^{\alpha}$ &  $ C_{\sigma',\psi_{3,1}}^{\alpha}$ \\
\hline

$\tilde{1}, \tilde{t}$ & $-\frac{\sqrt{\pi }}{8\Gamma ^{2}\left(\frac{3}{4} \right)}$ & $-\frac{\sqrt{\pi }}{8\Gamma ^{2}\left(\frac{3}{4} \right)}$ & $2^{\frac{1}{4}}$ & $0$ \\

$\tilde{\varepsilon''}, \tilde{\varepsilon}$ & $-\frac{\sqrt{\pi }}{8\Gamma ^{2}\left(\frac{3}{4} \right)}$ & $-\frac{\sqrt{\pi }}{8\Gamma ^{2}\left(\frac{3}{4} \right)}$ & $-2^{\frac{1}{4}}$ & $0$ \\

$\tilde{\sigma}, \tilde{\sigma'}$ & $-\frac{\sqrt{\pi }}{8\Gamma ^{2}\left(\frac{3}{4} \right)}$ &
$\frac{\sqrt{\pi }}{8\Gamma ^{2}\left(\frac{3}{4} \right)}$ & $0$ & $7^{\frac{1}{2}}2^{-\frac{5}{4}}$ \\ \hline
\end{tabular}
\end{center}

\subsubsection{Correlator $\langle \sigma \sigma'\rangle_{\alpha}$}
\begin{center}
\begin{tabular}{|c|c|c|c|c|}\hline
$\alpha$ &  $A$  & $B$ & $ C_{\sigma,\psi_{1,1}}^{\alpha}$ &  $ C_{\sigma,\psi_{3,1}}^{\alpha}$ \\
\hline

$\tilde{1}$ & $-\left(\frac{5+\sqrt{5}}{5-\sqrt{5}} \right)^{\frac{1}{4}}\frac{\sqrt{\pi }}{8\Gamma ^{2}\left(\frac{3}{4} \right)}$ & $-\left(\frac{5+\sqrt{5}}{5-\sqrt{5}} \right)^{\frac{1}{4}}\frac{\sqrt{\pi }}{8\Gamma ^{2}\left(\frac{3}{4} \right)}$ & $2^{\frac{1}{4}}\left(\frac{5+\sqrt{5}}{5-\sqrt{5}} \right)^{\frac{1}{4}}$ & $0$ \\

$\tilde{\varepsilon''}$ & $-\left(\frac{5+\sqrt{5}}{5-\sqrt{5}} \right)^{\frac{1}{4}}\frac{\sqrt{\pi }}{8\Gamma ^{2}\left(\frac{3}{4} \right)}$ & $-\left(\frac{5+\sqrt{5}}{5-\sqrt{5}} \right)^{\frac{1}{4}}\frac{\sqrt{\pi }}{8\Gamma ^{2}\left(\frac{3}{4} \right)}$ & $-2^{\frac{1}{4}}\left(\frac{5+\sqrt{5}}{5-\sqrt{5}} \right)^{\frac{1}{4}}$ & $0$ \\

$\tilde{\sigma'}$ & $\left(\frac{5+\sqrt{5}}{5-\sqrt{5}} \right)^{\frac{1}{4}}\frac{\sqrt{\pi }}{8\Gamma ^{2}\left(\frac{3}{4} \right)}$ & $-\left(\frac{5+\sqrt{5}}{5-\sqrt{5}} \right)^{\frac{1}{4}}\frac{\sqrt{\pi }}{8\Gamma ^{2}\left(\frac{3}{4} \right)}$ & $0$ & $-7^{-\frac{1}{2}}2^{-\frac{5}{4}}\left(\frac{5+\sqrt{5}}{5-\sqrt{5}} \right)^{\frac{1}{4}}$ \\

$\tilde{\varepsilon}$ & $\left(\frac{5-\sqrt{5}}{5+\sqrt{5}} \right)^{\frac{3}{4}}\frac{\sqrt{\pi }}{8\Gamma ^{2}\left(\frac{3}{4} \right)}$ & $\left(\frac{5-\sqrt{5}}{5+\sqrt{5}} \right)^{\frac{3}{4}}\frac{\sqrt{\pi }}{8\Gamma ^{2}\left(\frac{3}{4} \right)}$ & $2^{\frac{1}{4}}\left(\frac{5-\sqrt{5}}{5+\sqrt{5}} \right)^{\frac{3}{4}}$ & $0$ \\

$\tilde{t}$ & $\left(\frac{5-\sqrt{5}}{5+\sqrt{5}} \right)^{\frac{3}{4}}\frac{\sqrt{\pi }}{8\Gamma ^{2}\left(\frac{3}{4} \right)}$ & $\left(\frac{5-\sqrt{5}}{5+\sqrt{5}} \right)^{\frac{3}{4}}\frac{\sqrt{\pi }}{8\Gamma ^{2}\left(\frac{3}{4} \right)}$ & $-2^{\frac{1}{4}}\left(\frac{5-\sqrt{5}}{5+\sqrt{5}} \right)^{\frac{3}{4}}$ & $0$ \\

$\tilde{\sigma}$ & $-\left(\frac{5-\sqrt{5}}{5+\sqrt{5}} \right)^{\frac{3}{4}}\frac{\sqrt{\pi }}{8\Gamma
^{2}\left(\frac{3}{4} \right)}$ & $\left(\frac{5-\sqrt{5}}{5+\sqrt{5}} \right)^{\frac{3}{4}}\frac{\sqrt{\pi
}}{8\Gamma ^{2}\left(\frac{3}{4} \right)}$ & $0$ &
$7^{-\frac{1}{2}}2^{-\frac{5}{4}}\left(\frac{5-\sqrt{5}}{5+\sqrt{5}} \right)^{\frac{3}{4}}$ \\ \hline
\end{tabular}
\end{center}

\subsubsection{Correlator $\langle \varepsilon \sigma'\rangle_{\alpha}$}
\begin{center}
\begin{tabular}{|c|c|c|c|}\hline
$\alpha$ &  $A$  & $B$ & $ C_{\varepsilon,\psi_{1,1}}^{\alpha}$  \\
\hline

$\tilde{1}$ & $-2^{\frac{1}{4}}\left(\frac{5+\sqrt{5}}{5-\sqrt{5}} \right)^{\frac{1}{4}}\frac{\Gamma \left(\frac{1}{4} \right)}{8\sqrt{\pi }\Gamma \left(\frac{3}{4} \right)}$ & $0$ & $\left(\frac{5+\sqrt{5}}{5-\sqrt{5}} \right)^{\frac{1}{4}}$ \\

$\tilde{\varepsilon''}$ & $2^{\frac{1}{4}}\left(\frac{5+\sqrt{5}}{5-\sqrt{5}} \right)^{\frac{1}{4}}\frac{\Gamma \left(\frac{1}{4} \right)}{8\sqrt{\pi }\Gamma \left(\frac{3}{4} \right)}$ & $0$ & $\left(\frac{5+\sqrt{5}}{5-\sqrt{5}} \right)^{\frac{1}{4}}$ \\

$\tilde{\sigma'}$ & $0$ & $0$ & $-\left(\frac{5+\sqrt{5}}{5-\sqrt{5}} \right)^{\frac{1}{4}}$ \\

$\tilde{\varepsilon}$ & $-2^{\frac{1}{4}}\left(\frac{5-\sqrt{5}}{5+\sqrt{5}} \right)^{\frac{3}{4}}\frac{\Gamma
\left(\frac{1}{4} \right)}{8\sqrt{\pi }\Gamma \left(\frac{3}{4} \right)}$ & $0$ & $-\left(\frac{5-\sqrt{5}}{5+\sqrt{5}} \right)^{\frac{3}{4}}$ \\

$\tilde{t}$ & $2^{\frac{1}{4}}\left(\frac{5-\sqrt{5}}{5+\sqrt{5}} \right)^{\frac{3}{4}}\frac{\Gamma
\left(\frac{1}{4} \right)}{8\sqrt{\pi }\Gamma \left(\frac{3}{4} \right)}$ & $0$ & $-\left(\frac{5-\sqrt{5}}{5+\sqrt{5}} \right)^{\frac{3}{4}}$ \\

$\tilde{\sigma}$ & $0$ & $0$ & $\left(\frac{5-\sqrt{5}}{5+\sqrt{5}}
\right)^{\frac{3}{4}}$ \\
\hline
\end{tabular}
\end{center}

These results perfectly agree with the correlator $ \langle \sigma' \varepsilon \rangle _{\alpha}$ previously
found, because the different values obtained for $A$ give the same overall constant when multiplied by the two
different factors $\frac{\Gamma \left(2\alpha _{-}^{2}-2 \right)\Gamma \left(1-\alpha _{-}^{2} \right)}{\Gamma
\left(\alpha _{-}^{2}-1 \right)}$ (in the case $\langle\phi _{21}\phi _{12} \rangle _{\alpha}$) and $\frac{\Gamma
\left(2\alpha _{+}^{2}-2 \right)\Gamma \left(1-\alpha _{+}^{2} \right)}{\Gamma \left(\alpha _{+}^{2}-1 \right)}$
(in the case $\langle\phi _{12}\phi _{21} \rangle _{\alpha}$).

\subsubsection{Correlator $\langle t \sigma'\rangle_{\alpha}$}
\begin{center}
\begin{tabular}{|c|c|c|c|}\hline
$\alpha$ &  $A$  & $B$ & $ C_{t,\psi_{1,1}}^{\alpha}$  \\
\hline

$\tilde{1}$ & $-2^{\frac{1}{4}}\left(\frac{5+\sqrt{5}}{5-\sqrt{5}} \right)^{\frac{1}{4}}\frac{\Gamma \left(\frac{1}{4} \right)}{8\sqrt{\pi }\Gamma \left(\frac{3}{4} \right)}$ & $0$ & $\left(\frac{5+\sqrt{5}}{5-\sqrt{5}} \right)^{\frac{1}{4}}$ \\

$\tilde{\varepsilon''}$ & $2^{\frac{1}{4}}\left(\frac{5+\sqrt{5}}{5-\sqrt{5}} \right)^{\frac{1}{4}}\frac{\Gamma \left(\frac{1}{4} \right)}{8\sqrt{\pi }\Gamma \left(\frac{3}{4} \right)}$ & $0$ & $\left(\frac{5+\sqrt{5}}{5-\sqrt{5}} \right)^{\frac{1}{4}}$ \\

$\tilde{\sigma'}$ & $0$ & $0$ & $\left(\frac{5+\sqrt{5}}{5-\sqrt{5}}
\right)^{\frac{1}{4}}$ \\

$\tilde{\varepsilon}$ & $-2^{\frac{1}{4}}\left(\frac{5-\sqrt{5}}{5+\sqrt{5}} \right)^{\frac{3}{4}}\frac{\Gamma
\left(\frac{1}{4} \right)}{8\sqrt{\pi }\Gamma \left(\frac{3}{4} \right)}$ & $0$ & $-\left(\frac{5-\sqrt{5}}{5+\sqrt{5}} \right)^{\frac{3}{4}}$ \\

$\tilde{t}$ & $2^{\frac{1}{4}}\left(\frac{5-\sqrt{5}}{5+\sqrt{5}} \right)^{\frac{3}{4}}\frac{\Gamma
\left(\frac{1}{4} \right)}{8\sqrt{\pi }\Gamma \left(\frac{3}{4} \right)}$ & $0$ & $-\left(\frac{5-\sqrt{5}}{5+\sqrt{5}} \right)^{\frac{3}{4}}$ \\

$\tilde{\sigma}$ & $0$ & $0$ & $-\left(\frac{5-\sqrt{5}}{5+\sqrt{5}} \right)^{\frac{3}{4}}$ \\
\hline
\end{tabular}
\end{center}

\subsubsection{Correlator $\langle \varepsilon''\sigma'\rangle_{\alpha}$}
\begin{center}
\begin{tabular}{|c|c|c|c|}\hline
$\alpha$ &  $A$  & $B$ & $ C_{\varepsilon'',\psi_{1,1}}^{\alpha}$  \\
\hline

$\tilde{1}, \tilde{t}$ & $-2^{\frac{1}{4}}\frac{\Gamma \left(\frac{1}{4} \right)}{8\sqrt{\pi }\Gamma \left(\frac{3}{4} \right)}$ & $0$ & $1$ \\

$\tilde{\varepsilon''}, \tilde{\varepsilon}$  & $2^{\frac{1}{4}}\frac{\Gamma \left(\frac{1}{4} \right)}{8\sqrt{\pi }\Gamma \left(\frac{3}{4} \right)}$ & $0$ & $1$ \\

$\tilde{\sigma}, \tilde{\sigma'}$  & $0$ & $0$ & $-1$ \\ \hline
\end{tabular}
\end{center}

\chapter{Properties of Lie algebras}
\label{Liealg}

The following summaries list the essential information for all simple Lie algebras that are simply laced, which
can be classified as $A_{r}$, $D_{r}$ (with $r$ equal to the rank) or $E_{n}$  (with $n=6,7\;\textrm{or}\;8$ equal
to the rank). For each algebra, we give the extended Dynkin diagram and the finite Cartan matrix. The numbers
appearing beside the nodes of the Dynkin diagrams give the numbering of the corresponding simple roots and their
comarks; black dots refer to the affine extension of the diagrams. The list of properties includes the dimension
of the algebra, the dual Coxeter number $h$, the highest root $\theta$ (in Dynkin label notation), the exponents,
and the group ${\cal O}(\hat{g})$ of outer automorphisms of $\hat{g}$, with the action of its generators on an
arbitrary weight (in Dynkin label notation).

\vspace{2cm}

\begin{flushleft}
$\textbf{A}_{r\geq 2}(su(r+1))$
\end{flushleft}

\vspace{2cm}

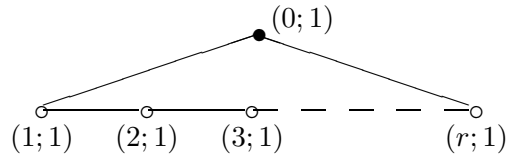
\setlength{\unitlength}{0.0125in}
\begin{picture}(40,0)(60,470)

\put(200,522){\line(3,1){90}}  \put(378,522){\line(-3,1){90}} \put(287,548){$ \bullet$}

\put(196,516){$ \circ$} \put(200,520){\line(1,0){40}} \put(240,516){$ \circ$} \put(244,520){\line(1,0){40}}
\put(284,516){$ \circ$} \put(288,520){\line(1,0){10}}\put(308,520){\line(1,0){10}}\put(328,520){\line(1,0){10}}
\put(348,520){\line(1,0){10}}\put(368,520){\line(1,0){10}}\put(378,516){$ \circ$} \put(186,505){\small$(1;1)$}
\put(230,505){\small$(2;1)$} \put(274,505){\small$(3;1)$} \put(368,505){\small$(r;1)$}
\put(295,555){\small$(0;1)$}
\end{picture}

\begin{center}
$A_{ij}=\left(\begin{array}{cccccc}
2 & -1 & 0 & \cdots & 0 & 0 \\
-1 & 2 & -1 & \cdots & 0 & 0\\
0 & -1 & 2 & \cdots & 0 & 0\\
\vdots & \vdots  & \vdots  & \vdots  & \vdots  & \vdots \\
0 & 0 & 0 & \cdots & 2 & -1\\
0 & 0 & 0 & \cdots & -1 & 2\\
\end{array}\right)$
\end{center}

\begin{eqnarray*}
&&\textrm{dim}\,A_{r}=r^{2}+2r \\
&& h=r+1 \\
&& \theta=(1,0,...,1) \\
&& \textrm{exponents} = 1,2,...,r\\
&& {\cal O}(\hat{A_{r}})=Z_{r+1}\,:\qquad
a[\lambda_{0},\lambda_{1},...,\lambda_{r-1}\lambda_{r}]=[\lambda_{r},\lambda_{0},...,\lambda_{r-2}\lambda_{r-1}]
\end{eqnarray*}

\vspace{2cm}

\begin{flushleft}
$\textbf{D}_{r\geq 4}(so(2r))$
\end{flushleft}

\vspace{2cm}

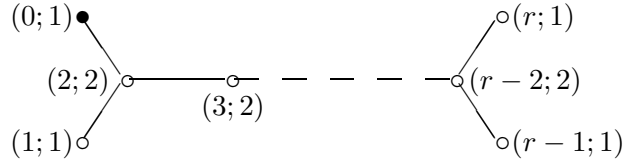
\setlength{\unitlength}{0.0125in}
\begin{picture}(40,0)(60,470)
\put(196,522){\line(-2,3){15}}\put(196,518){\line(-2,-3){15}} \put(177,543){$ \bullet$}\put(177,490){$ \circ$}
\put(196,516){$ \circ$} \put(200,520){\line(1,0){40}} \put(240,516){$ \circ$}
\put(244,520){\line(1,0){10}}\put(264,520){\line(1,0){10}}\put(284,520){\line(1,0){10}}
\put(304,520){\line(1,0){10}}\put(324,520){\line(1,0){10}} \put(334,516){$ \circ$}
\put(338,522){\line(2,3){15}}\put(338,518){\line(2,-3){15}} \put(353,543){$ \circ$}\put(353,490){$ \circ$}
\put(165,516){\small$(2;2)$} \put(150,543){\small$(0;1)$} \put(150,490){\small$(1;1)$}
\put(230,505){\small$(3;2)$} \put(342,516){\small$(r-2;2)$} \put(360,543){\small$(r;1)$}
\put(360,490){\small$(r-1;1)$}
\end{picture}

\begin{center}
$A_{ij}=\left(\begin{array}{ccccccc}

2 & -1 & 0 & \cdots & 0 & 0 & 0\\
-1 & 2 & -1 & \cdots & 0 & 0 & 0\\
0 & -1 & 2 & \cdots & 0 & 0 & 0\\
\vdots & \vdots  & \vdots  & \vdots  & \vdots  & \vdots & \vdots \\
0 & 0 & 0 & \cdots & 2 & -1 & -1\\
0 & 0 & 0 & \cdots & -1 & 2 & 0\\
0 & 0 & 0 & \cdots & -1 & 0 & 2\\
\end{array}\right)$
\end{center}

\begin{eqnarray*}
&&\textrm{dim}\,D_{r}=2r^{2}-r \\
&& h=2r-2 \\
&& \theta=(0,1,...,0) \\
&& \textrm{exponents} = 1,3,...,2r-3,r-1\\
&& {\cal O}(\hat{D_{2\ell}})=Z_{2}\times Z_{2}\,:\quad \, a[\lambda_{0},\lambda_{1}, \lambda_{2},
...,\lambda_{r-1},\lambda_{r}]=[\lambda_{1},\lambda_{0}, \lambda_{2},...,\lambda_{r},\lambda_{r-1}]\\
&& \qquad\qquad\qquad\qquad \quad \quad \tilde{a}[\lambda_{0},\lambda_{1}, \lambda_{2},
...,\lambda_{r-1},\lambda_{r}]=[\lambda_{r},\lambda_{r-1},
\lambda_{r-2},...,\lambda_{1},\lambda_{0}]\\
&& {\cal O}(\hat{D_{2\ell+1}})=Z_{4}\,:\qquad \; \; a[\lambda_{0},\lambda_{1}, \lambda_{2},
...,\lambda_{r-1},\lambda_{r}]=[\lambda_{r-1},\lambda_{r}, \lambda_{r-2},...,\lambda_{1},\lambda_{0}]\\
\end{eqnarray*}

\vspace{2cm}

\begin{flushleft}
$\textbf{E}_{6}$
\end{flushleft}

\vspace{4cm}

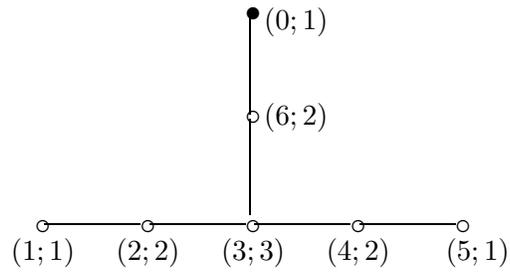
\setlength{\unitlength}{0.0125in}
\begin{picture}(40,0)(60,470)

\put(196,516){$ \circ$} \put(200,520){\line(1,0){40}} \put(240,516){$ \circ$} \put(244,520){\line(1,0){40}}
\put(284,516){$ \circ$} \put(288,520){\line(1,0){40}} \put(328,516){$ \circ$} \put(332,520){\line(1,0){40}}
\put(372,516){$ \circ$} \put(286,524){\line(0,1){40}} \put(284,562){$ \circ$} \put(286,566){\line(0,1){40}}
\put(284,605){$ \bullet$}

\put(186,505){\small$(1;1)$} \put(230,505){\small$(2;2)$} \put(274,505){\small$(3;3)$}
\put(318,505){\small$(4;2)$} \put(368,505){\small$(5;1)$} \put(292,562){\small$(6;2)$}
\put(292,602){\small$(0;1)$}
\end{picture}

\begin{center}
$A_{ij}=\left(\begin{array}{cccccc}

2 & -1 & 0 & 0 & 0 & 0 \\
-1 & 2 & -1 & 0 & 0 & 0\\
0 & -1 & 2 & -1 & 0 & -1\\
0 & 0 & -1 & 2 & -1 & 0 \\
0 & 0 & 0 & -1 & 2 & 0 \\
0 & 0 & -1 & 0 & 0 & 2\\
\end{array}\right)$
\end{center}

\begin{eqnarray*}
&&\textrm{dim}\,E_{6}=78 \\
&& h=12 \\
&& \theta=(0,0,...,1) \\
&& \textrm{exponents} = 1,4,5,7,8,11 \\
&& {\cal O}(\hat{E_{6}})=Z_{3}\,:\qquad  a[\lambda_{0},\lambda_{1},...,\lambda_{6}]=[\lambda_{1},\lambda_{5},
\lambda_{4}, \lambda_{3},\lambda_{6},\lambda_{0}, \lambda_{2}]
\end{eqnarray*}

\vspace{5cm}

\begin{flushleft}
$\textbf{E}_{7}$
\end{flushleft}

\vspace{2cm}

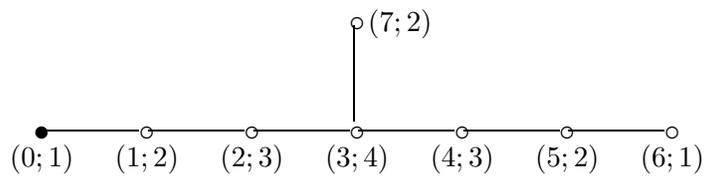
\setlength{\unitlength}{0.0125in}
\begin{picture}(40,0)(60,470)

\put(152,516){$ \bullet$} \put(156,520){\line(1,0){40}} \put(196,516){$ \circ$} \put(200,520){\line(1,0){40}}
\put(240,516){$ \circ$} \put(244,520){\line(1,0){40}} \put(284,516){$ \circ$} \put(288,520){\line(1,0){40}}
\put(328,516){$ \circ$} \put(332,520){\line(1,0){40}} \put(372,516){$ \circ$} \put(286,524){\line(0,1){40}}
\put(284,562){$ \circ$} \put(376,520){\line(1,0){40}} \put(416,516){$ \circ$}

\put(142,505){\small$(0;1)$} \put(186,505){\small$(1;2)$} \put(230,505){\small$(2;3)$}
\put(274,505){\small$(3;4)$} \put(318,505){\small$(4;3)$} \put(362,505){\small$(5;2)$}
\put(406,505){\small$(6;1)$} \put(292,562){\small$(7;2)$}
\end{picture}

\begin{center}
$A_{ij}=\left(\begin{array}{ccccccc}

2 & -1 & 0 & 0 & 0 & 0 & 0\\
-1 & 2 & -1 & 0 & 0 & 0 & 0\\
0 & -1 & 2 & -1 & 0 & 0 & -1\\
0 & 0 & -1 & 2 & -1 & 0 & 0\\
0 & 0 & 0 & -1 & 2 & -1 & 0 \\
0 & 0 & 0 & 0 & -1 & 2 & 0\\
0 & 0 & -1 & 0 & 0 & 0 & 2\\
\end{array}\right)$
\end{center}

\begin{eqnarray*}
&&\textrm{dim}\,E_{7}=133 \\
&& h=18 \\
&& \theta=(1,0,...,0) \\
&& \textrm{exponents} = 1,5,7,9,11,13,17\\
&& {\cal O}(\hat{E_{7}})=Z_{2}\,:\qquad  a[\lambda_{0},\lambda_{1},...,\lambda_{7}]=[\lambda_{6},\lambda_{5},
\lambda_{4}, \lambda_{3},\lambda_{2},\lambda_{1}, \lambda_{0}, \lambda_{7}]
\end{eqnarray*}

\vspace{2cm}

\begin{flushleft}
$\textbf{E}_{8}$
\end{flushleft}

\vspace{2cm}

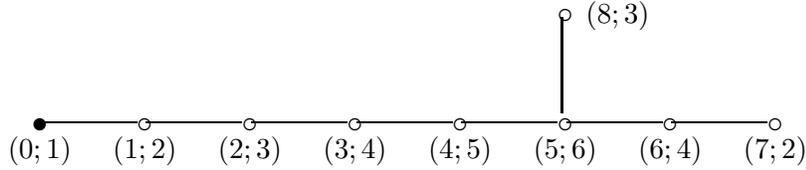
\setlength{\unitlength}{0.0125in}
\begin{picture}(40,0)(60,470)

\put(108,516){$ \bullet$} \put(112,520){\line(1,0){40}} \put(152,516){$ \circ$} \put(156,520){\line(1,0){40}}
\put(196,516){$ \circ$} \put(200,520){\line(1,0){40}} \put(240,516){$ \circ$} \put(244,520){\line(1,0){40}}
\put(284,516){$ \circ$} \put(288,520){\line(1,0){40}} \put(328,516){$ \circ$} \put(332,520){\line(1,0){40}}
\put(372,516){$ \circ$} \put(330,524){\line(0,1){40}} \put(328,562){$ \circ$} \put(376,520){\line(1,0){40}}
\put(416,516){$ \circ$}

\put(98,505){\small$(0;1)$} \put(142,505){\small$(1;2)$} \put(186,505){\small$(2;3)$} \put(230,505){\small$(3;4)$}
\put(274,505){\small$(4;5)$} \put(318,505){\small$(5;6)$} \put(362,505){\small$(6;4)$}
\put(406,505){\small$(7;2)$} \put(340,562){\small$(8;3)$}
\end{picture}

\begin{center}
$A_{ij}=\left(\begin{array}{cccccccc}

2 & -1 & 0 & 0 & 0 & 0 & 0 & 0\\
-1 & 2 & -1 & 0 & 0 & 0 & 0 & 0\\
0 & -1 & 2 & -1 & 0 & 0 & 0 & 0\\
0 & 0 & -1 & 2 & -1 & 0 & 0 & 0\\
0 & 0 & 0 & -1 & 2 & -1 & 0 & -1\\
0 & 0 & 0 & 0 & -1 & 2 & -1 & 0\\
0 & 0 & 0 & 0 & 0 & -1 & 2 & 0\\
0 & 0 & 0 & 0 & -1 & 0 & 0 & 2\\
\end{array}\right)$
\end{center}

\begin{eqnarray*}
&&\textrm{dim}\,E_{8}=248 \\
&& h=30 \\
&& \theta=(1,0,...,0) \\
&& \textrm{exponents} = 1,7,11,13,17,19,23,29
\end{eqnarray*}

\chapter{$E_{n}$ affine Toda field theories}
\label{Smatrlist}

In this appendix, we list the basic properties of the three affine Toda field theories related to the affine Lie
algebras $E_{6}$, $E_{7}$ and $E_{8}$. The $S$-matrix elements are given in the block form
$S_{ab}=\prod_{x}\{x\}_{\theta}$, with
\begin{displaymath}
\{x\}_{\theta}=\frac{s_{\frac{x+1}{h}}(\theta)s_{\frac{x-1}{h}}(\theta)}{s_{\frac{x+1-B}{h}}(\theta)s_{\frac{x-1+B}{h}}(\theta)}
, \qquad s_{\alpha}(\theta)=\frac{\sinh \left[\frac{1}{2}\left(\theta+i\pi \alpha\right)\right]}{\sinh
\left[\frac{1}{2}\left(\theta-i\pi\alpha\right)\right]}\,.
\end{displaymath}

\section{$E_{6}$ affine Toda field theory}

Masses are associated to the Dynkin diagram in the following way:

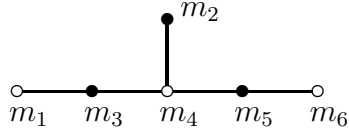
\setlength{\unitlength}{0.01cm}
\begin{picture}( 1000,300)(0,100)
\thicklines \put(492,190){$ \circ$} \put(510,200){\line( 1, 0){85}} \put(592,190){$ \bullet$}
\put(702,207){\line( 0, 1){85}} \put(692,285){$ \bullet$} \put(610,200){\line( 1, 0){85}} \put(692,190){$ \circ$}
\put(792,190){$ \bullet$} \put(892,190){$ \circ$} \put(710,200){\line( 1, 0){85}} \put(810,200){\line( 1, 0){85}}
\put(492,160){$ m_1$} \put(592,160){$ m_3$} \put(692,160){$ m_4$} \put(792,160){$ m_5$} \put(892,160){$ m_6$}
\put(720,300){$ m_2$}
\end{picture}

\vspace{1cm}

The two couples of particles (1,6) and (3,5) are mass degenerate and related by charge conjugation: $6=\bar{1}$
and $5=\bar{3}$. The mass spectrum is:

\begin{center}
\begin{tabular}{|cclc|l|} \hline
$m_1$ &=& $M$ & & 1   \\
$m_2$ &=& $M \, \sqrt{2}$ & & 1.41421  \\
$m_3$ &=& $M \,(\sqrt{3}+1)/\sqrt{2} $ & & 1.93185  \\
$m_4$ &=& $M \,(\sqrt{3}+1)$ & & 2.73205  \\\hline
\end{tabular}
\end{center}

\newpage

Each entry in the following table indicates the fusing angle $u_{ab}^{c}$ as a multiple of $\frac{i\pi}{12}$; the
subscript refers to the order of the corresponding $S$-matrix pole, if multiple.

\vspace{0.5cm}

\begin{flushleft}

\begin{tabular}{|c|c|c|c|c|c|c||c|c|c|c|c|c|c|}\hline
$\small \textbf{ab} \setminus \textbf{c}$  & \textbf{1} & \textbf{2} & \textbf{3} & \textbf{4} & \textbf{5} & \textbf{6} &  $\small \textbf{ab} \setminus \textbf{c}$  & \textbf{1} & \textbf{2} & \textbf{3} & \textbf{4} & \textbf{5} & \textbf{6}  \\
\hline

$\textbf{1}\:\textbf{1}$ & \hspace{6mm} & \hspace{6mm} & 2& \hspace{6mm} & \hspace{6mm} & 8 &  & \hspace{6mm} &
\hspace{6mm} & \hspace{6mm} & \hspace{6mm} & \hspace{6mm} & \hspace{6mm} \\ \hline

$\textbf{1}\:\textbf{2}$ & 9  &  & \hspace{6mm} & &5 & \hspace{6mm} & $\textbf{2}\:\textbf{2}$ &  & 8 & &2 & &
\\ \hline

$\textbf{1}\:\textbf{3}$&  & 9 & & 3 & &  & $\textbf{2}\:\textbf{3}$ &  &  & & & &10 \\ \hline

$\textbf{1}\:\textbf{4}$ &  &  & &  &10 &  & $\textbf{2}\:\textbf{4}$ &  & 11 & & $7_{3}$& & \\ \hline

$\textbf{1}\:\textbf{5}$ &  &  &7 &  & & 11 & $\textbf{2}\:\textbf{5}$ & 10 &  & & & & \\ \hline

$\textbf{1}\:\textbf{6}$ &  & 6 & &  & &  & $\textbf{2}\:\textbf{6}$ &  &  & 5& & &9 \\ \hline

\end{tabular}

\begin{tabular}{|c|c|c|c|c|c|c||c|c|c|c|c|c|c|}\hline
$\small \textbf{ab} \setminus \textbf{c}$  & \textbf{1} & \textbf{2} & \textbf{3} & \textbf{4} & \textbf{5} & \textbf{6} &  $\small \textbf{ab} \setminus \textbf{c}$  & \textbf{1} & \textbf{2} & \textbf{3} & \textbf{4} & \textbf{5} & \textbf{6} \\
\hline

$\textbf{3}\:\textbf{3}$ & 10 & \hspace{6mm}  & \hspace{6mm} & \hspace{6mm} & $8_{3}$& \hspace{6mm} &  &
\hspace{6mm} & \hspace{6mm} & \hspace{6mm} & \hspace{6mm} & \hspace{6mm} & \hspace{6mm} \\ \hline

$\textbf{3}\:\textbf{4}$ & \hspace{6mm} &  & $9_{3}$&  & &11 & $\textbf{4}\:\textbf{4}$ &  & $10_{3}$ & &
$8_{5}$& & \\ \hline

$\textbf{3}\:\textbf{5}$ &  &  & & $6_{3}$ & \hspace{6mm} &  & $\textbf{4}\:\textbf{5}$ & 11 & & & & $9_{3}$ & \\
\hline

$\textbf{3}\:\textbf{6}$ & 11 &  & &   & 7& & $\textbf{4}\:\textbf{6}$ &  &  & 10 & & & \\ \hline

\end{tabular}

\begin{tabular}{|c|c|c|c|c|c|c||c|c|c|c|c|c|c|c|c|}\hline
$\small \textbf{ab} \setminus \textbf{c}$  & \textbf{1} & \textbf{2} & \textbf{3} & \textbf{4} & \textbf{5} & \textbf{6} &  $\small \textbf{ab} \setminus \textbf{c}$  & \textbf{1} & \textbf{2} & \textbf{3} & \textbf{4} & \textbf{5} & \textbf{6}  \\
\hline

$\textbf{5}\:\textbf{5}$ & \hspace{6mm} & \hspace{6mm} & $8_{3}$ & \hspace{6mm} & \hspace{6mm} & 10  & & \hspace{6mm} & \hspace{6mm} & \hspace{6mm} & \hspace{6mm} & \hspace{6mm} & \hspace{6mm}  \\
\hline

$\textbf{5}\:\textbf{6}$ & & 9& \hspace{6mm} & 3  & & \hspace{6mm} & $\textbf{6}\:\textbf{6}$ & 8 & & & & 2& \\
\hline

\end{tabular}

\end{flushleft}

\vspace{1cm}

The $S$-matrix elements are:

\vspace{0.5cm}

\begin{center}
\begin{tabular}{|l|l|}\hline
$S_{11}=\{1\}\{7\}$ & $S_{22}=\{1\}\{5\}\{7\}\{11\}$     \\
$S_{12}=\{4\}\{8\}$ & $S_{23}=\{3\}\{5\}\{7\}\{9\}$   \\
$S_{13}=\{2\}\{6\}\{8\}$& $S_{24}=\{2\}\{4\}\{6\}^{2}\{8\}\{10\}$   \\
$S_{14}=\{3\}\{5\}\{7\}\{9\}$ &$S_{33}=\{1\}\{3\}\{5\}\{7\}^{2}\{9\}$    \\
$S_{15}=\{4\}\{6\}\{10\}$   & $S_{34}=\{2\}\{4\}^{2}\{6\}^{2}\{8\}^{2}\{10\}$       \\
$S_{16}=\{5\}\{11\}$    & $S_{35}=\{3\}\{5\}^{2}\{7\}\{9\}\{11\}$  \\ \hline
\end{tabular}

\begin{tabular}{|c|}
\hspace{1.5cm} $S_{44}=\{1\}\{3\}^{2}\{5\}^{3}\{7\}^{3}\{9\}^{2}\{11\}$ \hspace{1.5cm} \\ \hline
\end{tabular}

\end{center}

\newpage

\section{$E_{7}$ affine Toda field theory}

Masses are associated to the Dynkin diagram in the following way:

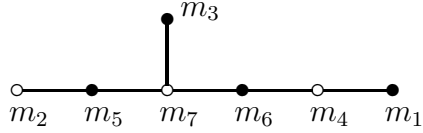
\setlength{\unitlength}{0.01cm}
\begin{picture}( 1000,300)(0,100)
\thicklines \put(492,190){$ \circ$} \put(510,200){\line( 1, 0){85}} \put(592,190){$ \bullet$}
\put(702,207){\line( 0, 1){85}} \put(692,285){$ \bullet$} \put(610,200){\line( 1, 0){85}} \put(692,190){$ \circ$}
\put(792,190){$ \bullet$} \put(892,190){$ \circ$} \put(710,200){\line( 1, 0){85}} \put(810,200){\line( 1, 0){85}}
\put(910,200){\line( 1, 0){85}} \put(992,190){$ \bullet$} \put(992,160){$ m_1$} \put(492,160){$ m_2$}
\put(592,160){$ m_5$} \put(692,160){$ m_7$} \put(792,160){$ m_6$} \put(892,160){$ m_4$} \put(720,300){$ m_3$}
\end{picture}

\vspace{0.5cm}

The mass spectrum is presented in the following table, where the parity assigned to each particle refers to the
$Z_{2}$ symmetry of the extended Dynkin diagram:

\vspace{0.5cm}

\begin{center}
\begin{tabular}{|cclc|l|l|} \hline
$m_1$ &=& $M$ & & 1  & \hspace{1mm} odd \\
$m_2$ &=& $2 M \cos({5\pi \over 18})$ & & 1.28557 & \hspace{1mm} even \\
$m_3$ &=& $2 M \cos({2\pi \over 18})$ & & 1.87938 & \hspace{1mm} odd \\
$m_4$ &=& $2 M \cos({\pi \over 18})$ & & 1.96961 & \hspace{1mm} even \\
$m_5$ &=& $4 M \cos({\pi \over 18}) \cos({5\pi \over 18})$ & & 2.53208 &
\hspace{1mm} even \\
$m_6$ &=& $4 M \cos({2\pi\over 18})\cos({4\pi \over 18}) $ & & 2.87938 &
\hspace{1mm} odd \\
$m_7$ &=& $4 M \cos({\pi \over 18}) \cos({2\pi \over 18})$ & & 3.70166 &
\hspace{1mm} even\\
\hline
\end{tabular}
\end{center}

\vspace{0.5cm}

Each entry in the table indicates the fusing angle $u_{ab}^{c}$ as a multiple of $\frac{i\pi}{18}$; the subscript
refers to the order of the corresponding $S$-matrix pole, if multiple.

\begin{center}
\begin{tabular}{|c||c|c|c|c|c|c|c|}\hline
$\textbf{ab} \setminus \textbf{c}$  & \textbf{1} & \textbf{2} & \textbf{3} & \textbf{4} & \textbf{5} & \textbf{6} & \textbf{7} \\
\hline \hline

$\textbf{1}\:\textbf{1}$ & \hspace{6mm} & 10 & & 2 & \hspace{6mm} & \hspace{6mm} & \hspace{6mm} \\
\hline

$\textbf{1}\:\textbf{2}$ & 13 & \hspace{6mm} & 7& \hspace{6mm} & & &  \\ \hline

$\textbf{1}\:\textbf{3}$&  & 14 & \hspace{6mm} & 10 & 6& &  \\ \hline

$\textbf{1}\:\textbf{4}$ & 17 &  &11 &  & & 3&  \\ \hline

$\textbf{1}\:\textbf{5}$ &  &  &14 &  & & 8&  \\ \hline

$\textbf{1}\:\textbf{6}$ &  &  & & 16 &12 & &4  \\ \hline

$\textbf{1}\:\textbf{7}$ &  &  & &  & & 15&  \\ \hline

\end{tabular}

\begin{tabular}{|c||c|c|c|c|c|c|c|}\hline
\hspace{1.5mm} $\textbf{2}\:\textbf{2}$ \hspace{1mm} & \hspace{6mm} & 12 & \hspace{6mm} &8 & 2& \hspace{6mm} & \hspace{6mm} \\
\hline

$\textbf{2}\:\textbf{3}$ & 15 & \hspace{6mm} & 11 & \hspace{6mm} & \hspace{6mm} & 5 & \\ \hline

$\textbf{2}\:\textbf{4}$ &  & 14 & & & 8& & \\ \hline

$\textbf{2}\:\textbf{5}$ &  & 17 & & 13& & &3 \\ \hline

$\textbf{2}\:\textbf{6}$ &  &  & 15& & & & \\ \hline

$\textbf{2}\:\textbf{7}$ &  &  & & & 16& &$10_{3}$ \\ \hline
\end{tabular}

\begin{tabular}{|c||c|c|c|c|c|c|c|}\hline
$\textbf{ab} \setminus \textbf{c}$  & \textbf{1} & \textbf{2} & \textbf{3} & \textbf{4} & \textbf{5} & \textbf{6} & \textbf{7}\\
\hline \hline

$\textbf{3}\:\textbf{3}$ &\hspace{6mm}  & 14 &\hspace{6mm} & \hspace{6mm} & \hspace{6mm} &\hspace{6mm} & 2 \\
\hline

$\textbf{3}\:\textbf{4}$ & 15& \hspace{6mm} & &  & && \hspace{6mm} \\
\hline

$\textbf{3}\:\textbf{5}$ & 16 &  & &  & & $10_{3}$&  \\
\hline

$\textbf{3}\:\textbf{6}$ &  & 16 & &  &$12_{3}$ & & $8_{3}$ \\
\hline

$\textbf{3}\:\textbf{7}$ &  &  & 17&  & & $13_{3}$&  \\ \hline
\end{tabular}

\begin{tabular}{|c||c|c|c|c|c|c|c|}\hline
\hspace{1.5mm} $\textbf{4}\:\textbf{4}$ \hspace{1mm} & \hspace{6mm} & \hspace{6mm} & \hspace{6mm} & 12& $10_{3}$&
\hspace{6mm} &4
\\ \hline

$\textbf{4}\:\textbf{5}$ &  & 15& & $13_{3}$& \hspace{6mm} & &$7_{3}$ \\ \hline

$\textbf{4}\:\textbf{6}$ &17  &  & & \hspace{6mm} & &$11_{3}$ & \hspace{6mm} \\ \hline

$\textbf{4}\:\textbf{7}$ &  &  & &16 & $14_{3}$& & \\ \hline
\end{tabular}

\begin{tabular}{|c||c|c|c|c|c|c|c|}\hline
\hspace{1.5mm} $\textbf{5}\:\textbf{5}$ \hspace{1mm} & \hspace{6mm} & \hspace{6mm} & \hspace{6mm} & \hspace{6mm} &
$12_{3}$& \hspace{6mm} & \hspace{6mm}
\\ \hline

$\textbf{5}\:\textbf{6}$ & 16 & & $14_{3}$&   & \hspace{6mm} & & \\ \hline

$\textbf{5}\:\textbf{7}$ &  &17  & & $15_{3}$ & & & $11_{5}$ \\ \hline
\end{tabular}

\begin{tabular}{|c||c|c|c|c|c|c|c|}\hline
\hspace{1.5mm} $\textbf{6}\:\textbf{6}$ \hspace{1mm} & \hspace{6mm} & \hspace{6mm} & \hspace{6mm} & $14_{3}$& \hspace{6mm} & \hspace{6mm} &$10_{5}$\\
\hline

$\textbf{6}\:\textbf{7}$ & 17 &  & $15_{3}$& \hspace{6mm} & & $13_{5}$& \hspace{6mm}  \\ \hline
\end{tabular}

\begin{tabular}{|c||c|c|c|c|c|c|c|}\hline
\hspace{1.5mm} $\textbf{7}\:\textbf{7}$ \hspace{1mm} & \hspace{6mm} &\hspace{0.1mm}$16_{3}$ &\hspace{6mm} &
\hspace{6mm} &$14_{5}$ & \hspace{6mm} &$12_{7}$\\ \hline
\end{tabular}

\end{center}

\vspace{1cm}

If we define $x:=\{x\}\{h-x\}$, the $S$-matrix elements are:

\begin{center}
\begin{tabular}{|l|l|l|l|}\hline
$S_{11}=1\;\{9\}$ & & &   \\
$S_{12}=6$ & $S_{22}=1\;7$ & &  \\
$S_{13}=5\;\{9\}$& $S_{23}=4\;8$ &$S_{33}=1\;5\;7\;\{9\}$ & \\
$S_{14}=2\;8$ & $S_{24}=5\;7$ &$S_{34}=4\;6\;8$ &  $S_{44}=1\;3\;7\;9$ \\
$S_{15}=5\;7$   & $S_{25}=2\;6\;8$ &$S_{35}=3\;5\;7\;9$ &  $S_{45}=4\;6^{2}\;8$  \\
$S_{16}=3\;7\;\{9\}$    & $S_{26}=4\;6\;8$ &$S_{36}=3\;5\;7^{2}\;\{9\}$ &  $S_{46}=2\;4\;6\;8^{2}$ \\
$S_{17}=4\;6\;8$     & $S_{27}=3\;5\;7\;9$ &$S_{37}=2\;4\;6^{2}\;8^{2}$ &  $S_{47}=3\;5^{2}\;7^{2}\;9$ \\
\end{tabular}

\begin{tabular}{|l|l|l|}\hline
\hspace{1mm} $S_{55}=1\;3\;5\;7^{2}\;9$ & &  \\
\hspace{1mm} $S_{56}=3\;5^{2}\;7^{2}\;9$ & \hspace{1mm} $S_{66}=1\;3\;5^{2}\;7^{2}\;9\;\{9\}$ &\\
\hspace{1mm} $S_{57}=2\;4^{2}\;6^{2}\;8^{3}$ & \hspace{1mm} $S_{67}=2\;4^{2}\;6^{3}\;8^{3}$& \hspace{1mm}
$S_{77}=1\;3^{2}\;5^{3}\;7^{4}\;9^{2}$ \hspace{0.5mm}  \\ \hline
\end{tabular}
\end{center}

\newpage

\section{$E_{8}$ affine Toda field theory}

Masses are associated to the Dynkin diagram in the following way:

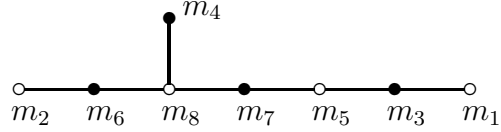
\setlength{\unitlength}{0.01cm}
\begin{picture}( 1100,300)(0,100)
\thicklines \put(492,190){$ \circ$} \put(510,200){\line( 1, 0){85}} \put(592,190){$ \bullet$}
\put(702,207){\line( 0, 1){85}} \put(692,285){$ \bullet$} \put(610,200){\line( 1, 0){85}} \put(692,190){$ \circ$}
\put(792,190){$ \bullet$} \put(892,190){$ \circ$} \put(710,200){\line( 1, 0){85}} \put(810,200){\line( 1, 0){85}}
\put(910,200){\line( 1, 0){85}} \put(992,190){$ \bullet$} \put(992,160){$ m_3$} \put(1010,200){\line( 1, 0){85}}
\put(1092,190){$ \circ$} \put(1092,160){$ m_1$} \put(492,160){$ m_2$} \put(592,160){$ m_6$} \put(692,160){$ m_8$}
\put(792,160){$ m_7$} \put(892,160){$ m_5$} \put(720,300){$ m_4$}
\end{picture}

The mass spectrum is:

\begin{center}
\begin{tabular}{|cclc|l|} \hline
$m_1$ &=& $M$ & & 1   \\
$m_2$ &=& $2 M \cos({6\pi \over 30})$ & & 1.61803  \\
$m_3$ &=& $2 M \cos({\pi \over 30})$ & & 1.98904  \\
$m_4$ &=& $4 M \cos({6\pi \over 30})\cos({7\pi \over 30})$ & & 2.40487  \\
$m_5$ &=& $4 M \cos({4\pi \over 30}) \cos({6\pi \over 30})$ & & 2.95629 \\
$m_6$ &=& $4 M \cos({\pi\over 30})\cos({6\pi \over 30}) $ & & 3.21834 \\
$m_7$ &=& $8 M \cos^{2}({6\pi \over 30}) \cos({7\pi \over 30})$ & & 3.89116 \\
$m_8$ &=& $8 M \cos({4\pi\over 30}) \cos^{2}({6\pi \over 30})$ & & 4.78339 \\ \hline
\end{tabular}
\end{center}

\vspace{0.5cm}

Each entry in the table indicates the fusing angle $u_{ab}^{c}$ as a multiple of $\frac{i\pi}{30}$; the subscript
refers to the order of the corresponding $S$-matrix pole, if multiple.

\begin{center}
\begin{tabular}{|c||c|c|c|c|c|c|c|c|}\hline
$\textbf{ab} \setminus \textbf{c}$  & \textbf{1} & \textbf{2} & \textbf{3} & \textbf{4} & \textbf{5} & \textbf{6} & \textbf{7} & \textbf{8} \\
\hline \hline

$\textbf{1}\:\textbf{1}$ & 20 & 12 & 2 & \hspace{6mm} & \hspace{6mm} & \hspace{6mm} & \hspace{6mm} & \hspace{6mm} \\
\hline

$\textbf{1}\:\textbf{2}$ & 24 & 18 & 14 & 8 & & & & \\ \hline

$\textbf{1}\:\textbf{3}$ & 29 & 21 & \hspace{6mm} & 13 & 3 & & & \\ \hline

$\textbf{1}\:\textbf{4}$ & \hspace{6mm} & 25 & 21 & 17 & 11 & 7 & & \\ \hline

$\textbf{1}\:\textbf{5}$ &  & \hspace{6mm} & 28 & 22 & & 16 & 4 & \\ \hline

$\textbf{1}\:\textbf{6}$ &  &  & & 25 & 19 & & 9 &  \\ \hline

$\textbf{1}\:\textbf{7}$ &  &  & &  & 27 & 23 & & 5 \\ \hline

$\textbf{1}\:\textbf{8}$ &  &  & &  & & & 26 & $16_{3}$\\ \hline
\end{tabular}

\begin{tabular}{|c||c|c|c|c|c|c|c|c|}\hline
\hspace{1.5mm} $\textbf{2}\:\textbf{2}$ \hspace{1mm} & 24 & 20  & \hspace{6mm} & 14  & 8 & 2 & \hspace{6mm} & \hspace{6mm} \\
\hline

$\textbf{2}\:\textbf{3}$ & 25 & \hspace{6mm} & 19 & \hspace{6mm} & \hspace{6mm} & 9 & & \\
\hline

$\textbf{2}\:\textbf{4}$ & 27 & 23 & & & & \hspace{6mm} & 5 &\\ \hline

$\textbf{2}\:\textbf{5}$ & \hspace{6mm} & 26 & & & & $16_{3}$ & & \\ \hline

$\textbf{2}\:\textbf{6}$ &  & 29 & 25 & & $19_{3}$ & & $13_{3}$ & 3 \\ \hline

$\textbf{2}\:\textbf{7}$ &  &  & & 27 & & $21_{3}$ & $17_{3}$ & $11_{3}$ \\ \hline

$\textbf{2}\:\textbf{8}$ &  &  & & & & 28 & $22_{3}$ & \\ \hline
\end{tabular}

\begin{tabular}{|c||c|c|c|c|c|c|c|c|}\hline
$\textbf{ab} \setminus \textbf{c}$  & \textbf{1} & \textbf{2} & \textbf{3} & \textbf{4} & \textbf{5} & \textbf{6} & \textbf{7}& \textbf{8} \\
\hline \hline

$\textbf{3}\:\textbf{3}$ & \hspace{6mm}  & 22  & $20_{3}$ & \hspace{6mm} & 14 & $12_{3}$ & 4 & \hspace{6mm}  \\
\hline

$\textbf{3}\:\textbf{4}$ & 26 & \hspace{6mm} & \hspace{6mm} &  & $16_{3}$ & \hspace{6mm} & \hspace{6mm} & \\
\hline

$\textbf{3}\:\textbf{5}$ & 29 &  & 23 & $21_{3}$ & \hspace{6mm} & & $13_{3}$ & 5 \\
\hline

$\textbf{3}\:\textbf{6}$ &  & 26 & $24_{3}$ &  &  & $18_{3}$ &  & $8_{3}$ \\
\hline

$\textbf{3}\:\textbf{7}$ &  &  & 28 &  & $22_{3}$ & & & \\ \hline

$\textbf{3}\:\textbf{8}$ &  &  &  &  & 27 & $25_{3}$ & & $17_{5}$ \\ \hline
\end{tabular}

\begin{tabular}{|c||c|c|c|c|c|c|c|c|}\hline
\hspace{1.5mm} $\textbf{4}\:\textbf{4}$ \hspace{1mm} & 26 & \hspace{6mm} & \hspace{6mm} & $20_{3}$ & \hspace{6mm}
& $16_{3}$ & $12_{3}$ & 2
\\ \hline

$\textbf{4}\:\textbf{5}$ & 27 & & $23_{3}$ & \hspace{6mm} & $19_{3}$ & \hspace{6mm} & \hspace{6mm} & $9_{3}$ \\
\hline

$\textbf{4}\:\textbf{6}$ & 28 &  & & $22_{3}$ & & & & \hspace{6mm} \\ \hline

$\textbf{4}\:\textbf{7}$ & \hspace{6mm} & 28 & & $24_{3}$ &  & & $18_{5}$ & $14_{5}$ \\ \hline

$\textbf{4}\:\textbf{8}$ &  &  & & 29 & $25_{3}$ & & $21_{5}$ & \\ \hline
\end{tabular}

\begin{tabular}{|c||c|c|c|c|c|c|c|c|}\hline
\hspace{1.5mm} $\textbf{5}\:\textbf{5}$ \hspace{1mm} & \hspace{6mm} & \hspace{6mm} & \hspace{6mm} & $22_{3}$ &
$20_{5}$ & \hspace{6mm} & \hspace{6mm} & $12_{5}$
\\ \hline

$\textbf{5}\:\textbf{6}$ & 27 & $25_{3}$ & & \hspace{6mm}  & \hspace{6mm} & & $17_{5}$ & \hspace{6mm} \\ \hline

$\textbf{5}\:\textbf{7}$ & 29 &  & $25_{3}$ &  & & $21_{5}$ & & \\ \hline

$\textbf{5}\:\textbf{8}$ &  &  & 28 & $26_{3}$ & $24_{5}$ & & & $18_{7}$ \\ \hline
\end{tabular}

\begin{tabular}{|c||c|c|c|c|c|c|c|c|}\hline
\hspace{1.5mm} $\textbf{6}\:\textbf{6}$ \hspace{1mm} & \hspace{6mm} & \hspace{6mm} & $24_{3}$ & \hspace{6mm} &
\hspace{6mm} & $20_{5}$ & \hspace{6mm} & $14_{5}$
\\ \hline

$\textbf{6}\:\textbf{7}$ & 28 & $26_{3}$ & \hspace{6mm} &  & $22_{5}$ & \hspace{6mm} & & $16_{7}$ \\ \hline

$\textbf{6}\:\textbf{8}$ &  & 29 & $27_{3}$ &  & & $23_{5}$ & $21_{7}$ & \hspace{6mm} \\ \hline
\end{tabular}

\begin{tabular}{|c||c|c|c|c|c|c|c|c|}\hline
\hspace{1.5mm} $\textbf{7}\:\textbf{7}$ \hspace{1mm} & \hspace{6mm} & $26_{3}$ & \hspace{6mm} & $24_{5}$ &
\hspace{6mm} & \hspace{6mm} & $20_{7}$ & \hspace{6mm} \\ \hline

$\textbf{7}\:\textbf{8}$ & 29 & $27_{3}$ & & $25_{5}$ & & $23_{7}$ & \hspace{6mm} & $19_{9}$ \\ \hline \hline

$\textbf{8}\:\textbf{8}$ & $28_{3}$ & \hspace{6mm} &$26_{5}$ & \hspace{6mm} & $24_{7}$ & & $22_{9}$ & \small $20_{11}$ \\
\hline
\end{tabular}

\end{center}

\newpage

If we define $x:=\{x\}\{h-x\}$, the $S$-matrix elements are:

\begin{center}

\begin{tabular}{|l|l|l|}\hline
\hspace{1mm} $S_{11}=1\;11$ & &   \\

\hspace{1mm} $S_{12}=7\;13$ & \hspace{1mm} $S_{22}=1\;7\;11\;13$ &   \\

\hspace{1mm} $S_{13}=2\;10\;12$& \hspace{1mm} $S_{23}=6\;8\;12\;14 $   & \hspace{1mm} $S_{33}=1\;3\;9\;11^{2}\;13$  \\

\hspace{1mm} $S_{14}=6\;10\;14$ & \hspace{1mm} $S_{24}=4\;8\;10\;12\;14$ & \hspace{1mm} $S_{34}=5\;7\;9\;11\;13\;15$ \\

\hspace{1mm} $S_{15}=3\;9\;11\;13$   & \hspace{1mm} $S_{25}=5\;7\;9\;11\;13\;15$ & \hspace{1mm} $S_{35}=2\;4\;8\;10^{2}\;12^{2}\;14$   \\

\hspace{1mm} $S_{16}=6\;8\;12\;14$    & \hspace{1mm} $S_{26}=2\;6\;8\;10\;12^{2}\;14$ & \hspace{1mm} $S_{36}=5\;7^{2}\;9\;11\;13^{2}\;15$  \\

\hspace{1mm} $S_{17}=4\;8\;10\;12\;14$     & \hspace{1mm} $S_{27}=4\;6\;8\;10^{2}\;12\;14^{2}$ & \hspace{1mm} $S_{37}=3\;5\;7\;9^{2}\;11^{2}\;13^{2}\;15$ \\

\hspace{1mm} $S_{18}=5\;7\;9\;11\;13\;15 $  & \hspace{1mm} $S_{28}=3\;5\;7\;9^{2}\;11^{2}\;13^{2}\;15$   & \hspace{1mm} $S_{38}=4\;6^{2}\;8^{2}\;10^{2}\;12^{2}\;14^{3}$ \\

\end{tabular}

\begin{tabular}{|l|l|}\hline
\hspace{5mm} $S_{44}=1\;5\;7\;9\;11^{2}\;13\;15$ &    \\

\hspace{5mm} $S_{45}=4\;6\;8^{2}\;10\;12^{2}\;14^{2}$ & \hspace{5mm} $S_{55}=1\;3\;5\;7\;9^{2}\;11^{3}\;13^{2}\;15$   \\

\hspace{5mm} $S_{46}=3\;5\;7\;9^{2}\;11^{2}\;13^{2}\;15$ & \hspace{5mm} $S_{56}=4\;6^{2}\;8^{2}\;10^{2}\;12^{2}\;14^{3}$  \\

\hspace{5mm} $S_{47}=3\;5\;7^{2}\;9^{2}\;11^{2}\;13^{3}\;15$ & \hspace{5mm} $S_{57}=2\;4\;6^{2}\;8^{2}\;10^{3}\;12^{3}\;14^{3}$  \\

\hspace{5mm} $S_{48}=2\;4\;6^{2}\;8^{2}\;10^{3}\;12^{3}\;14^{3}$ \hspace{12mm} & \hspace{5mm} $S_{58}=3\;5^{2}\;7^{3}\;9^{3}\;11^{3}\;13^{4}\;15^{2}$ \hspace{12mm} \\

\end{tabular}

\begin{tabular}{|l|l|}\hline
\hspace{5mm} $S_{66}=1\;3\;5\;7^{2}\;9^{2}\;11^{3}\;13^{3}\;15$ & \hspace{5mm} $S_{77}=1\;3\;5^{2}\;7^{3}\;9^{3}\;11^{4}\;13^{4}\;15^{2}$      \\

\hspace{5mm} $S_{67}=3\;5^{2}\;7^{2}\;9^{3}\;11^{3}\;13^{3}\;15^{2}$ & \hspace{5mm} $S_{78}=2\;4^{2}\;6^{3}\;8^{4}\;10^{4}\;12^{5}\;14^{5}$   \\

\hspace{5mm} $S_{68}=2\;4^{2}\;6^{2}\;8^{3}\;10^{4}\;12^{4}\;14^{4}$ \hspace{10mm} & \hspace{5mm}
$S_{88}=1\;3^{2}\;5^{3}\;7^{4}\;9^{5}\;11^{6}\;13^{6}\;15^{3}$ \hspace{7mm} \\ \hline

\end{tabular}

\end{center}

\newpage

\begin{flushleft}\Large
\textbf{Acknowledgements}
\end{flushleft}

\vspace{0.5cm}

I'm extremely grateful to Giuseppe Mussardo for having introduced me into the area of conformal field theories
and integrable systems with generous and profound explanations, for his excellent guide in choosing and developing
this specific subject of work, for his uncommon kindness and for having given me the opportunity of joining his
group at SISSA in Trieste.

A special thanks goes also to Vittorio Gorini, who was a precious scientific and human guide in my decisions
during the entire course of studies in physics, followed my work with interest and provided me financial support
to participate in some conferences.

I'm grateful to Gesualdo Delfino, Davide Fioravanti and Roberto Tateo for their generosity in numerous
explanations, for their kindness and for valuable suggestions. I thank my friend Paola Mosconi for having shared
with me her house and everyday life in Trieste, for interesting discussions and for her help with English.

I thank SISSA for a pre-graduate grant and for hospitality.

It is a pleasure to acknowledge the support I received from people who have been close to me during these years of
study. In particular, I'm grateful to my father for his deep interest in my activity, his important guide in my
life and his suggestions about English, and to my mother, my brother Alberto and my grandmother Mariuccia for
their encouragement and participation to my joys and difficulties. I would like to thank my boyfriend Andrea
Colombo for having shared with me all the different emotions of this work and for stimulating my interest in
other fields of physics. Finally, I thank three special friends, Chiara Cappellini, Giovanna Frigerio and Chiara
Sampietro, for having contributed to make so exciting and happy the years of the university.

\end{document}